\documentclass[onecolumn,useamsfonts]{pasj02}
\usepackage{natbib}
\usepackage{aas2pasj}
 
\usepackage{float}
\usepackage{amsmath}
\usepackage{tikz}
\usetikzlibrary{positioning, arrows.meta}
 
\usepackage{xr}
 
\newif\ifXRFROZEN \XRFROZENtrue
\ifXRFROZEN
  \input{xref_frozen_T_pasj}
\else
\fi
 
\providecommand{\ContinuedFloat}{\addtocounter{figure}{-1}}

\usepackage{xcolor}

\jyear{2026}
\Received{}
\Accepted{}

\graphicspath{{./}{figures/}}

\begin{document}
 
\title{B-fields In STar-forming Region Observations (BISTRO): \\ Technical Foundations for Quantitative 450/850\,$\mu$m JCMT SCUBA-2/POL-2 Polarimetry}
 
\author{%
Ray S.~\textsc{Furuya},\altaffilmark{1,2}\altemailmark\orcid{0000-0002-0786-7307} \email{rsfuruya@gmail.com}
David \textsc{Berry},\altaffilmark{3}\orcid{0000-0001-6524-2447}
Pierre \textsc{Bastien},\altaffilmark{4}\orcid{0000-0002-0794-3859}
Tetsuo \textsc{Hasegawa},\altaffilmark{5}\orcid{0000-0003-1853-0184}
Masafumi \textsc{Matsumura},\altaffilmark{6}\orcid{0000-0002-6906-0103}
Koji \textsc{Kawabata},\altaffilmark{7,8,9}\orcid{0000-0001-6099-9539}
Kazunari \textsc{Iwasaki},\altaffilmark{10}\orcid{0000-0002-9892-1881}
Per \textsc{Friberg},\altaffilmark{3}\orcid{0000-0002-8010-8454}
Simon \textsc{Coud\'{e}},\altaffilmark{5}\orcid{0000-0002-0859-0805}
Lapo \textsc{Fanciullo},\altaffilmark{11}\orcid{0000-0001-9930-9240}
Thiem \textsc{Hoang},\altaffilmark{12,13}\orcid{0000-0003-2017-0982}
Yasuo \textsc{Doi},\altaffilmark{14}\orcid{0000-0001-8746-6548}
Hua-bai \textsc{Li},\altaffilmark{15}\orcid{0000-0003-2641-9240}
Jia-Wei \textsc{Wang},\altaffilmark{3}\orcid{0000-0002-6668-974X}
Tao-Chung \textsc{Ching},\altaffilmark{16}\orcid{0000-0001-8516-2532}
Hiroko \textsc{Shinnaga},\altaffilmark{17}\orcid{0000-0001-9407-6775}
Hiroyuki \textsc{Nakanishi},\altaffilmark{17}
Sarah \textsc{Graves},\altaffilmark{3}
A-Ran \textsc{Lyo},\altaffilmark{12}
Junhao \textsc{Liu},\altaffilmark{18}\orcid{0000-0002-4774-2998}
Janik \textsc{Karoly},\altaffilmark{19}\orcid{0000-0001-5996-3600}
Jihye \textsc{Hwang},\altaffilmark{20,21}\orcid{0000-0001-7866-2686}
Glenn J. \textsc{White},\altaffilmark{22}\orcid{0000-0002-7126-691X}
Doris \textsc{Arzoumanian},\altaffilmark{20,21}\orcid{0000-0002-1959-7201}
Sarah \textsc{Sadavoy},\altaffilmark{23}\orcid{0000-0001-7474-6874}
Kohji \textsc{Tomisaka},\altaffilmark{24}\orcid{0000-0003-2726-0892}
Pham Ngoc \textsc{Diep},\altaffilmark{25}\orcid{0000-0002-2808-0888}
Mehrnoosh \textsc{Tahani},\altaffilmark{26}\orcid{0000-0001-8749-1436}
Nguyen Bich \textsc{Ngoc},\altaffilmark{25}\orcid{0000-0002-5913-5554}
Archana \textsc{Soam},\altaffilmark{27}\orcid{0000-0002-6386-2906}
Hsi-Wei \textsc{Yen},\altaffilmark{16}\orcid{0000-0003-1412-893X}
Chakali \textsc{Eswaraiah},\altaffilmark{28,29,30}\orcid{0000-0003-4761-6139}
Vera \textsc{K\"{o}nyves},\altaffilmark{31}\orcid{0000-0002-3746-1498}
Jeong-Eun \textsc{Lee},\altaffilmark{32}\orcid{0000-0003-3119-2087}
Valentin J.~M.\ \textsc{Le Gouellec},\altaffilmark{33,34}\orcid{0000-0002-5714-799X}
Ekta \textsc{Sharma},\altaffilmark{28}\orcid{0000-0002-4541-0607}
Le Ngoc \textsc{Tram},\altaffilmark{35}\orcid{0000-0002-6488-8227}
Fr\'{e}d\'{e}rick \textsc{Poidevin},\altaffilmark{36,37}\orcid{0000-0002-5391-5568}
Tie \textsc{Liu},\altaffilmark{38}\orcid{0000-0002-5286-2564}
Mark \textsc{Rawlings},\altaffilmark{3}\orcid{0000-0002-6529-202X}
Najimudeen \textsc{Bijas},\altaffilmark{39}\orcid{0000-0002-6398-7530}
Fumitaka \textsc{Nakamura},\altaffilmark{24,40}\orcid{0000-0001-5431-2294}
Rachel \textsc{Friesen},\altaffilmark{41}\orcid{0000-0001-7594-8128}
Jean-Fran\c{c}ois \textsc{Robitaille},\altaffilmark{42}\orcid{0000-0001-5079-8573}
Yoshiaki \textsc{Misugi},\altaffilmark{43}\orcid{0000-0002-4093-6925}
Raiga \textsc{Kashiwagi},\altaffilmark{43}\orcid{0000-0002-1461-3866}
Takashi \textsc{Onaka},\altaffilmark{44}\orcid{0000-0002-8234-6747}
Motohide \textsc{Tamura},\altaffilmark{5,44,45}\orcid{0000-0002-6510-0681}
Harriet \textsc{Parsons},\altaffilmark{3}\orcid{0000-0002-6327-3423}
Tyler L.\ \textsc{Bourke},\altaffilmark{46,47}\orcid{0000-0001-7491-0048}
David \textsc{Eden},\altaffilmark{48,49}\orcid{0000-0002-5881-3229}
Huei-Ru Vivien \textsc{Chen},\altaffilmark{50}\orcid{0000-0002-9774-1846}
Nagayoshi \textsc{Ohashi},\altaffilmark{16}\orcid{0000-0003-0998-5064}
Jongsoo \textsc{Kim},\altaffilmark{12,13}\orcid{0000-0002-1229-0426}
Shih-Ping \textsc{Lai},\altaffilmark{51,16}\orcid{0000-0001-5522-486X}
Keping \textsc{Qiu},\altaffilmark{52,53}\orcid{0000-0002-5093-5088}
Di \textsc{Li},\altaffilmark{28,54}\orcid{0000-0003-3010-7661}
Woojin \textsc{Kwon},\altaffilmark{55,56}\orcid{0000-0003-4022-4132}
Shu-ichiro \textsc{Inutsuka},\altaffilmark{57}\orcid{0000-0003-4366-6518}
Kate \textsc{Pattle},\altaffilmark{19}\orcid{0000-0002-8557-3582}
Derek \textsc{Ward-Thompson},\altaffilmark{31}\orcid{0000-0003-1140-2761}
and the BISTRO Collaboration}
\altaffiltext{1}{Institute of Liberal Arts and Sciences, Tokushima University, Minami Jousanjima-Cho 1-1, Tokushima, Tokushima 770-8502, Japan}
\altaffiltext{2}{Visiting Associate Professor, Center for Computational Astrophysics (CfCA), National Astronomical Observatory of Japan (FY~2023), Osawa 2-21-1, Mitaka, Tokyo 181-8588, Japan}
\altaffiltext{3}{East Asian Observatory, 660 N.\ A`oh\={o}k\={u} Place, University Park, Hilo, HI 96720, USA}
\altaffiltext{4}{Centre de recherche en astrophysique du Qu\'ebec \& D\'epartement de physique, Universit\'e de Montr\'eal, C.P.~6128, Succursale Centre-ville, Montr\'eal, QC H3C~3J7, Canada}
\altaffiltext{5}{National Astronomical Observatory of Japan, National Institutes of Natural Sciences, Osawa, Mitaka, Tokyo 181-8588, Japan}
\altaffiltext{6}{Faculty of Education \& Center for Educational Development and Support, Kagawa University, Saiwai-cho 1-1, Takamatsu, Kagawa 760-8522, Japan}
\altaffiltext{7}{Hiroshima Astrophysical Science Center, Hiroshima University, Kagamiyama 1-3-1, Higashi-Hiroshima, Hiroshima 739-8526, Japan}
\altaffiltext{8}{Department of Physics, Hiroshima University, Kagamiyama 1-3-1, Higashi-Hiroshima, Hiroshima 739-8526, Japan}
\altaffiltext{9}{Core Research for Energetic Universe (CORE-U), Hiroshima University, Kagamiyama 1-3-1, Higashi-Hiroshima, Hiroshima 739-8526, Japan}
\altaffiltext{10}{Center for Computational Astrophysics, National Astronomical Observatory of Japan, Mitaka, Tokyo 181-8588, Japan}
\altaffiltext{11}{Tamkang University, No.\ 151, Yingzhuan Rd., Tamsui Dist., New Taipei City 251301, Taiwan}
\altaffiltext{12}{Korea Astronomy and Space Science Institute, 776 Daedeokdae-ro, Yuseong-gu, Daejeon 34055, Republic of Korea}
\altaffiltext{13}{University of Science and Technology, Korea, 217 Gajeong-ro, Yuseong-gu, Daejeon 34113, Republic of Korea}
\altaffiltext{14}{Department of Earth Science and Astronomy, Graduate School of Arts and Sciences, The University of Tokyo, 3-8-1 Komaba, Meguro, Tokyo 153-8902, Japan}
\altaffiltext{15}{Department of Physics, The Chinese University of Hong Kong, Shatin, N.T., Hong Kong}
\altaffiltext{16}{Academia Sinica Institute of Astronomy and Astrophysics, No.1, Sec.\ 4., Roosevelt Road, Taipei 10617, Taiwan}
\altaffiltext{17}{Department of Physics and Astronomy, Graduate School of Science and Engineering, Kagoshima University, 1-21-35 Korimoto, Kagoshima, Kagoshima 890-0065, Japan}
\altaffiltext{18}{Division of ALMA, National Astronomical Observatory of Japan, Mitaka, Tokyo 181-8588, Japan}
\altaffiltext{19}{Department of Physics and Astronomy, University College London, WC1E 6BT London, UK}
\altaffiltext{20}{Institute for Advanced Study, Kyushu University, Japan}
\altaffiltext{21}{Department of Earth and Planetary Sciences, Faculty of Science, Kyushu University, Nishi-ku, Fukuoka 819-0395, Japan}
\altaffiltext{22}{School for Physical Sciences, The Open University, Walton Hall, Milton Keynes MK7 6AA, United Kingdom, and RALSpace, The Rutherford Appleton Laboratory, Chilton, Didcot, OX11 0QX, United Kingdom}
\altaffiltext{23}{Department for Physics, Engineering Physics and Astrophysics, Queen's University, Kingston, ON K7L 3N6, Canada}
\altaffiltext{24}{Division of Theoretical Astronomy, National Astronomical Observatory of Japan, Mitaka, Tokyo 181-8588, Japan}
\altaffiltext{25}{Vietnam National Space Center, Vietnam Academy of Science and Technology, 18 Hoang Quoc Viet, Hanoi, Vietnam}
\altaffiltext{26}{University of South Carolina, Columbia, SC 29208, USA}
\altaffiltext{27}{Indian Institute of Astrophysics, II Block, Koramangala, Bengaluru 560034, India}
\altaffiltext{28}{CAS Key Laboratory of FAST, National Astronomical Observatories, Chinese Academy of Sciences, People's Republic of China}
\altaffiltext{29}{National Astronomical Observatories, Chinese Academy of Sciences, A20 Datun Road, Chaoyang District, Beijing 100012, People's Republic of China}
\altaffiltext{30}{Indian Institute of Science Education and Research (IISER) Tirupati, Rami Reddy Nagar, Karakambadi Road, Mangalam (P.O.), Tirupati 517507, India}
\altaffiltext{31}{Jeremiah Horrocks Institute, University of Central Lancashire, Preston PR1 2HE, UK}
\altaffiltext{32}{Astronomy Program, Department of Physics and Astronomy, Seoul National University, 1 Gwanak-ro, Gwanak-gu, Seoul 08826, Republic of Korea}
\altaffiltext{33}{SOFIA Science Center, Universities Space Research Association, NASA Ames Research Center, Moffett Field, California 94035, USA}
\altaffiltext{34}{Universit\'e Paris-Saclay, CNRS, CEA, Astrophysique, Instrumentation et Mod\'elisation de Paris-Saclay, 91191 Gif-sur-Yvette, France}
\altaffiltext{35}{Leiden Observatory, Leiden University, PO Box 9513, 2300 RA Leiden, The Netherlands}
\altaffiltext{36}{Instituto de Astrof\'isica de Canarias, E-38205 La Laguna, Tenerife, Canary Islands, Spain}
\altaffiltext{37}{Universidad de La Laguna, Dpto.\ Astrof\'isica, E-38206 La Laguna, Tenerife, Spain}
\altaffiltext{38}{Key Laboratory for Research in Galaxies and Cosmology, Shanghai Astronomical Observatory, Chinese Academy of Sciences, 80 Nandan Road, Shanghai 200030, People's Republic of China}
\altaffiltext{39}{Jodrell Bank Centre for Astrophysics, School of Physics and Astronomy, The University of Manchester, Manchester M13 9PL, UK.}
\altaffiltext{40}{SOKENDAI (The Graduate University for Advanced Studies), Hayama, Kanagawa 240-0193, Japan}
\altaffiltext{41}{National Radio Astronomy Observatory, 520 Edgemont Road, Charlottesville, VA 22903, USA}
\altaffiltext{42}{Univ.\ Grenoble Alpes, CNRS, IPAG, 38000 Grenoble, France}
\altaffiltext{43}{Division of Science, National Astronomical Observatory of Japan, 2-21-1 Osawa, Mitaka, Tokyo 181-8588, Japan}
\altaffiltext{44}{Department of Astronomy, Graduate School of Science, The University of Tokyo, 7-3-1 Hongo, Bunkyo-ku, Tokyo 113-0033, Japan}
\altaffiltext{45}{Astrobiology Center, National Institutes of Natural Sciences, 2-21-1 Osawa, Mitaka, Tokyo 181-8588, Japan}
\altaffiltext{46}{SKA Observatory, Jodrell Bank, Lower Withington, Macclesfield SK11 9FT, UK}
\altaffiltext{47}{Jodrell Bank Centre for Astrophysics, School of Physics and Astronomy, University of Manchester, Oxford Road, Manchester M13 9PL, UK}
\altaffiltext{48}{Department of Physics, University of Bath, Claverton Down, Bath, BA2 7AY, UK}
\altaffiltext{49}{Armagh Observatory and Planetarium, College Hill, Armagh, BT61 9DB, UK}
\altaffiltext{50}{Institute of Astronomy and Department of Physics, National Tsing Hua University, Hsinchu 300044, Taiwan}
\altaffiltext{51}{Institute of Astronomy and Department of Physics, National Tsing Hua University, Hsinchu 30013, Taiwan}
\altaffiltext{52}{School of Astronomy and Space Science, Nanjing University, 163 Xianlin Avenue, Nanjing 210023, People's Republic of China}
\altaffiltext{53}{Key Laboratory of Modern Astronomy and Astrophysics (Nanjing University), Ministry of Education, Nanjing 210023, People's Republic of China}
\altaffiltext{54}{University of Chinese Academy of Sciences, Beijing 100049, People's Republic of China}
\altaffiltext{55}{Department of Earth Science Education, Seoul National University, 1 Gwanak-ro, Gwanak-gu, Seoul 08826, Republic of Korea}
\altaffiltext{56}{SNU Astronomy Research Center, Seoul National University, 1 Gwanak-ro, Gwanak-gu, Seoul 08826, Republic of Korea}
\altaffiltext{57}{Department of Physics, Graduate School of Science, Nagoya University, Furo-cho, Chikusa-ku, Nagoya 464-8602, Japan}
 
 
\maketitle
 
\begin{abstract}
Multiwavelength dust polarimetry probes interstellar magnetic fields, grain properties and alignment, and contributions from dust at different temperatures.
Combining measurements across wavelengths and facilities requires reliable calibration, uncertainty estimates, and compatible spatial responses.
We present five technical validations of SCUBA-2/POL-2 polarimetry on the James Clerk Maxwell Telescope at 450 and 850\,\micron, using 174 archival fields in 69 molecular cloud regions and subsets appropriate to each test.
Extended-source calibration avoids a deterministic shift of $\Delta\beta=-1.05$ in the dust emissivity index $\beta$ that would result from calibrating extended emission using peak brightness.
We modeled Stokes~$I$ noise with additive and signal-dependent components.
The measured median ratio of surface-brightness noise at 450/850\,\micron\ is 25.
Independent tests of Stokes~$Q,U$ noise support approximately equal additive variances and weak error correlations, providing a basis for propagating polarization uncertainties.
Corrections determined for individual fields reduce additive zero-level offsets of order $1$\,mJy\,arcsec$^{-2}$ in the 450\,\micron\ Stokes~$Q$ and~$U$ intensities.
The three-step regression leaves a median combined residual amplitude of $0.07$\,mJy\,arcsec$^{-2}$.
Independent comparisons with SOFIA/HAWC+ at 154 and 214\,\micron\ support the correction in strongly affected fields.
We matched the 450\,\micron\ maps to the modeled 850\,\micron\ beam and the common $12\farcs6$--$300\arcsec$ range of angular scales.
We compared three reductions of the same 450\,\micron\ observations across 30 sources.
Taking the median angular residual relative to the final product for each source, we find that beam matching reduces the median over sources from $6\fdg3$ to $4\fdg7$.
After matching, 29 of the 30 sources have median residuals below $10\degr$.
We also quantified how residual $Q/U$ offsets propagate in ratios of polarization fractions, differences in spectral indices, and polarization angles.
These validations provide procedures and quantitative references for interpreting POL-2 measurements and combining them with polarimetry at other wavelengths.
\end{abstract}

\section{Introduction}
\label{sec:intro}
 
\subsection{Background}
\label{ss:intro_physmot}
 
Polarized thermal emission from interstellar dust provides a view of the structures of magnetic fields associated with the formation of molecular clouds and stars \citep[for reviews, e.g.,][]{Andersson2015,Pattle2023PPVII}.
Measurements at several wavelengths also constrain grain properties and alignment because dust temperature, emissivity, and alignment efficiency determine the relative contributions of different emitting components along the line of sight.
The wavelength dependence of the polarization fraction and angle therefore carries information about both the dust and the magnetic field ($\vec{B}$-field).
To extract this information, we must characterize the calibration, noise, and spatial response of the measurements.
 
Observations with the Planck satellite, the High-resolution Airborne Wideband Camera Plus (HAWC+) on the Stratospheric Observatory for Infrared Astronomy (SOFIA), and interferometers such as the Submillimeter Array (SMA) and Atacama Large Millimeter/submillimeter Array (ALMA) have extended the range of wavelengths and spatial scales available for studying magnetic fields \citep[see the review by][]{Pattle2023PPVII}.
These facilities trace the field over successive ranges of spatial scale: Planck on the scales of entire molecular clouds, single-dish telescopes such as the James Clerk Maxwell Telescope (JCMT) down to $\sim$0.1\,pc dense cores, and interferometers such as ALMA from several $\times 1000$\,au down to a few $\times 10$\,au \citep{Hull2019}.

The JCMT has a long record of submillimeter imaging polarimetry.
Its first-generation camera, the Submillimetre Common-User Bolometer Array (SCUBA; \citealt{Holland1999}), was used with the SCUPOL polarimeter \citep{Greaves2003}.
The SCUPOL archive established an extensive record of 850\,\micron\ imaging polarimetry \citep{Matthews2009}.
SCUBA was replaced by the Submillimetre Common-User Bolometer Array 2 (SCUBA-2; \citealt{Holland2013}), which images at 450 and 850\,\micron\ simultaneously.
Its flux calibration, through flux conversion factors (FCFs), and its beam shapes have been established from observations of calibration sources \citep{Dempsey2013,Mairs2021}.
With the POL-2 polarimeter \citep{Friberg2016}, SCUBA-2/POL-2 measures polarization in both bands simultaneously.
The B-fields In STar-forming Region Observations (BISTRO) survey has made extensive use of the 850\,\micron\ capability to map polarized emission from nearby molecular clouds \citep{Ward-Thompson2017}.
Reliable measurements in both POL-2 bands can extend these studies and connect them to far-infrared and millimeter polarimetry.
 
The atmospheric window near 450\,\micron\ is valuable for studies of both dust continuum and molecular lines.
It includes CO $J=6$--$5$ near 434\,\micron, one of the transitions used in Herschel spectroscopy to investigate the warm molecular gas around embedded protostars \citep[e.g.,][]{Yang2018COPS}.
For dust polarimetry, 450\,\micron\ provides an intermediate wavelength between the far-infrared HAWC+ bands and 850\,\micron.
It thus allows us to compare the polarization spectrum (the polarization fraction as a function of wavelength, $P(\lambda)$) and the projected structure of the $\vec{B}$-field over a wider wavelength range.

Ground-based observations in this window are demanding because atmospheric transmission is lower than at 850\,\micron\ and is more sensitive to changes in the column of water vapor \citep{Archibald2002,Dempsey2013}.
Variations in atmospheric emission also introduce sky noise, while changing opacity modulates the astronomical signal \citep{Archibald2002}.
The two bands are known to have different beam profiles and calibration factors for surface brightness \citep{Mairs2021}.
Before interpreting differences in the measured polarization, we must characterize these effects together with the recovery of the extended structure during map making.
 
\label{ss:intro_gap}
POL-2 observations have demonstrated the potential of 450\,\micron\ polarimetry in bright regions such as OMC~1 \citep{Hwang2021}.
To extend such studies to a broader sample, we developed and tested a consistent procedure using archival POL-2 observations.
The procedure calibrates surface brightness, selects reliable measurements, and compares maps with different spatial and noise responses.
Our aim is to establish which wavelength-dependent signatures can be assessed reliably and the uncertainty scales needed to interpret them.

\subsection{Contribution of this paper}
\label{ss:intro_contribution}
 
The scientific aims outlined above motivate three observational requirements, which organize the analysis as three methodological pillars.
 
\emph{Pillar~1 -- Photometric consistency.}
We examine the calibration of extended emission and the conversion from observed spectral slopes to dust emissivity indices (technical validation~1, TV1; subsection~\ref{ss:flux}).
Relating these slopes to grain properties requires a consistent calibration of surface brightness and an interpretation that accounts for dust temperature.
 
\emph{Pillar~2 -- Noise characterization.}
We analyze the noise in Stokes~$I$ and Stokes~$Q,U$ to establish the basis for data selection and error propagation (TV2 and TV3; subsections~\ref{ss:noiseI} and \ref{ss:noisemodel}).
The measured Stokes uncertainties allow us to assess the significance of changes in polarization fraction and angle when investigating spatial variations in grain alignment and the orientation of the $\vec{B}$-field.
 
\emph{Pillar~3 -- Map-domain comparability.}
We develop a correction for additive zero-level offsets in the 450\,\micron\ Stokes~$Q$ and~$U$ intensities and prepare the maps for comparison.
We test the sensitivity of the recovered angles to beam matching and spatial filtering (TV4 and TV5; subsections~\ref{ss:offset} and \ref{ss:beamval}).
Compatible zero levels and spatial responses are required to compare the polarized dust emission and projected structures of the $\vec{B}$-field traced at different wavelengths.
 
\begin{figure*}[htp]
\centering
\includegraphics{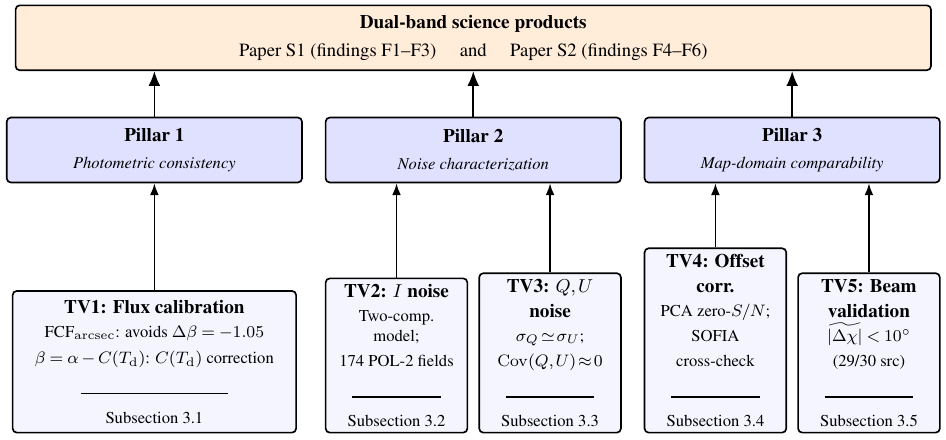}
\caption{
Relationship between the five technical validations and the three methodological pillars.
TV1 establishes photometric consistency.
TV2 and TV3 characterize the noise and polarization uncertainties.
TV4 and TV5 prepare comparable maps and assess the sensitivity of the recovered polarization angles to processing choices.
Abbreviations in the boxes: flux conversion factor (FCF) and principal component analysis (PCA).
``Offset corr.''\ denotes the correction of the 450\,$\mu$m Stokes-$Q/U$ zero-level offset.
Together, these validations provide the observational basis for the analyses of the two bands in Papers~S1 and~S2.
Section~\ref{sec:summary} summarizes the principal results and their scientific implications.
}
\label{fig:T_pillar_overview}
\end{figure*}

Figure~\ref{fig:T_pillar_overview} relates these three requirements to the five technical validations.
We then assess how the residual systematic effects propagate into ratios of polarization fractions, differences between spectral indices, and polarization angles (subsection~\ref{sec:uncertainty}).
The resulting calibration choices, noise diagnostics, and correction procedures are intended for both future observations and analyses of archival POL-2 data.
We discuss their applications to grain alignment, the structure of the $\vec{B}$-field, and comparisons among the environments of molecular clouds in the two science papers of this series (Papers~S1 and~S2).

\subsection{Organization of this paper}
\label{ss:intro_roadmap}
 
We organize the remainder of this paper as follows.
Section~\ref{sec:obs} describes the JCMT SCUBA-2/POL-2 observations, the seven nested samples (table~\ref{tbl:T_sample_inventory}), and the basic reduction parameters.

Subsection~\ref{ss:flux} presents Pillar~1: the framework for flux calibration, the choice of extended-source FCF, and its application to spectral indices with the $C(T_{\rm d})$ correction (equation~\ref{eq:planck_correction}).
We define the quantities for the two bands in sub-subsection~\ref{ss:fcf_dualbandquantities}: the spectral indices $\alpha$ and $\gamma$ of the total and polarized intensities and the polarization-fraction ratio $P_{450}/P_{850}$.
Subsections~\ref{ss:noiseI} and~\ref{ss:noisemodel} develop Pillar~2 in two parts: the characterization of additive noise in Stokes~$I$ (subsection~\ref{ss:noiseI}) and the two-component noise model for Stokes~$Q$ and~$U$ (subsection~\ref{ss:noisemodel}).
The latter concludes by examining how the spectral indices derived from the two bands depend on the signal-to-noise ratio ($S/N$).

Subsection~\ref{ss:offset} presents the construction stage of Pillar~3, in which we identify and investigate the additive zero-level offsets in the 450\,$\mu$m Stokes~$Q$ and~$U$ intensities and apply a correction based on principal component analysis (PCA).
This stage includes independent validation using SOFIA/HAWC+ observations.
Subsection~\ref{ss:beamval} provides the final validation of Pillar~3.
The comparisons cover a single source on the $12\arcsec$ grid, the same source on the $4\arcsec$ grid, and a subsample of 30 sources used for beam validation.
We conclude from these comparisons that beam harmonization preserves the local pattern of polarization angles within a median residual of $\widetilde{|\Delta\chi|} \sim 4\degr$--$6\degr$ in the angle $\chi$.
The median residual is $\widetilde{|\Delta\chi|} < 10\degr$ in 29 of the 30 sources at a Stokes-$I$ signal-to-noise ratio $I/\sigma_I > 30$.

Section~\ref{sec:discussion} combines the contributions to the systematic uncertainty from the three pillars.
We then propagate these contributions into the ratio of polarization fractions in the two bands, $P_{450}/P_{850}$, and the difference between the spectral indices of polarized and total intensity, $\gamma - \alpha$ (subsection~\ref{sec:uncertainty}).
Section~\ref{sec:summary} summarizes the framework and its applicability to future polarimetric analyses at 450\,$\mu$m with the JCMT.
Appendices~\ref{app:ricean}--\ref{app:mas_debias} collect the supporting derivations, diagnostics, and documentation of the reduction for Pillars~2 and~3, with references from the sections they support.
 
Figure~\ref{fig:T_pillar_overview} summarizes the five technical validations TV1--TV5 of section~\ref{sec:techresults} and their grouping into three methodological pillars.
Table~\ref{tbl:T_pillar_index} lists the major technical topics in this series, their corresponding pillars, and their locations in this paper or in Papers~S1 and~S2.
 
\begin{longrotatetable}
\begin{deluxetable*}{lcllc}
\tablecaption{Technical topics in Paper~T, their methodological pillars, and related material in Papers~S1 and~S2\label{tbl:T_pillar_index}}
\tablewidth{0pt}
\tablehead{
  \colhead{Topic} &
  \colhead{Pillar} &
  \colhead{Location in this paper series} &
  \colhead{Keywords} &
  \colhead{Note}
}
\startdata
\multicolumn{5}{c}{\textit{Paper~T --- Pillar~1: Photometric consistency}} \\
\hline
Flux Calibration $\&$ Surface Brightness Units & 1 & Subsection~\ref{ss:flux} & FCF$_{\rm arcsec}$; mJy\,arcsec$^{-2}$; $\beta$ & 1 \\
\hline
\multicolumn{5}{c}{\textit{Paper~T --- Pillar~2: Noise characterization}} \\
\hline
Stokes $I$ Noise Characterization & 2 & Subsection~\ref{ss:noiseI} & Two-component model; $\sigma_\lambda$; $k_\lambda$ & 2 \\
Stokes $Q,U$ Noise Model and Isotropy Check & 2 & Subsection~\ref{ss:noisemodel} & Polarization noise; $\sigma_Q\!\simeq\!\sigma_U$; $(Q,U)$ isotropy & 3 \\
Ricean Expectation Formulations & 2 (supp.) & Appendix~\ref{app:ricean} & Exact Bessel-function expectation; high-$S/N$ expansion & 4 \\
Spectral-Index $S/N$ Dependence & 2 & Sub-subsections~\ref{ss:spidx_snrbias}--\ref{ss:spidx_snrbias_sens} & Running medians; imposed $S/N(\gamma)$ threshold & 5 \\
\hline
\multicolumn{5}{c}{\textit{Paper~T --- Pillar~3: Map-domain comparability}} \\
\hline
Offset Recognition & 3 (constr.) & Subsection~\ref{ss:offset} & three-wavelength SOFIA/HAWC+ comparison; 450 isolated & 6 \\
Investigation of a Hardware Origin & 3 (constr.) & Subsection~\ref{ss:offset} & Pointing; IP model; brightness dependence & 7 \\
Correction of Offsets from Software & 3 (constr.) & Subsection~\ref{ss:offset} & Three-step regression; \texttt{pca.zero\_snr}; Kolmogorov--Smirnov (KS) test & 8 \\
Final Validation of Beam Harmonization & 3 (final) & Subsection~\ref{ss:beamval} & KS test; $\widetilde{|\Delta\chi|}$ floor; survey of 30 sources & 9 \\
\hline
\multicolumn{5}{c}{\textit{Paper~T --- Cross-pillar consolidation}} \\
\hline
Systematic Uncertainty Budget & 1, 2, 3 & Subsection~\ref{sec:uncertainty} & FCF; residual offset; beam mismatch; error propagation & 10 \\
\hline
\multicolumn{5}{c}{\textit{Material in Papers~S1 and~S2 (informational; cross-referenced from Paper~T)}} \\
\hline
$^{12}$CO($J$=3--2) Contamination & --- & Paper~S1 & 850\,$\mu$m; Orion~A; Heterodyne Array Receiver Program (HARP) & 11 \\
Spectral-Index Distance-Bias Analysis & --- & Paper~S2 & Beam-size effects on $\gamma - \alpha$ and $\beta$ & 12 \\
Stacked $P_{450}$ Histogram & --- & Paper~S2 & Log-normal; percentile statistics & 13 \\
Reference scales of $\mathcal{S}$ & --- & Paper~S2 & Random-orientation reference scale; noise response & 14 \\
$\vec{B}$-Field Maps Atlas & --- & Paper~S2 & 450/850\,$\mu$m; distance-ordered & 15 \\
Distance Determination & --- & Paper~S2 appendix~\ref{S2-app:distance} & Gaia; \citet{Zucker2020}; \citet{Prisinzano2022}; VLBI & 16 \\
\enddata
\tablecomments{
The Pillar column refers to the three methodological requirements described in subsection~\ref{ss:intro_contribution}: constr.\ denotes construction, final denotes the concluding validation, and supp.\ a supporting derivation.
Items~(11)--(16) describe material in Papers~S1 and~S2; the ``Location'' column identifies its home paper.
(1)~Extended-source calibration and the $\Delta\beta=-1.05$ bias from applying point-source FCFs to extended emission.
(2)~Additive and signal-dependent Stokes-$I$ noise, $\sigma_\lambda^2+k_\lambda^2 I^2$, validated by jackknifing across 82 POL-2 fields.
(3)~Two-component Stokes-$Q,U$ noise model and empirical tests of noise isotropy.
(4)~The exact Ricean expectation, compared with its high-$S/N$ expansion, which diverges as the polarized signal approaches the noise.
(5)~Dependence of $\alpha$ and $\gamma$ on $S/N$: diagnostic using running medians and sensitivity of the population result to an imposed $S/N(\gamma)$ threshold.
(6)~The $\sim$20\arcdeg--30\arcdeg\ discrepancies between the angles at 450\,$\mu$m and those from SOFIA/HAWC+ at 154\,$\mu$m that motivated the investigation of the zero-level offset in Stokes~$Q/U$ at 450\,$\mu$m.
(7)~Tests of telescope pointing, source brightness, and instrumental polarization (IP) as possible origins of the $Q/U$ offsets.
(8)~Three-step constrained regression, the response of the offsets to \texttt{pca.zero\_snr}, and SOFIA/HAWC+ validation.
(9)~Comparison of three products: maps with the native beam, maps after deconvolution--reconvolution, and final maps with matched beams and spatial-frequency coverage for L1688~Ophiuchus~A and the subsample of 30 sources.
This comparison establishes the reference for comparing angles in Papers~S1 and~S2.
(10)~Contributions from calibration and residual $Q/U$ zero-level offsets to $P_{450}/P_{850}$ and $\gamma-\alpha$, including cancellation of common multiplicative factors; residual effects of the beam response are considered separately.
(11)~Contamination by molecular lines at 850\,$\mu$m: median $\sim$10\% for pixels with $I_{850}/\Delta I_{850}>15$ and intensity-weighted mean $\sim$8\%.
(12)~Regression tests of beam-size effects versus intrinsic variations in $\gamma-\alpha$ and $\beta$.
(13)~Log-normal polarization fractions and the use of percentile statistics.
(14)~Random-orientation reference scale of $\sim$52\arcdeg\ for the angle-dispersion function $\mathcal{S}$ and the noise response of its raw estimator.
(15)~Atlas of $\vec{B}$-field orientations at both wavelengths for all sources detected at 450\,$\mu$m, ordered by distance.
(16)~Hierarchical distances from catalogs based on Gaia, superseded by very long baseline interferometry (VLBI) maser parallaxes where available.
These distances are used and documented in Papers~S1 and~S2 only.
}
\end{deluxetable*}
\end{longrotatetable}

\section{Observations, Data Retrieval, and Sample}
\label{sec:obs}
 
In this section, we describe the JCMT SCUBA-2/POL-2 observations and data retrieval, the seven nested samples
(subsection~\ref{ss:obs_sample}), and the basic reduction parameters.
Together they provide the observational basis for Pillars~1--3 in
figure~\ref{fig:T_pillar_overview} and table~\ref{tbl:T_pillar_index}.
Throughout the series, we distinguish samples of sources from samples of beams,
as defined in table~\ref{tbl:T_sample_inventory}.
 
\subsection{JCMT SCUBA-2/POL-2 observations and data retrieval}
\label{ss:obs_instrument}

On the JCMT, operated by the East Asian Observatory (EAO), POL-2 modulates incident polarization with a half-wave plate rotating at approximately 2~Hz.
A fixed wire-grid analyzer converts this polarization modulation into an intensity modulation at approximately 8~Hz, which the SCUBA-2 bolometer arrays in the focal plane record simultaneously at 450 and 850\,$\mu$m.
With the standard reduction settings, the \texttt{calcqu} program in the Starlink software suite fits the bolometer time series over each full rotation of the half-wave plate and produces Stokes $I$, $Q$, and $U$ time series sampled at approximately 2~Hz.
Stokes $Q$ and $U$ are recovered from the fourth-harmonic modulation, while Stokes $I$ is estimated from the fitted non-modulated component of the same data (equation~\ref{eq:pol2_signal_app} in appendix~\ref{app:QU_diagnostics}).
The 450\,$\mu$m polarization data analyzed here were therefore obtained in parallel with the 850\,$\mu$m data.

The POL-2 instrument and its observing mode for the two bands are described by \citet{Friberg2016} and \citet{Friberg2018}.
Here we address the data reduction and systematic calibration specific to the 450\,$\mu$m band.
We selected data from POL-2 observations available in the JCMT archive as of 2022 February.
For the 450 \, $\mu$m analyses, we restricted the sample to observations taken under the highest quality Weather Band (WB~1), for which the atmospheric opacity at the zenith satisfies $\tau_{225\,{\rm GHz}}^{\rm zenith} < 0.05$.
We did not apply a selection based on opacity to the 850\,$\mu$m data, for which the requirements on atmospheric transmission are less restrictive.
We downloaded the full raw time-series data set for both bands, totaling 54\,Tb, from the Canadian Astronomy Data Center (CADC) JCMT Science Archive to the Astronomy Data Center of the National Astronomical Observatory of Japan (NAOJ) for reduction.
 
\subsection{Sample inventory}
\label{ss:obs_sample}
We distinguish the archive fields used to characterize instrumental performance from the astronomical regions and matched sky samples used for the scientific comparisons.
A single molecular cloud region may require several POL-2 pointings, and not every pointing provides the measurements needed for each diagnostic that uses both bands.

We used seven samples, identified throughout this paper and Papers~S1 and~S2 by the number of sources or pixel pairs they contain.
In the {\em source-based} samples, each entry is a named astronomical source or a POL-2 field.
The {\em beam-based} sample consists of $12\arcsec$ pixel pairs at a matched resolution, each providing the smallest unit of independent measurement in the two bands.
Table~\ref{tbl:T_sample_inventory} summarizes their counts, short names, and primary uses.
We adopted the short names {\sc Census}, {\sc PolSrc}, {\sc SpIdx}, and {\sc DualBand} for the four main samples.
We refer to the three additional subsamples (36, 30, and 9) descriptively, without short names: for example, the subsample of 30 sources in {\sc SpIdx} used for beam validation.
The selection criteria for each sample are described below and in the notes to table~\ref{tbl:T_sample_inventory}.
 
\begin{deluxetable}{l c c l}
\tablecaption{Sample inventory across the BISTRO 450/850\,$\mu$m dual-wavelength paper series \label{tbl:T_sample_inventory}}
\tablewidth{0pt}
\tablehead{
 \colhead{Sample name} &
 \colhead{Short name} &
 \colhead{Count} &
 \colhead{Primary use}
}
\startdata
\multicolumn{4}{l}{\textit{Source-based samples (each entry is a named astronomical source or POL-2 field)}} \\
POL-2 archive fields                   & {\sc Census}   & 174 & Paper T subsections~\ref{ss:noiseI}--\ref{ss:noisemodel} (Pillar~2 noise characterization) \\
Unique star-forming regions            & {\sc PolSrc}   & 69  & Paper~S2 (complete dual-wavelength census; F4--F6) \\
\quad FCF/spectral-index subsample     & ---            & 36  & Paper T subsection~\ref{ss:flux} (Pillar~1 FCF verification) \\
\quad Statistical science subsample    & {\sc SpIdx}    & 35  & Paper~S1 (principal $\alpha$, $\gamma$, $R_{\rm p}$ statistics by source; F2--F3) \\
\quad Beam-validation subsample        & ---            & 30  & Paper T subsection~\ref{ss:beamval} (Pillar~3 final beam-harmonization validation) \\
\quad HGBS-overlap subsample           & ---            & 9   & Paper~S2 subsection~\ref{S2-ss:S2_HGBS} (F6 $N_{\rm H}$, $T_{\rm d}$ per-pixel correlations) \\
\hline
\multicolumn{4}{l}{\textit{Beam-based sample (matched-resolution $12\arcsec$ pixel-pair; the smallest unit of independent dual-band measurement)}} \\
Method-3 HGBS pixel-pair sample        & {\sc DualBand} & 417 & Paper~S1 subsections~\ref{S1-ss:Iratio}, \ref{S1-ss:S_dispersion}; Paper~S2 subsection~\ref{S2-ss:S2_HGBS} (F2/F3 per-pixel; F6)
\enddata
\tablecomments{The source-based samples are nested as follows.
The 174 POL-2 fields in the JCMT archive (2022 Feb., WB\,1) reduce to 69 \textit{unique} astronomical regions after mosaicking and quality selection.
Of those 69, 39 yield total-intensity and polarized-intensity spectral indices ($\alpha$, $\gamma$) that pass the quality assessment for individual fields: the 36 printed in table~\ref{S2-tbl:SpIdx_Global} of Paper~S2, plus three further fields that fall below the $N_{\rm samp} \ge 5$ floor and are therefore not tabulated.
The final subset used for the principal statistics for individual sources at both wavelengths in Paper~S1 comprises 35 of them.
The subsample used for the final beam validation in subsection~\ref{ss:beamval} contains 30 sources.
It consists of the 35 less Orion~A Field~4, Orion~A Field~5 and Serpens~South, which lack the native-beam member of the three-product set; Sagittarius~B2, which falls below the $N \ge 20$ requirement; and the Brick, whose native-beam pixels do not coincide with those of the final product.
Of the 69 regions, 9 fall within the sky coverage of the public \textit{Herschel} Gould Belt Survey (HGBS; column density $N_{\rm H}$ and dust temperature $T_{\rm d}$) used for the correlations between measurements for individual pixels in Paper~S2~subsection~\ref{S2-ss:S2_HGBS}.
We used the full set of 174 fields for Pillar~2 because each field provides an independent sampling of the noise conditions.
Pillar~1 uses the subsample of 36 sources for FCF/spectral-index analysis, and the final validation of Pillar~3 (subsection~\ref{ss:beamval}) uses the subsample of 30 sources for beam validation.
We constructed the beam-based {\sc DualBand} sample (417 pixel-pair measurements) on a common $12\arcsec$ grid at a matched effective resolution of $12\farcs6$ from the 9 regions that overlap with HGBS.
These pixel pairs are the units of independent measurement for the F6 correlations of Paper~S2 and for the $P$--$\mathcal{S}$ test of Paper~S1 for geometric depolarization.
Elsewhere in the science papers, ``$N_{\rm samp}$'' counts the matched pixel pairs used in the analysis in question and is not in general a subset of these 417.
For example, the F1 angle comparison uses the larger set of $N_{\rm samp} = 1649$ pixels across 21 regions.
We report scientific results from the two wavelengths either for the {\sc SpIdx} subset of 35 sources (Paper~S1, principal F2/F3) or for the full {\sc PolSrc} census of 69 regions, including fields with lower significance (Paper~S2, F4--F6).}
\end{deluxetable}
 
We selected the parent sample according to three criteria: (i) availability to the lead author as of 2022 February; (ii) 450\,$\mu$m observations obtained under Weather Band~1 ($\tau_{225\,{\rm GHz}}^{\rm zenith} < 0.05$); and (iii) a peak Stokes~$I$ signal-to-noise ratio $>100$ at 450\,$\mu$m, sufficient for polarization analysis in both bands.
We did not require a detection in polarization fraction or polarized intensity to define this parent sample.
The 174 fields used to characterize noise and the 36 sources used to verify the FCFs span the range of distances in the sample, from $\sim$100\,pc (low-mass cores) to $\sim$8\,kpc (Galactic Center clouds).
We therefore applied the methodology across the signal levels and conditions of the sky background represented in this survey.

We discuss the consequences of these selection criteria for the scientific results in Papers~S1 and~S2.
Each result applies to the subset that provides the required observable; it does not automatically apply to the entire census of regions.
This distinction preserves the connection between a reported trend in the population and the cloud material actually sampled.

\subsection{Basic reduction parameters and calibration}
\label{ss:obs_reduction}
 
We processed raw time-series data with {\tt pol2map.py} (version 2021 July 23) from the Sub-Millimetre User Reduction Facility ({\tt SMURF}) within {\tt Starlink}.
The program \texttt{pol2map} makes the $I$, $Q$, and $U$ maps with the iterative map maker of SMURF, \texttt{makemap}~\citep{Chapin2013}.
The release history of \texttt{pol2map} records a change on 2022 January 19 to use a weighted mean of the nominal FCFs for observations spanning different calibration epochs.
We used the development version of \texttt{pol2map.py} dated 2021 July 23 throughout this analysis.
Recent BISTRO publications have used \texttt{pol2map}~\citep[e.g.,][]{Choi2024,Karoly2025,Yang2025,Hwang2025}.
We follow the established three-step \texttt{pol2map} reduction procedure validated for POL-2 data (\citealt{Parsons2021}; e.g., \citealt{Ward-Thompson2017,Pattle2017,Kwon2018,Soam2018,Pattle2021}).

Reduction of the 450 $\mu$m data requires the {\tt normalization} parameter to compensate for differences in intensity scale between observations.
This is because the 450\,$\mu$m data are more sensitive than the 850\,$\mu$m data to variations in atmospheric opacity between observations.
This normalization is a prerequisite for correcting the zero-level offsets in the 450\,$\mu$m Stokes-$Q/U$ intensities in subsection~\ref{ss:offset}.

One of us (RSF) ran all reductions with \texttt{pol2map} described in this paper, from the raw time-series data to the Stokes $I$, $Q$, and $U$ maps, on the computers of the NAOJ Astronomy Data Center.
These reductions were completed before AI tools were introduced into this work.
The use of AI tools in the subsequent analysis and in the preparation of the text is described in the Acknowledgments.
 
To correct for instrumental polarization (IP), we applied the ``August 2019'' model \citep{EAO2019IPmodels}.
This model was derived from POL-2 observations of bright extended sources (OMC~1, DR~21, and G34) instead of Uranus \citep{Berry2019ipcor}.
Its accuracy near the zenith and at low elevations has not been established.

We calibrated all maps, including those of observations taken before 2018 June~30, with the arcsecond FCFs that \citet{Mairs2021} derived for observations after that date:
FCF$_{\rm arcsec,450} = 3.87 \pm 0.53$ and FCF$_{\rm arcsec,850} = 2.07 \pm 0.12$~Jy\,pW$^{-1}$\,arcsec$^{-2}$ (their table~4), with uncertainties of 14\% and 6\% in the stable part of the night and 17\% and 7\% in the evening and morning (their section~5).
We used this extended-source calibration to express the surface brightness of emission extended beyond the beam size in mJy \, arcsec$^{-2}$.
We used it in Papers~T, S1, and~S2 for all derived quantities that depend on wavelength.
We explain in subsection~\ref{ss:flux} why we use the arcsecond FCFs rather than the peak (point-source) FCFs and discuss the systematic implications of this choice.

We applied attenuation factors of $1.96 \pm 0.023$ at 450\,$\mu$m and $1.35 \pm 0.008$ at 850\,$\mu$m to account for the insertion of the POL-2 system in front of the focal plane of SCUBA-2.
We took the factors 1.96 and 1.35 from the POL-2 Efficiency Report (P.~Friberg 2016\footnote{POL2: Efficiency Report $\langle$\texttt{https://www.eao.hawaii.edu/pol2/Efficiency\%20Report}$\rangle$.}; see also \citealt{Mairs2021}; the 850\,$\mu$m factor was also adopted by \citealt{Sato2024}).
The report compared photometry of planets with and without POL-2 in the beam and gives the factors without uncertainties.
Each factor is half the measured ratio of the photometry without POL-2 to the photometry with POL-2.
The factor of 2 is the loss expected for an ideal analyzer, which transmits half of the unpolarized power.
RSF assigned the uncertainties from the photometry tables of the report, after checking these tables against the records and recollections of the joint meetings of the BISTRO and POL-2 teams.
The factor 1.96 is half the 450\,$\mu$m ratio of 2015 October~7, the night of Uranus observations that the report adopted.
The uncertainty 0.023 is half the statistical uncertainty of the same ratio.
The uncertainty 0.008 is half the statistical uncertainty of the 850\,$\mu$m ratio of 2015 October~1, the night that gives 1.35.
The value 0.008 is comparable to the corresponding uncertainty for October~7 (0.007) and to the difference between the 850\,$\mu$m factors of the two nights (0.008).
The ratios in the report scatter by 1\%--4\% (one standard deviation), both between pairs of scans within the night of 2015 October~7 and among the nights of 2012--2016.
This scatter is less than the uncertainties of the arcsecond FCFs (14\% and 6\%; sub-subsection~\ref{ss:fcf_calibration}).

We adopt the beam sizes of \citet{Mairs2021}, given as half-power beam widths (HPBWs): $\theta_{\rm HPBW}^{450} = 8\farcs6 \pm 1\farcs3$ and $\theta_{\rm HPBW}^{850} = 12\farcs6 \pm 1\farcs9$.
Throughout this paper and Papers~S1 and~S2, the two widths denote the \emph{effective} full widths at half maximum (FWHMs) of the two-component beam decomposition of \citet{Mairs2021}.
Their equation~(12) gives $\theta_{\rm eff} = (\alpha^{\rm Mairs}\,\theta_{\rm MB}^{2} + \beta^{\rm Mairs}\,\theta_{\rm S}^{2})^{1/2}$, where $\alpha^{\rm Mairs}$ and $\beta^{\rm Mairs}$ are their relative amplitudes $\alpha$ and $\beta$ of the main-beam and secondary components of widths $\theta_{\rm MB}$ and $\theta_{\rm S}$.
These effective widths do not describe the main-beam component alone, for which \citet{Mairs2021} gave $\theta_{\rm MB} = 6\farcs2$ and $11\farcs0$ (their table~6).
They also differ from the empirical effective widths $10\farcs0 \pm 0\farcs6$ and $14\farcs4 \pm 0\farcs3$ obtained from the ratio of peak to arcsecond FCFs (their equation~9), which include the far wings lying beyond the two-Gaussian model.
For the deconvolution--reconvolution in subsection~\ref{ss:beamval}, we require the beam {\em shape} rather than a single width and therefore retain both components explicitly.

We regridded the Stokes parameters onto a common 12\arcsec\ pixel grid for analyses of the overlap between the two bands.
The 12\arcsec\ grid samples the 850\,$\mu$m beam of 12\farcs6 at 1.05 pixels per FWHM, corresponding to one sample per beam.
For the 450\,$\mu$m band alone, we used a 4\arcsec\ pixel grid for analyses at the native resolution.
The 4\arcsec\ grid samples the 8\farcs6 beam at 2.15 pixels per FWHM.
We found empirically that this grid suppresses reduction artifacts in all Stokes parameters.
 
We debiased the polarized intensity and calculated the polarization fractions using the modified asymptotic estimator (MAS) of \citet{Plaszczynski2014} and \citet{Montier2015}.
We give the details of the implementation for the 450\,$\mu$m band in subsection~\ref{ss:offset}.
Appendix~\ref{app:ricean} presents the exact Ricean expectation that we use when modeling measured rather than debiased polarized intensities.
\section{Results and Analysis}
\label{sec:techresults}
 
This section presents the five technical validations (TV1--TV5) that establish the methodological basis of the BISTRO polarimetric survey at two wavelengths.
TV1 (subsection~\ref{ss:flux}) establishes the framework for photometry and surface brightness required to compare the two bands.
TV2 and TV3 (subsections~\ref{ss:noiseI}--\ref{ss:noisemodel}) develop the empirical and analytic noise models for Stokes~$I$, $Q$, and~$U$.
TV4 (subsection~\ref{ss:offset}) identifies and corrects additive zero-level offsets in the 450\,\micron\ Stokes~$Q$
and~$U$ intensities, while TV5 (subsection~\ref{ss:beamval}) quantifies the residual floor for polarization angles after beam harmonization.
Together, these five validations define the systematic precision floor of the survey used throughout Papers~S1 and~S2.
 
\subsection{Technical Validation 1: Flux Calibration, Surface Brightness Units, and Dust Spectral Indices}
\label{ss:flux}
\label{app:FCF_units}%
 
To compare the 450 and 850\,\micron\ polarization observables reliably, we must place both maps on a consistent surface-brightness scale before calculating spectral indices or polarization ratios.
We therefore re-examine the JCMT formalism for flux conversion and demonstrate that the commonly adopted convention of point-source calibration is not appropriate for the extended molecular cloud emission that dominates the present survey.
We establish the calibration of surface brightness used in all three papers in this series and quantify its effects on the dust spectral indices ($\alpha$, $\beta$, $\gamma$) used in F2--F5.
This analysis constitutes Pillar~1 of figure~\ref{fig:T_pillar_overview} and table~\ref{tbl:T_pillar_index}.
 
\subsubsection{Motivation and overview}
\label{ss:fcf_motivation}
 
In this work, we present the surface brightness in units of mJy\,arcsec$^{-2}$, rather than the commonly used mJy\,beam$^{-1}$.
As we demonstrate below, the use of point-source calibration for spatially extended molecular cloud emission introduces a bias that depends on wavelength and propagates directly into dust spectral indices.

\subsubsection{Dual-band derived quantities: $\alpha$, $\gamma$, and $P_{450}/P_{850}$}
\label{ss:fcf_dualbandquantities}
The spectrum of polarized dust emission does not necessarily follow the spectrum of total emission from the same line of sight.
Comparing their slopes quantifies how the polarization fraction changes between 450 and 850\,$\mu$m, without yet identifying the responsible grain properties or structure along the line of sight.
 
We use the spectral indices of total and polarized intensity, and their algebraic relation to the ratio of polarization fractions, in three parts of the analysis: the calibration of Pillar~1, the construction stage of Pillar~3 (subsection~\ref{ss:offset}), and the propagation of uncertainties from all three pillars (subsection~\ref{sec:uncertainty}).
We discuss their physical interpretation, including the spectral index of grain emissivity and alignment that depends on grain size, in Paper~S1.
The spectral indices of total intensity $I_\lambda$ and polarized intensity $PI_\lambda$ are defined as
\begin{equation}
\alpha \equiv \frac{\log(I_{450}/I_{850})}{\log(\nu_{450}/\nu_{850})}
\quad \text{and} \quad
\gamma \equiv \frac{\log(PI_{450}/PI_{850})}{\log(\nu_{450}/\nu_{850})},
\label{eq:spectral_indices}
\end{equation}
where $\nu_{450} = 665.0$\,GHz and $\nu_{850} = 349.5$\,GHz are the center frequencies of the SCUBA-2/POL-2 passbands, so that $\log(\nu_{450}/\nu_{850}) \approx 0.279$.
The polarization-fraction ratio $P_{450}/P_{850}$ follows directly from equation~(\ref{eq:spectral_indices}):
\begin{equation}
\frac{P_{450}}{P_{850}} = \left(\frac{\nu_{450}}{\nu_{850}}\right)^{\gamma - \alpha}.
\label{eq:Pratio_SpIdx}
\end{equation}
We use equation~(\ref{eq:Pratio_SpIdx}) in subsection~\ref{sec:uncertainty} to propagate the FCF systematics and the residual zero-level offsets in the 450\,$\mu$m Stokes-$Q/U$ intensities to $P_{450}/P_{850}$ and $\gamma - \alpha$.
This relation is an algebraic identity rather than a physical model.
We also estimate $P_{450}/P_{850}$ in Paper~S2 (subsection~\ref{S2-ss:S2_P_ratio}).
The formulae for the uncertainty budget below apply most directly to the intensity-weighted and median estimators.
The median estimator is the default for the ratio between the two bands in Paper~S1.

For the same measured line of sight, a negative $\gamma-\alpha$ means $P_{450}<P_{850}$: the polarization fraction is higher at the longer wavelength.
Summaries for individual sources do not necessarily obey the same numerical relation when they use different pixel selections or methods of aggregation.
 
\subsubsection{Point-source versus extended-source calibration}
\label{ss:fcf_calibration}

The Observatory has established the standard calibration scheme for the SCUBA-2 data through long-term observations of calibration sources.
Two types of flux conversion factors (FCFs) are recommended
\citep{Mairs2021}:
(i) the peak FCF (FCF$_{\rm peak}$), optimized for point-source photometry and expressed in Jy\,beam$^{-1}$\,pW$^{-1}$,
and (ii) the arcsecond-based FCF (FCF$_{\rm arcsec}$), expressed in Jy\,arcsec$^{-2}$\,pW$^{-1}$ and appropriate for spatially extended emission.
The standard \texttt{pol2map} pipeline with \texttt{JY=yes} applies
FCF$_{\rm peak}$ automatically, producing maps in mJy\,beam$^{-1}$.
This procedure is appropriate for point-source photometry,
but is less suitable for expressing the surface brightness
of molecular clouds, whose emission typically extends
beyond the beam.
Applying FCF$_{\rm peak}$ to extended emission effectively folds the beam solid angle into the intensity scale, obscuring the true surface brightness.
 
We therefore retained the detector output in pW during mapping (\texttt{JY = no}) and applied FCF$_{\rm arcsec}$ manually during post-processing.
For all observations, we adopt the following
arcsecond-based FCF values, scaled from those that \citet{Mairs2021} derived for observations obtained after 2018 June 30:
\begin{align}
{\rm FCF}_{\rm arcsec,450}^{\rm POL2} &= 7.59~{\rm Jy\,pW^{-1}\,arcsec^{-2}}, \\
{\rm FCF}_{\rm arcsec,850}^{\rm POL2} &= 2.79~{\rm Jy\,pW^{-1}\,arcsec^{-2}},
\end{align}
which include the POL-2 attenuation factors of $1.96 \pm 0.023$ at 450\,$\mu$m and $1.35 \pm 0.008$ at 850\,$\mu$m (subsection~\ref{ss:obs_reduction}).
Each factor multiplies all Stokes parameters of its band alike, so the factors cancel in polarization fractions and angles, in $P_{450}/P_{850}$, and in $\gamma-\alpha$.
An error in their ratio would shift $\alpha$, $\gamma$, and $\beta$ of all sources by the same amount, about 0.02 for the adopted uncertainties of the two factors.
We quote surface brightnesses derived from these factors in mJy\,arcsec$^{-2}$ throughout the paper series.

The adopted attenuation factors describe only the loss of total power caused by the insertion of the POL-2 system.
These factors do not include depolarization or cross-polarization within POL-2, which lower the modulation amplitude without lowering the transmitted power (the POL-2 Efficiency Report; \citealt{Parsons2021}).
The report describes pure depolarization as not very large but present and lists the measurements with sources of known polarization as pending.
To our knowledge, the modulation efficiency of POL-2 has not since been calibrated in either band.
A fixed difference in the modulation efficiency between the two bands would not cancel in $P_{450}/P_{850}$ or $\gamma-\alpha$.
Such a difference would scale $P_{450}/P_{850}$ by a common factor and shift $\gamma-\alpha$ by a common amount for all sources.
We have not corrected our results for this effect.

We could not apply FCFs for individual calibration epochs to the observations taken before 2018 June~30, because we converted the coadded maps from pW manually.
For those observations, our surface brightnesses are lower than with the FCFs of their epochs by 16\% at 450\,$\mu$m and by 3\%--8\% at 850\,$\mu$m \citep[table~4 of][]{Mairs2021}.
These differences are comparable to the uncertainties of the arcsecond FCFs (14\% and 6\%, respectively).

\begin{figure}[htbp]
\centering
\includegraphics[width=0.95\columnwidth]{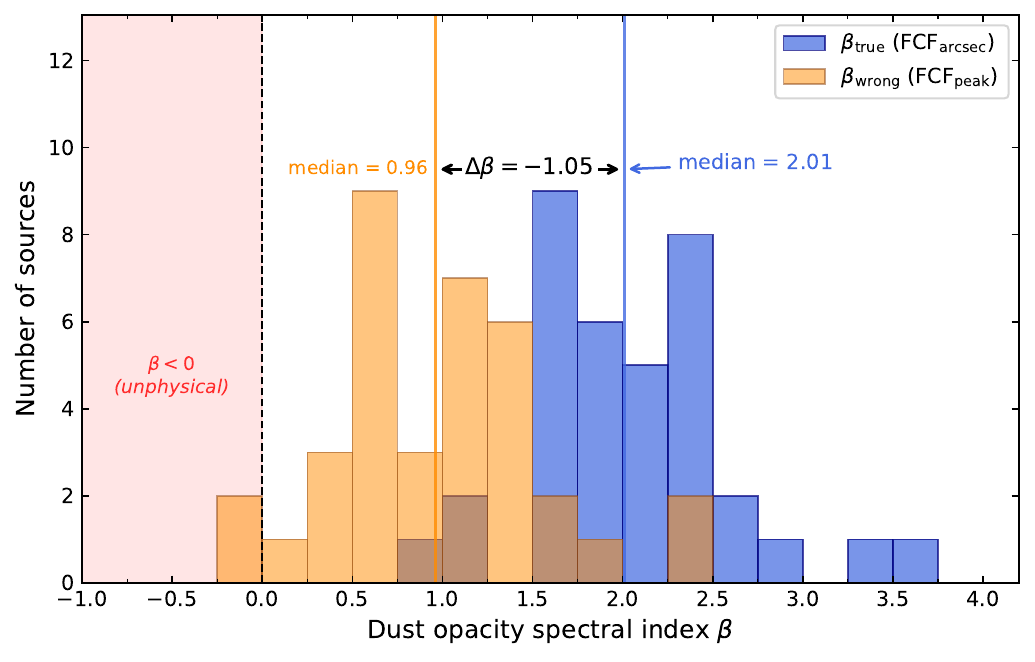}
\caption{
Distribution of the dust opacity spectral index $\beta$ for the sources in the table of spectral indices in Paper~S2 subsection~\ref{S2-ss:S2_SpIdx}, computed with the full Planck function and dust temperatures for individual sources.
The blue histogram shows the results with extended-source calibration (FCF$_{\rm arcsec}$; median $\beta=2.0\pm0.1$).
The orange histogram shows the values obtained by incorrectly applying point-source calibration (FCF$_{\rm peak}$) to the same emission, producing the deterministic shift $\Delta\beta=-1.05$ of equation~\ref{eq:beta_bias}, independent of $T_{\rm d}$.
The shaded region ($\beta<0$, labeled ``unphysical'') marks values inconsistent with the adopted dust model.
With point-source calibration (orange histogram), two of the 36 entries, G202.3$+$2.5 and the Rosette Molecular Cloud, have formally negative inferred $\beta$ values.
We do not plot the uncertainties of the individual $\beta$ values.
The uncertainty in temperature contributes through the $C(T_{\rm d})$ correction (equation~\ref{eq:planck_correction}) and broadens the inferred distributions without changing the relative shift caused by the calibration.
This contribution is typically $\sigma_{\beta,T}\lesssim0.3$--$0.4$ for $T_{\rm d}=15\pm5$~K in low-mass star-forming regions and cold infrared dark clouds, and for $25\pm10$~K in high-mass star-forming regions and the Central Molecular Zone (Paper~S2 sub-subsection~\ref{S2-sss:SpIdxFramework}).
}
\label{afig:beta_histogram}
\end{figure}
 
\subsubsection{Wavelength-Dependent Calibration Bias}
\label{ss:fcf_bias}
The two calibration approaches differ in their wavelength dependence.
For a point-source calibrator, FCF$_{\rm peak}$ is the known flux density divided by the measured peak signal in pW.
FCF$_{\rm arcsec}$ is the same flux density divided by the signal integrated in the calibration aperture, in pW\,arcsec$^{2}$.
Their ratio is therefore the integral of the beam response normalized to the unit peak, which defines the
effective beam solid angle at each wavelength:
\begin{align}
\Omega^{\rm FCF}_{450} &\equiv
\frac{{\rm FCF}_{\rm peak,450}}{{\rm FCF}_{\rm arcsec,450}}
\approx 122~{\rm arcsec^2},
\label{eq:R450}\\
\Omega^{\rm FCF}_{850} &\equiv
\frac{{\rm FCF}_{\rm peak,850}}{{\rm FCF}_{\rm arcsec,850}}
\approx 239~{\rm arcsec^2}.
\label{eq:R850}
\end{align}
Thus, $\Omega^{\rm FCF}_\lambda$ is the solid angle of the {\em compound} JCMT beam, including the narrow main beam and the broad error beam.
It differs from the Gaussian area $\Omega^{\rm G}_\lambda$ implied by the adopted effective FWHM (table~\ref{tab:omega_conventions}).
Equating the two would give
$\theta_{\rm FWHM} = (\Omega^{\rm FCF}_\lambda/1.133)^{1/2} = 10\farcs4$ and
$14\farcs5$ at 450 and 850\,$\mu$m, respectively, rather than the two-component effective widths of $8\farcs6$ and $12\farcs6$.
These inferred widths agree within $1\sigma$ with the empirical effective FWHMs of $10\farcs0 \pm 0\farcs6$ and $14\farcs4 \pm 0\farcs3$ that \citet{Mairs2021} obtained from the same FCF ratio.
This agreement provides an internal check on the consistency of the two calibration conventions.
Approximating $\Omega^{\rm FCF}_\lambda$ by a Gaussian beam area therefore neglects part of the compound beam, whose solid angle sets the calibration rather than the nominal HPBW.

We use the repeatability of the beam solid angle as the tolerance, rather than the full FCF tolerance.
The absolute flux scale cancels and only the uncertainty in the effective beam solid angle remains, because $\Omega^{\rm FCF}_\lambda$ is a ratio of two FCFs derived from calibration observations toward a common source.
With $\Omega \propto
\theta_{\rm eff}^{2}$ and the dispersions of the empirical effective FWHM measured on stable nights and quoted above, the fractional uncertainties are
$\epsilon_{450} = 12\%$ and $\epsilon_{850} = 4.2\%$.
These give an uncertainty in the shift $\Delta\beta$ of the dust emissivity index $\beta$ (equation~\ref{eq:beta_bias}) of
$\sigma(\Delta\beta) = [\ln(\nu_{450}/\nu_{850})]^{-1}\,
(\epsilon_{450}^{2} + \epsilon_{850}^{2})^{1/2} = 0.20$.
Propagating instead the full FCF tolerances of 17\% (450\,$\mu$m)
and 7\% (850\,$\mu$m) adopted in sub-subsection~\ref{sss:discussion_fcf} would give the more conservative upper bound $\sigma(\Delta\beta) \lesssim 0.29$.
We use the former below and retain the latter as a bound.
 
The compound-beam solid angles at 450 and 850\,$\mu$m differ by a factor of two, so the two calibration conventions are not interchangeable for extended emission.
However, the difference is not the one that a diffraction-limited beam would predict.
For a Gaussian beam $\Omega \propto \lambda^{2}$, which would give
$\Omega^{\rm FCF}_{450}/\Omega^{\rm FCF}_{850} = (\nu_{850}/\nu_{450})^{2}
\approx 0.28$;
the measured ratio is $0.51$, much closer to the scaling of $\lambda^{1}$.
The excess arises because the JCMT 450\,$\mu$m beam is compound: a substantial fraction of the power lies in the broad error beam, which inflates $\Omega^{\rm FCF}_{450}$ above its diffraction-limited value.
The convention for beam area changes the predicted bias in the spectral index: a $\lambda^{2}$ ratio would imply $\Delta\beta = -2.00$, while the ratio for the compound beam implies the $\Delta\beta = -1.05$ derived below.
 
Using FCF$_{\rm arcsec}$ accounts for this wavelength dependence and gives a true surface brightness.
If the beam were perfectly known, the bias could equally well be absorbed by convolving both bands to a common resolution and rescaling by the common beam solid angle, and the choice of calibration convention would be immaterial.
However, the 450\,$\mu$m beam is compound, and its effective solid angle differs from the diffraction-limited value.
The conversion from Jy\,beam$^{-1}$ to surface-brightness units therefore requires a correction for the beam area whose value is uncertain for the compound 450\,$\mu$m beam.
Working in surface-brightness units from the outset removes the need to assume this correction.
Applying FCF$_{\rm peak}$ to extended emission without accounting for the beam solid angle instead biases the spectral analysis.

Table~\ref{tab:omega_conventions} summarizes the three conventions for beam solid angle that we use for different purposes throughout this paper.

\begin{deluxetable}{lccc}
\tablecaption{Beam solid-angle conventions used in this paper\label{tab:omega_conventions}}
\tablehead{
\colhead{Symbol} & \colhead{Convention} &
\colhead{450\,$\mu$m} & \colhead{850\,$\mu$m}
}
\startdata
$\Omega^{\rm FCF}_\lambda$ & native (compound) & 122 & 239 \\
$\Omega^{\rm mat}_\lambda$ & Gaussian reference & 180 & 180 \\
$\Omega^{\rm G}_\lambda$   & Gaussian equivalent & 84 & 180 \\
\enddata
\tablecomments{
All areas are in arcsec$^{2}$.
The areas $\Omega^{\rm FCF}_\lambda$ derived from calibration are defined by equations~(\ref{eq:R450})--(\ref{eq:R850}) and used to convert archival beam-averaged data to surface brightness (sub-subsection~\ref{ass:NoiseEstimation}).
The Gaussian-equivalent areas are $\Omega^{\rm G}_\lambda=1.133\,\theta_{\rm eff}^{2}$, with the two-component effective FWHMs $\theta_{\rm eff}=8\farcs6$ and $12\farcs6$ from equation~(12) of \citet{Mairs2021}.
These areas enter the scaling test of the Noise Equivalent Flux Density (NEFD) in sub-subsection~\ref{ass:NEFDTransmission}.
They are 31\% and 25\% smaller than $\Omega^{\rm FCF}_\lambda$, respectively, because the two-Gaussian representation omits far wings included in the aperture measurement used to derive FCF$_{\rm arcsec}$.
For the common $12\farcs6$ effective resolution (sub-subsection~\ref{sss:discussion_beam}), we adopt the Gaussian reference area $\Omega^{\rm mat}_\lambda=\pi(12\farcs6)^2/[4\ln2]\simeq180=\Omega^{\rm G}_{850}$, distinct from $\Omega^{\rm FCF}_{850}=239$.
Harmonization matches the modeled 450\,$\mu$m response to the modeled 850\,$\mu$m compound beam; it does not make either beam a single Gaussian.
Thus $\Omega^{\rm mat}_\lambda$ is a reference convention, not a solid angle measured after harmonization.
Conversions from Jy\,beam$^{-1}$ must use an area consistent with the normalization of the input flux calibration.
}
\end{deluxetable}
 
\subsubsection{Analytical Propagation into Spectral Indices}
\label{ss:fcf_propagation}
If FCF$_{\rm peak}$ is incorrectly applied to extended emission,
the intensity ratio between 450 and 850\,$\mu$m is biased by a factor
$\Omega^{\rm FCF}_{450}/\Omega^{\rm FCF}_{850}$.
The bias propagates directly into the total-intensity spectral index
$\alpha \equiv \log(I_{450}/I_{850}) / \log(\nu_{450}/\nu_{850})$:
\begin{equation}
\alpha_{\rm wrong}
= \alpha_{\rm true}
+ \frac{\log(\Omega^{\rm FCF}_{450}/\Omega^{\rm FCF}_{850})}{\log(\nu_{450}/\nu_{850})}
\approx \alpha_{\rm true} - 1.05.
\label{eq:alpha_bias}
\end{equation}
For modified-blackbody emission $S_\nu \propto \nu^{\beta} B_\nu(T_{\rm d})$, with flux density $S_\nu$, Planck function $B_\nu$, and dust temperature $T_{\rm d}$, the spectral index $\alpha$ of total intensity differs from the dust opacity spectral index $\beta$ by the temperature-dependent Planck shape factor $C(T_{\rm d})$:
\begin{equation}
\beta = \alpha - C(T_{\rm d}),
\qquad
C(T_{\rm d}) \equiv
\frac{\log\left[B_{\nu_{450}}(T_{\rm d})/B_{\nu_{850}}(T_{\rm d})\right]}
     {\log(\nu_{450}/\nu_{850})},
\label{eq:planck_correction}
\end{equation}
identical to Paper~S2 equation~(\ref{S2-eq:planck_correction}).
The observed slope of the intensity spectrum, $\alpha$, is thus the opacity slope, $\beta$, plus the slope of the Planck function, $C(T_{\rm d})$.
The slope of the Planck function approaches 2 in the Rayleigh--Jeans limit and falls well below 2 at the dust temperatures of molecular clouds ($C \approx 1.01$ at $T_{\rm d} = 15$\,K and $1.45$ at $25$\,K).
We refer to obtaining $\beta$ by subtracting $C(T_{\rm d})$ from $\alpha$ as the $C(T_{\rm d})$ correction.
Because the calibration error does not affect $C(T_{\rm d})$, the bias in $\alpha$ propagates unchanged to $\beta$:
\begin{equation}
\beta_{\rm wrong} = \beta_{\rm true} - 1.05.
\label{eq:beta_bias}
\end{equation}
 
\subsubsection{Application to the Observed Sample}
\label{ss:fcf_application}
We illustrate the predicted bias using the sources listed in the table of spectral indices in subsection~\ref{S2-ss:S2_SpIdx} of Paper~S2.
The two distributions have median values of $\beta_{\rm true} = 2.0 \pm 0.1$ and
$\beta_{\rm wrong} = 1.0 \pm 0.1$.
Their unrounded medians are $2.008$ and $0.962$, respectively, giving
$\beta_{\rm wrong}-\beta_{\rm true}=-1.046$.
The predicted offset is $\Delta\beta = -1.05 \pm 0.20$ (figure~\ref{afig:beta_histogram}); its quoted uncertainty propagates the repeatability of the effective beam solid angles given in sub-subsection~\ref{ss:fcf_bias}.
Because point-source calibration shifts $\beta$ rigidly, this comparison illustrates the magnitude of the bias rather than providing an independent test of it.
We quote $\beta$ to two significant figures because systematic uncertainty limits precision.

We evaluated the $C(T_{\rm d})$ correction using an adopted dust temperature.
We assumed $15$\,K for low-mass star-forming regions and $25$\,K for high-mass star-forming and Central Molecular Zone (CMZ) regions, with nine exceptions.
For six sources, we adopted temperatures from the \textit{Herschel} Gould Belt Survey \citep[HGBS;][]{Andre2010}, taking the median over the $12\arcsec$ pixels with positive $\alpha$ in each source.
We used the HGBS Ophiuchus (L1688) map for L1688 Ophiuchus~A \citep{Ladjelate2020} and the HGBS Orion~A map for OMC~1, OMC~2, OMC~3, and Orion~A Fields~4 and~5, obtained from the HGBS archive (earlier HGBS maps of the central and southern parts of Orion~A were presented by \citealt{Roy2013} and \citealt{Polychroni2013}).
For the other three, we used published temperatures instead: $15$\,K for the two cold infrared-dark clouds and $20$\,K for the Brick.

Changing the adopted temperatures by $\pm 3$\,K moves the median $\beta$ by $\mp 0.1$.
The sensitivity is strongly asymmetric: $\partial\beta/\partial T_{\rm d} = -0.078$\,K$^{-1}$ at $15$\,K, compared to $-0.025$\,K$^{-1}$ at $25$\,K.
The quoted $\pm 0.1$ is this systematic term; the bootstrap statistical uncertainty on the median is comparable ($0.13$).

The median is $2.06$ for the statistical sample of 35 sources and $2.01$ for the 36 sources listed in table~\ref{S2-tbl:SpIdx_Global} of Paper~S2; the two samples agree at this precision.
The distributions obtained with the correct and incorrect calibrations have the same median absolute deviation (MAD), ${\rm MAD}(\beta) = 0.33$, because the calibration error enters $\beta$ as a rigid offset.
The measured shift is therefore $3.2\,{\rm MAD}$, well outside the dispersion among sources, while the $\pm0.20$ tolerance on the predicted offset corresponds to $0.61\,{\rm MAD}$.
Even with the $C(T_{\rm d})$ correction evaluated from the full Planck function, this shift drives two sources (an infrared dark cloud of G202.3+2.5, and the Rosette Molecular Cloud) into the formally unphysical regime $\beta < 0$.
This shift also produces systematically low values of $\beta$ that would bias the physical interpretation of the dust properties: six sources ($17\%$) fall below $\beta_{\rm wrong} = 0.5$.

The correctly calibrated distribution is centered on values broadly consistent with thermal emission from grains in dense molecular clouds, including coagulated grains with thin ice mantles \citep{Ossenkopf1994,Juvela2015,Kohler2015}.
Nevertheless, we interpret the high-$\beta$ tail with caution, because values of $\beta \gtrsim 3$ lie at the limit supported by the present measurement uncertainties (see also Paper~S2 subsection~\ref{S2-ss:S2_SpIdx}).
The biased distribution instead shifts toward values characteristic of protoplanetary disks ($\beta \lesssim 1$; \citealt{Testi2014}).
Calibration artifacts could therefore be mistaken for physical grain evolution.

We must therefore distinguish FCF$_{\rm peak}$ from FCF$_{\rm arcsec}$ to interpret the measurements reliably, because the bias may remain undetected without independent checks of the calibration.

We calculated all statistics in this paper from full-precision values and rounded them only for presentation.
The medians quoted here differ slightly ($\sim$0.01) from those in the table of spectral indices in subsection~\ref{S2-ss:S2_SpIdx} of Paper~S2.
The differences arise because the table rounds $\alpha$ to two decimal places before calculating $\beta$.

We conclude that the use of FCF$_{\rm arcsec}$ and mJy\,arcsec$^{-2}$ units is required for a consistent calibration of the surface brightness of extended emission.
Incorrect application of FCF$_{\rm peak}$ introduces a deterministic wavelength-dependent bias that
(1) changes the spectral indices by $\Delta\beta = -1.05$,
(2) systematically lowers the inferred values of $\beta$, and
(3) can lead to physically misleading interpretations even when the results appear reasonable.
We therefore recommend that the adopted calibration be taken into account when deriving and interpreting physical properties from such observations.

\subsubsection{Summary of Technical Validation 1}
\label{ss:fcf_summary}
 
\textit{This section establishes Pillar~1 of the technical framework.}
We calibrated all data products used in the scientific analyses presented in Papers~S1 and~S2 onto a common surface-brightness scale.
The common scale avoids the deterministic shift of $\Delta\beta = -1.05$ that applying FCF$_{\rm peak}$ without the required beam-area conversion would introduce for extended emission.
The remaining uncertainty in the photometric calibration enters the budget of systematic uncertainties in subsection~\ref{sec:uncertainty} through the FCF tolerances for each band.
These tolerances cancel in the quantities $P_{450}/P_{850}$ and $\gamma - \alpha$ derived from the two bands.
In conclusion, we regard the quantities $\alpha$, $\gamma$, and $P_{450}/P_{850}$ defined in sub-subsection~\ref{ss:fcf_dualbandquantities} as physically meaningful for the molecular cloud emission targeted by Papers~S1 and~S2.

\subsection{Technical Validation 2: Noise Characteristics and Selection of Polarization Data Based on Stokes $I$ Intensity}
\label{ss:noiseI}
\label{as:IinvDI}%
 
\subsubsection{Motivation and overview}
\label{ss:noiseI_motivation}
 
To identify the polarization measurements that we can use reliably in the comparison between the bands, we next quantify the noise properties after establishing the calibration in subsection~\ref{ss:flux}.
In this subsection and the next, we construct and validate a noise model that links Stokes-$I$ intensity, polarized intensity, and the precision of the polarization angles.
We use this model to define the Stokes-$I$ selection thresholds adopted in Papers~S1 and~S2.

\begin{figure*}[htp]
\centering
\includegraphics{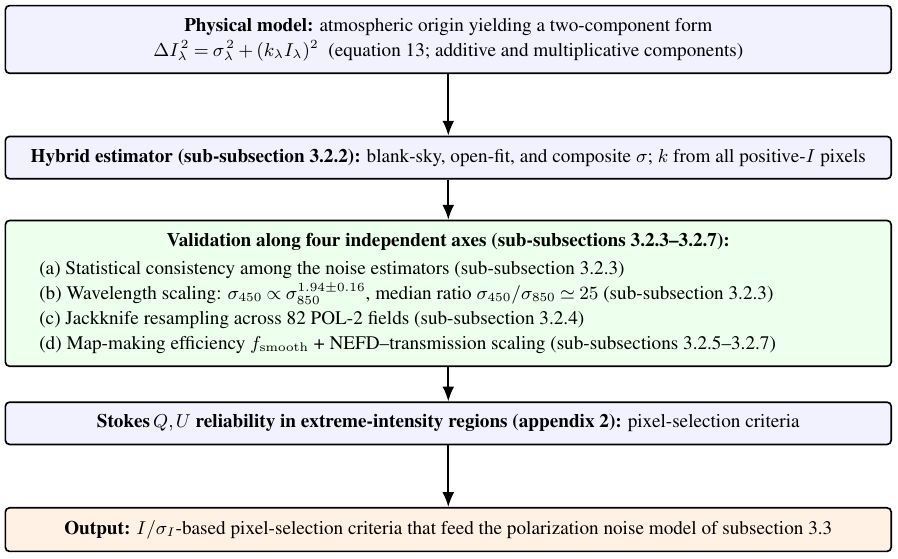}
\caption{
\textbf{Roadmap for the Stokes\,$I$ noise characterization (subsection~\ref{ss:noiseI}, Pillar~2a).}
The atmospheric origin of POL-2 noise leads to a two-component model with additive ($\sigma_\lambda$) and multiplicative ($k_\lambda I_\lambda$) components.
We estimate $\sigma_\lambda$ from distinct intensity regimes and $k_\lambda$ from a fit to all pixels with positive intensity (sub-subsection~\ref{ass:NoiseEstimation}).
We validate both parameters through four independent tests: consistency among the estimators, wavelength scaling, jackknife stability across 82 POL-2 fields, and map-making/atmospheric diagnostics (sub-subsections~\ref{ass:NoiseResults}--\ref{ass:NEFDTransmission}).
We then define a criterion based on $I/\sigma_I$ for selecting the pixels used in the polarization noise model of subsection~\ref{ss:noisemodel}.
Appendix~\ref{app:QU_diagnostics} provides supporting diagnostics of the reliability of Stokes\,$Q,U$ for two exceptionally bright 850\,\micron\ fields, W51 and DR21.
}
\label{fig:T_S4_roadmap}
\end{figure*}

We develop Pillar~2, {\em Noise characterization}, in two complementary parts (figure~\ref{fig:T_pillar_overview} and table~\ref{tbl:T_pillar_index}).
Here we characterize additive Stokes\,$I$ noise across the full archive of 174 POL-2 fields.
In subsection~\ref{ss:noisemodel}, we develop the two-component noise model for Stokes\,$Q$ and $U$.
Figure~\ref{fig:T_S4_roadmap} shows the organization of this subsection.
Together, the two parts establish that the retained pixels are statistically reliable for the subsequent analyses of the two bands.
 
Above a threshold that depends on the source, the uncertainty $\Delta I_\lambda$ of the Stokes~$I$ intensity $I_\lambda$ at wavelength $\lambda$ increases monotonically with $I_\lambda$.
This increase marks a transition from the regime dominated by atmospheric noise to the regime in which the noise depends on the signal.
This behavior originates from small fluctuations in atmospheric transmission along the line of sight.
The observed intensity toward a source at a ground-based submillimeter telescope can be written as
\begin{equation}
I_{\rm obs}(t) = I_0\,e^{-\tau(t)} + \int_0^s B_\nu(T_{\rm atm})\,e^{-\tau(z,t)}\,dz,
\label{eq:Iobs}
\end{equation}
where $I_0$ is the intrinsic astronomical signal, $\tau(t)$ is the time-variable atmospheric opacity, $B_\nu(T_{\rm atm})$ is the Planck function at atmospheric temperature, $s$ is the length of the path through the atmosphere, and $z$ is the line-of-sight depth coordinate (with $z=0$ at the telescope aperture and $z=s$ at the top of the atmosphere).
In the optically thin limit ($\tau \ll 1$), the exponential term can be linearized as
\begin{equation}
I_{\rm obs}(t) \simeq I_0\,[1 - \tau(t)] + I_{\rm atm}(t),
\label{eq:Iobs_linear}
\end{equation}
where $I_{\rm atm}(t)$ represents the additive component of atmospheric thermal emission.
We argue that the residual fluctuations in the POL-2 maps arise primarily from imperfect correction of the time-variable transmission term $e^{-\tau(t)}$.
The common-mode part of the additive atmospheric emission $I_{\rm atm}(t)$ is largely removed during map making with \texttt{pol2map}, although residual atmospheric emission may still contribute to $\epsilon(t)$.
After applying an extinction correction using the estimated zenith opacity $\hat{\tau}(t)$, the measured intensity becomes
\begin{equation}
I_{\rm corr}(t) = I_{\rm obs}(t)\,e^{\hat{\tau}(t)} 
  \simeq I_0 \,[1 + \delta g(t)] + \epsilon(t),
\label{eq:Icorr}
\end{equation}
where $\delta g(t) \equiv e^{\hat{\tau}(t)-\tau(t)} - 1$ represents the residual multiplicative fluctuations of the gain, and $\epsilon(t)$ denotes the additive noise of photon statistics, detector readout, and residual sky emission.
The corresponding total uncertainty can be written as
\begin{equation}
\Delta I_\lambda^2 = 
 {\rm Var}[\epsilon(t)] + {\rm Var}[\delta g(t)]\,I_\lambda^2
 = \sigma_\lambda^2 + (k_\lambda I_\lambda)^2,
\label{eq:DI_model}
\end{equation}
which naturally reproduces the empirical model used throughout this study.
The first term, $\sigma_\lambda$, includes all additive contributions in $\epsilon(t)$---photon statistics, detector readout, and residual sky emission---and gives an approximately constant $\Delta I_\lambda$ at low $I_\lambda$.
At 450\,\micron\ it is dominated by the atmospheric terms, whereas at 850\,\micron\ a component that does not depend on atmospheric transmission is larger (sub-subsection~\ref{ass:NEFDTransmission}).
\citet{Holland2013} found the on-sky sensitivity of the SCUBA-2 transition-edge-sensor (TES) arrays in good weather to be at least a factor of two worse than predicted from a model of the instrument, telescope, and sky.
A major contributor was dark noise in the detectors, which was higher than expected.
We measure the two contributions in the Noise Equivalent Flux Density (NEFD)--transmission test of sub-subsection~\ref{ass:NEFDTransmission}.
This test fits the instrumental floor $B$ of equation~(\ref{eq:nefd_model}) separately from the sky-limited term $A/T$, where $T$ is the atmospheric transmission.

The second term introduces a linear increase in $\Delta I_\lambda$ with $I_\lambda$ once the residual gain fluctuation $(k_\lambda I_\lambda)$ exceeds $\sigma_\lambda$.
The transition occurs at
\begin{equation}
I_{\rm trans} \simeq \frac{\sigma_\lambda}{k_\lambda},
\label{eq:Itrans}
\end{equation}
which varies among sources depending on their intrinsic brightness and atmospheric stability.
At low intensities, the uncertainty is dominated by additive atmospheric noise ($\sigma_\lambda$).
At high intensities, the uncertainty is dominated by multiplicative transmission fluctuations ($k_\lambda I_\lambda$), which act as an effective gain-noise term.
The fitted parameters $\sigma_\lambda$ and $k_\lambda$ thus provide quantitative diagnostics of atmospheric stability and calibration precision.
From the observed consistency of this model across all targets, we conclude that it is applicable to the noise characterization of the POL-2 data.

Throughout this paper, we use the term noise floor for the asymptotic minimum noise level that remains as the integration time and the averaging of signals increase.
We distinguish the noise floor from a raw root-mean-square (RMS) estimate and identify the source of each floor as {\em additive}, {\em instrumental}, {\em systematic}, {\em empirical}, {\em effective}, or {\em map-domain}, as appropriate.

\subsubsection{Implementation of the Hybrid Noise Estimation}
\label{ass:NoiseEstimation}
We implemented the hybrid estimator shown in box~2 of figure~\ref{fig:T_S4_roadmap} by combining estimates of $\sigma_\lambda$ from different intensity ranges with a fit for $k_\lambda$ in the two-component model of equation~(\ref{eq:DI_model}).
Our aim is to characterize the uncertainty in Stokes~$I$ across both faint extended emission and bright cloud structures.
The amount of emission-free sky differs between POL-2 fields, so we cannot estimate the scale of the additive noise in the same way for every target.

First, we estimated the additive noise floor $\sigma_\lambda$ from emission-free pixels with Stokes-$I$ signal-to-noise ratios $0 < S/N < 2$.
These regions of low intensity represent the regime in which atmospheric noise dominates and $\Delta I_\lambda$ remains approximately constant.
We excluded outliers through iterative $\sigma$ clipping.
At each iteration, we excluded pixels $3\sigma$ or more from the median $\Delta I_\lambda$, where $\sigma$ is 1.48 times the median absolute deviation.
We repeated the procedure three times, stopping early when fewer than 20 pixels remained.
The median of the surviving values $\Delta I_\lambda$ provides a robust estimate $\sigma_{\rm blank}$.
When fewer than 20 emission-free pixels are available, the procedure instead uses the second percentile of the full distribution of $\Delta I_\lambda$.
However, every field has at least 20 such pixels at both wavelengths.

The composite estimator, denoted $\sigma_{\rm comb}$, is the median of four estimates made on the same map.
The four estimates are $\sigma_{\rm blank}$; the median of three determinations that do not use the blank-sky mask; the estimate $\sigma_{\rm open}$ from the transition regime described below; and the intercept of a straight-line fit $\Delta I_\lambda = a\,I_\lambda + b$ to the pixels with $0.5 < S/N < 3$.
Estimates that are not finite and positive are excluded from the median.
The three determinations are (i) the second percentile of the full distribution of $\Delta I_\lambda$, (ii) a two-component fit of equation~(\ref{eq:DI_model}) restricted to $S/N > 5$, and (iii) the modal $\Delta I_\lambda$ among pixels below the 30th percentile of $I_\lambda$.
When 100 or fewer pixels met the criterion in (iii), we replaced that determination with the 10th percentile of the full distribution of $\Delta I_\lambda$.
Taking the median of several estimates rather than adopting any one of them keeps $\sigma_{\rm comb}$ usable for fields that contain too few emission-free pixels to support $\sigma_{\rm blank}$.

Second, we estimated the multiplicative noise coefficient $k_\lambda$ by fitting equation~(\ref{eq:DI_model}) to all pixels with positive $I_\lambda$, which span both the additive and the signal-dependent regimes.
We minimized a Huber loss of the residuals with the bounded quasi-Newton method L-BFGS-B, imposing $\sigma_\lambda > 0$ and $0 < k_\lambda \leq 1$.
We instead adopted the posterior medians of a Markov chain Monte Carlo (MCMC) fit of the same model with \texttt{emcee} \citep{Foreman-Mackey2019} for six fields: Cepheus~A, G202.3+2.5 (mosaicked field), Perseus~IC348, Monoceros~R2, NGC~891, and Serpens~Main (mosaicked field).

Third, we turned to the transition regime, where the two noise components contribute comparably.
We fitted the same functional form to pixels with $5 < {\rm S/N} < 30$ without prior constraints, obtaining the estimates $\sigma_{\rm open}$ and $k_{\rm open}$ from this open fit.
This fit at intermediate intensities enters $\sigma_{\rm comb}$ and is the preferred estimate when blank-sky coverage is insufficient [case~(i) below].

Fourth, we selected among the candidate estimates separately for each field and band.
We base this selection on three quantities measured from the map itself, because the reliability of an estimator depends on the available blank sky and the contribution of multiplicative noise.
The blank-sky fraction $f_{\rm blank}$ is the fraction of pixels with $I_\lambda/\Delta I_\lambda < 2$, and the high-$S/N$ fraction $f_{\rm hiSN}$ is the fraction with $I_\lambda/\Delta I_\lambda > 10$.
The multiplicative ratio is $r_{\rm mult} = k_\lambda \widetilde{I}_{\lambda,80}/\sigma_{\rm blank}$, where $\widetilde{I}_{\lambda,80}$ is the median intensity of pixels above the 80th percentile of $I_\lambda$.

We evaluate the following cases in order and adopt the first that applies.
(i) If $f_{\rm blank} < 0.1$, the extended source fills the field, and we prefer the estimate $\sigma_{\rm open}$ from the transition regime because the field contains too little blank sky to determine a blank-sky floor.
(ii) If $r_{\rm mult} > 0.5$ and $k_\lambda > 0.01$, multiplicative noise dominates, and we prefer the full two-component fit of equation~(\ref{eq:DI_model}).
(iii) If $f_{\rm hiSN} < 0.05$, additive noise dominates and we prefer $\sigma_{\rm blank}$.
(iv) If $f_{\rm blank} > 0.3$ and $r_{\rm mult} < 0.1$, the source is compact, with ample blank sky and negligible multiplicative noise, and we again prefer $\sigma_{\rm blank}$.
(v) Otherwise, we classify the field as hybrid and prefer the composite estimate $\sigma_{\rm comb}$.

We accept a candidate only if it is finite and positive.
If the preferred estimate is not finite and positive, we adopt the first finite and positive estimate among alternative estimates, taken in a fixed order for each regime (for a hybrid field, $\sigma_{\rm open}$ and then $\sigma_{\rm blank}$).
We do not impose absolute limits on $\sigma_\lambda$ during this selection.
Apart from positivity, the fits bound $\sigma_\lambda$ only by values taken from the field's own $\Delta I_\lambda$ distribution, and the open fit is unbounded.
 
For the 174 POL-2 fields, this rule selects $\sigma_{\rm comb}$ for 107 fields at 850\,\micron\ and 89 fields at 450\,\micron, and $\sigma_{\rm blank}$ for the remaining $67$ and $85$ fields, respectively.
When two candidate estimates coincide numerically with $\sigma_{\rm comb}$, we count the field in the category for the composite estimate.
The additive noise scale $\sigma_{\rm final}$ adopted by the selection rule spans $0.0024$--$0.059$\,mJy\,arcsec$^{-2}$ at 850\,\micron\ and $0.010$--$6.1$\,mJy\,arcsec$^{-2}$ at 450\,\micron, with medians of $0.0079$ and $0.21$\,mJy\,arcsec$^{-2}$.
These ranges are results of the procedure, not imposed constraints.
We do not reject a field or replace an estimate solely because of its absolute noise level.

To visualize the fit, we mark four classes of pixels, Conditions~A--D.
Condition~A retains pixels whose noise is at or below the 99.99th percentile of the map's $\Delta I_\lambda$ distribution.
This extreme cut excludes the few catastrophic outliers, such as bad bolometer columns and map-edge pixels with almost no integration time.
It leaves the physical noise distribution unchanged.
Condition~B retains pixels within $1'$ of the center of the map, where the integration time is the longest and most uniform.
Condition~C retains the brightest 5\% of pixels in $I_\lambda$, and Condition~D retains pixels with $I_\lambda/\Delta I_\lambda \geq 30$.
These two conditions isolate the regime in which the multiplicative term dominates.
The scatter panels highlight the intersection A$\cap$C, and the noise maps highlight the intersection A$\cap$B$\cap$C$\cap$D.
The fit itself uses all pixels with positive $I_\lambda$.
Condition~A pixels can lie above the fitted $\sigma_\lambda$ because the estimator describes the additive {\em floor}, not the mean of the distribution.
The scatter of individual low-$I$ pixels above this floor is therefore expected and does not indicate a failed fit.

\begin{figure*}[tp]
\centering
\includegraphics[width=0.95\textwidth]{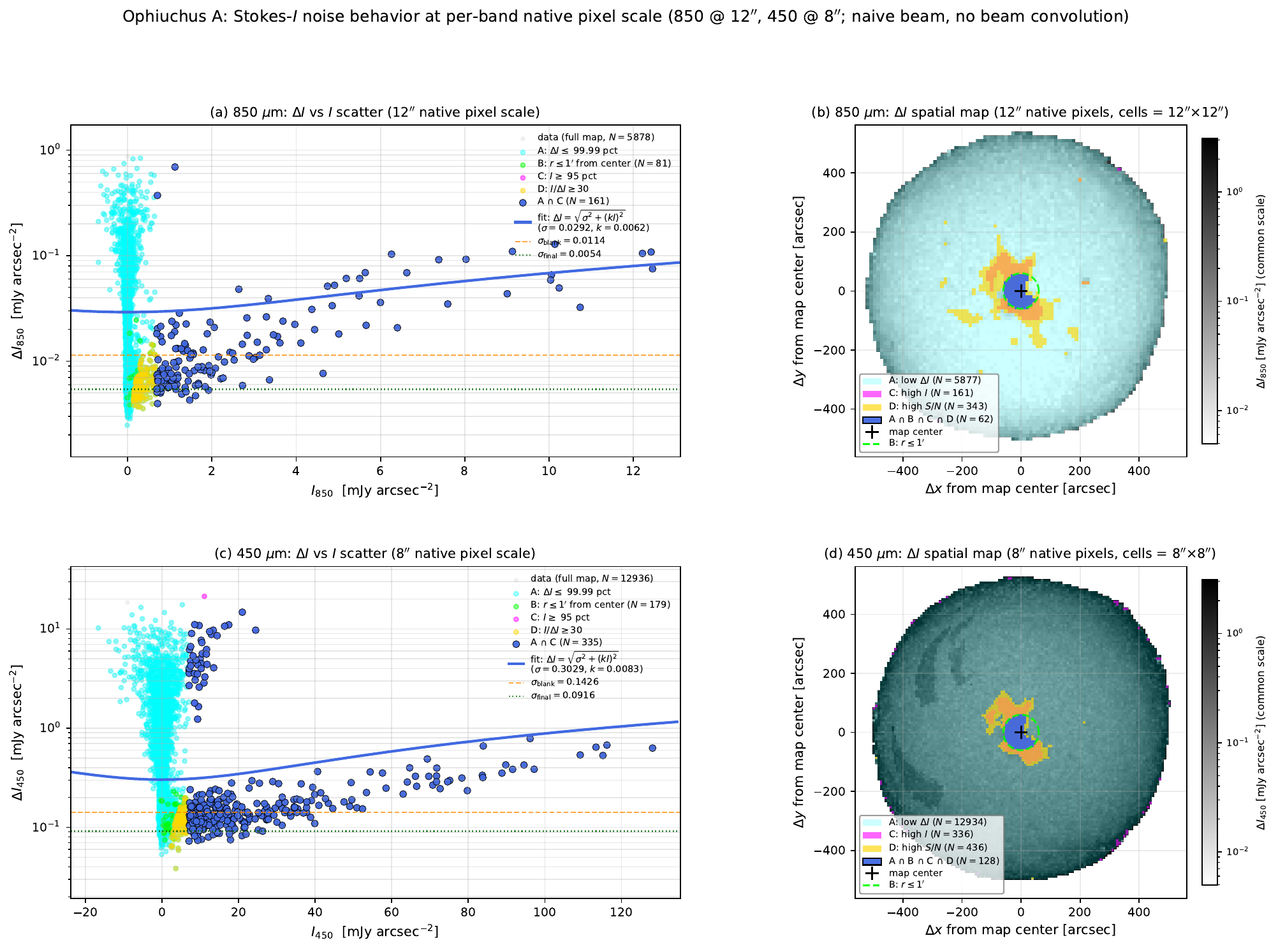}
\caption{
Hybrid Stokes-$I$ noise analysis for Ophiuchus~A using native beams without beam convolution and native pixels of $12\arcsec$ at 850\,$\mu$m and $8\arcsec$ at 450\,$\mu$m.
\textbf{(a,c)} Noise scatter $\Delta I_\lambda$ versus $I_\lambda$ at 850 and 450\,$\mu$m, respectively.
\textbf{(b,d)} Corresponding noise maps, with one rendered cell for each native pixel and a shared $\Delta I$ color scale.
Gray points in the scatter panels show all finite pixels, including pixels with negative $I$ and the tail from the bright source.
Conditions~A--D of sub-subsection~\ref{ass:NoiseEstimation} are highlighted in cyan, lime, magenta, and gold, respectively.
Royal-blue points show A$\cap$C: high-$I$ pixels with $\Delta I$ at or below the 99.99th percentile.
The heavy royal-blue curve is the two-component model $\Delta I_\lambda=\sqrt{\sigma_\lambda^2+(k_\lambda I_\lambda)^2}$, fitted with a robust Huber loss to all pixels with positive $I$ (equation~\ref{eq:DI_model}).
We discuss the relation of this curve to the bimodal $\Delta I$ distribution after figure~\ref{afig:IvsDI_OMC1}.
Horizontal dashed and dotted lines mark the blank-sky floor $\sigma_{\rm blank}$ and the adopted additive noise scale $\sigma_{\rm final}$.
In panels~(b,d), the conditions are overlaid in the same colors on the grayscale noise maps.
Royal-blue cells mark the pixels that satisfy all four conditions (A$\cap$B$\cap$C$\cap$D).
The dashed green circle bounds Condition~B ($r\leq1'$), and the cross marks the map center.
Paper~S2 figure~\ref{S2-afig:Bmap_OphiuA} shows the corresponding map of polarization segments.
}
\label{afig:IvsDI}
\end{figure*}

\begin{figure*}[tp]
\centering
\includegraphics[width=0.95\textwidth]{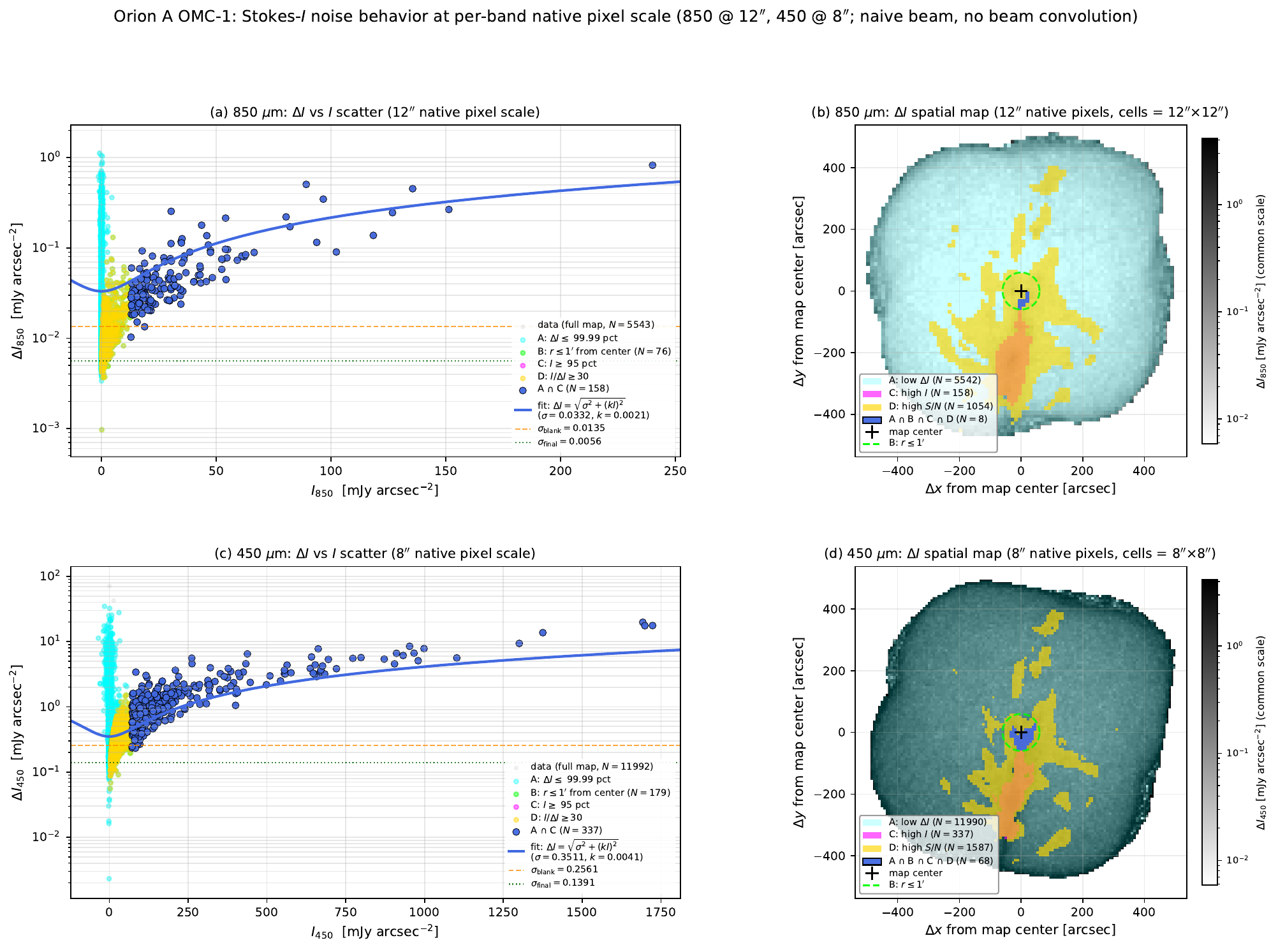}
\caption{
Same as figure~\ref{afig:IvsDI} but for the high-mass star-forming region Orion~A OMC-1.
The brighter central source produces a substantially larger population of pixels with high $S/N$ (Condition~D, gold cells in panels~b,~d) extending over a wider spatial area than in Ophiuchus~A.
However, the order of $\sigma$ values at both wavelengths remains comparable across the two sources.
Paper~S2 figure~\ref{S2-afig:Bmap_OriAS1} shows the corresponding map of polarization segments.
}
\label{afig:IvsDI_OMC1}
\end{figure*}

Figures~\ref{afig:IvsDI} and~\ref{afig:IvsDI_OMC1} illustrate hybrid noise estimates for two bright sources in the hybrid regime, Ophiuchus~A and Orion~A OMC-1.
Both examples use the native beam and pixel scale in each band (12$''$ at 850\,$\mu$m and 8$''$ at 450\,$\mu$m).
The map panels share a $\Delta I$ color scale to show the higher noise at 450\,$\mu$m directly.
These examples demonstrate that the two-component model captures the qualitative $\Delta I_\lambda$--$I_\lambda$ behavior from blank-sky pixels through source pixels with high $S/N$.
We discuss the conservative choice of $\sigma$ in the fit shown in the scatter panels after the figures.

To compare the two wavelengths, we used consistent flux calibration for all measurements of intensity and noise.
We converted raw detector units (pW) to surface brightness (mJy\,arcsec$^{-2}$) with the arcsecond FCFs of the post-2018 JCMT calibration scale \citep{Mairs2021}.
We multiplied these FCFs by the POL-2 attenuation factors of $1.96 \pm 0.023$ at 450 \micron\ and $1.35 \pm 0.008$ at 850 \micron\ to obtain 7.59 and 2.79~Jy\,pW$^{-1}$\,arcsec$^{-2}$, respectively (sub-subsection~\ref{ss:fcf_calibration}).
To maintain internal consistency among data from different preprocessing pipelines, we converted archival data in beam-averaged units to surface brightness using the solid angles of the native compound beams, $\Omega^{\rm FCF}_{450} = 122$ and $\Omega^{\rm FCF}_{850} = 239$ arcsec$^2$ (table~\ref{tab:omega_conventions}).

We scaled the noise measured at the native pixel size to standardized pixel sizes using $\sigma_{\rm scaled} = \sigma_{\rm native} \times \sqrt{A_{\rm target}/A_{\rm native}}$, where $A$ is the pixel area.
The statistics for 174 fields in sub-subsection~\ref{ass:NoiseResults} use maps made with 12\arcsec\ pixels at both wavelengths, so they do not need such scaling.
The square-root scaling assumes uncorrelated noise.
This assumption is valid in the regime dominated by atmospheric noise, which characterizes the majority of our polarization measurements.
The resulting parameters $(\sigma_\lambda, k_\lambda)$ quantitatively characterize the noise behavior of each map and provide the empirical basis for the subsequent analysis.
The selected additive scale and signal-dependent term describe different limits on the Stokes-$I$ measurement.
They inform the selection on intensity and the uncertainty estimates.
The uncertainties of the polarization components require the separate treatment developed below.

\paragraph{Interpretation of the conservative-$\sigma$ fit}
We use conservative curves of the noise model in panels~(a)~and~(c) of figures~\ref{afig:IvsDI} and~\ref{afig:IvsDI_OMC1}.
The fitted envelope lies above the empirical blank-sky floor $\sigma_{\rm blank}$ and the adopted additive noise scale $\sigma_{\rm final}$, shown by the horizontal lines.
The fitted curve does not follow the lower envelope of the high-$S/N$ points in the centers of the bright sources.

The well-covered center of a POL-2 field and its sparsely sampled outskirts can have different uncertainties, even where both contain little astronomical emission.
Panels~(a)~and~(c) show that the distribution of $\Delta I_\lambda$ at low $I_\lambda$ is heterogeneous.
A tight cluster with low $\Delta I$ from pixels in the central source coexists with a sparse cluster with high $\Delta I$ from map edges and areas of low coverage.
A single pair of parameters in $\Delta I_\lambda = \sqrt{\sigma_\lambda^2 + (k_\lambda I_\lambda)^2}$ cannot describe both populations separately.
The robust fit to all pixels with positive $I$ therefore gives a $\sigma_\lambda$ that reflects the median of the bimodal $\Delta I$ distribution rather than either floor alone.
The fitted envelope and the noise scale used to select pixels in which astronomical emission is detected above the adopted $S/N$ threshold therefore serve different purposes and must be reported separately.
We interpret faint structure using the uncertainty appropriate to its map coverage, and do not treat the conservative fitted envelope as a uniform sensitivity across the field.

We retained the resulting $(\sigma_\lambda, k_\lambda)$ as the standard pair of noise parameters for each field throughout the paper series.
We treated $\sigma_{\rm blank}$ and $\sigma_{\rm final}$, shown as horizontal reference lines, as the relevant baselines at low $I$ for the populations of blank-sky and high-quality pixels, respectively.
Sources in other noise regimes, such as filaments dominated by additive noise (Taurus~B211/B213) and fields limited by mosaic processing (Serpens~Main), show the same offset of the fitted $\sigma_\lambda$ under our uniform classification.
We use these parameters unchanged in the Stokes-$Q,U$ analysis of subsection~\ref{ss:noisemodel} and the survey statistics of sub-subsection~\ref{ass:NoiseResults}.

Estimating image noise in ground-based polarimetric maps made with TES bolometers is difficult.
We therefore assess the method by comparison with the official JCMT SCUBA-2/POL-2 sensitivity calculator \citep{EAOSCUBA2ITC} and find general agreement, using independently inferred degradation factors for map making and smoothing (appendix~\ref{app:itc_noise_comparison}).

\begin{figure*}[tp]
\centering
\includegraphics{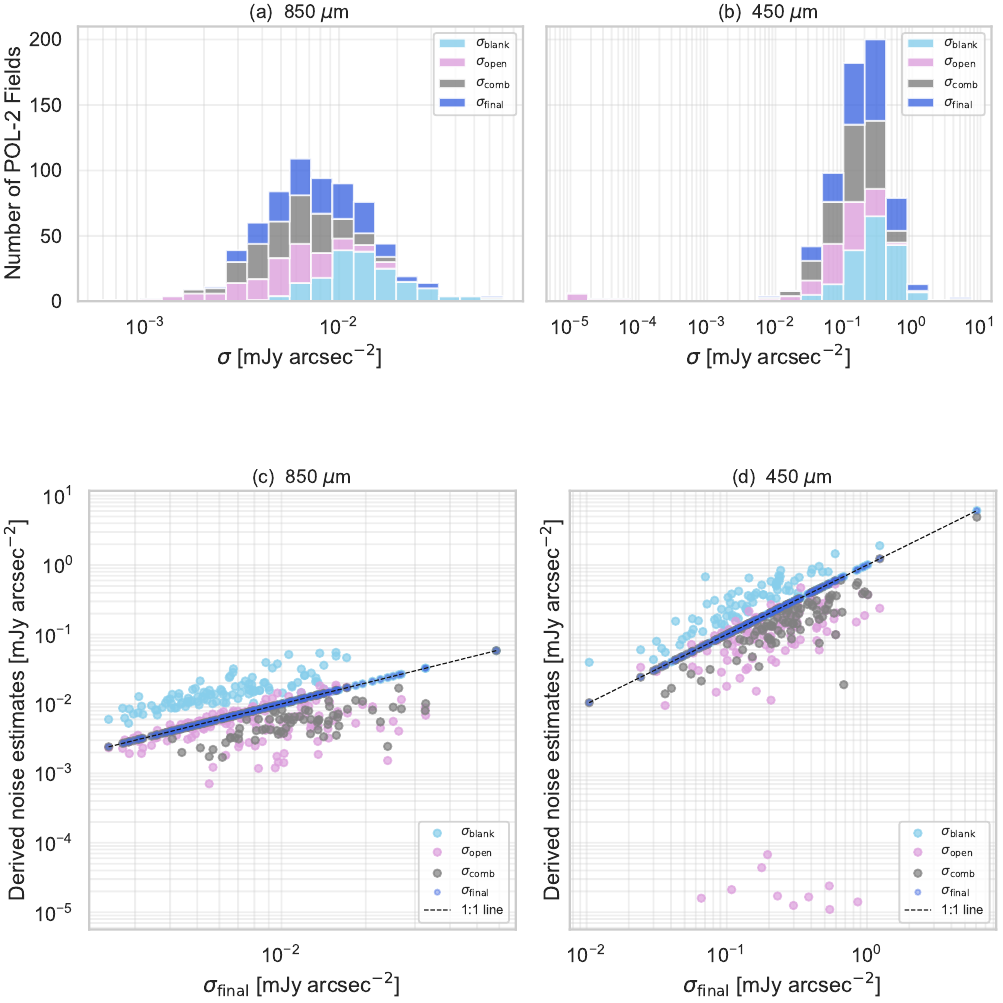}
\caption{
Additive noise estimates for 174 POL-2 fields.
\textbf{(a,b)} Stacked histograms of $\sigma_{\rm blank}$, $\sigma_{\rm open}$, $\sigma_{\rm comb}$, and the adopted $\sigma_{\rm final}$ at 850 and 450\,$\mu$m, respectively.
Bar heights count determinations for each combination of field and estimator, not individual fields.
The blank-sky, composite, and adopted estimates occupy similar narrow ranges.
Fits for the estimator $\sigma_{\rm open}$ in the transition regime return negative values or values below $10^{-5}$\,mJy\,arcsec$^{-2}$ in 21 fields at 850\,$\mu$m and 49 at 450\,$\mu$m.
We exclude these determinations from all panels.
\textbf{(c,d)} Each estimator compared with $\sigma_{\rm final}$ at the corresponding wavelength.
Where they are not adopted, $\sigma_{\rm blank}$ lies above the one-to-one line and $\sigma_{\rm comb}$ below it, in most fields by less than about a factor of three.
The $\sigma_{\rm final}$ series lies on this line by construction.
The other estimators provide the consistency check (sub-subsection~\ref{ass:NoiseEstimation}).
}
\label{afig:IvsDIstats}
\end{figure*}
 
\begin{figure*}[tp]
\centering
\includegraphics[width=0.6\textwidth]{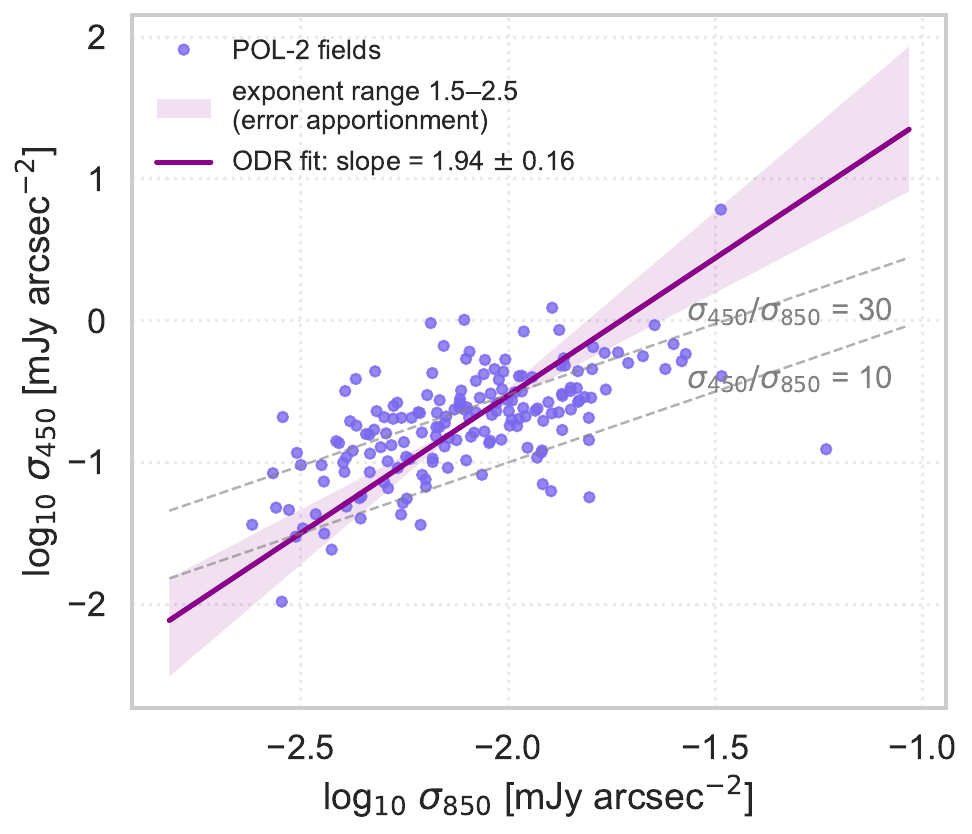}
\caption{
Correlation between the additive noise levels derived at 450~$\mu$m and 850~$\mu$m across the 174 POL-2 fields.
Each point represents one field.
We do not weight the regression or draw error bars because the estimation procedure returns a single $\sigma_\lambda$ for each field and band without an uncertainty for the individual estimate.
The magenta line shows the orthogonal distance regression (ODR) fit in log-log space, yielding
a slope of $m = 1.94 \pm 0.16$ and an intercept of $b = 3.35 \pm 0.33$.
The fit corresponds to the empirical relation
$\sigma_{450} = 2.25\times10^{3}\,\sigma_{850}^{1.94}$,
indicating that the 450~$\mu$m noise increases more steeply than a linear scaling with $\sigma_{850}$.
The shaded wedge spans the exponents recovered when the uncertainty is apportioned differently between the two axes ($m \simeq 1.5$--$2.5$).
The fitted exponent is therefore indicative rather than tightly determined (sub-subsection~\ref{ass:NoiseResults}).
Dashed gray lines mark fixed $\sigma_{450}/\sigma_{850}$ ratios of 10 and 30, illustrating the typical dynamic range.
The median ratio across the sample is $\sigma_{450}/\sigma_{850} = 25$, with an interquartile range of 16.9--35.7.
}
\label{afig:sig450vs850}
\end{figure*}
 
\begin{figure*}[tp]
\centering
\includegraphics[width=0.75\textwidth]{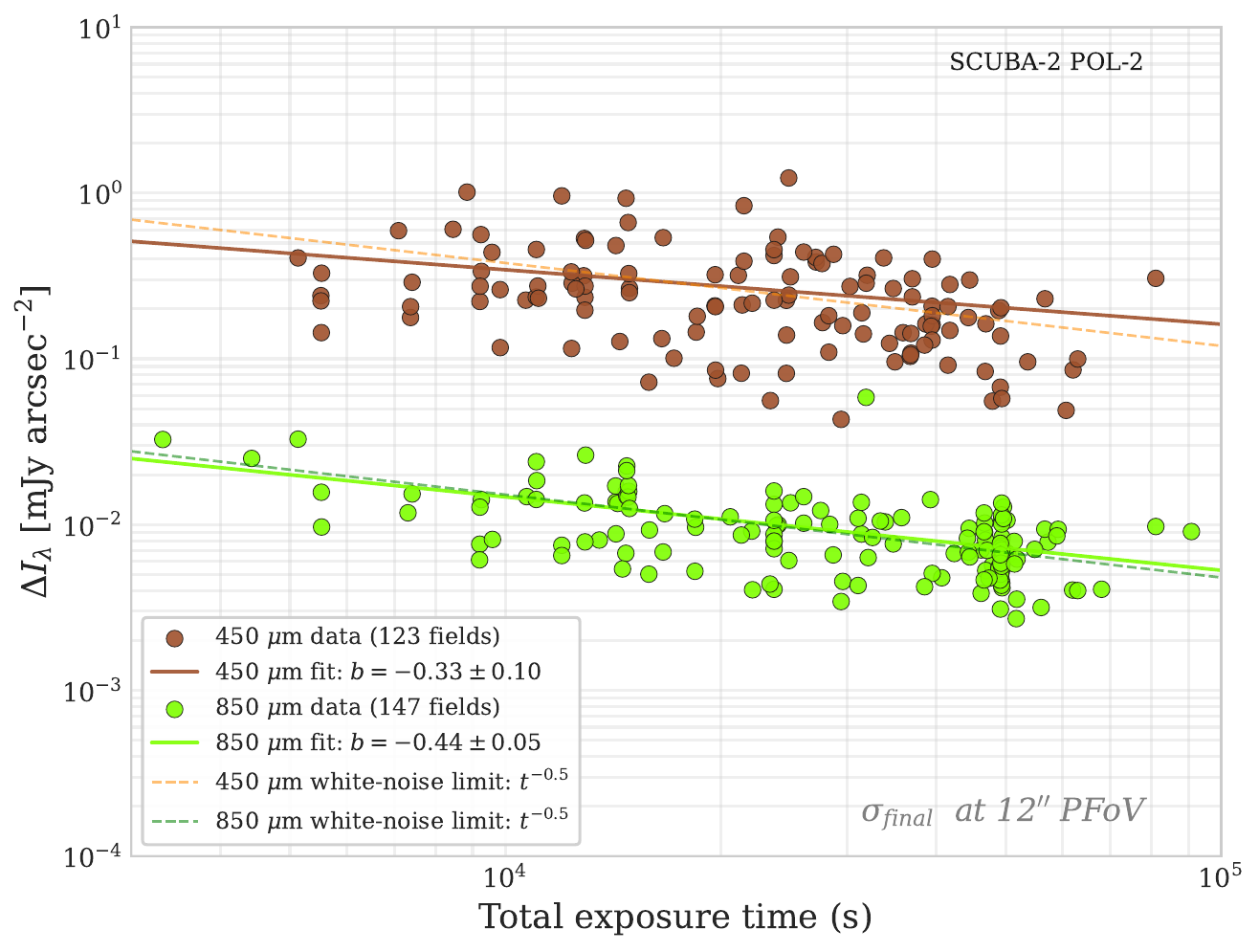}
\caption{
Dependence of the additive noise level $\Delta I_\lambda$ on the total exposure time
for the POL-2 observations at 450~$\mu$m (sienna) and 850~$\mu$m (chartreuse).
Each point represents a single mapped field, and solid lines show
the best-fit power-law relations $\Delta I_\lambda = a\,t^{b}$,
yielding slopes of $b = -0.33 \pm 0.10$ for 450~$\mu$m and
$b = -0.44 \pm 0.05$ for 850~$\mu$m, based on 123 and 147 fields, respectively.
The number of fields analyzed to characterize the NEFD exceeds the number of star-forming regions in the main scientific sample because the noise analysis incorporates all POL-2 observations available in the JCMT archive.
This data set includes not only surveys of star-forming clouds, but also observations of T~Tauri stars, late-type stars, supernova remnants, galaxies, and quasi-stellar objects obtained during the same instrument era.
The plotted $\Delta I_\lambda$ values correspond to the final adopted additive noise estimates
$\sigma_{\rm final}$ introduced in figure~\ref{afig:IvsDIstats}.
Dashed lines indicate the expected white-noise limits ($\propto t^{-0.5}$), normalized to intersect the empirical fits near the median exposure time.
Both fitted slopes are consistent with the $t^{-0.5}$ white-noise prediction (dashed lines), within $1.1\sigma$ at 850~$\mu$m and $1.8\sigma$ at 450~$\mu$m.
All values correspond to noise levels evaluated at a pixel size of 12$^{\prime\prime}$.}
\label{afig:sensitivity_time}
\end{figure*}
 
\subsubsection{Statistical Validation and Wavelength-Dependent Characteristics}
\label{ass:NoiseResults}
We compare the available estimates from the maps and examine how the adopted Stokes-$I$ noise scale varies between wavelengths and with total exposure time.
We address two tests in box~3 of figure~\ref{fig:T_S4_roadmap}: (a) consistency among the noise estimators and (b) wavelength scaling, $\sigma_{450} \propto \sigma_{850}^{1.94 \pm 0.16}$.
These comparisons allow us to assess the noise scale in regions with different source morphologies and observing conditions.

The hybrid procedure for noise estimation (sub-subsection~\ref{ass:NoiseEstimation}) yields multiple independent determinations of the additive noise floor $\sigma_\lambda$ for each field.
To validate the consistency and robustness of these estimates, we performed a statistical analysis on the entire sample of 174 POL-2 fields.
The adopted estimates describe a much larger uncertainty in surface brightness at 450\,$\mu$m than at 850\,$\mu$m, with substantial variation among fields.
Their dependence on wavelength and exposure time characterizes the analyzed data.
This dependence does not imply that every cloud will improve along the same sensitivity curve when more data are added.
 
\paragraph{Consistency of the Noise Estimators}
We compare noise estimates from different intensity ranges to test how sensitive the adopted noise level is to the selection of pixels used to estimate it.
Extended emission can contaminate map regions nominally identified as blank sky, whereas compact sources may leave ample emission-free area for a direct noise estimate.

In figure~\ref{afig:IvsDIstats}, we compared the distributions of four noise estimates: $\sigma_{\rm blank}$ from emission-free pixels, $\sigma_{\rm open}$ from the open fit in the transition regime (sub-subsection~\ref{ass:NoiseEstimation}), $\sigma_{\rm comb}$ from the median of four estimates that include $\sigma_{\rm blank}$ and $\sigma_{\rm open}$, and $\sigma_{\rm final}$ adopted by the selection rule.
Panels (a) and (b) show stacked histograms at 850 \micron\ and 450 \micron\, respectively.
Three of the four estimates converge within narrow ranges.
At 850 \micron\ the distributions of $\sigma_{\rm blank}$, $\sigma_{\rm comb}$, and $\sigma_{\rm final}$ have 5th--95th percentiles between $0.003$ and $0.03$\,mJy\,arcsec$^{-2}$.
At 450 \micron\ the corresponding range is $0.03$ to $0.85$\,mJy\,arcsec$^{-2}$.
Part of this agreement is built into the estimators and does not by itself confirm the noise floor, because $\sigma_{\rm comb}$ includes $\sigma_{\rm blank}$ among its inputs and $\sigma_{\rm final}$ equals one of the two in every field.

The exception is $\sigma_{\rm open}$, which returns a negative value or a value below $10^{-5}$\,mJy\,arcsec$^{-2}$ in 21 fields at 850 \micron\ and 49 at 450 \micron.
These fits did not converge to usable noise estimates; their results do not measure unusually low noise floors, and we excluded them from the histograms.
The behavior of $\sigma_{\rm open}$ follows the selection rule in sub-subsection~\ref{ass:NoiseEstimation}.
None of the fields with unconverged $\sigma_{\rm open}$ reaches case~(i), the only case that prefers $\sigma_{\rm open}$.
The requirement that an estimate be finite and positive excludes the unconverged estimates from $\sigma_{\rm final}$.

Panels (c) and (d) provide direct one-to-one comparisons between $\sigma_{\rm final}$ and the other three estimators. 
Where they are not adopted, $\sigma_{\rm blank}$ exceeds $\sigma_{\rm final}$ by a median factor of 2.3 at 850 \micron\ and 1.9 at 450 \micron, and $\sigma_{\rm comb}$ falls below it by factors of 2.1 and 1.7.
As noted above, these comparisons are partially built into the estimators.
The adopted $\sigma_{\rm final}$ spans 1.4 orders of magnitude at 850 \micron\ and 2.8 at 450 \micron.

The hybrid procedure also separates the additive and multiplicative components.
We used valid estimates from the maps as an internal check on the adopted additive noise scale, including in fields with limited blank sky.
We evaluated the observational calibration of the adopted noise scale using repeated data because agreement with $\sigma_{\rm final}$ does not constitute an independent measurement of noise.

\paragraph{Wavelength Dependence of Additive Noise}
 
Figure~\ref{afig:sig450vs850} shows the correlation between additive noise levels at 450 \micron\ and 850 \micron\ in 174 POL-2 fields.
Each point represents a single POL-2 field.
We fit the relation without weights in either coordinate because the estimation procedure does not supply uncertainties for individual fields.
The orthogonal distance regression (ODR) fit in log--log space yields:
\begin{equation}
\log_{10}\sigma_{450} = (1.94 \pm 0.16)\,\log_{10}\sigma_{850} + (3.35 \pm 0.33),
\label{eq:sigma_powerlaw}
\end{equation}
corresponding to $\sigma_{450} = (2.25 \times 10^{3})\,\sigma_{850}^{1.94}$.
Because the relation is non-linear, the intercept $3.35$ and equivalently the prefactor $2.25 \times 10^{3}$ are specific to $\sigma_\lambda$ expressed in mJy\,arcsec$^{-2}$ and must be rescaled if other units of surface brightness are used.
The exponent $m$ is independent of the choice of units.
The exponent $m = 1.94 \pm 0.16$ exceeds unity, indicating that 450 \micron\ noise increases more rapidly than the linear scaling with $\sigma_{850}$.
We argue that this superlinear dependence reflects the combined influence of wavelength-dependent fluctuations in atmospheric transmission, detector characteristics, and differences in beam solid angle.
 
We quote the exponent as indicative, with $m \simeq 1.5$--$2.5$ the range spanned by the choices of uncertainty described below, and note that $m > 1$ holds throughout.
The formal fit error in $m$ understates the true uncertainty of the exponent.
The two bands are correlated at Spearman $r_s = 0.62$ ($p \approx 1 \times 10^{-19}$).
At that level of scatter, the recovered slope depends on how the uncertainty is apportioned between the two axes.
The unweighted fit adopted here returns $m = 1.94 \pm 0.16$.
Weighting each axis by the measured dispersion of $\log_{10}\sigma_\lambda$ among fields (0.25 and 0.37~dex at 850 and 450 \micron, respectively) returns $m = 1.53 \pm 0.11$.
Assigning the uncertainty entirely to $\sigma_{850}$ returns $m = 2.49$.
Only in the degenerate case in which $\sigma_{850}$ is treated as exact does the fit return $m \simeq 1$, an apportionment that is inconsistent with both bands being determined by the same estimation procedure.
The noise ratio, which we quote next, is insensitive to this choice.

The median ratio $\sigma_{450}/\sigma_{850} = 25$ (interquartile range 16.9--35.7) demonstrates that 450 \micron\ maps are typically a few tens of times noisier in surface brightness than the corresponding 850 \micron\ maps. 
We interpret this broad distribution as a measure of how strongly 450 \micron\ observations depend on atmospheric conditions.

\paragraph{Noise Scaling with Integration Time}
The archive allows us to compare the sensitivity achieved in the maps with the total exposure time across the observed fields.
In general, longer exposure is the most straightforward way to reach fainter cloud envelopes at 450\,$\mu$m, but its benefit may be limited by atmospheric conditions.
 
We present the dependence of the final noise floor on the total exposure time in figure~\ref{afig:sensitivity_time}.  
Each point shows the adopted estimate of the additive noise, $\sigma_{\rm final}$, for a field as a function of its total integration time at the corresponding wavelength.
Solid lines indicate the best-fit power-law relations of the form $\Delta I_\lambda = a_\lambda\,t^{b_\lambda}$, with $t$ expressed in seconds, which yield
$\Delta I_{450} = (7.1 \pm 6.6)\,t^{-0.33 \pm 0.10} \quad \text{mJy\,arcsec}^{-2}$ and $\Delta I_{850} = (0.87 \pm 0.42)\,t^{-0.44 \pm 0.05} \quad \text{mJy\,arcsec}^{-2}$.
The dashed curves show the ideal white-noise limits normalized to intersect the empirical fits near the median integration time.

The 5th--95th percentiles of the scatter about the best-fit relations span a factor of about 4 at 850\,\micron\ and 7 at 450\,\micron.
We suggest that this scatter reflects field-to-field variations in atmospheric opacity, telescope elevation, and the number of nights contributing to each field.
The best-fit exponent is steeper at 850\,\micron\ than at 450\,\micron, but the two differ by only $1.1\sigma$.
The normalization at 450\,\micron\ ($a_{450} = 7.1 \pm 6.6$) is poorly constrained.
These fits alone do not establish which band is more strongly affected by atmospheric fluctuations.
We assess the relative atmospheric contribution at the two wavelengths using the noise scaling between bands in sub-subsection~\ref{sss:QU_isotropy}.

The best-fit power-law indices $b_{450} = -0.33 \pm 0.10$ and $b_{850} = -0.44 \pm 0.05$ are shallower than the ideal radiometer prediction $b_{\rm ideal} = -0.5$ for uncorrelated white noise (shown as dashed lines), by $1.1\sigma$ at 850\,\micron\ and by $1.8\sigma$ at 450\,\micron.
Neither departure is significant.
If the departure at 450\,\micron\ is real, it may reflect the combination of imperfect atmospheric correction and time-variable transmission fluctuations.
At 850\,\micron, where atmospheric stability is generally better, the smaller departure is consistent with the expectation for photon noise.

Across the analyzed data, fields with ten times longer integration have noise levels lower by a factor of $\sim 2.8$ at 850~$\mu$m and $\sim 2.1$ at 450~$\mu$m, compared with the factor of $\sim 3.2$ expected for ideal $t^{-0.5}$ scaling.
Because this comparison mixes fields observed under different conditions, it constrains the sensitivity achieved by the archive rather than the gain from extending any one observation.
We consider these factors a practical guide for sensitivity forecasts in future surveys.

\begin{figure*}[tp]
\centering
\includegraphics[width=0.98\textwidth]{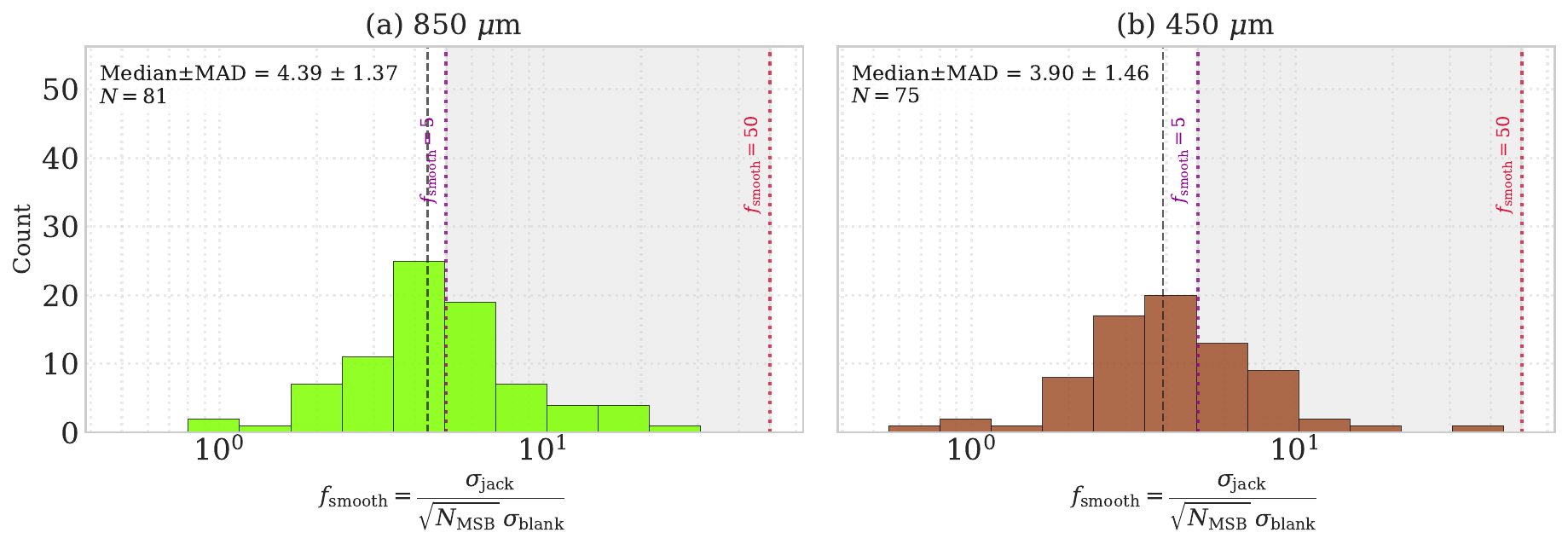}
\caption{
Distributions of $f_{\mathrm{smooth}}$, the efficiency factor for map making, over (a) 81 fields at 850~$\mu$m and (b) 75 fields at 450~$\mu$m where successful jackknife analysis yielded valid noise estimates for individual observations.
The shaded gray region ($5 \le f_{\mathrm{smooth}} \le 50$) marks the range chosen empirically, by visual comparison with the maps, to represent normal map-making performance.
In this range, the RMS of the final map ($\sigma_{\mathrm{blank}}$) is consistent with expectations from the noise estimated by the jackknife analysis for individual observations and the number of coadded MSBs.
Values below this range ($f_{\mathrm{smooth}}<5$) indicate overestimated blank-sky RMS levels.
No field lies above the range.
For 850~$\mu$m, the distribution has $\mathrm{median}\pm\mathrm{MAD}=4.4\pm1.4$, with 58\% of fields below the nominal range.
The 450~$\mu$m data give $3.9\pm1.5$, 
with 65\% below the nominal range. 
The two distributions are not statistically distinguishable (KS test: $D=0.16$, $p=0.25$).
}
\label{afig:f_smooth}
\end{figure*}

In appendix~\ref{app:itc_noise_comparison}, we compare our estimates with predictions from the JCMT SCUBA-2/POL-2 sensitivity calculator provided by the observatory.
On a common mJy\,arcsec$^{-2}$ scale, the SCUBA-2 Integration Time Calculator (ITC) provided by the EAO gives the sensitivities predicted by the model and used to prepare JCMT observation proposals.
The noise measured in the BISTRO science maps exceeds this baseline by factors of $\mathcal{R}_{\rm ITC}\simeq 6$ at 850\,$\mu$m and $\simeq 13$ at 450\,$\mu$m.
These factors are comparable in scale to the independently inferred degradation factors for map making and smoothing (figure~\ref{afig:f_smooth}).

\subsubsection{Jackknife Validation of the Additive Noise Model}
\label{ass:JackknifeTest}
We next assess test~(c) in box~3 of figure~\ref{fig:T_S4_roadmap}: jackknife resampling across 82 POL-2 fields.
Differencing maps from disjoint observing blocks tests the repeatability of the recovered emission without relying solely on a blank-sky noise estimate.
The intensity distribution of a cloud should remain unchanged between repeated observations, whereas atmospheric and detector noise varies between observations.
 
To assess whether hybrid noise estimates accurately capture the intrinsic instrumental and atmospheric noise floor, we performed a jackknife analysis based on the individual \texttt{Step~1} maps produced by the \texttt{pol2map.py} pipeline.
We treat each Minimum Schedulable Block (MSB), the basic scheduling unit for JCMT observations, as an independent noise realization.
Each MSB in the analyzed POL-2 data set typically has an integration time of $\sim$40~minutes.
The jackknife test compares two independent subsets of the data, as in the odd/even split that \citet{Ward-Thompson2017} used to test the repeatability of their POL-2 polarization maps.
Here we used it to evaluate the reproducibility of the measured surface-brightness noise and thus directly measure the noise level reached in a single MSB before any coaddition.
 
For each target, we sorted the observation files chronologically and assigned alternating files to two non-overlapping groups: even-indexed files to Group~A and odd-indexed files to Group~B.
We analyzed a field at a given wavelength only when at least four observation files were available.
We pooled the pixels from the maps of individual observations in each group without averaging, so that each matched pixel pair below compares one observation from Group~A with one from Group~B:
\begin{equation}
M_A = S + N_A, \quad M_B = S + N_B ,
\end{equation}
where $S$ represents the astronomical signal and $N_A$, $N_B$ denote independent noise realizations.  
The difference between the two maps removes the coherent sky emission while preserving the stochastic noise component:
\begin{equation}
D = M_A - M_B = N_A - N_B .
\end{equation}
We matched the two data sets in sky coordinates with a positional tolerance of 3\farcs0, consistent with the pointing accuracy of the JCMT \citep{Mairs2021}.
For each matched pixel pair, we estimated the jackknife noise as
\begin{equation}
\sigma_{\mathrm{Jack}} = \frac{1}{\sqrt{2}}\,\mathrm{std}(D),
\end{equation}
where $\mathrm{std}(D)$ is the standard deviation of the intensity differences between pixels.
We assigned all observations equal statistical weight, without weighting by coverage or $S/N$.
We applied the same flux calibration to both subsets, following sub-subsection~\ref{ass:NoiseEstimation}.
 
We analyzed the entire mapped field for each target, including the outer regions with lower coverage and higher noise.
Using the entire field provides conservative but representative estimates of the noise in individual observations.
This choice also allows direct comparison between targets, even for extended sources that fill the field of view (e.g., DR21, Galactic Center).
 
Across the analyzed fields (81 at 850~\micron\ and 75 at 450~\micron; sub-subsection~\ref{ass:fSmooth}), the 16th--84th percentile ranges of the jackknife noise for individual observations are $\sigma_{\mathrm{Jack}}(I) \simeq 0.15$--$0.35$~mJy\,arcsec$^{-2}$ at 850~\micron\ and $\simeq 3.4$--$9.6$~mJy\,arcsec$^{-2}$ at 450~\micron.
The noise of an individual observation depends on atmospheric opacity, integration time per file, and source brightness.
For example, Ophiuchus~A at 850~\micron\ has $\sigma_{\mathrm{Jack}}(I) \simeq 0.23$~mJy\,arcsec$^{-2}$ in a single MSB.
With $N_{\mathrm{obs}}=20$ independent observations, the expected coadded noise scales as
\begin{equation}
\sigma_{\mathrm{coadd}} \approx \frac{0.23}{\sqrt{20}} \simeq 0.051~\mathrm{mJy\,arcsec^{-2}},
\end{equation}
which exceeds the blank-sky noise of the same field ($\sigma_{\rm blank} \simeq 0.011$~mJy\,arcsec$^{-2}$) by a factor of 4.4.
This ratio is the factor $f_{\rm smooth}$ defined in sub-subsection~\ref{ass:fSmooth}.
At 850~\micron, the jackknife noise for individual observations is typically 13--45 times larger (median 26) than the noise of the final map, measured in the coadded, spatially filtered product (e.g., $\sigma_{\mathrm{Jack}} \sim 0.2$ versus $\sigma_{\mathrm{final}} \sim 0.009$~mJy\,arcsec$^{-2}$).
We attribute this difference to the cumulative effects of temporal averaging over $N_{\mathrm{obs}} = 4$--$34$ observations, spatial filtering that suppresses large-scale atmospheric emission, and common-mode subtraction that removes correlated sky fluctuations.

The jackknife analysis thus provides a noise estimate for each observation that is independent of the blank-sky selection.
With the averaging convention and map coverage specified, we use the difference map to quantify the noise that varies between the two observing subsets.
We cannot constrain all errors in the final cloud map with this test alone, because systematic components shared by both subsets can cancel along with the source emission.
In sub-subsection~\ref{ass:fSmooth}, we compare the jackknife and blank-sky estimates and define $f_{\mathrm{smooth}}$ to quantify the efficiency with which map making with \texttt{pol2map} converts the noise in individual scans into the final RMS noise level estimated in the coadded map.

\subsubsection{Quantifying Map-Making Efficiency: The $f_{\mathrm{smooth}}$ Metric}
\label{ass:fSmooth}
We compare the noise inferred from repeated observations with the blank-sky noise in the combined map to identify fields where the two estimates disagree.
We use $f_{\rm smooth}$, the efficiency metric for map making, in this comparison to assess test~(d) in box~3 of figure~\ref{fig:T_S4_roadmap}.
An accurate estimate of the noise in the final combined map is needed to analyze faint, extended emission reliably in Stokes-$I$ maps.
 
To compare map-making efficiency across the POL-2 observations, including the BISTRO survey, we define a dimensionless smoothing factor
\begin{equation}
    f_{\mathrm{smooth}} = 
    \frac{\sigma_{\mathrm{Jack}}}{
      \sqrt{N_{\mathrm{MSB}}}\,\sigma_{\mathrm{blank}} } ,
\end{equation}
which represents the ratio between the noise expected from the jackknife analysis of individual observations (sub-subsection~\ref{ass:JackknifeTest}) and the residual noise level measured in the final coadded map.
In this form, $f_{\mathrm{smooth}}$ effectively measures how efficiently the \texttt{pol2map} pipeline converts raw, sky-limited time streams into spatially filtered maps suitable for scientific analysis.
Values near unity would imply purely statistical averaging of the noise, whereas larger values indicate additional suppression of correlated noise through common-mode subtraction, high-pass filtering, and beam convolution.

Figure~\ref{afig:f_smooth} shows the distributions of $f_{\mathrm{smooth}}$ for the 82 fields with jackknife estimates (81 at 850\,$\mu$m, 75 at 450\,$\mu$m).
The other fields in the sample of 174 fields were not included in the reduction of individual observations (84 at 850\,$\mu$m and 91 at 450\,$\mu$m) or have fewer than four observation files (9 and 8).

The shaded gray band ($5 \le f_{\mathrm{smooth}} \le 50$) marks the empirically defined range corresponding to ``normal'' map-making performance.
In this range, the RMS of the final map ($\sigma_{\mathrm{blank}}$) is consistent with expectations based on the noise estimated by the jackknife analysis for individual MSBs and the number of coadded observations.
Values below the empirical range ($f_{\mathrm{smooth}} < 5$) may reflect overestimated RMS levels in the blank sky, which would lead to overly conservative contour thresholds.
Values above the range ($f_{\mathrm{smooth}} > 50$) may reflect underestimated RMS levels in the blank sky or excessive filtering of genuine emission.
We chose these boundaries empirically, by visual comparison with the maps, because we have not established a method to derive them.
The boundaries have no fundamental physical significance.

At 850\,$\mu$m, the $f_{\mathrm{smooth}}$ values have $\mathrm{median}\pm\mathrm{MAD}=4.4\pm1.4$, with 42\% of fields in the nominal range and 58\% below ($f_{\mathrm{smooth}} < 5$).
At 450\,$\mu$m, $\mathrm{median}\pm\mathrm{MAD}=3.9\pm1.5$, with 35\% of the fields in the nominal range and 65\% below.
No field at either wavelength lies above the range.
A two-sample Kolmogorov-Smirnov test comparing the two distributions yields $D=0.16$ and $p=0.25$.
The 450\,$\mu$m and 850\,$\mu$m distributions are therefore not statistically distinguishable.
The ratio of the two noise estimates provides a check for each field on how blank-sky contours should be displayed, once both estimates have been placed on the same averaging and surface-brightness scale.
It does not replace the polarization uncertainties or the selection criteria used for the scientific measurements.

\subsubsection{Practical Application of the $f_{\mathrm{smooth}}$ Metric}
\label{ass:fSmoothApplication}
To display faint structure in the Stokes-$I$ maps, we compared noise estimates from repeated observations with those from blank sky and flagged contour levels that need further inspection.
We now interpret the $f_{\rm smooth}$ diagnostic (sub-subsection~\ref{ass:fSmooth}), continuing test~(d) in box~3 of figure~\ref{fig:T_S4_roadmap}.
The interpretation also depends on map coverage and the amount of emission-free sky.

We adopt an empirical scheme in which $f_{\mathrm{smooth}}$ guides the choice of contour thresholds in the Stokes~$I$ maps.
 
For fields with $f_{\mathrm{smooth}} < 5$, the blank-sky RMS is likely overestimated because of residual faint emission in the putatively blank regions.
This situation affects 58\% of our 850\,$\mu$m fields and 65\% of our 450\,$\mu$m fields.
To distinguish faint residual emission from a source that fills most of the map, we use the fraction of each map that survives the blank-sky selection.
A low $f_{\mathrm{smooth}}$ alone cannot distinguish these cases.
When a source fills most of the map, the only unmasked pixels lie at the noisy edges, and $\sigma_{\mathrm{blank}}$ is not representative for a different reason.

For fields in this low range, contour thresholds can be relaxed using a scaling factor $\sigma_{\mathrm{effective}} = \sigma_{\mathrm{blank}} \times \sqrt{f_{\mathrm{smooth}}/5}$.
We use $\sigma_{\mathrm{effective}}$ only for the definition and display of the contours.
It does not enter the polarization noise model of subsection~\ref{ss:noisemodel}, the criteria for selecting pixels, or any error bar reported in this series.
 
The fields with $5 \le f_{\mathrm{smooth}} \le 50$ show map-making performance consistent with the expectations of the jackknife analysis.
These fields are 42\% of the fields at 850\,$\mu$m and 35\% at 450\,$\mu$m.
For these fields, standard contour levels (3, 5, 10$\sigma$, etc.) based directly on $\sigma_{\mathrm{blank}}$ are appropriate without modification.
 
No field in our sample has $f_{\mathrm{smooth}} > 50$.
Such a value would indicate an underestimated blank-sky RMS.
For such fields, we recommend more conservative contours, with $\sigma_{\mathrm{effective}} \approx \sigma_{\mathrm{blank}} \times \sqrt{f_{\mathrm{smooth}}/10}$.

The $f_{\mathrm{smooth}}$ metric thus provides a quantitative measure of map-making efficiency and a uniform check on the consistency of noise calibration across the POL-2 data collected over several years.
At both wavelengths, more than half of the fields fall below the nominal range, and the two distributions do not differ significantly.
The ratio diagnoses the agreement between two noise estimates under their stated conventions; it is not a universal measure of map-making efficiency.
Its interpretation is most useful alongside field coverage, blank-sky selection, and observing conditions.

We use these empirical scalings to select contour levels that reveal faint cloud emission without overstating the significance of structure near the noise level.
Fields with little emission-free area require a different interpretation of the diagnostic from that for fields whose blank-sky regions are well characterized.
The scalings guide the display of the maps; they do not describe fundamental noise properties.

\begin{figure*}[tp]
\centering
\includegraphics{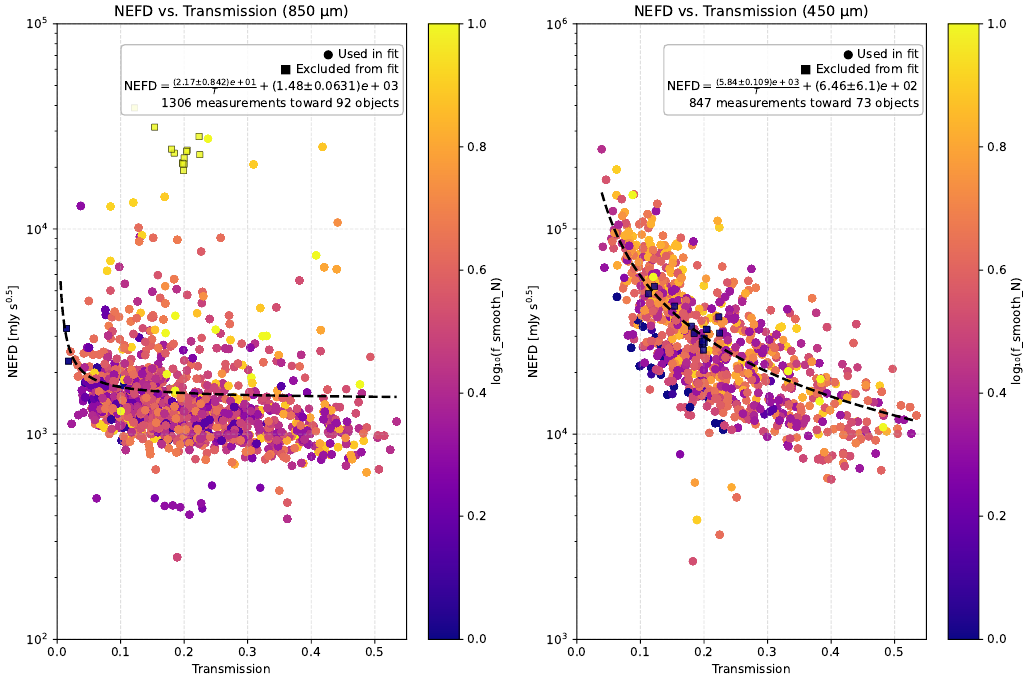}
\caption{
Noise Equivalent Flux Density (NEFD) as a function of atmospheric transmission at 666\,GHz
for (left) 850\,$\mu$m and (right) 450\,$\mu$m POL-2 observations.
Each point represents a measurement for one MSB.
Filled circles show the data used in the fit, and open squares mark excluded outliers (transmission $<0.02$).
The color of each point indicates $\log_{10}(f_{\rm smooth})$ (sub-subsection~\ref{ass:fSmooth}).
The points do not show a correlation between map-making efficiency and atmospheric conditions.
The dashed black curves indicate the best-fit relation ${\rm NEFD} = A/T + B$
(equation~\ref{eq:nefd_model}), with parameters from table~\ref{tab:nefd_fit}.
The dependence on inverse transmission is much stronger at 450\,$\mu$m, where $A/T$ exceeds $B$ throughout the sampled transmission range, than at 850\,$\mu$m, where the constant term $B$ dominates (table~\ref{tab:nefd_fit}).
}
\label{afig:nefd_transmission}
\end{figure*}

\subsubsection{Validation with Observational Conditions: NEFD--Transmission Scaling}
\label{ass:NEFDTransmission}
We compare the NEFD of individual MSBs with atmospheric transmission at 666\,GHz during each observation.
The comparison provides an independent physical test of the additive and multiplicative noise models by relating the sensitivity of individual observations to sky opacity.
This comparison completes test~(d) in box~3 of figure~\ref{fig:T_S4_roadmap}, which combines the $f_{\rm smooth}$ and NEFD--transmission tests (sub-subsections~\ref{ass:fSmooth}--\ref{ass:NEFDTransmission}).

\paragraph{Physical Motivation and Empirical Model}
The NEFD characterizes the observing sensitivity through the flux density that yields a signal-to-noise ratio of unity in one second of integration.
For ground-based submillimeter telescopes, NEFD is expected to vary inversely with atmospheric transmission $T = e^{-\tau_{\lambda} \cdot {\rm AM}}$, where $\tau_{\lambda}$ is the zenith opacity at wavelength $\lambda$ and AM is the airmass. 
In the low-opacity regime typical of Mauna Kea ($T \gtrsim 0.2$), the dominant noise sources include:  
(1)~\textit{sky photon noise}, proportional to $\sqrt{I_{\rm sky}}$, where $I_{\rm sky}$  increases with atmospheric emission (i.e., decreases with $T$); and  
(2)~\textit{instrument readout noise}, arising from TES bolometer electronics, cryogenic  amplifiers, and digitization.  
These considerations motivate a simple two-parameter model:
\begin{equation}
{\rm NEFD}(T) = \frac{A}{T} + B.
\label{eq:nefd_model}
\end{equation}
In this model, $A$ parameterizes the contribution that depends on transmission, while $B$ represents the instrumental noise floor.
We adopt this form as an empirical description of the dependence of NEFD on transmission, as in the POL-2 sensitivity relations provided by the observatory \citep{EAOPOL2obs}.
We do not attempt a detailed radiative-transfer calculation.

\paragraph{Data and Methodology}
For each MSB, we measured the spread of $\Delta I$ over the map in three ways: 1.4826 times the median absolute deviation, the interquartile range divided by 1.349, and the standard deviation.
We adopted the largest of these three estimates as the noise $\sigma_{\rm MSB}$ in that observation.
We derived the NEFD from this estimate using the adopted beam solid angle and the elapsed observing time of the MSB, and associated it with the opacity and airmass of the corresponding observation.
Specifying the conventions for flux density and averaging allows the resulting sensitivity measures to be compared between wavelengths.
For each MSB, we converted $\sigma_{\rm MSB}$ from mJy\,arcsec$^{-2}$ to NEFD in mJy\,s$^{1/2}$ as
\begin{equation}
{\rm NEFD} = \sigma_{\rm MSB} \, \Omega^{\rm G}_{\rm beam} \, \sqrt{t_{\rm int}},
\end{equation}
where $\Omega^{\rm G}_{\rm beam}$ is the nominal Gaussian-equivalent beam solid angle (84\,arcsec$^2$ at 450\,$\mu$m and 180\,arcsec$^2$ at 850\,$\mu$m; table~\ref{tab:omega_conventions}), and $t_{\rm int}$ is the elapsed observing time of each MSB (median $\approx 2450$\,s at 450\,$\mu$m and $\approx 2250$\,s at 850\,$\mu$m).

We calculated atmospheric transmission in the 666\,GHz band from the 225\,GHz zenith optical depth ($\tau_{225}$) derived from the JCMT water-vapor monitor (WVM).
For each observation, we took the geometric mean of the optical depths at the start and end, and used the empirical relation \citep{Dempsey2013}:
\begin{equation}
\tau_{666} = 26.0 \, (\tau_{225} - 0.012),
\end{equation}
to obtain the opacity.
We then accounted for the airmass AM through $T = \exp(-\tau_{666} \cdot {\rm AM})$, again taking the geometric mean of the values at the start and end of each observation.

After excluding L1689B (see the note to table~\ref{tab:nefd_fit}), we fitted measurements from 847 MSBs toward 73 fields at 450\,$\mu$m and 1302 measurements toward 92 fields at 850\,$\mu$m.
These span $0.05 \lesssim T \lesssim 0.5$.
We excluded MSBs with $T < 0.02$ from the fits.
We estimated the parameters by Markov Chain Monte Carlo (MCMC) sampling with \texttt{emcee} \citep{Foreman-Mackey2019}, using 50 walkers, 2000 iterations, and a burn-in of 500 steps.
We used posterior distributions to estimate $A$ and $B$ and their uncertainties.

The fitted terms, one depending on transmission and one constant, describe how this noise measure varies with the observing conditions and allow us to assess, on the same observational basis, how the two bands respond to weather.
Separating these terms does not identify a unique detector or atmospheric mechanism.

\paragraph{Results} In Figure~\ref{afig:nefd_transmission}, we present the NEFD--$T$ scaling at both wavelengths, color-coded by $\log_{10}(f_{\rm smooth})$ to highlight the independence between map-making efficiency and atmospheric transmission. 
Table~\ref{tab:nefd_fit} summarizes the best-fit parameters obtained from the MCMC sampling.
 
\begin{deluxetable}{lccc}
\tablecaption{Best-fit NEFD--Transmission Parameters for figure \ref{afig:nefd_transmission}\label{tab:nefd_fit}}
\tablehead{
\colhead{Band} & \colhead{$A$ [mJy\,s$^{0.5}$]} & \colhead{$B$ [mJy\,s$^{0.5}$]} & \colhead{$N_{\rm MSB}$}
}
\startdata
450\,$\mu$m & $(5.84 \pm 0.11) \times 10^3$ & $(6.5 \pm 6.1) \times 10^2$ & 847 \\
850\,$\mu$m & $21.7 \pm 8.4$ & $(1.48 \pm 0.06) \times 10^3$ & 1302 \\
\enddata
\tablecomments{Best-fit parameters of the model ${\rm NEFD} = A/T + B$ (equation~\ref{eq:nefd_model}) obtained by Markov-chain Monte Carlo sampling using measurements from individual MSBs.
$N_{\rm MSB}$ is the number of MSBs retained in the fit after the exclusion of L1689B (Paper~S2 table~\ref{S2-tab:surv}) and the requirement $T > 0.02$.
For each band, the fit uses only the MSBs observed in that band, with the measurements matched one-to-one by observation identifier.}
\end{deluxetable}
 
We found the following physical trends.
First, the atmospheric term $A$ depends strongly on wavelength.
The term $A_{450} \simeq 5.8\times10^{3}$ exceeds $A_{850} \simeq 22$ by more than two orders of magnitude, and the dependence on transmission is correspondingly much stronger at 450\,$\mu$m
(Spearman $r_s = -0.74$, versus $-0.37$ at 850\,$\mu$m).
The ordering is as expected.
The 450\,$\mu$m (666\,GHz) band lies in a far more opaque and more variable part of the submillimeter window than 850\,$\mu$m (345\,GHz).
A given fractional change in $T$ therefore modulates the contribution of sky noise much more strongly.

Second, the values of the instrumental noise floor $B$ are $B_{450}\simeq6.5\times10^{2}$ (formally compatible with zero at $1\sigma$) and $B_{850}\simeq1.5\times10^{3}$\,mJy\,s$^{0.5}$.
At 450\,\micron\ the atmospheric term therefore dominates the fit almost entirely over the sampled range of transmission, whereas at 850\,\micron\ the constant term dominates.
For the latter band, $A/T \simeq 1.3\times10^{2}$ at the median $T = 0.17$, less than a tenth of $B$.
The stability of $B$ over nearly an order of magnitude in NEFD, together with $A/T \gg B$ at 450\,\micron\ throughout the sampled range of transmission, indicates that atmospheric effects dominate the performance limit at 450\,\micron.

Finally, we attribute the increase in the NEFD scatter at low transmission ($T\lesssim0.2$) to sky fluctuations on short time scales and turbulence.
The color coding does not show a clear systematic variation of $f_{\rm smooth}$ with atmospheric transmission.
Other factors, such as scan strategy and field structure, may also influence $f_{\rm smooth}$.
 
\paragraph{Comparison with Previous Results} \citet{Archibald2002} presented the NEFD of the original SCUBA instrument as a function of normalized sky transmission (their figure~15), and \citet{Holland2013} showed that the NEFD of SCUBA-2 decreases with increasing sky transmission in both bands (their figure~13), qualitatively as in figure~\ref{afig:nefd_transmission}.
For POL-2, the observatory provides the relations ${\rm NEFD}_{450} = 981.5/T_{450} - 87.3$ and ${\rm NEFD}_{850} = 310/T_{850} - 26$\,mJy\,s$^{1/2}$, which have the form of equation~(\ref{eq:nefd_model}) \citep{EAOPOL2obs}.
These relations were derived from the Stokes~$Q$ and $U$ maps of 27 observations of four calibrators, excluding the central point source, using the exposure time per pixel and the transmission in each band.
Our coefficients are therefore not directly comparable with those of the observatory: we derived the NEFD of Stokes~$I$ from the elapsed time of each MSB and used the 666\,GHz transmission for both bands.

The validated NEFD--transmission relation has three applications to JCMT observing and data calibration.
First, the relation enables sensitivity forecasts in real time from current $\tau_{225}$ values.
These forecasts support dynamic scheduling of the queue.
Second, MSBs that deviate more than $3\sigma$ from the model flag potential problems, such as pointing errors, transient clouds, or incomplete reductions.
Third, the relation allows direct conversion between the expected atmospheric transparency and the integration time required to reach a desired $S/N$ ratio for polarization measurements.

The NEFD scaling for individual MSBs agrees with the noise in the final coadded map ($\sigma_{\rm final}$; sub-subsection~\ref{ass:NoiseEstimation}).
We conclude from this agreement that the hybrid noise estimates consistently describe atmospheric and instrumental components across the intensity range of the BISTRO data.
Based on this validation, we also consider the model useful for predicting the sensitivity of future submillimeter polarimetric observations with SCUBA-2 and successor instruments, such as SCUBA-3 \citep{Li2024SCUBA3}.

\paragraph{Comparison with the SCUBA-2 Integration Time Calculator}
We find that the instrument noise floors $B_{450}$ and $B_{850}$ derived from the NEFD--transmission analysis (table~\ref{tab:nefd_fit}) are qualitatively consistent with the predictions of the JCMT SCUBA-2 Integration Time Calculator (ITC; \citealt{EAOSCUBA2ITC,Friberg2016noise}; appendix~\ref{app:itc_noise_comparison}) for Grade~1--2 weather conditions, after conversion to the same surface-brightness scale.

We interpret this agreement qualitatively because two differences limit a direct quantitative comparison.
First, the ITC predicts point-source sensitivity in units of mJy\,beam$^{-1}$, while our analysis characterizes noise in extended-source maps in mJy \, arcsec$^{-2}$.
Second, the ITC assumes idealized conditions for a single observation, while our measurements reflect the co-addition of multiple MSBs obtained under heterogeneous weather grades spanning several years.
We consider the scaling of noise with exposure time given by the radiometer equation, on which the ITC relies, to be an incomplete description of the sensitivity achieved in the maps when correlated fluctuations cannot be neglected.
The observed scaling provides an empirical constraint on the sensitivity achievable in these observations.
We recommend taking this constraint into account when planning exposure times for similar observations.

\subsubsection{Summary of Technical Validation 2}
\label{ss:noiseI_summary}

We formulated a two-component Stokes-$I$ model to separate the background uncertainty from the signal-dependent error toward bright emission and relate intensity cuts to the measured noise conditions of each field.
The usable range of surface brightness in a cloud map depends on both contributions.
\textit{This section establishes the Stokes-$I$ part of Pillar~2 of the technical framework.}
We validated the two-component Stokes-$I$ model and the hybrid estimator across the POL-2 archive of 174 fields through four independent tests (sub-subsections~\ref{ass:NoiseResults}--\ref{ass:NEFDTransmission}).
The adopted additive noise scale $\sigma_{\rm final}$ of sub-subsection~\ref{ass:NoiseEstimation} characterizes the noise floor and is distinct from the uncertainty of each pixel, which can also depend on intensity and map coverage.
For the dual-band parent samples in Papers~S1 and~S2, the denominator $\sigma_I$ in the stated Stokes-$I$ thresholds is the per-pixel uncertainty recorded in the adopted map catalogs.
Reproducing these selections requires the same catalog uncertainties for the relevant field, wavelength, and map product.
They should not be replaced by the additive noise floor, an independently estimated rms, or a noise scale used to display contours (sub-subsection~\ref{ass:fSmoothApplication}).

Sub-subsection~\ref{ss:noisemodel_summary} presents the combined result for Pillar~2, joining the Stokes-$I$ result above with the Stokes-$Q,U$ result of subsection~\ref{ss:noisemodel}.
Next, we assess the reliability of the polarization fractions and angles using the corresponding Stokes~$Q$ and~$U$ measurements.
 
\subsection{Technical Validation 3: Noise Model for Stokes \texorpdfstring{$Q$}{Q} and \texorpdfstring{$U$}{U}}
\label{ss:noisemodel}
\label{as:QU_noise_model}%
 
\subsubsection{Motivation and overview}
\label{ss:noisemodel_motivation}
We complete Pillar~2 by extending the Stokes\,$I$ noise characterization of subsection~\ref{ss:noiseI} to the polarization Stokes parameters (TV3 of the five technical validations TV1--TV5).
A bright dust-continuum detection does not necessarily provide a precise polarization measurement, because the Stokes $Q$ and $U$ components can remain small even when Stokes $I$ is well measured.
We must therefore characterize their uncertainties separately before interpreting faint polarization or differences in field orientation.
The two-component model for Stokes\,$Q$ and $U$ separates the regime dominated by atmospheric noise from that dominated by residual systematic effects, with empirical tests of isotropy and cross-validation using jackknife estimates (figure~\ref{fig:T_pillar_overview} and table~\ref{tbl:T_pillar_index}).
 
We proceed in five steps that parallel the Stokes-$I$ characterization of subsection~\ref{ss:noiseI}.
Sub-subsection~\ref{sss:QU_physical_origin} derives the physical origin of polarization noise from fluctuations in atmospheric transmission.
Sub-subsection~\ref{sss:QU_two_component} formulates the two-component additive$+$multiplicative model for Stokes~$Q$ and~$U$.
Sub-subsection~\ref{sss:QU_regimes} identifies the observational regimes in which each component dominates and connects the model to the strategy for masking the two bands adopted in Papers~S1 and~S2.
Sub-subsection~\ref{sss:QU_validation} presents the implementation and cross-validation strategy.
Sub-subsection~\ref{sss:QU_isotropy} empirically tests the isotropy assumptions---equal additive noise in Stokes $Q$ and $U$ ($\sigma_Q \simeq \sigma_U$) and uncorrelated errors (${\rm Cov}(Q,U)\approx0$)---across wavelength and beam configurations.
Sub-subsections~\ref{ss:spidx_snrbias}--\ref{ss:spidx_snrbias_sens} then apply the noise treatment to the dual-band spectral indices of sub-subsection~\ref{ss:fcf_dualbandquantities}, before these quantities enter the scientific analyses of Papers~S1 and~S2.
The subsection closes with sub-subsection~\ref{ss:noisemodel_summary}, the combined result for Pillar~2.

The relevant noise regime depends on the polarized signal relative to its additive uncertainty, not on total intensity or polarization fraction alone.
This distinction determines when a simplified uncertainty model can be used to compare the polarization seen through the two submillimeter bands.
 
\subsubsection{Physical Origin of Polarization Noise}
\label{sss:QU_physical_origin}
Noise in polarized emission shares the properties established for Stokes $I$ in subsection~\ref{as:IinvDI}, with additional effects from polarized radiative transfer through the terrestrial atmosphere.
Atmospheric transmission fluctuations affect all Stokes parameters identically: opacity variations along the line of sight multiply the signal by a scalar and preserve its polarization state.
We treat atmospheric thermal emission as unpolarized and atmospheric attenuation as independent of polarization.
Under these assumptions, atmospheric emission contributes to Stokes $I$, but not to mean Stokes $Q$ or $U$.
Following equation~(\ref{eq:Iobs}), we express the Stokes parameters observed through an atmosphere with opacity that varies in time as
\begin{equation}
\left[\begin{array}{c} 
I_{\rm obs}(t) \\ 
Q_{\rm obs}(t) \\ 
U_{\rm obs}(t) 
\end{array}\right]
= e^{-\tau(t)} 
\left[\begin{array}{c} 
I_0 \\ 
Q_0 \\ 
U_0 
\end{array}\right]
+ \left[
\begin{array}{c} 
I_{\rm atm}(t) \\
0 \\
0
\end{array}\right],
\label{eq:stokes_propagation}
\end{equation}
where atmospheric emission contributes only to Stokes $I$ because thermal emission from randomly oriented atmospheric molecules is unpolarized. 
Multiplicative attenuation therefore affects all Stokes parameters, whereas additive atmospheric emission affects only $I$.
This difference determines the noise properties of polarimetric observations.
 
\subsubsection{Derivation of the Two-Component Model}
\label{sss:QU_two_component}
The extinction correction multiplies the observed signal by $e^{\hat{\tau}(t)}$, while the true atmospheric attenuation
is $e^{-\tau(t)}$.
The corrected source signal therefore retains a multiplicative factor $e^{\hat{\tau}(t)-\tau(t)}$.
We write this factor as $1+\delta g(t)$, where $\delta g(t)\equiv e^{\hat{\tau}(t)-\tau(t)}-1$.
At a given wavelength, the same residual gain factor applies to all Stokes parameters under the assumption that the attenuation is independent of polarization.
After extinction correction with estimated opacity $\hat{\tau}(t)$, analogous to equation~(\ref{eq:Icorr}), the corrected Stokes parameters become
\begin{equation}
Q_{\rm corr}(t) = Q_{\rm obs}(t)\,e^{\hat{\tau}(t)} \simeq Q_0\,[1+\delta g(t)] + n_Q(t),
\label{eq:Qcorr}
\end{equation}
\begin{equation}
U_{\rm corr}(t) = U_{\rm obs}(t)\,e^{\hat{\tau}(t)} \simeq U_0\,[1+\delta g(t)] + n_U(t),
\label{eq:Ucorr}
\end{equation}
where $\delta g(t)$ represents the fractional error in gain that remains after the extinction correction and varies in time.
The additive terms $n_Q(t)$ and $n_U(t)$ arise from the readout noise of the SCUBA-2 detector, photon statistics, and residual instrumental effects after calibration.
Assuming that additive noise is not correlated with residual gain fluctuations, we take the variance of equations~(\ref{eq:Qcorr})--(\ref{eq:Ucorr}) to obtain the two-component model:
\begin{equation}
(\Delta Q_\lambda)^2 = \sigma_{Q,\lambda}^2 + k_g^2\,Q_\lambda^2, \qquad
(\Delta U_\lambda)^2 = \sigma_{U,\lambda}^2 + k_g^2\,U_\lambda^2,
\label{eq:QU_twocomponent}
\end{equation}
where $\sigma_{Q,\lambda}$ and $\sigma_{U,\lambda}$ quantify the additive noise floor, and $k_g^2 \equiv {\rm Var}[\delta g(t)]$ is the gain variance coefficient shared with Stokes $I$.
 
\subsubsection{Observational Regimes and Physical Interpretation}
\label{sss:QU_regimes}
The applicability and interpretation of equation~(\ref{eq:QU_twocomponent}) depend on the polarization
properties of the observed field.
\paragraph{Low Polarization Regime ($P \lesssim 5\%$)}
In molecular clouds where typical polarization fractions toward the denser parts are a few percent \citep{Crutcher2004, Planck2020-XII}, the polarized intensities scale linearly with the total intensity:
\begin{equation}
Q \approx p\,I\cos(2\chi), \qquad U \approx p\,I\sin(2\chi),
\label{eq:lowpol_approx}
\end{equation}
where $p=P/100$ is the fractional linear polarization with $P$ expressed in percent, and $\chi$ is the polarization position angle on the plane of the sky.
For approximately constant $p$ and $\chi$, both $Q$ and $U$ scale linearly with $I$.
For isotropic field orientations, $\langle p^2\cos^2(2\chi)\rangle \approx \langle p^2\sin^2(2\chi)\rangle \approx \langle p\rangle^2/2$.
Here $\langle\cdot\rangle$ denotes an ensemble average over the orientation angle $\chi$.
We specify the meaning of this symbol locally when we use it elsewhere.
Substituting into equation~(\ref{eq:QU_twocomponent}), we obtain
\begin{equation}
(\Delta Q_\lambda)^2 \approx \sigma_{Q,\lambda}^2 + \left(\frac{k_g\langle p\rangle}{\sqrt{2}}\,I_\lambda\right)^2 \equiv \sigma_{Q,\lambda}^2 + (k'_{Q,\lambda} I_\lambda)^2,
\label{eq:lowpol_model}
\end{equation}
where $k'_{Q,\lambda} = k_g\langle p\rangle/\sqrt{2}$ represents an effective coefficient that links polarization noise to total intensity.
The factor $1/\sqrt{2}$ arises from the orientation average $\langle\cos^2(2\chi)\rangle = \langle\sin^2(2\chi)\rangle = 1/2$.
Replacing $\langle p^2\rangle$ by $\langle p\rangle^2$ requires two additional assumptions: that the fractional polarization varies little over the pixels that are averaged, ${\rm Var}(p) \ll \langle p\rangle^2$, and that $p$ and $\chi$ are not correlated. 
This formulation allows us to fit the uncertainties in $Q$ and $U$ with the functional form used for Stokes $I$, using $I$ as an independent variable.
 
\paragraph{High Polarization Regime ($P \gtrsim 10\%$)}
In regions where polarization can exceed 10\%, such as the edges of some star-forming cores observed with an interferometer \citep{Hull2014}, equation~(\ref{eq:QU_twocomponent}) must be applied without approximation.
The direct coupling between $\Delta Q$ and $Q$ (similarly for $U$) becomes dominant in these regions.
A sufficient dynamic range in the polarized signal is required to constrain $k_g$ independently from the total intensity.
 
\paragraph{Transition Intensity and Observable Predictions}
Analogously to equation~(\ref{eq:Itrans}) for Stokes $I$, the transition from the regime dominated by atmospheric noise to the regime dominated by gain fluctuations occurs at
\begin{equation}
Q_{\rm trans} \simeq \frac{\sigma_{Q,\lambda}}{k_g}, \qquad U_{\rm trans} \simeq \frac{\sigma_{U,\lambda}}{k_g}.
\label{eq:QU_trans}
\end{equation}
For the low polarization approximation, this transition corresponds to the total intensity
\begin{equation}
I_{\rm trans}^{\rm (pol)} \simeq \frac{\sqrt{2}\,\sigma_{Q,\lambda}}{k_g\langle p\rangle} \approx \frac{\sqrt{2}\,I_{\rm trans}}{\langle p\rangle},
\label{eq:I_trans_pol}
\end{equation} 
where $I_{\rm trans}$ is the transition intensity for Stokes $I$ from equation~(\ref{eq:Itrans}).
Under the assumption of comparable additive noise floors in Stokes $Q$ and $I$, $\sigma_{Q,\lambda} \simeq \sigma_\lambda$, the second approximation implies that polarization measurements become limited by gain fluctuations at intensities $\sim$30--140 times higher than total-intensity measurements (for $P \sim 1$--5\%).
The higher onset of the regime dominated by gain fluctuations explains why most POL-2 polarization measurements remain dominated by atmospheric noise even when Stokes $I$ shows significant gain noise.
The present data do not test the assumed equality; we will assess the isotropy of the $Q$ and $U$ noise floors in sub-subsection~\ref{sss:QU_isotropy}.

\begin{figure*}[tp]
\centering
\includegraphics{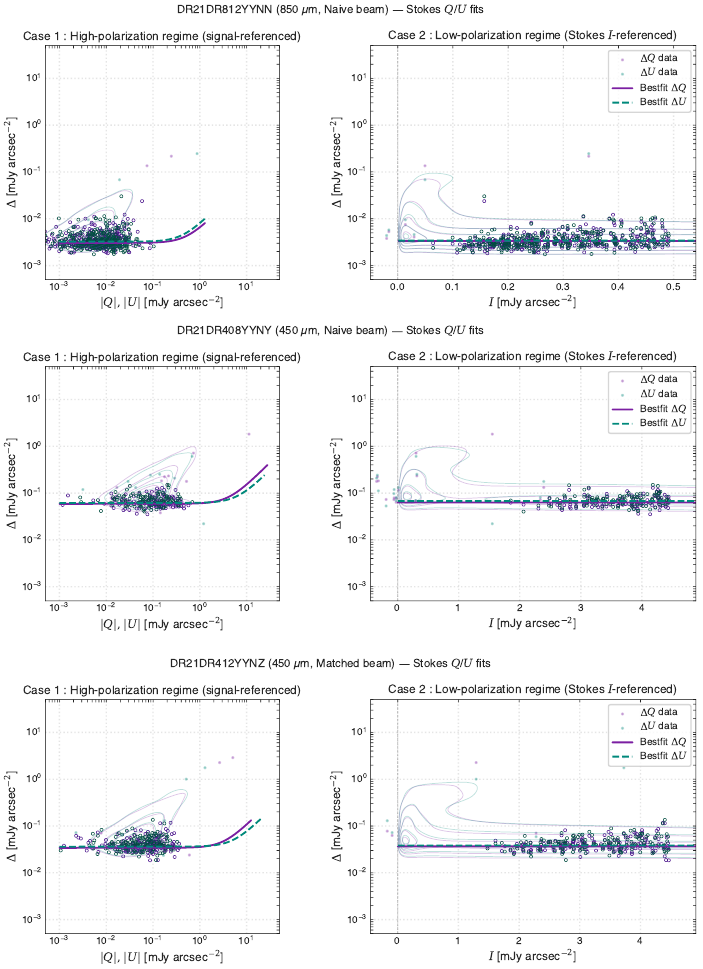}
\caption{
Two-component Stokes-$Q,U$ noise analysis for DR21: native 850\,$\mu$m (top), native 450\,$\mu$m (middle), and 450\,$\mu$m data matched to the 850\,$\mu$m beam (bottom) (Paper~S1 section~\ref{S1-sec:obs}).
Following the Stokes-$I$ analysis of figure~\ref{afig:IvsDI}, the left panels show Case~1, the high polarization regime, with $\Delta Q$ and $\Delta U$ versus $|Q|$ and $|U|$ on logarithmic axes.
The right panels show Case~2, the low polarization regime, versus Stokes~$I$ on a linear horizontal axis.
The vertical dashed line marks $I=0$.
Crimson and royal-blue points and kernel-density contours represent $Q$ and $U$, respectively.
Open circles mark the subset of pixels chosen for display: those with $I/\Delta I > 30$ between the 5th and 90th percentiles of $I$.
The fits use the pixel selections of sub-subsection~\ref{sss:QU_validation}.
Solid ($Q$) and dashed ($U$) curves show fits of $\Delta^2=\sigma^2+(kS)^2$, where $S=|Q|$, $|U|$, or $I$ as appropriate.
The model describes both wavelengths despite the complex emission structure of DR21, with consistent fit parameters between the native and matched beams.
Only pixels meeting the fitting criteria enter the fits.
Appendix~\ref{ass:qcmask} describes the quality mask used throughout this series and the map-making artifacts that motivate it.
}
\label{afig:QUnoise_DR21}
\end{figure*}
 
\subsubsection{Validation and Implementation Strategy}
\label{sss:QU_validation}
We tested the model by comparing the uncertainties in $Q$ and $U$ with the corresponding Stokes signals and estimating the additive noise where the signals are small.
The same noise model should describe uncertainties in both weak and strong polarized emission.
 
The two-component model makes several testable predictions:
\begin{enumerate}
\item The ratio $\sigma_{Q,\lambda}/\sigma_{U,\lambda} \approx 1$ due to instrumental isotropy;
\item The gain coefficient $k_g$ extracted from $Q$ and $U$ should match that of Stokes $I$;
\item The wavelength scaling $\sigma_{Q,450}/\sigma_{Q,850}$ should follow the trends in atmospheric opacity, depending on observing conditions;
\item Pixels with $I < I_{\rm trans}^{\rm (pol)}$ should have constant $\Delta Q$ and $\Delta U$, while those above should show linear scaling.
\end{enumerate}
 
We implemented the model in the following order.
First, we determined $\sigma_{Q,\lambda}$ and $\sigma_{U,\lambda}$ from emission-free regions with $S/N(I) < 2$.
Second, we fitted the model expressed in terms of $I$ [equation~(\ref{eq:lowpol_model})] by MCMC to the brightest 5\% of the pixels in $I$.
We excluded pixels above the 99.99th percentile of $\Delta I$ and required at least 200 selected pixels in each field.
Third, we fitted equation~(\ref{eq:QU_twocomponent}) directly by MCMC to pixels with $S/N(PI) > 10$, requiring at least 200 such pixels in each field.
Here $PI = \sqrt{Q^2+U^2}$ is the non-debiased polarized intensity.
The sequence uses as many pixels as each field allows.
We applied the second fit without checking that a field lies in the low polarization regime.

The three steps test whether the uncertainty model follows the measured polarized signal across the sampled intensity range.
They provide separate checks on additive noise and errors associated with gain fluctuations.
Agreement in one check does not establish all of the assumptions needed for polarization inference.
 
\begin{figure*}[tp]
\centering
\includegraphics{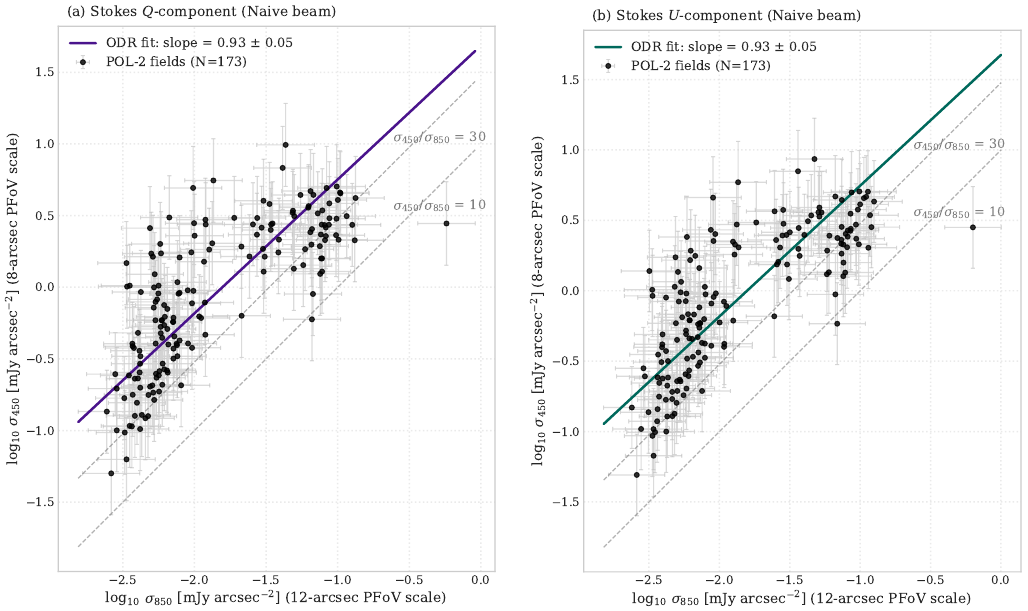}
\caption{
Additive polarization-noise levels at 450 versus 850\,$\mu$m for 173 POL-2 fields at the native beam resolutions: \textbf{(a)} Stokes~$Q$ and \textbf{(b)} Stokes~$U$.
Each point represents one field, with $1\sigma$ uncertainties from fits of the noise model in $\log_{10}\sigma$ space.
Magenta lines show orthogonal distance regression in log--log space.
Both components give a slope of $m=0.93\pm0.05$, indicating similar, nearly proportional scaling between bands.
Gray dashed lines mark $\sigma_{450}/\sigma_{850}=10$ and 30.
The median ratios are 58.7 for $Q$ and 59.5 for $U$, compared with 25 for Stokes~$I$ (figure~\ref{afig:sig450vs850}), reflecting their different noise contributions.
This nearly proportional scaling of the polarization noise is consistent with prediction~(3) of subsection~\ref{as:QU_noise_model}.
}
\label{afig:sigQU_450vs850}
\end{figure*}
 
\begin{figure*}[tp]
\centering
\includegraphics[width=0.72\textwidth]{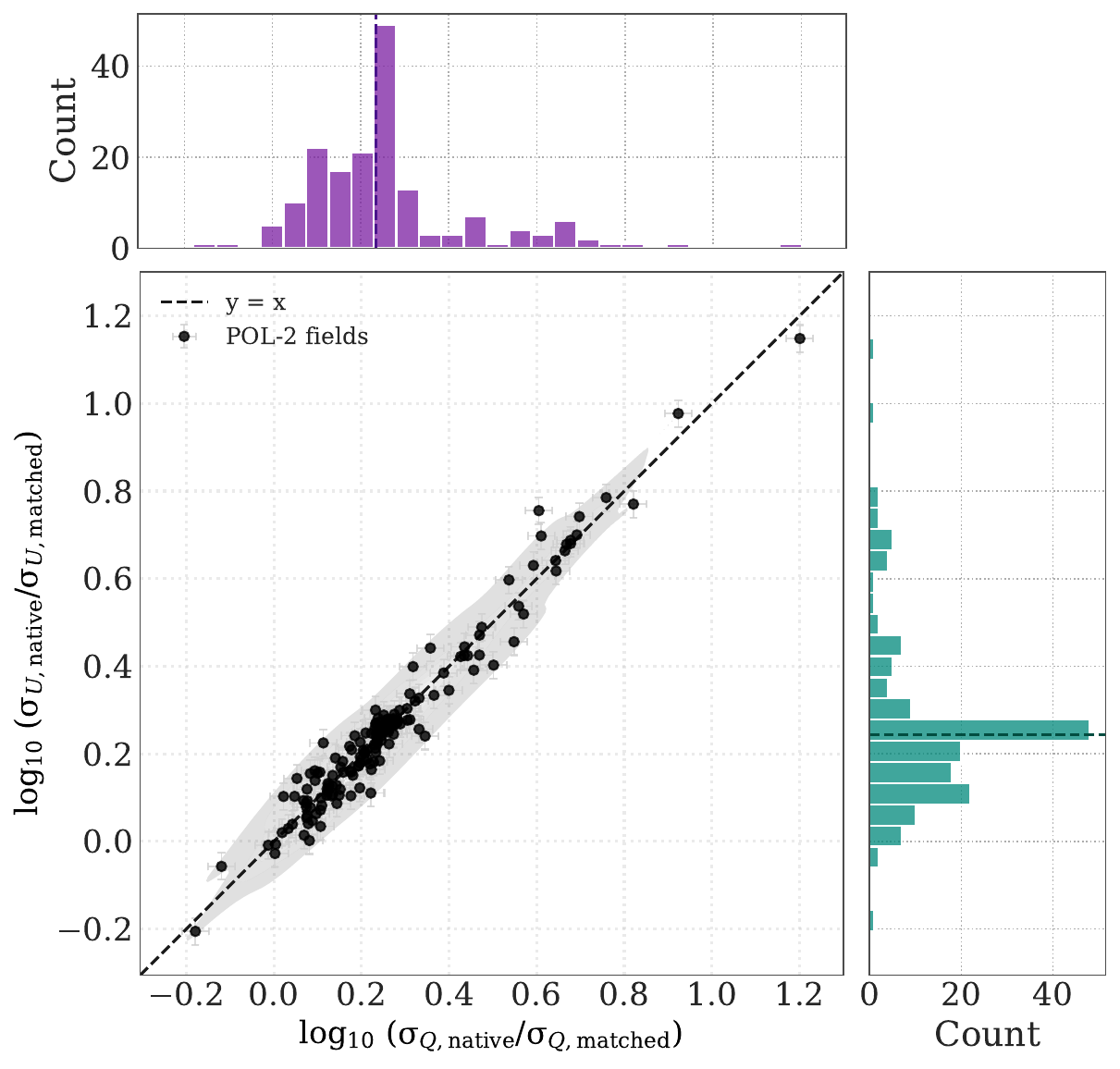}
\caption{
Additive noise in the 450\,$\mu$m reductions with the native beam and with the beam matched to that at 850\,$\mu$m for 172 POL-2 fields.
Define $R_Q=\sigma_{Q,\rm native}/\sigma_{Q,\rm matched}$ and $R_U=\sigma_{U,\rm native}/\sigma_{U,\rm matched}$.
The central panel plots $\log_{10}R_Q$ horizontally and $\log_{10}R_U$ vertically, with one point for each field and error bars propagated from the assumed 5\% fractional uncertainties in the noise estimates.
Purple contours show the kernel-density estimate, and the black dashed line marks $y=x$.
The marginal histograms show $Q$ at the top and $U$ at the right.
Orange dashed lines mark medians of $\log_{10}R_Q=0.235$ and $\log_{10}R_U=0.243$, or linear ratios of approximately 1.7.
The diagonal concentration indicates similar scaling of $Q$ and $U$ noise, preserving the isotropy described in subsection~\ref{as:QU_noise_model}.
The factor of 1.7 combines beam convolution and the change in pixel scale (matched-beam preparation: Paper~S1 section~\ref{S1-sec:obs}).
}
\label{afig:sigRatio_408vs412}
\end{figure*}

\subsubsection{Empirical Verification of Polarization-Noise Isotropy}
\label{sss:QU_isotropy}
\label{ass:QU_isotropy_validation}
 
Following the two-component formulation in equation~(\ref{eq:QU_twocomponent}), we examine the assumptions about additive noise underlying the simplified expressions for noise propagation used in submillimeter polarimetry \citep[e.g.,][]{Ward-Thompson2017}.
The polarization angle is $\chi=\tfrac{1}{2}\operatorname{atan2}(U,Q)$.
Anisotropic noise in the Stokes $Q$--$U$ plane can therefore make the uncertainty in angle and the bias in polarization amplitude depend on the polarization orientation.
Equal marginal noise scales, negligible covariance, and joint Gaussianity are distinct conditions.
Together, these conditions yield a circular Gaussian error distribution centered on the true Stokes values.
 
Let $q$ and $u$ denote measured values of Stokes $Q$ and $U$, and $q_0$ and $u_0$ their true values.
For Gaussian additive errors with $\sigma_Q=\sigma_U=\sigma$ and zero covariance, the joint probability density is
\begin{equation}
f(q,u)=\frac{1}{2\pi\sigma^2}
\exp\!\left[-\frac{(q-q_0)^2+(u-u_0)^2}{2\sigma^2}\right].
\label{eq:PDFxy_equal}
\end{equation}
This model yields the Rice distribution for the measured polarized intensity $\sqrt{q^2+u^2}$.
For jointly Gaussian errors with unequal variances or nonzero covariance, the corresponding expression is
\begin{equation}
f(q,u)=
\frac{1}{2\pi\sigma_Q\sigma_U\sqrt{1-\rho_{QU}^2}}
\exp\!\left[
-\frac{1}{2(1-\rho_{QU}^2)}
\left\{
\frac{(q-q_0)^2}{\sigma_Q^2}
+\frac{(u-u_0)^2}{\sigma_U^2}
-\frac{2\rho_{QU}(q-q_0)(u-u_0)}{\sigma_Q\sigma_U}
\right\}\right],
\label{eq:PDFxy_general}
\end{equation}
where $\rho_{QU}$ is the correlation coefficient of the additive errors in $Q$ and $U$.
Comparing $\sigma_Q$ and $\sigma_U$ in each field constrains the marginal noise scales but leaves $\rho_{QU}$ free.
We therefore also measured $\rho_{QU}$ in the same fields by two independent methods.

\paragraph{Assessment of Stokes-$Q,U$ Noise Isotropy across Fields}
For each observed field, we compared the fitted additive noise scales of Stokes $Q$ and $U$ through their logarithmic ratio.
This comparison tests for a systematic imbalance between the two components.
We first derived the additive noise parameters $\sigma_{Q,\lambda}$ and $\sigma_{U,\lambda}$ for all POL-2 fields using the two-component model (equations~\ref{eq:QU_twocomponent} and \ref{eq:lowpol_model}).
We evaluated $\log_{10}(\sigma_Q/\sigma_U)$ for each data set: the data with the native beam at 450 and 850\,$\mu$m and the data with matched beams at 450\,$\mu$m (table~\ref{tab:QU_isotropy_summary}).
An ideal isotropic system has a median of zero.

To assess whether the median differs significantly from zero, we performed the Wilcoxon signed-rank test and describe the results below.
The size of the noise imbalance between $Q$ and $U$ is more directly relevant to a polarization measurement than its statistical significance alone.
 
We measured the error correlation $\rho_{QU}$ using two different methods.
The first uses the blank-sky pixels, $S/N(I) < 2$, from which we obtained $\sigma_Q$ and $\sigma_U$ above.
In these pixels, the true Stokes $Q$ and $U$ vanish, so the measured values represent noise and their correlation estimates $\rho_{QU}$.
For the second method, we differenced pairs of independent observations at positions that match within $3\arcsec$.
The difference cancels the astrophysical signal exactly and leaves the noise.

We applied a circular $5\sigma$ clip in the Stokes $Q$--$U$ plane for both methods.
Clipping $Q$ and $U$ separately would truncate the joint distribution along its axes and bias $\rho_{QU}$ toward zero.
A radial clip is rotationally symmetric and leaves $\rho_{QU}$ unbiased.
We quote the Pearson coefficient, confirm it with the Spearman rank coefficient, and calibrate the estimator by recomputing $\rho_{QU}$ after randomly permuting $U$.
 
\paragraph{Results}
Given the small typical differences in noise scale and the Wilcoxon $p$-values above 0.2, we do not detect a systematic signed imbalance between the fitted noise scales of $Q$ and $U$ in any of the configurations.
Table~\ref{tab:QU_isotropy_summary} summarizes the statistics.
The mean absolute values of $\log_{10}(\sigma_Q/\sigma_U)$ are 0.024, 0.032, and 0.028 for the three data sets (450\,$\mu$m with the native beam, 450\,$\mu$m with matched beams, and 850\,$\mu$m with the native beam, respectively).
These values indicate typical deviations of 6\%--8\% from perfect isotropy.
The mean of $|\log_{10}(\sigma_Q/\sigma_U)|$ measures the typical absolute difference in noise scale, whereas the signed median and MAD in table~\ref{tab:QU_isotropy_summary} measure any systematic displacement of the ratio distribution and its scatter.
We list all three statistics together for direct comparison.
The corresponding fractions of fields satisfying $|\log_{10}(\sigma_Q/\sigma_U)| < 0.05$ are 90\% (157/175), 81\% (148/182), and 85\% (172/203), respectively, further confirming that the fitted noise scales of Stokes~$Q$ and~$U$ are closely matched in most fields.

The distributions are sharply centered near zero, with median offsets $|\mathrm{median}|\!<\!0.005$ and scatter ${\rm MAD}\!\simeq\!0.03$.
The scatter corresponds to only 7\% fractional variations ($10^{0.03}\!\approx\!1.07$).
We identified outliers with $|\log_{10}(\sigma_Q/\sigma_U)|\!>\!0.1$, including the Rosette Mosaic field and Ophiuchus~C.
The outliers account for less than 5\% of the sample.
 
\begin{table}[t]
\caption{
Statistical validation of noise isotropy across POL-2 configurations.
}
\label{tab:QU_isotropy_summary}
{\centering\footnotesize\setlength{\tabcolsep}{4pt}
\begin{tabular}{lccccccccl}
\hline
 & & \multicolumn{4}{c}{$\log_{10}(\sigma_Q/\sigma_U)$} & \multicolumn{3}{c}{$\rho_{QU}$\tablenotemark{d}} & \\
\cline{3-6}\cline{7-9}
Data Set & $N_{\rm field}$ &
Median &
Mean $|{\cdot}|$\tablenotemark{a} &
MAD\tablenotemark{b} & Wilcoxon $p$\tablenotemark{c} &
Median & Mean $|{\cdot}|$ & Wilcoxon $p$ & Notable outliers\tablenotemark{e}\\
\hline
450\,$\mu$m native compound beam        & 175 & $+0.0014$ & $0.024$ & $0.031$ & $0.45$ & $+0.0018$ & $0.021$ & $0.24$ & \parbox[t]{2.5cm}{\raggedright NGC~1333\strut} \\
450\,$\mu$m beam-matched ($12\farcs6$)  & 182 & $-0.0018$ & $0.032$ & $0.038$ & $0.90$ & $+0.0016$ & $0.025$ & $0.65$ & \parbox[t]{2.5cm}{\raggedright Rosette~N, Serpens~Main, Oph~C\strut} \\
850\,$\mu$m native beam                 & 203 & $+0.0048$ & $0.028$ & $0.031$ & $0.29$ & $+0.0072$ & $0.020$ & $<0.001$ & \parbox[t]{2.5cm}{\raggedright Oph~C, SDC~24.489, IRDC~G35.39$-$0.33\strut} \\
\hline
\end{tabular}
\par}
\tablenotetext{a}{Mean of the $|\log_{10}(\sigma_Q/\sigma_U)|$ values for individual fields; this is the statistic quoted in the text and is not the same as the signed median in the preceding column.}
\tablenotetext{b}{Median absolute deviation of the distribution of $\log_{10}(\sigma_Q/\sigma_U)$ over fields.}
\tablenotetext{c}{The $p$-value is from the two-sided Wilcoxon signed-rank test
against the null hypothesis of zero median offset.}
\tablenotetext{d}{Pearson correlation of Stokes $Q$ and $U$ over the blank-sky pixels ($S/N(I)<2$) of each field, after a circular $5\sigma$ clip in the $Q$--$U$ plane.
A clip on $Q$ and $U$ separately would bias $\rho_{QU}$ toward zero, while a radial clip does not.
The largest $|\rho_{QU}|$ are $0.39$, $0.13$ and $0.07$ for the three data sets.
The two fields above $0.2$ are the two 450\,$\mu$m reductions of Sagittarius~B2, where the blank-sky cut does not isolate empty sky.
An independent estimate from differences of repeated observations, in which the astrophysical signal cancels, gives median $\rho_{QU}$ of $0.000$ over 84 targets at 450\,$\mu$m and $+0.004$ over 101 targets at 850\,$\mu$m.}
\tablenotetext{e}{Representative fields with $|\log_{10}(\sigma_Q/\sigma_U)|>0.1$ (less than 5\,\% of each sample), sorted by $|\log_{10}|$ descending; cell entries show the top three science outliers.
Full listings for each data set: 450\,$\mu$m native compound beam (175 fields, 1 science outlier $=$ 0.6\,\%): NGC~1333 (mosaicked field; Paper~S2 figure~\ref{S2-afig:Bmap_N133MO}).
450\,$\mu$m beam-matched (182 fields, 3 science outliers $=$ 1.6\,\%): Rosette~North (mosaic component field; Paper~S2 figure~\ref{S2-afig:Bmap_RoseMO}), Serpens~Main (mosaicked field; Paper~S2 figure~\ref{S2-afig:Bmap_SerpMO}), L1688 Ophiuchus~C (Paper~S2 figure~\ref{S2-afig:Bmap_OphiuC}).
The non-BISTRO compact sources CTA~102 and IK~Tauri and the field G16 excluded for data quality (Paper~S2 table~\ref{S2-tab:surv}) also exceed the threshold and are excluded for this paper.
850\,$\mu$m native compound beam (203 fields, 5 science outliers $=$ 2.5\,\%): L1688 Ophiuchus~C (Paper~S2 figure~\ref{S2-afig:Bmap_OphiuC}), SDC~24.489 (Paper~S2 figure~\ref{S2-afig:Bmap_SD24P4}), IRDC~G035.39$-$00.33 (mosaicked field), B211/213 North (mosaic component field; Paper~S2 figure~\ref{S2-afig:Bmap_B211MO}), IC~1396E (Paper~S2 figure~\ref{S2-afig:Bmap_IC139E}).}
\end{table}
 
The DR21 case study in figure~\ref{afig:QUnoise_DR21} illustrates that the two-component model recovers the same additive noise levels for both Stokes components, even in bright, spatially complex fields.
The outlying fields remain relevant to the uncertainty estimates for individual sources.

\paragraph{Error Correlation between $Q$ and $U$}
We verified that the measured correlations are small.
The blank-sky method gives median $\rho_{QU}$ values of $+0.002$, $+0.002$ and $+0.007$ for the 450\,$\mu$m data with the native beam, the 450\,$\mu$m data with matched beams, and the 850\,$\mu$m data with the native beam, respectively.
The corresponding mean $|\rho_{QU}|$ values are $0.021$, $0.025$ and $0.020$.
In those three configurations $99\%$, $99\%$ and $100\%$ of fields have $|\rho_{QU}| < 0.1$.

The method using differences between observations gives median $\rho_{QU}$ values of $0.000$ for 84 targets with repeated 450\,$\mu$m observations and $+0.004$ for 101 targets with repeated 850\,$\mu$m observations.
The corresponding mean $|\rho_{QU}|$ are $0.020$ and $0.017$, and every target lies below $|\rho_{QU}| = 0.06$.
Randomly permuting $U$ gives mean $|\rho_{QU}| = 0.005$--$0.014$.
The measured values therefore resolve the correlation rather than only the scatter of the estimator.
The Pearson and Spearman coefficients agree within $5\%$ of each other.

At 850\,$\mu$m, the positive median is significant at $p < 0.001$ with the blank-sky method and $p = 0.04$ with the method using differences between observations.
We thus detected a correlation, but its magnitude is negligible.
 
Neglecting the cross term in equation~(\ref{eq:varPhi_full}) therefore has only a small effect on the propagated uncertainty in polarization angle.
With $\sigma_Q \simeq \sigma_U \equiv \sigma$ and $\mathrm{Cov}(Q,U) = \rho_{QU}\sigma^2$, the omitted term is $2\rho_{QU} QU\sigma^2$, compared with the retained term $(Q^2+U^2)\sigma^2$.
The fractional error, $2\rho_{QU}|QU|/(Q^2+U^2)$, is largest at $|Q| = |U|$, where it equals $\rho_{QU}$.
For $|\rho_{QU}| \lesssim 0.02$ the resulting error in the uncertainty of the polarization angle $\sigma_\chi$ is at most $1\%$.
 
The blank-sky estimates exceed $|\rho_{QU}| = 0.2$ for two fields, both reductions of Sagittarius~B2 at 450\,$\mu$m: the single-field reduction and the reduction of the CMZ mosaic component.
In these fields, an $S/N(I) < 2$ cut does not isolate empty sky.
At 850\,$\mu$m the same source can be differenced across observations, and the two methods give $\rho_{QU} = +0.028$ and $\rho_{QU} = +0.001$ over $9.4\times10^{3}$ matched positions.
We attribute the difference between these estimates to residual sky in the blank-sky selection rather than to a correlation of the errors.

\paragraph{Wavelength and Beam Effects}
The comparison between wavelengths in figure~\ref{afig:sigQU_450vs850} further supports the isotropy established above: $\sigma_{Q,450}$ and $\sigma_{U,450}$ scale with their 850\,$\mu$m counterparts as $\sigma_{450}\!\propto\!\sigma_{850}^{0.93}$.
This slope near unity is what equation~(\ref{eq:stokes_propagation}) requires of $Q$ and $U$.
Lacking the additive atmospheric emission term that enters Stokes~$I$, the two components carry only the multiplicative attenuation, so their noise scales almost linearly between bands.
The noise in total intensity, which does carry that term, scales superlinearly over the same fields ($m = 1.94 \pm 0.16$, equation~\ref{eq:sigma_powerlaw}).
The two exponents are therefore not expected to agree.
We regard their difference as a consequence of the asymmetry set out in sub-subsection~\ref{sss:QU_physical_origin} rather than as a discrepancy.

Figure~\ref{afig:sigRatio_408vs412} shows that matching the beam at 450\,$\mu$m uniformly reduces both $\sigma_Q$ and $\sigma_U$ by a factor of $\approx 1.7$, without altering their ratio.
We conclude from these tests that the noise isotropy and scaling are preserved across the wavelength and beam configurations.
We quantify the preservation of polarization angles under the same procedure for matching the beams in subsection~\ref{ss:beamval}.

\paragraph{Theoretical Consistency and Approximate Treatment}
To compare $\vec{B}$-field orientations between wavelengths, we must estimate angle uncertainties for the selected lines of sight.
For the angle, the relevant assumptions concern the errors in $Q$ and $U$, their covariance, and the signal-to-noise regime.
The uncertainty in intensity enters the polarization fraction separately.
We consider the measured equality $\sigma_Q\simeq\sigma_U$ and the measured $|\rho_{QU}|\!\lesssim\!0.02$ to justify the use of the simplified Gaussian model [equation~(\ref{eq:PDFxy_equal})] instead of the general elliptical form [equation~(\ref{eq:PDFxy_general})].
In the following expressions, $P$ and $\sigma_P$ are expressed as dimensionless fractions rather than percentages.
Under these isotropic and uncorrelated conditions, the variance propagation for $P$ simplifies to
\begin{equation}
\sigma_P^2 \simeq \frac{\sigma^2}{I^2},
\qquad
\sigma_Q^2\simeq\sigma_U^2\equiv\sigma^2,\;
\mathrm{Cov}(Q,U)\approx0,
\label{aeq:CovQUzero}
\end{equation}
or equivalently, $\sigma_P\simeq\sigma/I$ in terms of the standard deviation.
Likewise, the variance of the polarization angle,
\begin{equation}
\sigma_\chi^2
=\frac{1}{(2P^2I^2)^2}
\left[
Q^2\sigma_U^2
+U^2\sigma_Q^2
-2QU\,\mathrm{Cov}(Q,U)
\right],
\label{eq:varPhi_full}
\end{equation}
simplifies to the standard approximation:
\begin{equation}
\sigma_\chi\simeq\frac{1}{2}\frac{\sigma_P}{P}.
\label{eq:varPhi_simplified}
\end{equation}
These expressions require that both $|\rho_{QU}|$ and $|\log_{10}(\sigma_Q/\sigma_U)|$ be small.
The measurements meet that requirement.
We find $|\rho_{QU}| < 0.1$ in $99\%$--$100\%$ of fields, and the scatter of $\log_{10}(\sigma_Q/\sigma_U)$ is ${\rm MAD} \simeq 0.03$ (table~\ref{tab:QU_isotropy_summary}).
At a low signal-to-noise ratio in polarization, the linearized standard error is not adequate, and the angle distribution itself must be evaluated.

The agreement between the additive noise scales of $Q$ and $U$, together with the measured $|\rho_{QU}|\!\lesssim\!0.02$, supports the circular error model.
We therefore characterize the $(Q,U)$ error distribution by its two variances rather than using a full covariance matrix in routine processing.
We adopted Gaussian errors following standard practice in submillimeter polarimetry.
Under this further assumption of jointly Gaussian errors, the measured polarized intensity follows the Rice distribution.
For sufficiently well measured polarization, we estimate angle uncertainties with equation~(\ref{eq:varPhi_simplified}).
We use a von Mises approximation for the doubled angle $2\chi$ to remove the $180\arcdeg$ ambiguity of polarization orientation.

A weak polarization amplitude can reflect measurement noise as well as a small polarized contribution from the cloud.
Similarly, an apparent angular dispersion contains both sky structure and uncertainty in the measured Stokes vector.
The noise treatment specifies the measurement contribution that must be considered when interpreting polarization amplitudes and angular structure.
It does not by itself separate imperfect grain alignment from cancellation along the line of sight, nor does it establish that every remaining feature is astrophysical.
 
\subsubsection{Signal-to-Noise Dependence of the Dual-Band Spectral Indices}
\label{ss:spidx_snrbias}
We apply the noise treatment above to the spectral indices $\alpha$ and $\gamma$ of equation~(\ref{eq:spectral_indices}), which we use in Papers~S1 and~S2 to compare polarized and total emission (Paper~S2 subsection~\ref{S2-ss:S2_SpIdx}).
A spurious change in spectral slope near the detection limit could be mistaken for a change in the properties of the aligned dust.
Polarized intensity is intrinsically weaker than total intensity, so $\gamma$ is more susceptible than $\alpha$ to bias that depends on $S/N$.
At low $S/N$, debiasing of $PI$ with the modified asymptotic estimator (appendix~\ref{app:mas_debias}) does not completely remove the Ricean bias.
Fluctuations in $PI_{450}$ and $PI_{850}$ therefore bias $\gamma$ downward.
 
For a spectral index $X = \log(F_{450}/F_{850}) / \log(\nu_{450}/\nu_{850})$, where $F$ represents either Stokes~$I$ or polarized intensity $PI$, the uncertainty is
\begin{equation}
\sigma_X = \frac{1}{\ln(\nu_{450}/\nu_{850})} \sqrt{\left(\frac{\sigma_{F_{450}}}{F_{450}}\right)^2 + \left(\frac{\sigma_{F_{850}}}{F_{850}}\right)^2},
\label{eq:sigma_spectral_index}
\end{equation}
and the formal $S/N$ ratio of each index is $S/N(X) = |X|/\sigma_X$.
 
To examine how the spectral indices vary with measurement precision, we selected pixels in each source with $I/\sigma_I > 30$, $0 < P < 100\%$, and a defined uncertainty in polarization angle at both wavelengths, and required $\alpha > 0$.
We divided the logarithmic ranges of the signal-to-noise ratios of the two spectral indices, $S/N(\alpha)$ and $S/N(\gamma)$, separately into five equal bins.
For each bin containing at least three pixels, we calculated the median index and its uncertainty, $1.4826\,\mathrm{MAD}/\sqrt{N}$, where MAD is the unscaled median absolute deviation and $N$ is the pixel count.
We compared these running medians with the median of all selected pixels, whose uncertainty is $1.253\times1.4826\,\mathrm{MAD}/\sqrt{N}$.
For the medians of individual sources in Paper~S2, we instead use the Maritz--Jarrett estimator.
We recorded as a diagnostic the $S/N$ above which the running median becomes approximately constant, but did not use this value to select pixels.
 
A running median that does not change with $S/N$ indicates that the measured index does not vary systematically with measurement precision.
A systematic tilt, particularly at low $S/N$, indicates that measurements dominated by noise can bias the ensemble statistics.

Neither result is conclusive on its own, because each index enters both axes of the comparison.
Because $S/N(X) = |X|/\sigma_X$, the bins with the lowest $S/N$ preferentially contain pixels with small $|X|$.
Part of a downturn at low $S/N$ therefore follows from the axis definition rather than noise bias.
A constant running median alone does not establish an unbiased estimate.
It also does not show that the selected pixels represent the full cloud.
 
For $\alpha$, every plotted pixel has $I/\sigma_I > 30$ at both wavelengths, so noise bias is not expected.
The uncertainty $\sigma_\alpha$ depends on the $S/N$ of Stokes~$I$ in both bands (equation~\ref{eq:sigma_spectral_index}), so $S/N(\alpha)$ orders pixels mainly by brightness.
The running median of $\alpha$ therefore traces changes with brightness within the source.
In the following examples, this running median changes by up to about $1$ across the $S/N(\alpha)$ range.
We do not interpret this variation as a bias.

For $\gamma$, the running median rises steeply through the bins at low $S/N$, where the axis definition and the residual Ricean bias act together.
The $S/N(\gamma)$ at which the running median levels off varies among sources.
This variation reflects differences in data quality, intrinsic polarization strength, and the distribution of $S/N$ values across the map.
This diagnostic shows how the summaries of the spectral indices change with formal precision.
We first illustrate this diagnostic for representative sources below.
We then examine how imposing an $S/N(\gamma)$ threshold affects the inferred spectral differences in sub-subsection~\ref{ss:spidx_snrbias_sens}.
 
\begin{figure}[htbp]
\centering
\includegraphics[width=0.72\columnwidth]{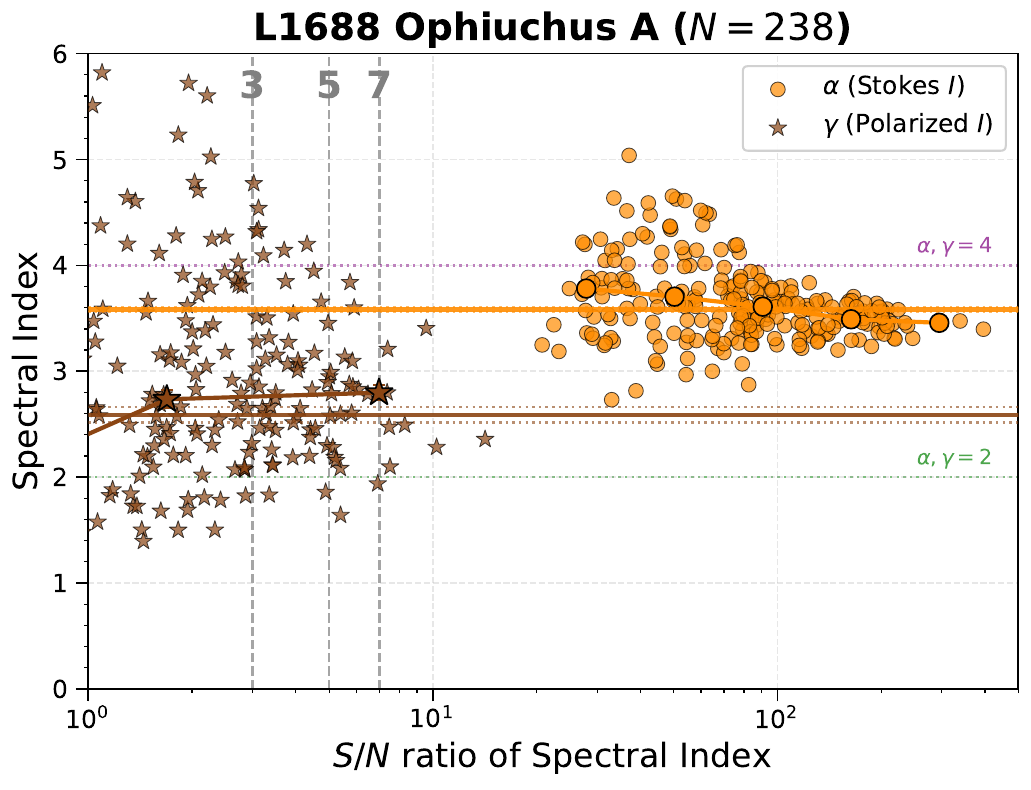}
\caption{Dependence of the spectral indices on their formal $S/N$ in L1688 Ophiuchus~A.
The spectral index $\alpha$ of Stokes~$I$ (orange circles) and the spectral index $\gamma$ of polarized intensity (brown stars) are plotted against their respective $S/N$ ratios.
Connected symbols with error bars show running medians in logarithmic bins of $S/N$ (sub-subsection~\ref{ss:spidx_snrbias}).
Horizontal solid lines indicate the medians of all pixels drawn ($\alpha = 3.58$, $\gamma = 2.59$), with dotted lines showing the standard error of the median.
Vertical dashed lines mark $S/N = 3$, 5, and 7 for reference.
A total of $N = 238$ pixels satisfy the basic selection criteria ($I/\sigma_I > 30$ at both wavelengths).
In the underlying maps, the pixel centers are spaced by $12\farcs0$, nearly one beam width ($12\farcs6$ effective FWHM).
Pixels and bins of the running median with $S/N < 1$ lie outside the plotted range.
For $\alpha$, the running median decreases by about $0.3$ across the $S/N(\alpha)$ range.
We do not interpret this trend as a bias, because $\alpha$ enters both axes.
For $\gamma$, the running median rises steeply through the bins with the lowest $S/N$, most of which lie below the plotted range, and levels off over the two bins with the highest $S/N$.
Only 8 pixels have $S/N(\gamma) > 7$, the display threshold of the spatially resolved maps of Paper~S2 (figure~\ref{S2-fig:SpIdx_maps_OphA}).
We do not apply a cut on $S/N(\gamma)$ when calculating the values for individual sources in Paper~S2 table~\ref{S2-tbl:SpIdx_Global}.
}
\label{fig:SpIdx_BiasCheck_OphA}
\end{figure}
 
\subsubsection{Diagnostics for Representative Sources}
\label{ss:spidx_snrbias_examples}
\begin{figure*}[htbp]
\centering
\includegraphics{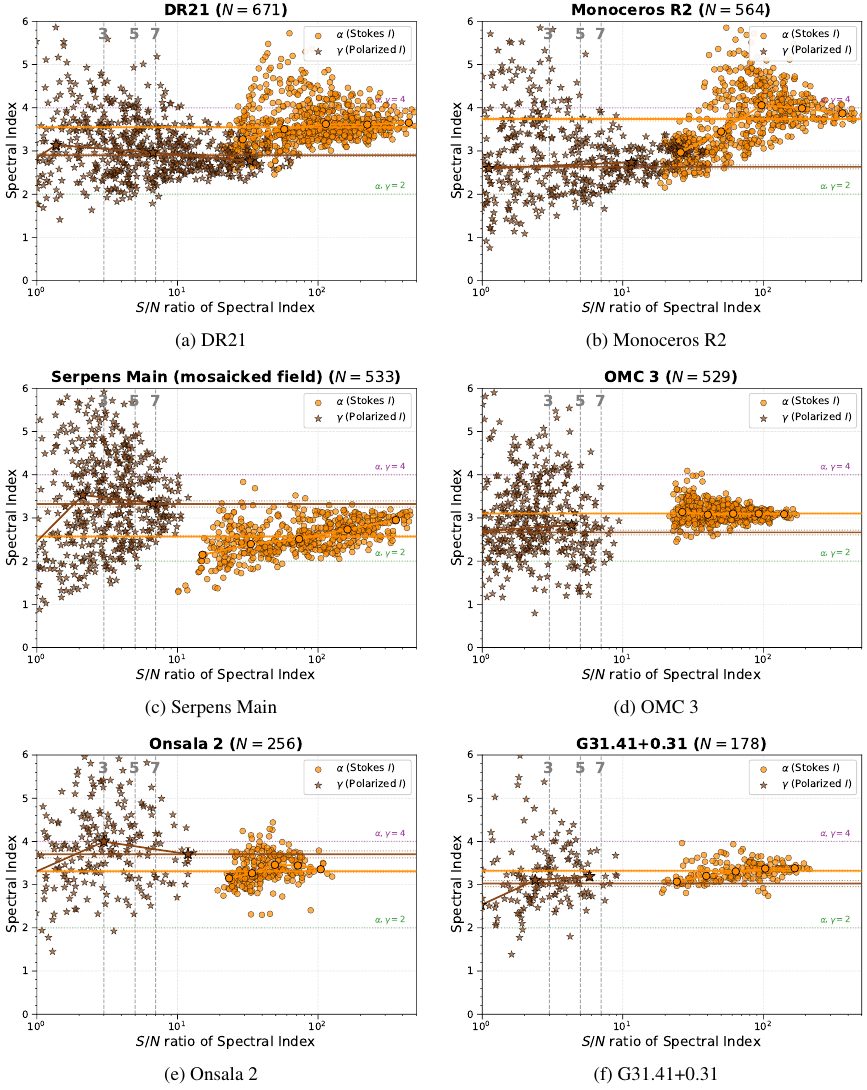}
\caption{
Dependence of the spectral indices on their formal $S/N$ for six representative sources spanning the range of data quality and behavior of the spectral indices in the statistical sample of Paper~S2.
Each panel shows the spectral index $\alpha$ of Stokes~$I$ (orange circles) and the spectral index $\gamma$ of polarized intensity (brown stars) as a function of their respective $S/N$ ratios.
Connected symbols with error bars show running medians in logarithmic bins of $S/N$.
Horizontal solid lines indicate the medians of all pixels drawn.
Vertical dashed lines mark $S/N = 3$, 5, and 7 for reference.
Pixels and bins of the running median with $S/N < 1$ lie outside the plotted range.
The panel title gives $N$, the number of plotted pixels with $I/\sigma_I>30$, $0<P<100\%$, a defined uncertainty in polarization angle at both wavelengths, and $\alpha>0$.
The plotted number $N$ exceeds $N_{\rm samp}$ in the text and Paper~S2 table~\ref{S2-tbl:SpIdx_Global}, which counts the smaller set selected for the scientific analysis of the spectral indices in Paper~S2 sub-subsection~\ref{S2-sss:SpIdx_Global}.
The median lines refer to the plotted pixels and do not necessarily reproduce the tabulated $\gamma-\alpha$.
Neither selection imposes a cut on $S/N(\gamma)$.
(a)~DR21: high-quality source with broad $S/N$ coverage and clear $\gamma<\alpha$.
(b)~Monoceros~R2: a source with a strongly negative $\gamma-\alpha=-1.04$, illustrating the lower-bound regime of the F2 distribution.
(c)~Serpens Main: $\gamma-\alpha=+0.01$, consistent with $\gamma\approx\alpha$ at $1\sigma$.
(d)~OMC~3: a field requiring correction of additive zero-level offsets in Stokes $Q/U$ at 450\,$\mu$m.
(e)~Onsala~2: a small but significant offset, $\gamma-\alpha=-0.38$.
(f)~G31.41+0.31: a JCMT calibration standard with small $N_{\rm samp}$.
The effect of imposing an $S/N(\gamma)$ threshold is quantified in sub-subsection~\ref{ss:spidx_snrbias_sens}.
}
\label{fig:SpIdx_BiasCheck_examples}
\end{figure*}

We examine whether the procedure for the spectral indices remains interpretable in bright compact regions, extended mosaics, and clouds with only modest contrast in polarized emission.
Figure~\ref{fig:SpIdx_BiasCheck_OphA} shows the diagnostic for L1688 Ophiuchus~A, our reference source in Papers~S1 and~S2 (Paper~S2 sub-subsection~\ref{S2-sss:SpIdx_OphA}).
In figure~\ref{fig:SpIdx_BiasCheck_examples}, we show six additional sources spanning the range of data quality and behavior of the spectral indices in the Paper~S2 statistical sample.
 
DR21 (number of sampled pixels $N_{\rm samp} = 86$; $\gamma - \alpha = -0.94$) represents a high-quality source with broad $S/N$ coverage that extends to $S/N(\gamma) > 50$.
The running median of $\gamma$ decreases slightly toward the highest $S/N$, so the separation between the $\alpha$ and $\gamma$ median lines is not produced by the pixels at low $S/N$.
Monoceros~R2 ($N_{\rm samp} = 73$; $\gamma - \alpha = -1.04$) has one of the most negative $\gamma - \alpha$ values in the statistical sample.
L1688 Ophiuchus~A has the most negative value, $-1.14$.
Serpens Main ($N_{\rm samp} = 58$; $\gamma - \alpha = +0.01$) is one of the sources consistent with $\gamma \approx \alpha$ at $1\sigma$, corresponding to $P_{450}/P_{850} \approx 1$.
In this mosaicked field, the running median of $\alpha$ increases steadily with $S/N(\alpha)$, by approximately $0.8$.
OMC~3 ($N_{\rm samp} = 72$; $\gamma - \alpha = -0.91$) is a field where additive zero-level offsets in Stokes $Q/U$ at 450\,$\mu$m required correction (subsection~\ref{ss:offset}).
Its bin with the highest $S/N(\gamma)$ is centered at $S/N(\gamma) \approx 4$.
Onsala~2 ($N_{\rm samp} = 28$; $\gamma - \alpha = -0.38$) represents a source with an intermediate pixel count and a modest wavelength offset.
G31.41+0.31 ($N_{\rm samp} = 16$; $\gamma - \alpha = -0.41$) is a frequently observed JCMT standard for flux calibration and has the fewest pixels of these six examples.
In all seven sources, the running median of $\gamma$ rises steeply through the bins at low $S/N$ and changes by less than $0.3$ between the two bins with the highest $S/N(\gamma)$ in each source.
We need to consider the data quality and how each source's measurements of the spectral indices depend on $S/N$ before interpreting differences in $\gamma-\alpha$ among clouds in terms of the emitting dust.
 
\subsubsection{Sensitivity of the Population Result to an Imposed $S/N(\gamma)$ Threshold}
\label{ss:spidx_snrbias_sens}
We do not apply a cut on $S/N(\gamma)$ to the statistical sample of Paper~S2 (Paper~S2 sub-subsection~\ref{S2-sss:SpIdx_Global}).
To assess the robustness of the population result to selection by $S/N(\gamma)$, we imposed a lower threshold on the pixels of that sample and recomputed the medians of $\alpha$ and $\gamma$ for each source using the remaining pixels.
We repeated this calculation with threshold values spanning $4 \le S/N(\gamma) \le 8$.
Between 28 and 34 of the 35 sources retain at least one qualifying pixel in the tested range.
The sample median $\gamma - \alpha$ changes from $-0.51$, the value without a threshold from the measured medians in Paper~S2 table~\ref{S2-tbl:SpIdx_Global}, to values between $-0.40$ and $-0.26$.
The fraction of sources showing $\gamma < \alpha$ lies between 65\% and 79\% over the same range, compared with 71\% (25/35) at the measured medians reported in Paper~S2 sub-subsection~\ref{S2-sss:SpIdx_Global}.
The sign of the median offset is preserved at each threshold tested, so the finding $P_{450}/P_{850} < 1$ does not depend on the threshold prescription, although its precise amplitude does.

Because $S/N(\gamma) = |\gamma|/\sigma_\gamma$, increasing a uniform threshold preferentially excludes pixels with small $|\gamma|$.
Such a cut is therefore not a neutral quality criterion.

\subsubsection{Summary of Technical Validation 3}
\label{ss:noisemodel_summary}
\textit{Together with subsection~\ref{ss:noiseI}, this section establishes Pillar~2 of the technical framework.}
We use a consistent budget of statistical errors for the polarization measurements in Papers~S1 and~S2.
We combine the characterization of additive noise in Stokes~$I$ from the 174-field POL-2 archive with a two-component model of Stokes~$Q,U$ in the $Q$--$U$ plane to construct the integrated noise model.
We tested the assumptions of the model, $\sigma_Q\!\simeq\!\sigma_U$ and $|\rho_{QU}|\!\lesssim\!0.02$, using two independent methods in the wavelength and beam configurations.
We find that the additive noise scales in Stokes $Q$ and $U$, $\sigma_Q$ and $\sigma_U$, agree to within $0.03$ in $|\log_{10}(\sigma_Q/\sigma_U)|$, with an error correlation of $|\rho_{QU}|\lesssim 0.02$.
The simplified error propagation used here is therefore accurate to $1\%$ in $\sigma_\chi$ wherever the polarization is well measured.
We therefore cannot attribute outliers among individual pixels in the polarized intensity products of Papers~S1 and~S2 to statistical noise that has not been adequately quantified.
Any apparent discrepancy between the two bands above the noise envelope quantified here is, by construction, a candidate astrophysical signal rather than a residual statistical artifact.

For the dual-band spectral indices, we find that the sign of the population offset $\gamma-\alpha$ is preserved when an $S/N(\gamma)$ threshold is imposed, whereas its amplitude depends on the selection of the pixels (sub-subsection~\ref{ss:spidx_snrbias_sens}).

\subsection{Technical Validation 4: Identification and Correction of the 450\,$\mu$m Stokes-$Q/U$ Zero-Level Offset}
\label{ss:offset}
 
\subsubsection{Motivation and overview}
\label{ss:offset_motivation}
 
To compare polarization at 450 and 850\,\micron, we check the zero levels of the Stokes-$Q$ and Stokes-$U$ maps.
We identify and correct, where required, systematic additive zero-level offsets in the 450\,\micron\ Stokes~$Q$ and~$U$ intensities.
Although invisible in analyses at a single wavelength, these offsets would otherwise appear as spurious differences between the bands in the polarization observables.
Throughout this subsection, 450\,\micron\ offset correction denotes this additive correction to the Stokes-$Q$ and Stokes-$U$ maps.
We do not apply the corresponding additive correction to Stokes-$I$ and do not correct all sources uniformly.
 
\begin{figure*}[htp]
\centering
\includegraphics{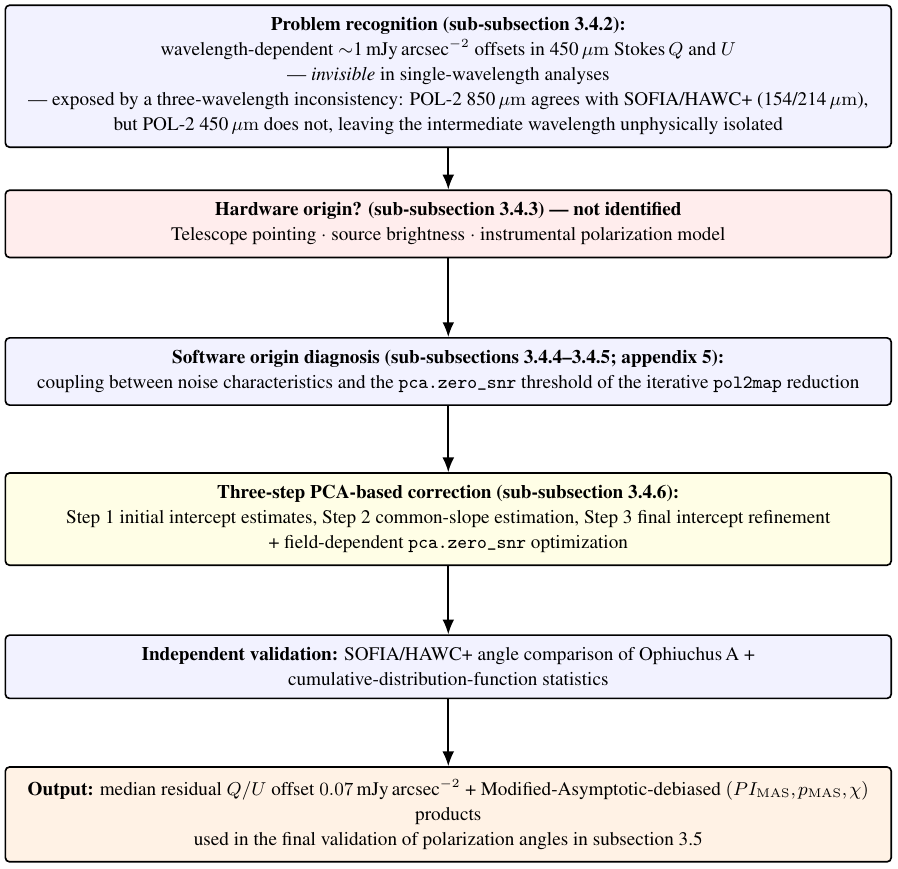}
\caption{
\textbf{Roadmap for the Pillar~3 construction stage (subsection~\ref{ss:offset}).}
From top to bottom, the boxes show problem recognition (sub-subsection~\ref{ass:Recognition}), tests of hardware origins (sub-subsection~\ref{as:HardwareInvestigation}; appendix~\ref{app:hardware_nulltests}), investigation of software origins and the three-step PCA-based correction (sub-subsections~\ref{ass:PhysicalEvidence}--\ref{ass:3step}), and independent SOFIA/HAWC+ validation (sub-subsection~\ref{ass:SOFIA_validation}).
The resulting products, debiased with the modified asymptotic estimator, enter the final validation of Pillar~3 in subsection~\ref{ss:beamval}.
}
\label{fig:T_S6_roadmap}
\end{figure*}

We develop Pillar~3, {\em Map-domain comparability}, in two stages (figure~\ref{fig:T_pillar_overview} and table~\ref{tbl:T_pillar_index}).
Here we describe the {\em construction} stage, in which we identify and correct these offsets and match the beam and spatial-frequency responses of both bands to produce the final maps at $12\farcs6$ resolution.
In subsection~\ref{ss:beamval}, we present the {\em final validation} stage and quantitatively demonstrate that these maps preserve the information on polarization angles required for the scientific analyses of the two bands.
Figure~\ref{fig:T_S6_roadmap} outlines the procedure in this subsection.
 
\subsubsection{Recognition of Systematic Inconsistency in the 450\,$\mu$m Data}
\label{ass:Recognition}
 
We begin with the comparison shown in box~1 of figure~\ref{fig:T_S6_roadmap}.
We recognized an inconsistency in the 450\,\micron\ data by comparing the $\vec{B}$-field orientations from POL-2 with those from independent SOFIA/HAWC+ Band~D (154\,\micron) polarimetry of $\rho$~Ophiuchus~A from \citet{Santos2019}.
The SOFIA/HAWC+ polarimetry has an angular resolution ($\theta_{\rm FWHM} \approx 13\farcs6$) similar to that of the POL-2 data.
The 850\,\micron\ POL-2 segments agreed with the 154\,\micron\ SOFIA orientations throughout the mapped region, but the 450\,\micron\ POL-2 segments exhibited systematic angular discrepancies of $\sim$20\arcdeg--30\arcdeg\ relative to both reference wavelengths.
We regard this pattern, in which only the intermediate wavelength is discrepant in an otherwise consistent sequence of three wavelengths, as a systematic artifact specific to the 450\,\micron\ measurements rather than an intrinsic astrophysical effect that depends on wavelength.
All three wavelengths should trace the same underlying field for optically thin dust emission.

This pattern exposes the wavelength-dependent zero-level offset in Stokes $Q/U$ at 450\,\micron\ addressed below.
We present a quantitative comparison with SOFIA/HAWC+ data at 154/214\,\micron, including Kolmogorov--Smirnov tests, in sub-subsection~\ref{ass:SOFIA_validation}.
There, we assess the discrepancy in the 450\,\micron\ data before the $Q/U$ zero-level offset correction and test whether that correction resolves it.

\subsubsection{Null Tests for Hardware Origins}
\label{as:HardwareInvestigation}
 
We tested the hypothesis of a hardware origin shown in box~2 of the section roadmap (figure~\ref{fig:T_S6_roadmap}).
Dedicated null tests examined three categories of instrumental effects as possible origins of the zero-level offset in Stokes $Q/U$ at 450\,\micron: telescope pointing and attitude parameters, dependence on source brightness, and the elevation-dependent IP model of \citet{Friberg2018}.
None of the three tests reproduces the observed $\gtrsim 1$\,mJy\,arcsec$^{-2}$ offsets.
The full diagnostic figures, methodology for each test, and quantitative tolerances are presented in appendix~\ref{app:hardware_nulltests}.
These multiple independent null results redirect the investigation to the data-reduction pipeline.

Boxes~3--6 of figure~\ref{fig:T_S6_roadmap} summarize the investigation of software origins.
The next three sub-subsections (\ref{ass:PhysicalEvidence}--\ref{ass:3step}) present the evidence for the offsets, the astrophysical constraints requiring zero intercepts in comparisons of Stokes parameters between the bands, and the three-step correction, which estimates and subtracts the intercepts (see Paper~S1 section~\ref{S1-sec:obs}).
We then present the independent SOFIA/HAWC+ validation and the products debiased with the modified asymptotic estimator that are used in subsection~\ref{ss:beamval}.

\subsubsection{Observational Evidence for Systematic Stokes-$Q/U$ Zero-Level Offsets}
\label{ass:PhysicalEvidence}
 
Appendix~\ref{app:hardware_nulltests} demonstrates that significant zero-level offsets in Stokes $Q/U$ at 450\,$\mu$m are detectable in data from individual MSBs, before coaddition.
These offsets in individual MSBs show no dependence on telescope azimuth, elevation, parallactic angle, or tracking geometry (appendix figure~\ref{afig:AzEl_offset_maps}).
For the seven bright fields examined there, we therefore infer that they are unlikely to originate from effects related to pointing or attitude (see also appendix~\ref{ass:HardwareSummary}).

\begin{figure*}[tp]
\centering
\includegraphics[width=0.96\textwidth]{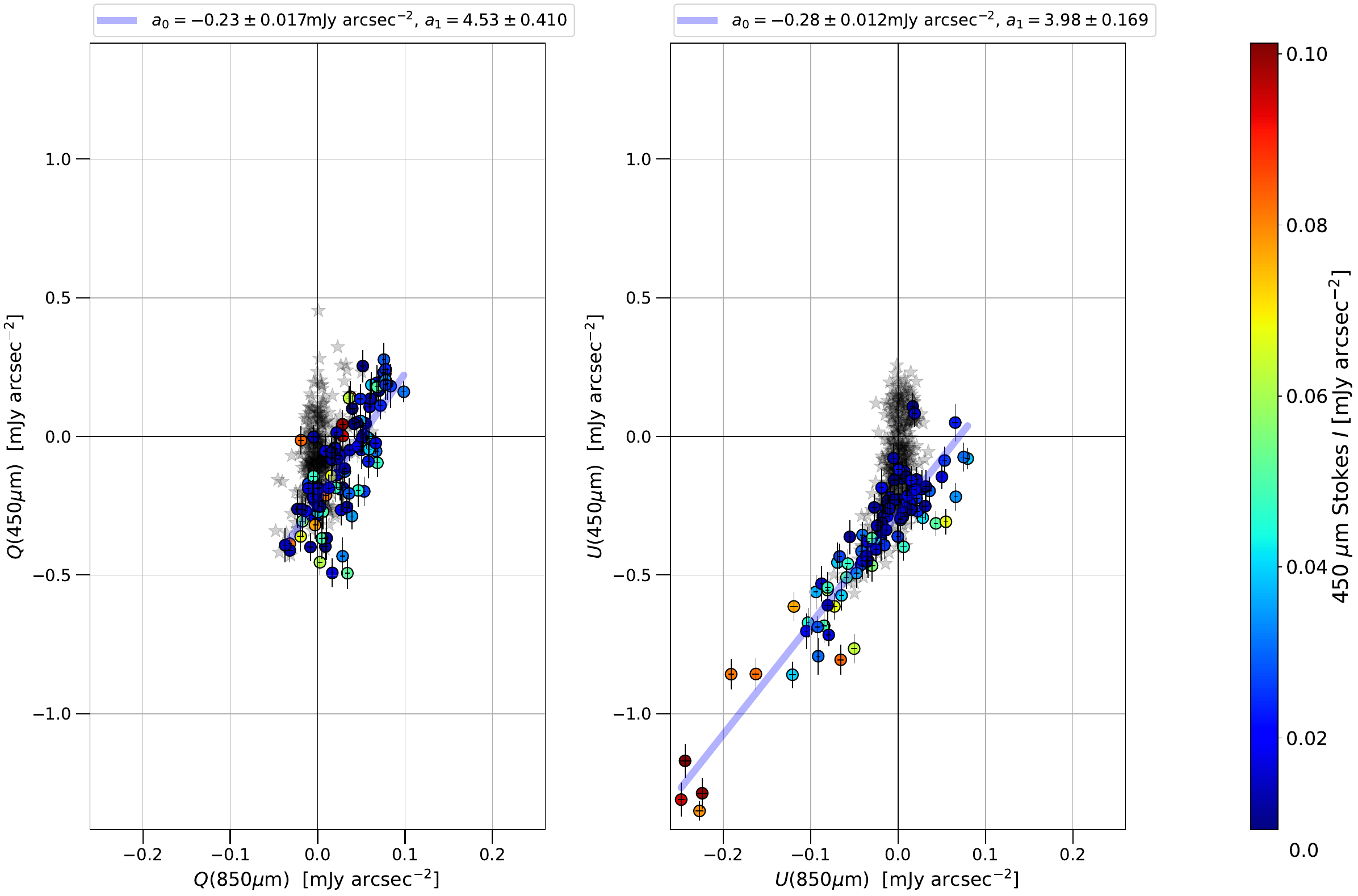}
\caption{Initial regression analysis for Ophiuchus~A (Step~1 of the correction for zero-level offsets in Stokes $Q/U$ at 450\,$\mu$m).
The panels show the relations between $Q_{450}$ and $Q_{850}$ (left) and between $U_{450}$ and $U_{850}$ (right) after matching the beams, with colors indicating the 450\,$\mu$m intensities.
In the underlying maps, the pixel centers are spaced by $12\farcs0$, nearly one beam width ($12\farcs6$ effective FWHM).
Weighted fits (\texttt{curve\_fit}) yield significant non-zero intercepts: $a_0^Q = -0.23\pm0.02\ \mathrm{mJy\ arcsec^{-2}}$ and $a_0^U = -0.28\pm0.01\ \mathrm{mJy\ arcsec^{-2}}$ (equation~\ref{aeq:OphA_intercepts}), indicating raw zero-level offsets at 450\,$\mu$m.
These non-zero intercepts depart from the proportional-emission model described in appendix~\ref{app:zero_intercept_assumptions} and motivate the investigation of the offsets and the three-step correction in sub-subsection~\ref{ass:3step}.
}
\label{afig:Q8U8vsQ4U4_initialEstimate}
\end{figure*}

After matching the beam size and spatial-frequency range of the two bands with \texttt{pol2map} (\citealt{Parsons2021}; appendix~\ref{app:reduction_workflow}), systematic $Q/U$ offsets become evident in the coadded maps.
For every bright field, linear regressions of $(Q_{450}, Q_{850})$ and $(U_{450}, U_{850})$ reveal significant, repeatable non-zero intercepts.
Figure~\ref{afig:Q8U8vsQ4U4_initialEstimate} illustrates this behavior for Ophiuchus~A, one of the regions with the highest signal-to-noise ratios in our sample.

We compared regression estimators based on unweighted least squares, bootstrap resampling of the residuals, a principal-axis (eigenvector) fit, and orthogonal distance regression.
We adopted weighted least squares for the offset estimates presented here.
For Ophiuchus~A, we used \texttt{scipy.curve\_fit} to obtain the intercepts of the regressions of $Q_{450}$ on $Q_{850}$ and of $U_{450}$ on $U_{850}$ (equation~\ref{aeq:Qfit_initial}),
\begin{equation}
a_0^Q = -0.23 \pm 0.02\ \mathrm{mJy\ arcsec^{-2}}, \qquad
a_0^U = -0.28 \pm 0.01\ \mathrm{mJy\ arcsec^{-2}},
\label{aeq:OphA_intercepts}
\end{equation}
with corresponding slopes
\begin{equation}
a_1^Q = 4.53 \pm 0.41, \qquad
a_1^U = 3.98 \pm 0.17.
\label{aeq:OphA_slopes}
\end{equation}
Both intercepts differ from zero at $\gtrsim 10\sigma$.
This result demonstrates wavelength-dependent zero-level shifts in the 450\,$\mu$m Stokes parameters relative to 850\,$\mu$m.
We discuss the physical interpretation of these non-zero intercepts in sub-subsection~\ref{ass:PhysicalConstraints}.

\subsubsection{Physical Basis of the Zero-Intercept Criterion}
\label{ass:PhysicalConstraints}
The zero-intercept condition relies on assumptions about both the polarized dust emission and its spatial weighting.
We assume that the emission is optically thin at both wavelengths and that the directions of grain alignment do not depend on wavelength.
We also require a sufficiently coherent polarization geometry along each line of sight and within the beam, with a compatible relative weighting of the emitting components at the two wavelengths.
 
If the ratio of polarized intensities is also approximately constant over the fitted region, the Stokes~$Q,U$ maps of the source emission are approximately proportional between the two bands.
Within this model, $Q_{850}=U_{850}=0$ implies $Q_{450}=U_{450}=0$, and the regressions of $Q_{450}$ versus $Q_{850}$ and of $U_{450}$ versus $U_{850}$ are expected to pass through the origin.
The zero-intercept condition is therefore a diagnostic based on these assumptions, rather than a consequence of optical thinness alone.
 
The observed non-zero intercepts (equation~\ref{aeq:OphA_intercepts}) indicate a departure from this proportional model.
We regard a non-astrophysical origin as the most plausible interpretation, given their recurrence across many fields.
This interpretation, together with the hardware tests in sub-subsection~\ref{as:HardwareInvestigation}, motivates our investigation of residual processing effects and the correction in sub-subsection~\ref{ass:3step}.
Appendix~\ref{app:zero_intercept_assumptions} discusses the astrophysical assumptions, potential sources of contamination, and limitations of this interpretation, together with the supporting references.

\subsubsection{Three-Step Offset Correction Procedure}
\label{ass:3step}
 
We implemented a three-step procedure to isolate and remove the additive $Q/U$ offsets at 450\,$\mu$m identified in sub-subsection~\ref{ass:PhysicalEvidence}.
To improve statistical precision, we require the polarization angles to be independent of wavelength (constraint~\ref{assumption:alignmentdirection} in appendix~\ref{app:zero_intercept_assumptions}).
We applied the MAS debiasing of \citet{Plaszczynski2014} and \citet{Montier2015} to the polarized intensity after this correction (appendix~\ref{app:mas_debias}).
 
\paragraph{Step 1: Initial Intercept Estimates}
 
We carried out a weighted least-squares regression of the following form:
\begin{equation}
Q_{450} = a_0^Q + a_1^Q Q_{850} + \epsilon_Q, \qquad
U_{450} = a_0^U + a_1^U U_{850} + \epsilon_U,
\label{aeq:Qfit_initial}
\end{equation}
where, at each pixel, $\epsilon_Q$ and $\epsilon_U$ are the residuals in $Q_{450}$ and $U_{450}$ that are not explained by the linear model.
We assume that they have zero mean and are dominated by measurement noise.
We use inverse-variance weights $1/\sigma_{Q,450}^2$ and $1/\sigma_{U,450}^2$.
The regression yields signed intercepts $(a_0^Q, a_0^U)$ and slopes $(a_1^Q, a_1^U)$ for each field, as illustrated in figure~\ref{afig:Q8U8vsQ4U4_initialEstimate} for Ophiuchus~A.
We interpret the signed intercepts as raw zero-level offsets at 450\,$\mu$m that violate the astrophysical constraints in appendix~\ref{app:zero_intercept_assumptions}.
 
\paragraph{Step 2: Estimation of a Common Slope}
 
\begin{figure*}[tp]
\centering
\includegraphics[width=0.48\textwidth]{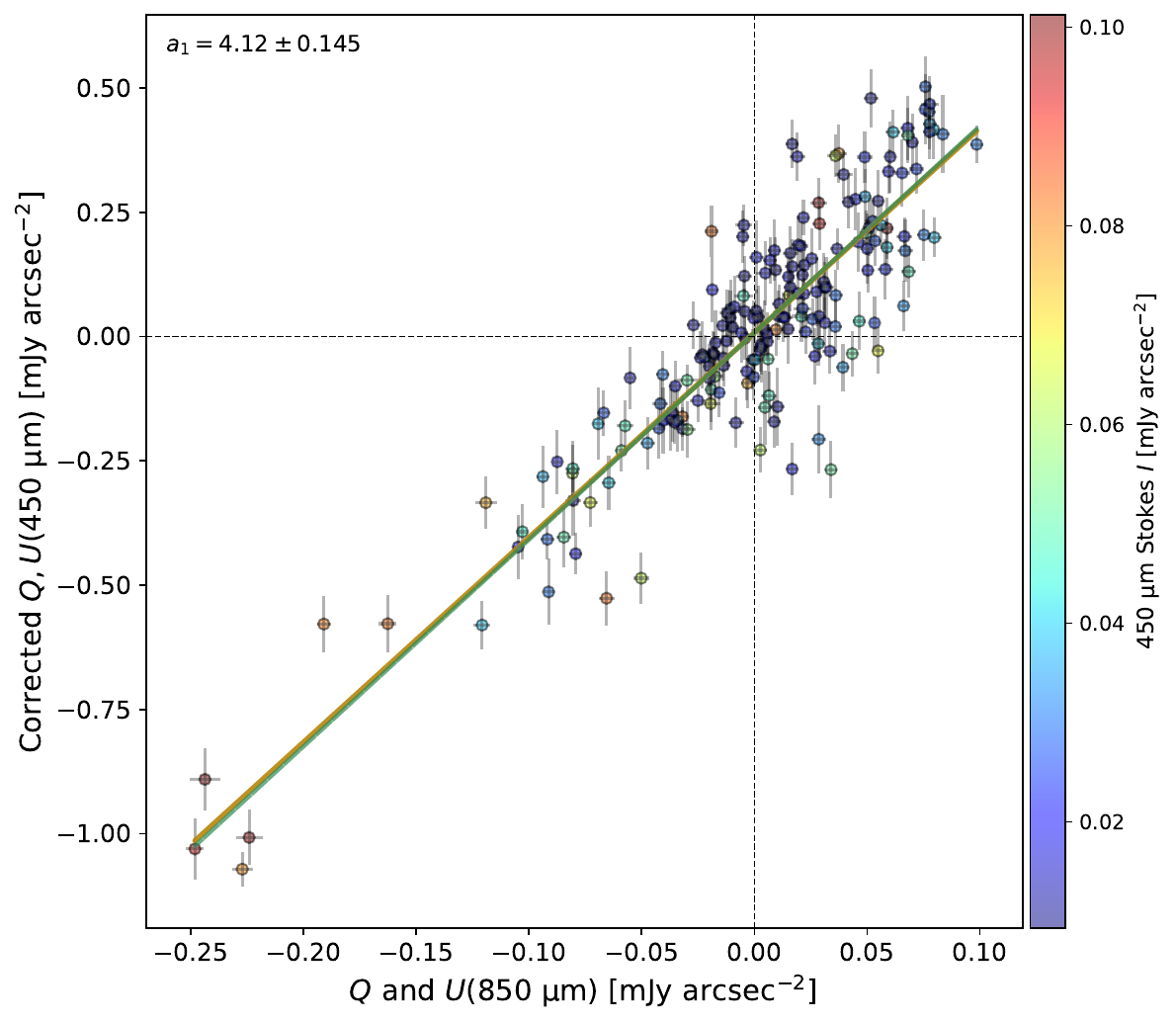}
\caption{
Combined regression of the Stokes parameters after subtraction of the intercepts for Ophiuchus~A (Step~2 of the three-step procedure).
We merged the $(Q_{450}^{\rm corr}, Q_{850})$ and $(U_{450}^{\rm corr}, U_{850})$ samples after subtracting the Step~1 intercepts (equation~\ref{aeq:QUcorr_step2}).
The combined regression yields a common slope $a_1^{\rm common} = 4.120 \pm 0.145$, a reduction in the slope uncertainty by factors of $2.8$ and $1.2$ compared to the individual slopes from Step~1 (equation~\ref{aeq:OphA_slopes}: $a_1^Q = 4.53 \pm 0.41$, $a_1^U = 3.98 \pm 0.17$).
This improvement in precision reflects the doubled sample size from combining the $Q$ and $U$ domains, which encode the same underlying polarization pattern (constraint~\ref{assumption:alignmentdirection} in appendix~\ref{app:zero_intercept_assumptions}) with independent noise realizations.
}
\label{afig:QU8vsQU4combined}
\end{figure*}

Constraint~(\ref{assumption:alignmentdirection}) in appendix~\ref{app:zero_intercept_assumptions} requires that the polarization position angle be independent of wavelength.
The constraint implies that the proportionality between the Stokes parameters at 450 and 850\,$\mu$m must be identical for the $Q$ and $U$ components: $a_1^Q = a_1^U \equiv a_1^{\rm common}$.
After subtracting the Step~1 intercepts:
\begin{equation}
Q_{450}^{\rm corr} = Q_{450} - a_0^Q, \qquad 
U_{450}^{\rm corr} = U_{450} - a_0^U,
\label{aeq:QUcorr_step2}
\end{equation}
we combined the $(Q_{450}^{\rm corr}, Q_{850})$ and $(U_{450}^{\rm corr}, U_{850})$ samples to derive a single common slope $a_1^{\rm common}$ through joint regression (figure~\ref{afig:QU8vsQU4combined}).
Combining the samples with the slope constrained to be the same reduces the slope uncertainty by factors of $2.8$ and $1.2$ relative to the separate Step~1 estimates of $a_1^Q$ and $a_1^U$, respectively (equation~\ref{aeq:OphA_slopes}).
 
\paragraph{Step 3: Final Intercept Refinement}
 
Holding $a_1^{\rm common}$ fixed, we re-estimated the intercepts using one-parameter fits:
\begin{equation}
Q_{450} = a_0^{Q,\rm final} + a_1^{\rm common} Q_{850}, \qquad
U_{450} = a_0^{U,\rm final} + a_1^{\rm common} U_{850}.
\label{aeq:Qfinal}
\end{equation}
Fixing the slope reduces the number of fitted parameters and tightens the uncertainties in the intercepts, yielding the final offset corrections $(a_0^{Q,\rm final}, a_0^{U,\rm final})$ applied to the 450\,$\mu$m maps.
The corrected Stokes parameters used for scientific analysis are then:
\begin{equation}
Q_{450}^{\rm corr} = Q_{450} - a_0^{Q,\rm final}, \qquad
U_{450}^{\rm corr} = U_{450} - a_0^{U,\rm final}.
\label{aeq:Q450corr_final}
\end{equation}
These corrected parameters satisfy the astrophysical requirement (derived in sub-subsection~\ref{ass:PhysicalConstraints}) that the regression intercepts be zero, enabling robust polarization analysis that combines the two bands.

\subsubsection{Application to Ophiuchus A}
\label{ass:OphA_application}
\begin{figure*}[tp]
\centering
\includegraphics[width=0.96\textwidth]{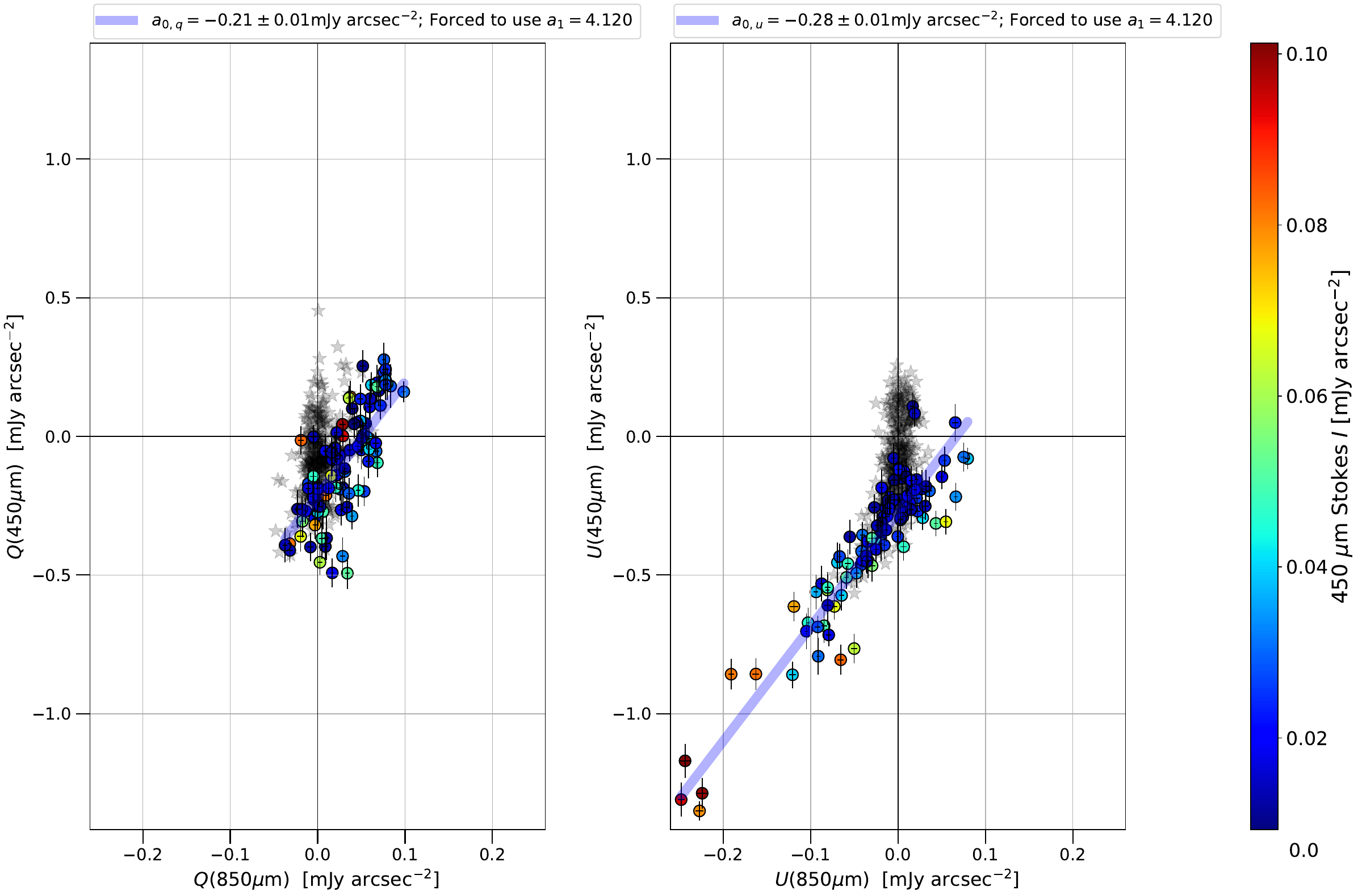}
\caption{Final intercept refinement for Ophiuchus~A (Step~3 of the three-step procedure), using the common slope $a_1^{\rm common} = 4.120$ derived in Step~2 (equation~\ref{aeq:OphA_commonslope}; figure~\ref{afig:QU8vsQU4combined}).
With the slope fixed, regressions reduce to one-parameter fits yielding $a_0^{Q,\rm final} = -0.21 \pm 0.01\ \mathrm{mJy\ arcsec^{-2}}$ and $a_0^{U,\rm final} = -0.28 \pm 0.01\ \mathrm{mJy\ arcsec^{-2}}$ (equation~\ref{aeq:OphA_finalintercepts}).
Subtracting these intercepts (equation~\ref{aeq:Q450corr_final}) leaves residual intercepts of $|a_0^{\rm sc}|\lesssim0.02$\,mJy\,arcsec$^{-2}$ (sub-subsection~\ref{ass:ResidualOffsets}).
}
\label{afig:Q8U8vsQ4U4_forcedSlope}
\end{figure*}

For Ophiuchus~A, one of the regions with the highest $S/N$ in the sample, the three-step procedure converges as follows.
Step~1 gives raw intercepts $a_0^Q = -0.23 \pm 0.02$\,mJy\,arcsec$^{-2}$ and $a_0^U = -0.28 \pm 0.01$\,mJy\,arcsec$^{-2}$, with separate slopes $a_1^Q = 4.53 \pm 0.41$ and $a_1^U = 3.98 \pm 0.17$ (equations~\ref{aeq:OphA_intercepts}, \ref{aeq:OphA_slopes}; figure~\ref{afig:Q8U8vsQ4U4_initialEstimate}).
Step~2 gives the more precise common slope $a_1^{\rm common} = 4.120 \pm 0.145$ (equation~\ref{aeq:OphA_commonslope}; figure~\ref{afig:QU8vsQU4combined}).
Step~3 gives the final intercept corrections $a_0^{Q,\rm final} = -0.21 \pm 0.01$ and $a_0^{U,\rm final} = -0.28 \pm 0.01$\,mJy\,arcsec$^{-2}$ (equation~\ref{aeq:OphA_finalintercepts}; figure~\ref{afig:Q8U8vsQ4U4_forcedSlope}).
We have subtracted these corrections from the 450\,\micron\ Stokes parameters obtained with the default threshold using equation~(\ref{aeq:Q450corr_final}).
The subtraction demonstrates the procedure.
The scientific analyses of Ophiuchus~A, a Category~1 field, use the map reduced with its optimized threshold, without subtraction (sub-subsection~\ref{ass:ResidualOffsets}).
The corrected maps of Ophiuchus~A have residual intercepts $|a_0^{\rm sc}| \lesssim 0.02$\,mJy\,arcsec$^{-2}$ in regressions with a constrained slope (sub-subsection~\ref{ass:ResidualOffsets}), consistent with zero at the $\lesssim 2\sigma$ level relative to the statistical uncertainty per fit.
The procedure preserves the wavelength-independent polarization pattern required by the astrophysical constraints of sub-subsection~\ref{ass:PhysicalConstraints}.
In particular, $Q_{450}$ and $U_{450}$ vanish wherever their 850\,\micron\ counterparts do (equation~\ref{aeq:Q450corr_final}).
 
\begin{equation}
a_1^{\rm common} = 4.120 \pm 0.145 ,
\label{aeq:OphA_commonslope}
\end{equation}
 
\begin{equation}
a_0^{Q,\rm final} = -0.21 \pm 0.01\ \mathrm{mJy\ arcsec^{-2}}, \qquad
a_0^{U,\rm final} = -0.28 \pm 0.01\ \mathrm{mJy\ arcsec^{-2}}.
\label{aeq:OphA_finalintercepts}
\end{equation}

Appendix~\ref{app:pca_grid_atlas} describes how we optimize the PCA threshold.
It discusses the empirical origin of the offsets in the masking and Fourier-domain interpolation in the filtering (FLT), principal-component analysis (PCA), and common-mode subtraction (COM) stages of \texttt{makemap} \citep{Chapin2013}.
It also gives the formalism for the grid search (equations~\ref{eq:a0tot_def}--\ref{eq:sigmaeff_def}), the parameter grid (appendix table~\ref{tab:PCAparams}), the mosaicking limitations that restrict application to single fields, and the distribution of the optimal thresholds.
Figures~\ref{afig:PCAcutoff_cat1_p1}--\ref{afig:PCAcutoff_cat3} show representative fields of Categories~1--3.
We summarize the resulting field categories below.

\begin{figure*}[tp]
\centering
\includegraphics{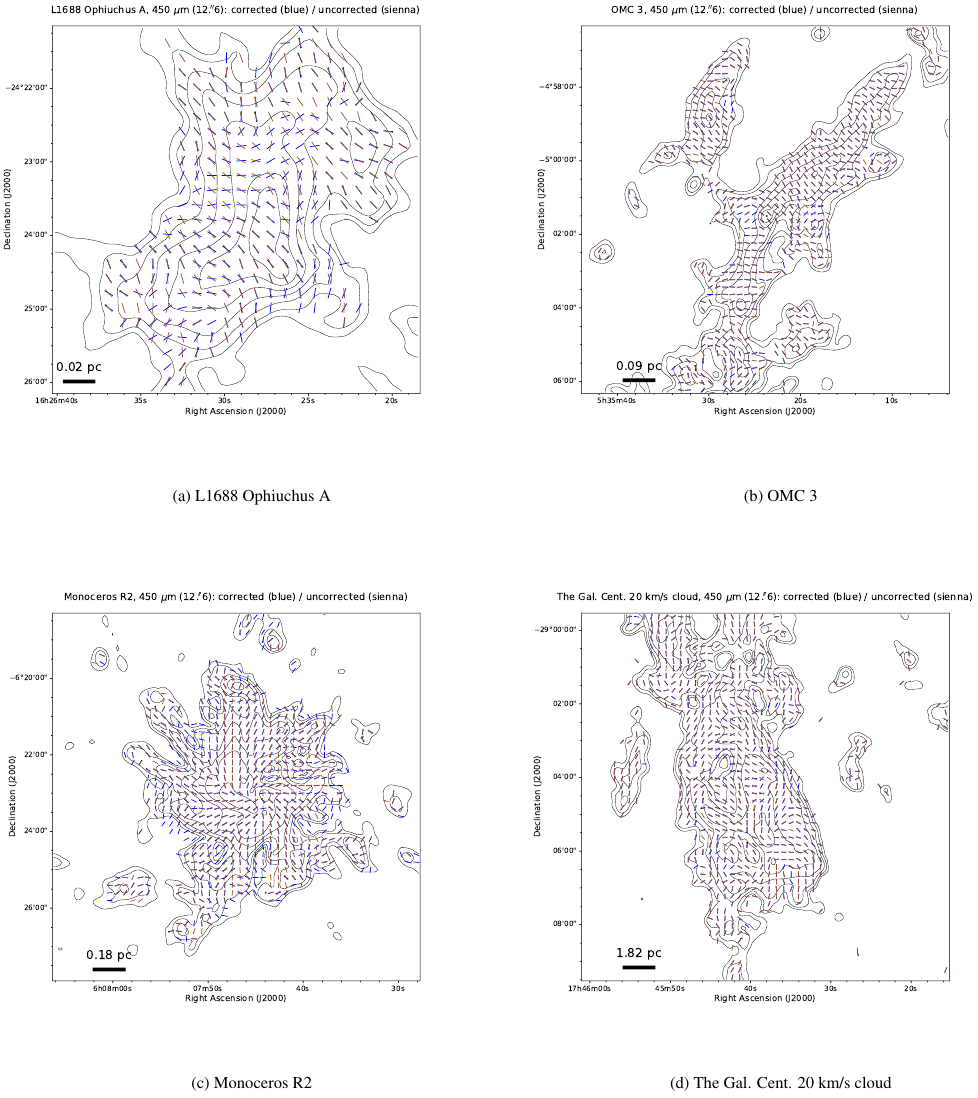}
\caption{
Comparison of $\vec{B}$-field orientations at 450\,$\mu$m before the correction for zero-level offsets in Stokes $Q/U$ (brown) and after the correction procedure (blue), for four representative targets observed as single fields.
Magnetic field segments are shown only where $I/\sigma_I > 30$, overlaid on 450\,$\mu$m Stokes $I$ contours at $10 \times 2^n \sigma_I$ for $n = 0, 1, \dots, 10$.
The corrected maps of these Category~1 fields use optimized \texttt{pca.zero\_snr} thresholds, without subtraction of the intercepts from the three-step procedure (sub-subsection~\ref{ass:3step}).
Offset correction primarily affects regions of low intensity while preserving the $\vec{B}$-field structure in bright emission.
}
\label{fig:450_singlefield_before_after}
\end{figure*}

\begin{figure*}[tp]
\ContinuedFloat
\centering
\includegraphics{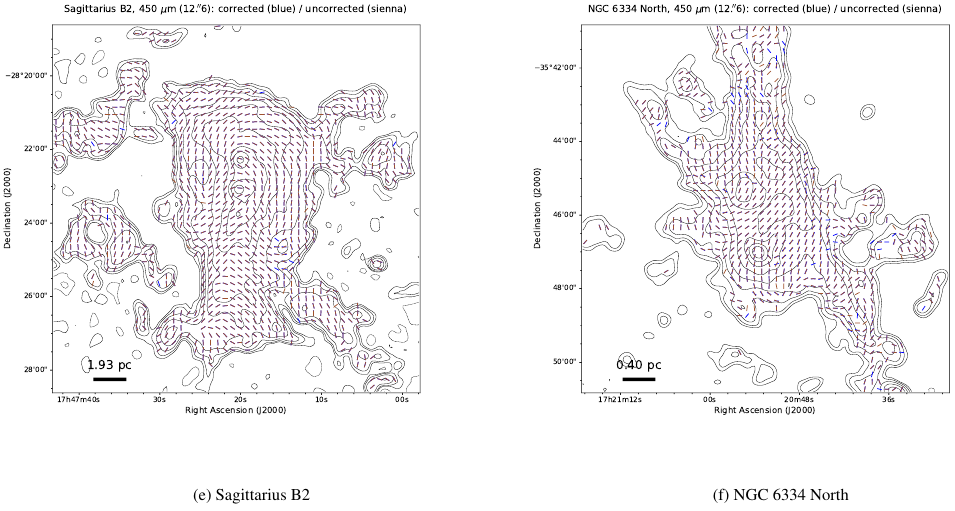}
\caption{
(Continued.)
Same as figure~\ref{fig:450_singlefield_before_after}, but for two distant regions with high dynamic range.
Offset correction produces the largest fractional changes in the faint outskirts of the clouds ($I/\sigma_I \lesssim 100$), while dense cores remain essentially unchanged.
These examples show the intensity regime in which the 450\,$\mu$m zero-level correction affects polarimetry in the two bands.
}
\end{figure*}

\subsubsection{Offset Significance and Field Categorization}
\label{ass:offset_significance}
 
\paragraph{Offset Diagnostic and Reduction Threshold}
We assess additive offsets in Stokes $Q/U$ relative to the adopted noise scale and classify fields by their response to the tested reduction thresholds.
A small additive baseline in $Q$ or $U$ can rotate the inferred field in a faint cloud envelope while making little difference to a bright polarized peak.
We compared the combined offset $a_0^{\rm tot} = [(a_0^Q)^2 + (a_0^U)^2]^{1/2}$ with the effective noise $\sigma_{\rm eff} = (\sigma_Q\,\sigma_U)^{1/2}$ (equations~\ref{eq:a0tot_def}--\ref{eq:sigmaeff_def} in appendix~\ref{app:pca_grid_atlas}) across the threshold grid.
The comparison classifies the 119 fields surveyed into three categories (representative fields in figures~\ref{afig:PCAcutoff_cat1_p1}--\ref{afig:PCAcutoff_cat3}).

The parameter \texttt{pca.zero\_snr} sets the signal-to-noise ratio below which \texttt{pol2map} treats a map pixel as blank and zeros it before the principal-component background is estimated.
The default threshold of three treats pixels below $3\sigma$ as free of sources.
We attribute the 450\,\micron\ zero-level offset to this low threshold rather than to PCA itself.
 
\paragraph{Field Classification}
We first require the offsets at the default threshold $\texttt{pca.zero\_snr}=3$ to satisfy $|a_0^Q| \leq \sigma_{\rm eff}$ and $|a_0^U| \leq \sigma_{\rm eff}$.
We assign fields that fail this test, such as Ophiuchus~A and OMC~3, to Category~1 regardless of their response at higher thresholds.
Among the remaining fields, we also assign Category~1 when the minimum $a_0^{\rm tot}$ across the grid exceeds $\sigma_{\rm eff}$.
For the other fields, whose minimum $a_0^{\rm tot}$ does not exceed $\sigma_{\rm eff}$, we assign Category~2 when the minimum occurs at a threshold other than three.
We assign Category~3 when the minimum occurs at the default threshold of three.
Applying these criteria places 43 fields in Category~1, 51 in Category~2, and 25 in Category~3.
The categories describe the response of a map baseline to the tested reduction choices, not the intrinsic classes of molecular clouds.
They identify where a correction or a separate assessment of the uncertainty is needed before the faint polarization structure is compared between the bands.
 
\paragraph{Coaddition and Scope of Application}
These results agree with the azimuth--elevation diagnostics for individual MSBs in appendix figure~\ref{afig:AzEl_offset_maps}.
Several fields that appear free of offsets in individual MSBs, including OMC~1, NGC~6334 field~2 as a single field, and NGC~7538, develop measurable offsets after coaddition.
We interpret this behavior as evidence that the residuals of the PCA interpolation accumulate nonlinearly.

The offsets of the mosaicked and mosaic-component products depend only weakly on the PCA threshold.
This weak dependence restricts the reliable use of the optimization framework to reductions of individual fields, as documented in appendix~\ref{app:pca_grid_atlas}.
We nevertheless classify these products with the same diagnostics and select their thresholds by the same rule.
Sub-subsection~\ref{ass:ResidualOffsets} states how each product is corrected.
We present diagnostics of these failure modes for OMC~1, NGC~1333, and NGC~2264 in appendix~\ref{app:mosaicking_diagnostics}.
 
We produced the final maps for scientific analysis with the standard \texttt{pol2map} three-step pipeline and two modifications chosen for each field.
We set \texttt{flt.filt\_edge\_largescale=300}, together with \texttt{smooth450=yes}, to enforce the common range of angular scales from 12\farcs6 to 300\arcsec.
For Category~1 fields, we adopted the \texttt{pca.zero\_snr}/\texttt{pca.zero\_snrlo} pair from the grid in appendix table~\ref{tab:PCAparams} that gives the smallest $a_0^{\rm tot}$.
Category~2 and~3 fields keep the default thresholds.
The complete \texttt{pol2map} commands needed to reproduce these maps, together with the choices for debiasing (sub-subsection~\ref{ass:ResidualOffsets}) and flux calibration with \texttt{JY=no}, are documented in appendix~\ref{app:reduction_workflow}.

\subsubsection{The Corrected 450\,$\mu$m Stokes Parameters and Their Residual Offsets}
\label{ass:ResidualOffsets}

We suppressed the dominant wavelength-dependent zero-level biases of the 450\,\micron\ data by matching the beams and spatial-frequency responses.
We then corrected the remaining offsets, where possible, either by using maps reduced with optimized thresholds or by subtracting the final intercepts of the three-step regression (sub-subsection~\ref{ass:3step}).
For maps corrected by intercept subtraction, the corrected Stokes parameters are
\begin{equation}
Q_{450}^{\rm corr} = Q_{450} - a_0^{Q,\rm final}, \qquad
U_{450}^{\rm corr} = U_{450} - a_0^{U,\rm final} ,
\label{eq:Q450corr_definition}
\end{equation}
where $(a_0^{Q,\rm final}, a_0^{U,\rm final})$ are the final intercepts of equation~(\ref{aeq:Qfinal}).

In the 35-source {\sc SpIdx} sample (table~\ref{tbl:T_sample_inventory}), we subtracted these intercepts in 17 fields.
The 17 fields comprise 14 Category~2 fields, 2 Category~3 fields, and 1 Category~1 field.
The mosaicked field G28.34+0.06 is one of these 17 fields.
We used maps reduced with optimized thresholds, without intercept subtraction, for the 14 Category~1 fields whose smallest $a_0^{\rm tot}$ occurs at a nondefault threshold (appendix table~\ref{tab:PCAparams}).
For the remaining four mosaicked fields without a three-step fit, we used the reduced maps without intercept subtraction.
The four fields are PGCC~G203.21$-$11.20, Serpens~Main, G202.3$+$2.5, and the Rosette Molecular Cloud.

The regression with the slope constrained produces small additional residual intercepts $(a_0^{Q,\rm sc}, a_0^{U,\rm sc})$ that we document but do not subtract.
For the 42-source single-field sample, we summarize the residual intercepts by their sample means, median absolute amplitudes, and mean uncertainties from the fits:
\begin{equation}
\langle a_0^{Q,\rm sc} \rangle = -0.033, \quad
\langle a_0^{U,\rm sc} \rangle = -0.071 \quad ({\rm mJy\,arcsec^{-2}}) ,
\label{eq:residual_mean}
\end{equation}
\begin{equation}
\mathrm{median}\,|a_0^{Q,\rm sc}| = 0.037, \quad
\mathrm{median}\,|a_0^{U,\rm sc}| = 0.053 \quad ({\rm mJy\,arcsec^{-2}}) ,
\label{eq:residual_median}
\end{equation}
\begin{equation}
\langle \Delta a_0^{Q,\rm sc} \rangle \simeq 0.036, \quad
\langle \Delta a_0^{U,\rm sc} \rangle \simeq 0.037 \quad ({\rm mJy\,arcsec^{-2}}) .
\label{eq:residual_uncertainty}
\end{equation}
We define the combined residual amplitude as $a_0^{\rm res} = \sqrt{(a_0^{Q,\rm sc})^2 + (a_0^{U,\rm sc})^2}$.
The median of $a_0^{\rm res}$ is $6.7 \times 10^{-2}$\,mJy\,arcsec$^{-2}$, with an interquartile range of $(4.4$--$13) \times 10^{-2}$\,mJy\,arcsec$^{-2}$.
The median absolute residual amplitudes are comparable to the statistical uncertainties of individual fits (equation~\ref{eq:residual_uncertainty}), consistent with a substantial contribution from statistical noise.

As a representative case, Ophiuchus~A yields $a_0^{Q,\rm sc} = 0.009 \pm 0.013$ and $a_0^{U,\rm sc} = 0.013 \pm 0.008$\,mJy\,arcsec$^{-2}$.
Both residual intercepts are consistent with zero within $2\sigma$ and are an order of magnitude below the raw intercepts of equation~(\ref{aeq:Qfit_initial}).
We therefore use the maps corrected according to equation~(\ref{eq:Q450corr_definition}), the maps reduced with optimized thresholds, and the four mosaicked maps without a three-step fit for the scientific analyses.
We do not subtract the additional residual intercepts from the regression with a constrained slope.
 
Figure~\ref{fig:450_singlefield_before_after} shows how optimizing the PCA threshold affects the maps of the $\vec{B}$-field for six representative Category~1 targets observed as single fields, spanning Ophiuchus~A (adopted distance $d = 139$\,pc) to Sgr~B2 (adopted distance $d = 8340$\,pc).
We obtain segments showing the $\vec{B}$-field by rotating the polarization angle by $90\degr$, assuming optically thin thermal dust emission \citep{Hildebrand1988, Andersson2015}:
\begin{equation}
\chi = \tfrac{1}{2}\arctan\!\left(\frac{U_{450}^{\rm corr}}{Q_{450}^{\rm corr}}\right) , \qquad
\theta_{\rm B} = \chi + \tfrac{\pi}{2} ,
\label{eq:chi_def}
\end{equation}
with $(Q_{450}^{\rm corr}, U_{450}^{\rm corr})$ defined in equation~(\ref{eq:Q450corr_definition}) (or, for maps reduced with optimized thresholds, the Stokes parameters as reduced) and segments shown only at $S/N \equiv I/\sigma_I > 30$.
Both $\chi$ and $\theta_{\rm B}$ are defined modulo $180\degr$, so we plot segments rather than vectors.
We compare maps made with the default $\texttt{pca.zero\_snr} = 3$ (brown segments) with maps corrected using the threshold optimized for each field (blue segments).
The contours of Stokes~$I$ at $10 \times 2^n\,\sigma_I$ ($n = 0$--10) show the intensity structure.
The correction mainly affects regions of low intensity ($\lesssim 100\sigma_I$), where the zero-level offset contributes polarization comparable to or greater than the intrinsic signal.
With the optimized PCA threshold, the patterns and azimuthal distortions seen in faint regions at the default threshold are absent, and cloud cores with high $S/N$ ($\gtrsim 160\sigma_I$) are essentially unchanged.
This behavior is consistent across all six fields, including the Galactic-Center 20\,km\,s$^{-1}$ cloud and Sgr~B2 at kiloparsec distances.
We independently validate the Ophiuchus~A result against SOFIA/HAWC+ 154\,\micron\ polarimetry in sub-subsection~\ref{ass:SOFIA_validation}.
 
We derive the final polarimetric products using the Modified Asymptotic Estimator of \citet{Plaszczynski2014} and \citet{Montier2015}: the debiased polarized intensity $PI_{\rm MAS}$, the debiased polarization fraction $p_{\rm MAS}$, and the polarization angle $\chi$.
We use the Stokes parameters adopted above (equation~\ref{eq:Q450corr_definition}, the maps reduced with optimized thresholds, or the four mosaicked maps without a three-step fit).
The noise term weighted by the Stokes components, the explicit MAS expression, and the corresponding debiased polarization fraction $p_{\rm MAS}$ and angle $\chi$ are documented in appendix~\ref{app:mas_debias}.

\subsubsection{Independent Validation Using SOFIA/HAWC+ Polarimetry}
\label{ass:SOFIA_validation}
 
To verify the corrected 450\,$\mu$m data independently, we compared the $\vec{B}$-field orientations with those obtained from SOFIA/HAWC+ observations at 154\,$\mu$m (Band~D) and 214\,$\mu$m (Band~E).
The preceding sub-subsection compared the maps before and after the 450\,$\mu$m offset correction.
The maps differ in regions of low intensity, and the $\vec{B}$-field structures in cloud cores with high $S/N$ are preserved.
The 450\,$\mu$m wavelength lies between the POL-2 850\,$\mu$m and SOFIA bands.
This comparison therefore tests directly whether the offset correction resolves the discrepancy between the 450\,$\mu$m $\vec{B}$-field orientations and those at the neighboring wavelengths, first identified in sub-subsection~\ref{ass:Recognition}.

\begin{figure*}[tp]
\centering
\includegraphics{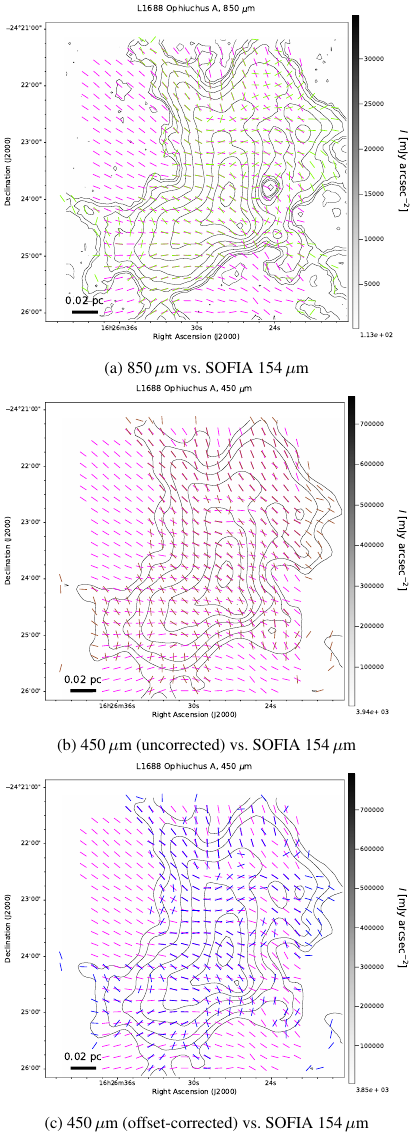}
\caption{
Comparison of $\vec{B}$-field orientations in the central region of L1688 Ophiuchus~A at 850\,$\mu$m (green), at 450\,$\mu$m before (brown) and after (blue) the Stokes-$Q/U$ zero-level offset correction, and from SOFIA/HAWC+ Band~D at 154\,$\mu$m (magenta).
The POL-2 segments are shown where $I/\sigma_I > 10$ and $PI/\sigma_{PI} > 1$, and the SOFIA segments where $I/\sigma_I > 200$, $PI/\sigma_{PI} > 3$, and $P < 50\%$.
These thresholds apply to the displayed segments only.
All segments are displayed over the corresponding POL-2 Stokes~$I$ contours, which begin at $10\sigma$ with $\sigma$ estimated from blank-sky regions.
Panel~(a) shows the good agreement between the 850\,$\mu$m and 154\,$\mu$m orientations.
Panel~(b) reveals a systematic difference in polarization angle between the 450\,$\mu$m segments before the correction and the 154\,$\mu$m segments across the bright ridge.
In panel~(c), the orientations in the 450\,$\mu$m map after the offset correction are consistent with those at 154\,$\mu$m and 850\,$\mu$m (sub-subsection~\ref{ass:3step}).
Figure~\ref{afig:SOFIA_CDF} and table~\ref{tbl:SOFIA_validation} show a quantitative comparison of the angular differences.
}
\label{afig:OphA_multiwavelength_Bfield}
\end{figure*}

\begin{figure*}[ht!p]
\centering
\includegraphics{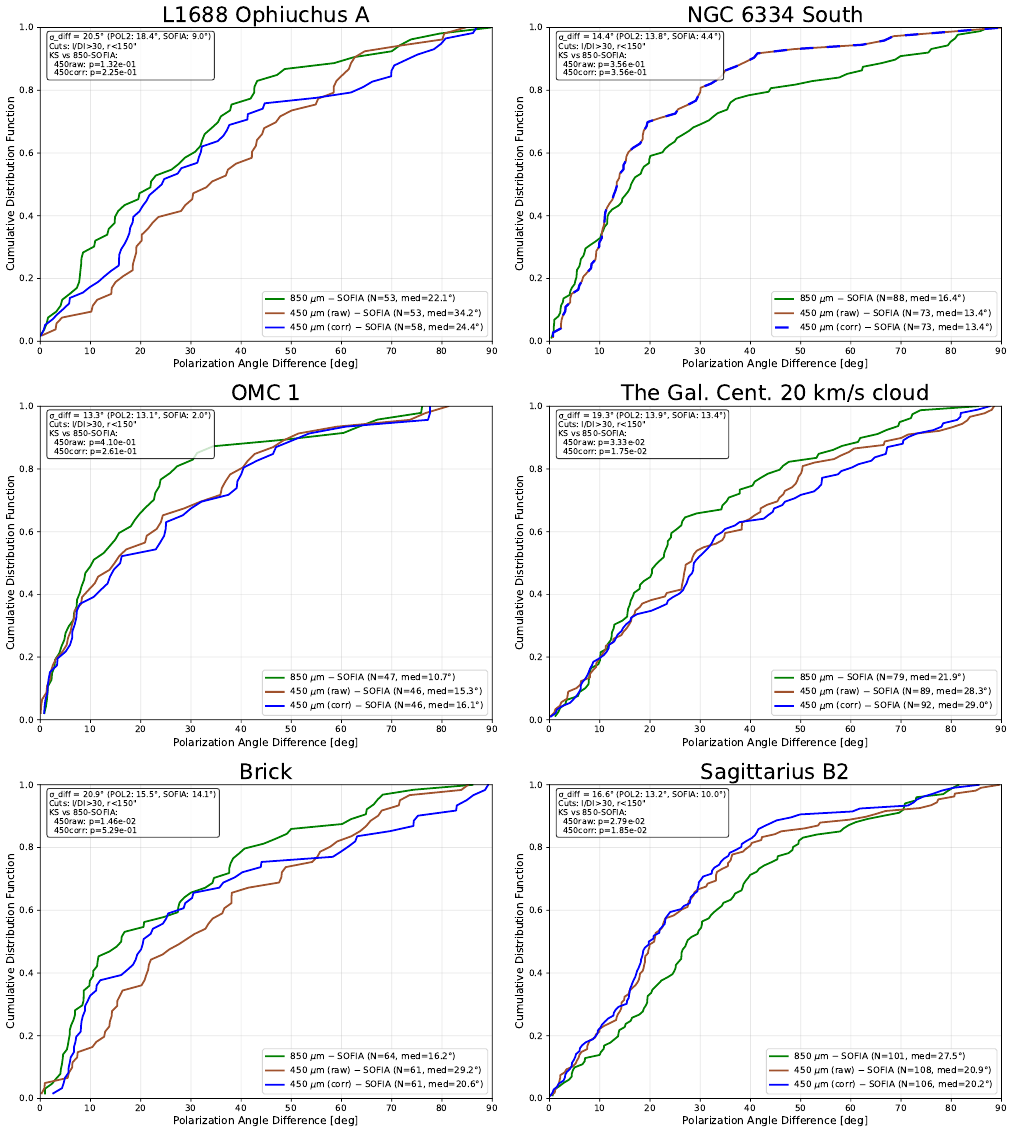}
\caption{
Cumulative distribution functions (CDFs) of absolute differences in polarization angle $|\Delta\chi|$ between SOFIA/HAWC+ and POL-2 for the six targets retained for quantitative comparison.
Green, brown, and blue curves show comparisons with 850\,$\mu$m data and with 450\,$\mu$m data before and after the Stokes-$Q/U$ zero-level offset correction, respectively.
The maps before and after correction use the default threshold and the threshold selected for each field, respectively; neither has the intercepts from the three-step procedure subtracted.
The 850\,$\mu$m comparison provides a reference for wavelength-dependent differences.
For NGC~6334 South, a Category~2 source that keeps the default threshold (sub-subsection~\ref{ass:offset_significance}), the maps before and after the correction are the same, so the dashed blue curve coincides with the brown curve.
Table~\ref{tbl:SOFIA_validation} gives the corresponding statistics and sample sizes.
\label{afig:SOFIA_CDF}
}
\end{figure*}

\paragraph{SOFIA/HAWC+ Data and Target Selection}
We used archival SOFIA/HAWC+ polarimetric observations of eight targets also observed in BISTRO, in Band~D ($\theta_{\rm FWHM} \approx 13\farcs6$ at 154\,\micron) and Band~E ($\theta_{\rm FWHM} \approx 18\farcs2$ at 214\,\micron; \citealt{Harper2018}).
The Band~D targets are L1688 Ophiuchus~A (Paper~S2 figure~\ref{S2-afig:Bmap_OphiuA}) and NGC~6334 South (Paper~S2 figure~\ref{S2-afig:Bmap_N633MO}).
The Band~E targets are OMC~1 (Paper~S2 figure~\ref{S2-afig:Bmap_OriAS1}), the Galactic Center 20~km~s$^{-1}$ cloud (Paper~S2 figure~\ref{S2-afig:Bmap_GC20KC}), the Brick (Paper~S2 figure~\ref{S2-afig:Bmap_GCBRCK}), Sgr~B2 (Paper~S2 figure~\ref{S2-afig:Bmap_GSGRB2}), DR~21 (Paper~S2 figure~\ref{S2-afig:Bmap_DR21DR}), and Monoceros~R2 (Paper~S2 figure~\ref{S2-afig:Bmap_MonoR2}) (principal investigators: D.~Dowell, D.~Chuss, T.~Pillai, R.~Pokhrel).
The SOFIA maps are Level~4 data products retrieved from the NASA/IPAC Infrared Science Archive (IRSA).
These maps were produced by the HAWC+ data reduction pipeline (\texttt{hawc\_drp}~v3.2.0, as recorded in the product headers) within the SOFIA Redux framework \citep{Clarke2023}.
We reprojected the SOFIA Stokes $I$, $Q$, and $U$ maps onto the 12\arcsec\ POL-2 pixel grid and converted them into the same format for polarization measurements as used for the POL-2 data.
We debiased the SOFIA measurements following \citet{Wardle1974}.
We did not apply beam convolution and compared the SOFIA polarization angles as measured.
For the 450\,\micron\ data before and after the offset correction, we used the reduction with the default threshold and the reduction with the threshold selected for each field (appendix table~\ref{tab:PCAparams}), respectively.
We do not subtract any intercepts obtained with the three-step procedure from either.
For quantitative comparison (figure~\ref{afig:SOFIA_CDF} and table~\ref{tbl:SOFIA_validation}), we used pixels with positive uncertainties and $0 < P < 100\%$ in every data set, and POL-2 pixels with $I_{850}/\sigma_{I,850} \geq 30$.
We applied the standard HAWC+ quality cuts $I/\sigma_I > 200$, $PI/\sigma_{PI} > 3$, and $P < 50\%$ \citep[e.g.,][]{Harper2018, Butterfield2024} only when selecting the SOFIA segments displayed in figure~\ref{afig:OphA_multiwavelength_Bfield}.
We restricted the comparison to pixels within 150\arcsec\ of the field center to avoid the edge introduced by the spatial filtering of the POL-2 data (Paper~S1 section~\ref{S1-sec:obs}).
We excluded DR~21 and Monoceros~R2 from the quantitative comparison because the median uncertainties of their SOFIA polarization angles ($\sigma_{\rm SOFIA} = 90\fdg4$ and $102\fdg5$, respectively; table~\ref{tbl:SOFIA_validation}) exceed half a cycle in polarization angle.
The remaining six targets have combined uncertainties $\sigma_{\rm diff} = \sqrt{\sigma_{\rm POL2}^2 + \sigma_{\rm SOFIA}^2}$ of $13\fdg3$ -- $20\fdg9$, suitable for statistical comparison.

\paragraph{Qualitative Comparison: Ophiuchus~A as a Representative Case}
We use Ophiuchus~A as a representative case because it has the largest 450\,\micron\ offset correction among the Category~1 targets with SOFIA comparison data and the best SOFIA Band~D data quality.
Figure~\ref{afig:OphA_multiwavelength_Bfield} compares the inferred orientations of the $\vec{B}$-field in the plane of the sky from POL-2 850\,\micron\ (green), 450\,\micron\ before (brown) and after (blue) the $Q/U$ zero-level offset correction, and SOFIA/HAWC+ Band~D 154\,\micron\ (magenta).
The 850\,\micron--SOFIA pair agrees to within $\lesssim 10\degr$--$15\degr$ along the bright ridge (panel a).
Before correction, the 450\,\micron\ segments deviate from the SOFIA orientations by $\sim 20\degr$--$30\degr$ (panel b).
After correction, the 450\,\micron--SOFIA angle differences are $\lesssim 10\degr$--$15\degr$ (panel c), comparable to those of the 850\,\micron--SOFIA reference.
These visual comparisons apply only to the bright ridge.
The medians for the whole field in table~\ref{tbl:SOFIA_validation} use all $N = 53$ matched pixels within $150\arcsec$ of the field center and are correspondingly larger: $34\fdg2$ before correction and $24\fdg4$ after correction, compared with $22\fdg1$ at 850\,\micron.

We performed an additional check on Ophiuchus~A with cumulative distribution functions (CDFs) separated by $S/N$, using subsets of low and high intensity.
The comparison supports the applicability of the qualitative result across the range of surface brightness sampled.
We did not find a statistically significant dependence on intensity, although the sample size in each bin is limited.

\paragraph{Quantitative Comparison Using Cumulative Distribution Functions}
We assess whether the correction reduces the observed 450\,\micron--SOFIA angle differences, without requiring identical polarization orientations at all wavelengths.
Far-infrared polarization provides an external view of the projected field against which the 450\,$\mu$m correction can be checked.
Different wavelengths can give different weights to dust components along the line of sight.
 
For all comparison targets, we calculated CDFs of the absolute differences in polarization angle $|\Delta\chi|$ between SOFIA/HAWC+ and three POL-2 data sets: 850\,\micron, and 450\,\micron\ before and after the $Q/U$ zero-level offset correction.
Each CDF uses all matched pixels for which the angle difference is defined, so the sample size $N$ may vary between wavelength pairs.
Figure~\ref{afig:SOFIA_CDF} presents the CDFs for the six targets retained for quantitative comparison, and table~\ref{tbl:SOFIA_validation} summarizes the median angle differences and sample sizes.
 
We found three patterns among the targets retained for quantitative comparison.
First, in Ophiuchus~A and the Brick, both Category~1 fields with significant raw $Q/U$ offsets, the 450\,\micron--SOFIA angle differences are smaller after correction.
The median 450\,\micron--SOFIA difference decreases from $34\fdg2$ to $24\fdg4$ in Ophiuchus~A, compared to the median angle difference of $22\fdg1$ between 850\,\micron\ and SOFIA.
In the Brick, the median difference decreases from $29\fdg2$ to $20\fdg6$, compared to the baseline of $16\fdg2$.
The two-sample Kolmogorov--Smirnov tests show that the 450\,\micron\ CDFs after correction are statistically consistent with the 850\,\micron\ reference in both cases.

Second, NGC~6334 South, a Category~2 field where the measured $Q/U$ offsets are consistent with noise (sub-subsection~\ref{ass:offset_significance}), keeps the default threshold.
Its maps before and after the correction are therefore the same, and its $|\Delta\chi|$ medians are identical ($13\fdg4$).

Third, OMC~1, the Galactic Center 20~km~s$^{-1}$ cloud, and Sgr~B2 show more complex behavior.
OMC~1 has a baseline of $10\fdg7$ and 450\,\micron\ medians of $15\fdg3$/$16\fdg1$ before/after the correction.
The Galactic Center 20~km~s$^{-1}$ cloud has a baseline of $21\fdg9$ and corresponding medians of $28\fdg3$/$29\fdg0$.
For Sgr~B2, the angle differences at 450\,\micron\ are smaller than the 850\,\micron\ baseline of $27\fdg5$: their medians are $20\fdg9$/$20\fdg2$ before/after the correction.

In general, where systematic biases were the largest, the 450\,\micron\ angles after the offset correction are closer to the independent far-infrared (FIR) measurements.
For the reported samples, the median discrepancies decrease in Ophiuchus A and the Brick, whereas the other targets show smaller or different responses.
These comparisons support checking the correction for individual sources.
Some of the remaining angle differences between bands may arise because dust components with different temperatures and $\vec{B}$-field orientations are weighted differently at each wavelength \citep[e.g.,][]{TassisPavlidou2015, PohDodelson2017}.
The CDF comparisons alone do not establish the physical origin of these differences.
Non-rejection by the Kolmogorov--Smirnov (KS) test does not establish identical field geometry or eliminate wavelength-dependent weighting of the dust.

\begin{deluxetable*}{lccrcccccc}
\tablecaption{Summary of POL-2 and SOFIA/HAWC+ Polarization Angle Comparison\label{tbl:SOFIA_validation}}
\tablewidth{0pt}
\tablehead{
\colhead{Target} &
\colhead{$\lambda_{\rm SOFIA}$} &
\colhead{Category} &
\colhead{$N$} &
\colhead{$\sigma_{\rm POL2}$} &
\colhead{$\sigma_{\rm SOFIA}$} &
\colhead{$\sigma_{\rm diff}$} &
\colhead{850--SOFIA} &
\colhead{450raw--SOFIA} &
\colhead{450corr--SOFIA} \\
\colhead{} &
\colhead{($\mu$m)} &
\colhead{} &
\colhead{} &
\colhead{(deg)} &
\colhead{(deg)} &
\colhead{(deg)} &
\colhead{(deg)} &
\colhead{(deg)} &
\colhead{(deg)}
}
\startdata
\multicolumn{10}{c}{\textit{Valid Comparisons}} \\
\hline
Ophiuchus~A      & 154 & 1 &  53 & 18.4 &  9.0 & 20.5 & 22.1 & 34.2 & 24.4 \\
NGC~6334 South   & 154 & 2 &  88 & 13.8 &  4.4 & 14.4 & 16.4 & 13.4 & 13.4 \\
OMC~1            & 214 & 1 &  47 & 13.1 &  2.0 & 13.3 & 10.7 & 15.3 & 16.1 \\
GC~20~km~s$^{-1}$& 214 & 1 &  79 & 13.9 & 13.4 & 19.3 & 21.9 & 28.3 & 29.0 \\
The Brick        & 214 & 1 &  64 & 15.5 & 14.1 & 20.9 & 16.2 & 29.2 & 20.6 \\
Sgr~B2           & 214 & 1 & 101 & 13.2 & 10.0 & 16.6 & 27.5 & 20.9 & 20.2 \\
\hline
\multicolumn{10}{c}{\textit{Excluded from Quantitative Analysis}} \\
\hline
DR~21            & 214 & 1 & \nodata & 15.7 & 90.4 & \nodata & \nodata & \nodata & \nodata \\
Monoceros~R2     & 214 & 1 & \nodata & 17.0 & 102.5 & \nodata & \nodata & \nodata & \nodata \\
\enddata
\tablecomments{
Column descriptions:
$\lambda_{\rm SOFIA}$ is the SOFIA/HAWC+ filter wavelength (154\,$\mu$m corresponds to Band~D; 214\,$\mu$m corresponds to Band~E);
Category refers to the classification of the zero-level offset in Stokes $Q/U$ at 450\,$\mu$m, as defined in sub-subsection~\ref{ass:offset_significance} (1: offset exceeding $\sigma_{\rm eff}$; 2: offset within $\sigma_{\rm eff}$; see appendix figures~\ref{afig:PCAcutoff_cat1_p1}--\ref{afig:PCAcutoff_cat3});
$N$ is the number of matched pixels satisfying $I_{850}/\sigma_{I,850} \geq 30$ within 150\arcsec\ of the field center;
$\sigma_{\rm POL2}$ and $\sigma_{\rm SOFIA}$ are the median uncertainties of the polarization angles for POL-2 (850\,$\mu$m) and SOFIA/HAWC+, respectively;
$\sigma_{\rm diff} = \sqrt{\sigma_{\rm POL2}^2 + \sigma_{\rm SOFIA}^2}$ is the combined angle uncertainty;
the final three columns give the median absolute differences in polarization angle $|\Delta\chi|$ between SOFIA and the three POL-2 data sets: 850\,$\mu$m, and 450\,$\mu$m before (450raw) and after (450corr) the $Q/U$ zero-level offset correction (maps reduced with the default and selected thresholds).
We exclude DR~21 and Monoceros~R2 because the median uncertainties of their SOFIA polarization angles exceed 90\degr.
These uncertainties make a quantitative comparison impossible.
NGC~6334 South is a Category~2 source that keeps the default threshold, so ``450raw'' and ``450corr'' are the same map and their values are identical.
}
\end{deluxetable*}

\subsubsection{Summary of Technical Validation 4}
\label{ss:offset_summary}
 
\textit{This section completes the construction stage of Pillar~3 of the technical framework.}
We applied two steps of spatial harmonization to the 450\,\micron\ Stokes~$I,Q,U$ maps used in Papers~S1 and~S2 and corrected additive zero-level offsets in Stokes~$Q$ and~$U$ where required.
First, we matched the 450\,\micron\ angular response to the modeled 850\,\micron\ compound beam (effective FWHM $12\farcs6$).
Second, we restricted the recovered angular scales to the common $12\farcs6$--$300\arcsec$ range of the standard 850\,\micron\ \texttt{pol2map} reduction.
We then corrected additive zero-level offsets of $\sim 1$\,mJy\,arcsec$^{-2}$ in the 450\,\micron\ Stokes~$Q$ and~$U$ intensities, where required.
After the correction, the combined residual offset amplitude has a median of $0.07$\,mJy\,arcsec$^{-2}$.
We independently checked the corrected polarization orientations against SOFIA/HAWC+ polarimetry for six targets, with Ophiuchus~A as the representative case.
The median angle discrepancies decrease after the correction in Ophiuchus~A and the Brick, whereas the other targets show smaller or different responses.
In subsection~\ref{ss:beamval}, we provide the final quantitative validation by testing whether these processing steps preserve the {\em morphology of the polarization angles} required for the scientific analyses of the two bands.

\subsection{Technical Validation 5: Validation of Beam-Harmonized Polarization Angles}
\label{ss:beamval}
 
\subsubsection{Motivation and overview}
\label{ss:beamval_motivation}
We isolate the sensitivity to beam treatment and to the recovery of extended emission by comparing reductions of the same 450\,$\mu$m observations, before the maps enter an interpretation based on two wavelengths.
The compound 450\,$\mu$m beam and the recovery of extended emission can alter the $\vec{B}$-field structure represented in a polarization map.
 
We now assess the geometric floor of residual errors in polarization angle that remains after the reduction of the two bands, using the corrected and beam-matched 450\,\micron\ maps from subsection~\ref{ss:offset}.
In the final 450 and 850\,\micron\ maps, we found localized residuals in polarization angle of several to ${\sim}10\degr$ in a few regions.
L1688 Ophiuchus~A provides the clearest example (the analysis F1 in Paper~S1).
We ask whether these residuals reflect an intrinsic dependence of the $\vec{B}$-field geometry on wavelength or arise mainly from matching the beam and spatial-frequency responses of the 450\,\micron\ data.
To distinguish these possibilities, we compared three independently processed 450\,\micron\ products derived from the same POL-2 observations:
(i)~the {\em direct native-beam} product (\texttt{pol2map} flag \texttt{smooth450=no}, default 450\,\micron\ spatial-frequency coverage), which uses the native compound 450\,\micron\ point-spread function with two components.
Equation~(10) of \citet{Mairs2021} models this function as $G_{\rm total} = \alpha^{\rm Mairs}\,G_{\rm MB} + \beta^{\rm Mairs}\,G_{\rm S}$ with main-beam FWHM $\theta_{\rm MB} = 6\farcs2$ and amplitude $\alpha^{\rm Mairs} = 0.89$, secondary-beam FWHM $\theta_{\rm S} = 18\farcs8$ and amplitude $\beta^{\rm Mairs} = 0.11$ (\citealt{Mairs2021}, their table~6);
(ii)~an analytically {\em deconvolved--reconvolved} product (\texttt{smooth450=yes}, default 450\,\micron\ spatial-frequency coverage), in which the 450\,\micron\ $I$, $Q$ and $U$ maps are smoothed to the 850\,\micron\ beam before coaddition.
\texttt{pol2map} forms the kernel by deconvolving the expected 850\,\micron\ beam with the expected 450\,\micron\ beam, both represented as two-component Gaussians \citep{Parsons2021}.
The 450\,\micron\ map-making retains its native largest recoverable scale of $\sim 600\arcsec$; and
(iii)~the {\em final science product} (\texttt{smooth450=yes} together with \texttt{flt.filt\_edge\_largescale=300}), in which the 450\,\micron\ data are smoothed to the 850\,\micron\ beam in the same way and their spatial-frequency coverage is truncated.
Both wavelengths then sample the {\em common} range of angular scales $12\farcs6$--$300\arcsec$ matched to the standard 850\,\micron\ \texttt{pol2map} reduction.
Product~(iii) is the reduction at the default PCA threshold (\texttt{pca.zero\_snr}=3).
For Category~1 fields whose scientific analyses use a map reduced with an optimized threshold (sub-subsection~\ref{ass:ResidualOffsets}), this comparison therefore refers to the map reduced with the default threshold.
 
All three products come from the same 450\,\micron\ observations.
Comparison between the data from (i) and (iii) measures the combined effects of beam matching and the change in recovered angular scales.
Products (ii) and (iii) have the same beam treatment, so their comparison measures the effect of restricting the recovered angular scales.
Throughout this section, we use the median absolute difference in polarization angle for each source, $\widetilde{|\Delta\chi|}$, as the residual statistic.
To describe the scatter, we report in parentheses the MAD of the underlying distribution of $\Delta\chi$.

This analysis provides the {\em final quantitative validation of Pillar~3}.
Together with the calibration and noise analyses in subsections~\ref{ss:flux}--\ref{ss:noisemodel}, it defines the floor of the uncertainty in polarization angle used in Papers~S1 and~S2.
Subsection~\ref{ss:offset} describes the beam construction.
Here we test empirically how well the resulting maps preserve the angle information used in the scientific analyses (figure~\ref{fig:T_pillar_overview} and table~\ref{tbl:T_pillar_index}).
 
We proceed in two stages.
Sub-subsection~\ref{ss:beam_validation_L1688} examines L1688~Ophiuchus~A as a representative bright source, using polarization measurements on a $12\arcsec$ grid and a supplementary comparison on the $4\arcsec$ map grid.
Sub-subsection~\ref{ss:beam_validation_survey} extends the same comparison to the eligible sources in the dual-band sample.
The distributions of angle differences for individual sources are shown in figure~\ref{fig:beam_validation_persource}.
The Ophiuchus~A example illustrates how beam matching changes the inferred $\vec{B}$-field orientations.
Comparisons among sources establish how often this behavior occurs and quantify the remaining angle differences.

The resulting differences measure how the recovered 450\,$\mu$m angle field responds to the specified processing changes.
They are neither a direct measurement of the angle difference between 450 and 850\,$\mu$m nor an independent observation of the true field geometry.
 
\subsubsection{Representative high-$S/N$ example: L1688 Ophiuchus~A}
\label{ss:beam_validation_L1688}
 
\begin{figure*}[htp]
\centering
\includegraphics[width=0.95\textwidth]{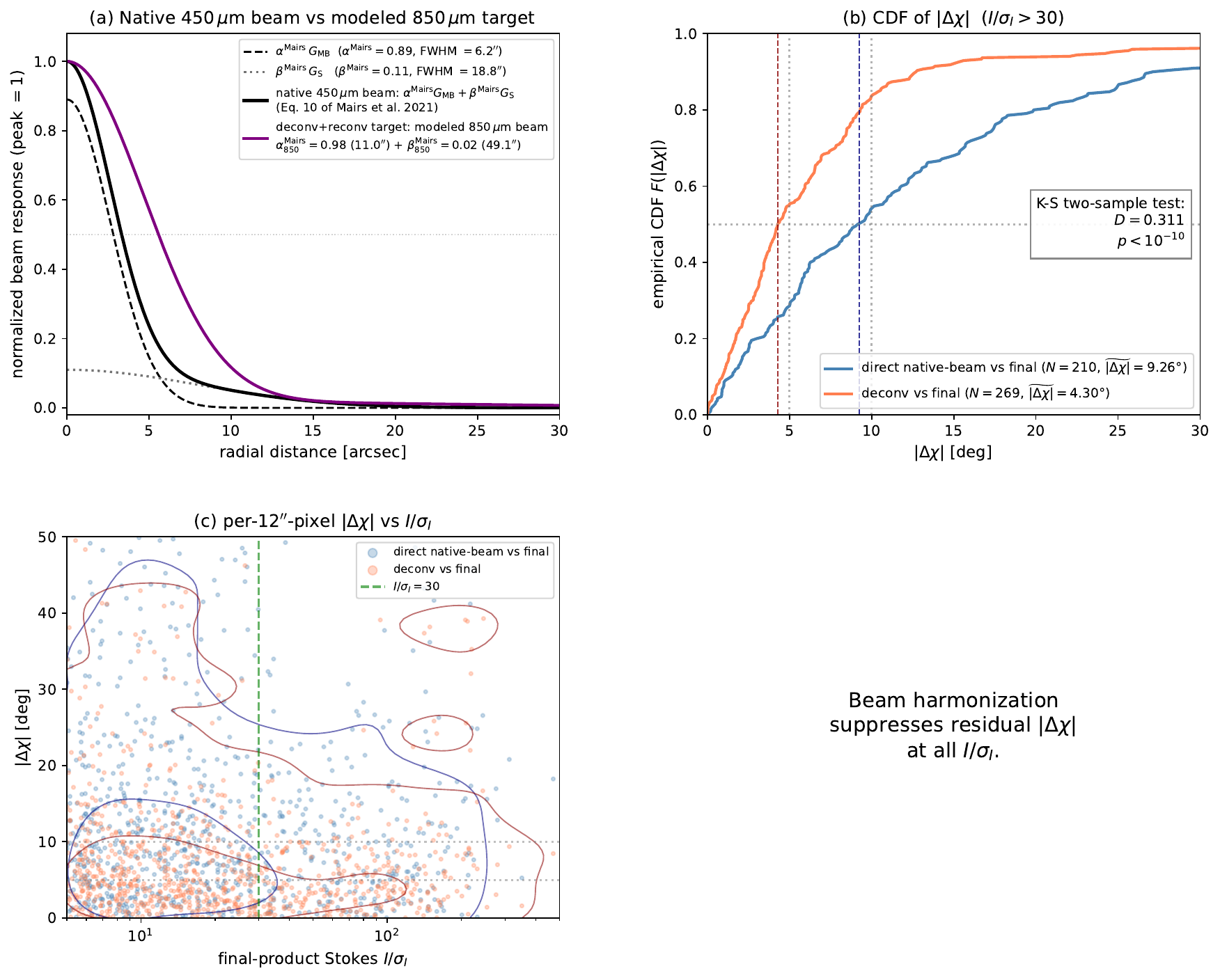}
\caption{
Representative validation of beam harmonization using the final polarization segments for L1688~Ophiuchus~A.
The final product is sampled at pixel centers spaced by $12\farcs0$, nearly one beam width ($12\farcs6$ effective FWHM).
The same three 450\,\micron\ products described in the text are compared at $I/\sigma_I > 30$ (in both products).
\textbf{(a)}~Decomposition of the native 450\,\micron\ beam from equation~(10) and table~6 of \citet{Mairs2021}: main-beam component $\alpha^{\rm Mairs}\,G_{\rm MB}$ (FWHM $\theta_{\rm MB} = 6\farcs2$, $\alpha^{\rm Mairs} = 0.89$; black dashed), secondary-beam component $\beta^{\rm Mairs}\,G_{\rm S}$ (FWHM $\theta_{\rm S} = 18\farcs8$, $\beta^{\rm Mairs} = 0.11$; gray dotted), and the combined two-component profile (black solid).
The panel also shows the deconvolution--reconvolution target, the modeled 850\,\micron\ compound beam (purple solid; main-beam FWHM $11\farcs0$ with amplitude 0.98 and secondary-beam FWHM $49\farcs1$ with amplitude 0.02, table~6 of \citealt{Mairs2021}; effective FWHM $12\farcs6$).
\textbf{(b)}~CDFs of $|\Delta\chi|$ for individual pixels in the {\em native-beam versus final} comparison (blue) and the {\em deconvolve--reconvolve versus final} comparison (orange).
Both comparisons use the same final science product.
Vertical dashed lines mark the medians, and light dotted lines indicate $5\degr$ and $10\degr$.
The inset gives the two-sample Kolmogorov--Smirnov statistic $D$ and $p$-value for the null hypothesis of a common parent distribution.
\textbf{(c)}~$|\Delta\chi|$ for individual pixels as a function of Stokes $I/\sigma_I$ in the final product for both comparison pairs [individual pixel values plus 50/90 percentile kernel-density estimate (KDE) contours].
The deconvolve--reconvolve step alone reduces the median absolute angle residual from $\widetilde{|\Delta\chi|} = 9\fdg3$ to $4\fdg3$ at $I/\sigma_I > 30$.
This comparison isolates the contribution of the beam morphology to the local mismatch in polarization angle from the contribution of truncating the spatial-frequency coverage that the subsequent (ii)-to-(iii) step would add.
}
\label{fig:beam_validation_main}
\end{figure*}

We analyzed polarization measurements toward L1688~Ophiuchus~A sampled on a grid with $12\arcsec$ spacing, comparable to the $12\farcs6$ beam FWHM.
Figure~\ref{fig:beam_validation_main} presents the comparison for positions satisfying $I/\sigma_I > 30$ in both products.
Panel~(a) shows the two-component decomposition of the native 450\,\micron\ beam, $\alpha^{\rm Mairs}\,G_{\rm MB} + \beta^{\rm Mairs}\,G_{\rm S}$, adopted from equation~(10) and table~6 of \citet{Mairs2021} (main-beam FWHM $6\farcs2$, $\alpha^{\rm Mairs} = 0.89$; secondary-beam FWHM $18\farcs8$, $\beta^{\rm Mairs} = 0.11$).
Panel~(a) also shows the modeled 850\,\micron\ compound beam used as the deconvolution--reconvolution target.
Comparing the map with the native beam to the final science product (blue in panels~b and~c), we found a median signed angle difference of only $+2\fdg5$, but $\widetilde{|\Delta\chi|} = 9\fdg3$ (MAD\,$=8\fdg7$) in the positions sampled.
After deconvolution--reconvolution (orange), this median decreases to $\widetilde{|\Delta\chi|} = 4\fdg3$ (MAD\,$=4\fdg0$).
The median residual for the source thus decreases by $5\fdg0$, or 54\% of its original value.
The CDFs for comparisons using the native and matched beams differ by up to 31 percentage points in the fraction of positions with $|\Delta\chi|$ below a given threshold.
This maximum separation gives a KS statistic $D = 0.31$ and a nominal $p$-value of $p < 10^{-10}$.
We interpret this statistically robust reduction by a factor of two at fixed $S/N$ as evidence that {\em the majority of the apparent local angular mismatch in L1688 originates from the morphology of the native 450\,\micron\ compound beam rather than from intrinsic changes in the polarization-angle field.}

We quote the KS statistics as supporting evidence only.
A KS $p$-value depends on the sample size and indicates whether the distributions differ, rather than the size of that difference.
We therefore assess harmonization using the absolute residuals and their reduction, which are the quantities relevant to the analyses of Papers~S1 and~S2.
 
\begin{figure*}[htp]
\centering
\includegraphics[width=0.95\textwidth]{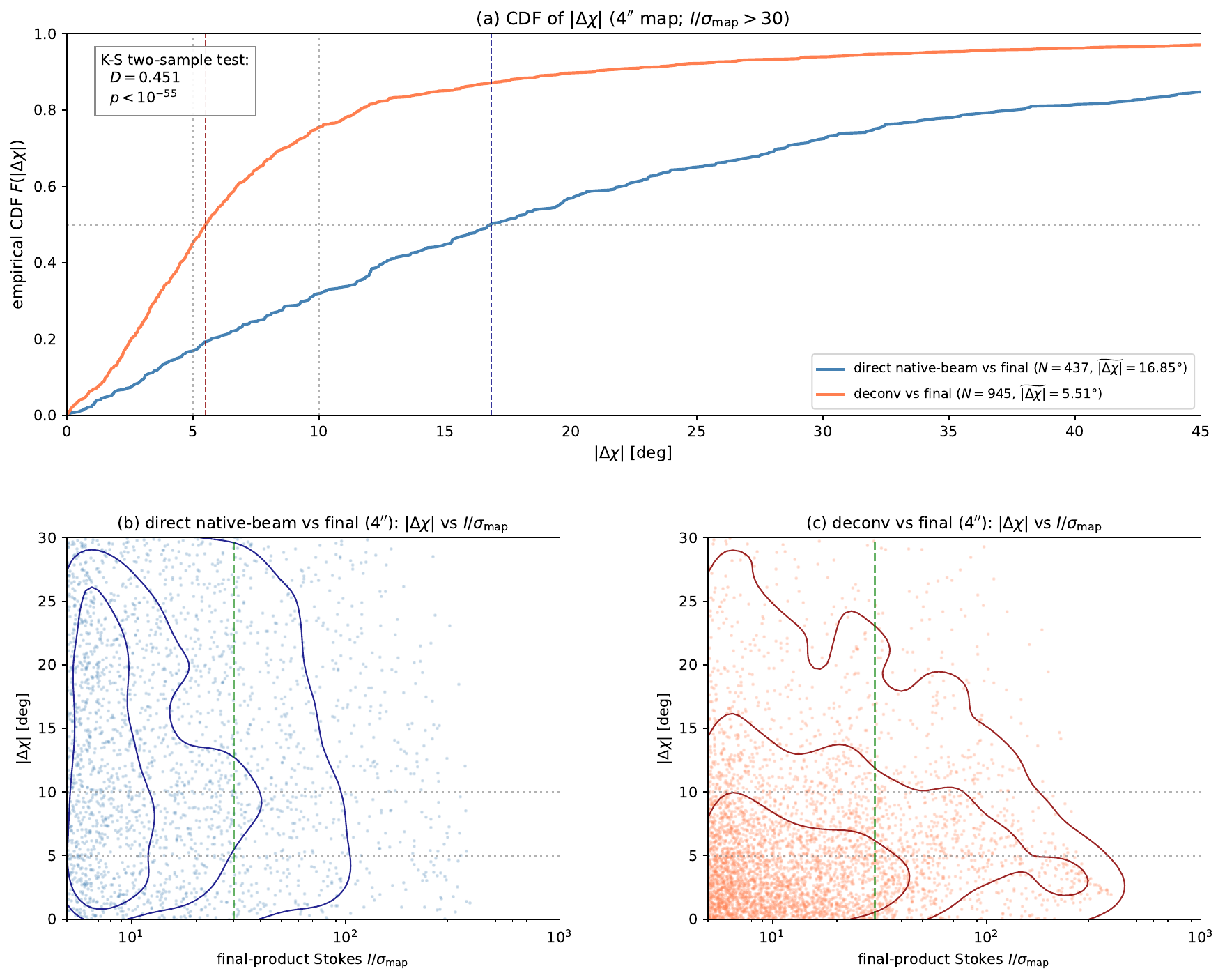}
\caption{
Supplementary comparison of three 450\,\micron\ products for L1688~Ophiuchus~A on the original $4\arcsec$ grid of the Stokes maps, before sampling at $12\arcsec$: the map with the native beam, the map matched to the modeled 850\,\micron\ compound beam, and the final science product with matched beams and spatial-frequency responses.
Here $\sigma_{\rm map}=1.4826\,\mathrm{MAD}(I)$ is a single measure of scatter over the map that includes source structure.
The ratio $I/\sigma_{\rm map}$ is therefore a selection proxy rather than the statistical intensity signal-to-noise ratio used for the $12\arcsec$ comparison.
\textbf{(a)} CDFs of $|\Delta\chi|$ relative to the final product for the native-beam (blue) and deconvolve--reconvolve (orange) maps at $I/\sigma_{\rm map}>30$, with medians and $5\degr/10\degr$ guides.
The inset gives the KS distance and a nominal $p$-value.
Spatial correlations and the shared final product preclude interpreting that value as a calibrated measure of significance for independent samples.
\textbf{(b,c)} Corresponding pixel values versus $I/\sigma_{\rm map}$ in the final product, with 50/75/90 percentile kernel-density contours.
The median $|\Delta\chi|$ decreases from $16\fdg9$ to $5\fdg5$ at $I/\sigma_{\rm map}>30$, and from $10\fdg1$ to $4\fdg1$ at $I/\sigma_{\rm map}>100$.
The reduction on both grids shows that the improvement is already present in the Stokes maps and does not arise solely from sampling at $12\arcsec$.
The different selections do not necessarily yield equal numerical residuals.
}
\label{fig:beam_validation_4arcsec}
\end{figure*}

We also compared the polarization angles directly on the $4\arcsec$ grid of the Stokes-$I,Q,U$ maps, before resampling onto the $12\arcsec$ grid.
We present this supplementary comparison in figure~\ref{fig:beam_validation_4arcsec}.
For this comparison, we selected the positions using $I/\sigma_{\rm map}$, where $\sigma_{\rm map}=1.4826\,{\rm MAD}(I)$ is a single measure of the scatter evaluated on the Stokes-$I$ map.
This scatter includes the structure of the source emission and differs from the statistical uncertainty $\sigma_I$ used for selection on the grid with $12\arcsec$ spacing.
We therefore use the $4\arcsec$ comparison to check whether the decrease in angle differences persists at finer sampling, without requiring numerical agreement between the two grids.
For $I/\sigma_{\rm map}>30$, comparing the native-beam map with the final product gives $\widetilde{|\Delta\chi|}\approx16\fdg9$.
After smoothing, this median decreases to $\approx5\fdg5$.
For $I/\sigma_{\rm map}>100$, the corresponding values are $\approx10\fdg1$ and $\approx4\fdg1$.
The angle differences from the final 450\,\micron\ map decrease after beam matching on both the $12\arcsec$ and $4\arcsec$ grids.
The decrease therefore persists when the maps are examined at finer sampling and cannot be attributed solely to the $12\arcsec$ sampling.

The same test on the sample from the $4\arcsec$ map gives $D = 0.45$ at $I/\sigma_{\rm map} > 30$.
We do not quote its nominal $p$-value.
Measurements in $4\arcsec$ pixels within the same beam are correlated, and the two distributions of angle differences share the same final map.
The independence assumption of the two-sample KS test therefore does not hold for this comparison.
 
\subsubsection{Validation across the dual-band sample}
\label{ss:beam_validation_survey}
To assess the broader applicability and robustness of the validation procedure established for L1688 Ophiuchus~A, we repeated the comparison of the three products for the sources in the Paper~S1 {\sc SpIdx} sample.
A correction that improves one bright molecular cloud core may produce different results in a filamentary cloud or a complex field in the Galactic center.

\begin{figure*}[htp]
\centering
\includegraphics[width=0.98\textwidth]{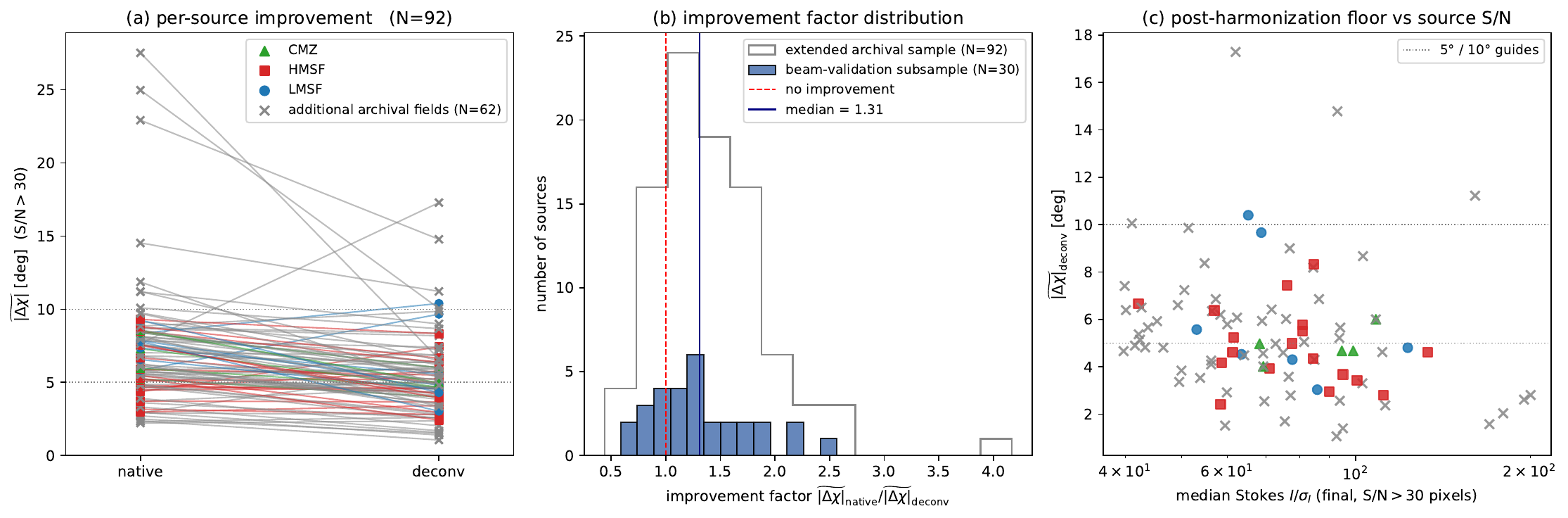}
\caption{
Validation of beam harmonization for the 30 Paper~S1 sources with all three 450\,$\mu$m products, $I/\sigma_I>30$, and $N_{\rm samp}\geq20$.
\textbf{(a)} Median absolute angle differences $\widetilde{|\Delta\chi|}$ for individual sources relative to the final science product, comparing native-beam (left) and deconvolve--reconvolve (right) maps.
Tracks are colored by environment: low-mass star formation (LMSF), high-mass star formation (HMSF), and the Central Molecular Zone (CMZ).
\textbf{(b)} Distributions of the improvement factor $\widetilde{|\Delta\chi|}_{\rm native}/\widetilde{|\Delta\chi|}_{\rm deconv}$ for the 30 sources (filled) and the extended archival sample of 92 fields (gray outline).
The extended sample comprises the 30 sources and 62 additional fields.
Only the additional fields are shown in gray in panels~(a,c).
\textbf{(c)} Residuals after harmonization versus the median Stokes $I/\sigma_I$ of eligible pixels in the final product.
Horizontal guides mark $5\degr$ and $10\degr$.
L1688~Ophiuchus~A (figure~\ref{fig:beam_validation_main}) has a larger residual with the native beam than the sample median ($9\fdg3$ versus $6\fdg3$), but a typical residual after harmonization.
}
\label{fig:beam_validation_survey}
\end{figure*}

We required all three products (i)--(iii) and $\geq 20$ matched pixels with $I/\sigma_I > 30$ in both compared products.
We pair pixels of products (i) and (iii) within $6\arcsec$ and pixels of products (ii) and (iii) within $2\arcsec$.
This selection retains $30$ of the $35$ sources of the dual-band statistical sample.
Three sources (Orion~A Field~4, Orion~A Field~5 and Serpens~South) have no native-beam reduction (i).
Sgr~B2 provides too few pixels at high $S/N$ for a robust statistic for that source, and the native-beam pixels of the Brick lie about $8\arcsec$ from those of product~(iii).
Figure~\ref{fig:beam_validation_survey} summarizes the comparisons for individual sources.
 
The main results for the comparison sample of 30 sources defined in table~\ref{tbl:T_sample_inventory} are as follows:
\begin{enumerate}
\item the median of the $\widetilde{|\Delta\chi|}$ values for individual sources decreases from $6\fdg3$ to $4\fdg7$ after beam matching;
\item $77\%$ of sources have smaller median angle differences after beam matching;
\item after harmonization, $29$ of the $30$ sources ($97\%$) reach $\widetilde{|\Delta\chi|}$ {\em below} the scale of $10\degr$ adopted as the physical tolerance in Paper~S1 F1; the single exception, L1689~North, lies at $10\fdg4 \pm 1\fdg2$ and is therefore consistent with that scale within its uncertainty; and
\item after harmonization, $63\%$ of sources reach $\widetilde{|\Delta\chi|}$ below $5\degr$.
\end{enumerate}
In each of the three environmental subsets, the median of the $\widetilde{|\Delta\chi|}$ values for individual sources also decreases after beam matching: from $7\fdg8$ to $4\fdg8$ in low-mass star formation (LMSF), from $7\fdg3$ to $4\fdg7$ in the CMZ, and from $5\fdg2$ to $4\fdg6$ in high-mass star formation (HMSF).
The fractional improvement is similar in LMSF and the CMZ (38\% and 36\%) and smaller in HMSF (12\%).
This comparison shows that harmonization is effective even in spatially complex fields in the Galactic Center.
We attribute the smaller improvement in the HMSF to the dominance of bright, compact polarized peaks in this sample.
The native 450\,\micron\ compound beam already gives modest local angle differences in these peaks, leaving less room for improvement.
 
\begin{figure*}[htp]
\centering
\includegraphics[width=\textwidth]{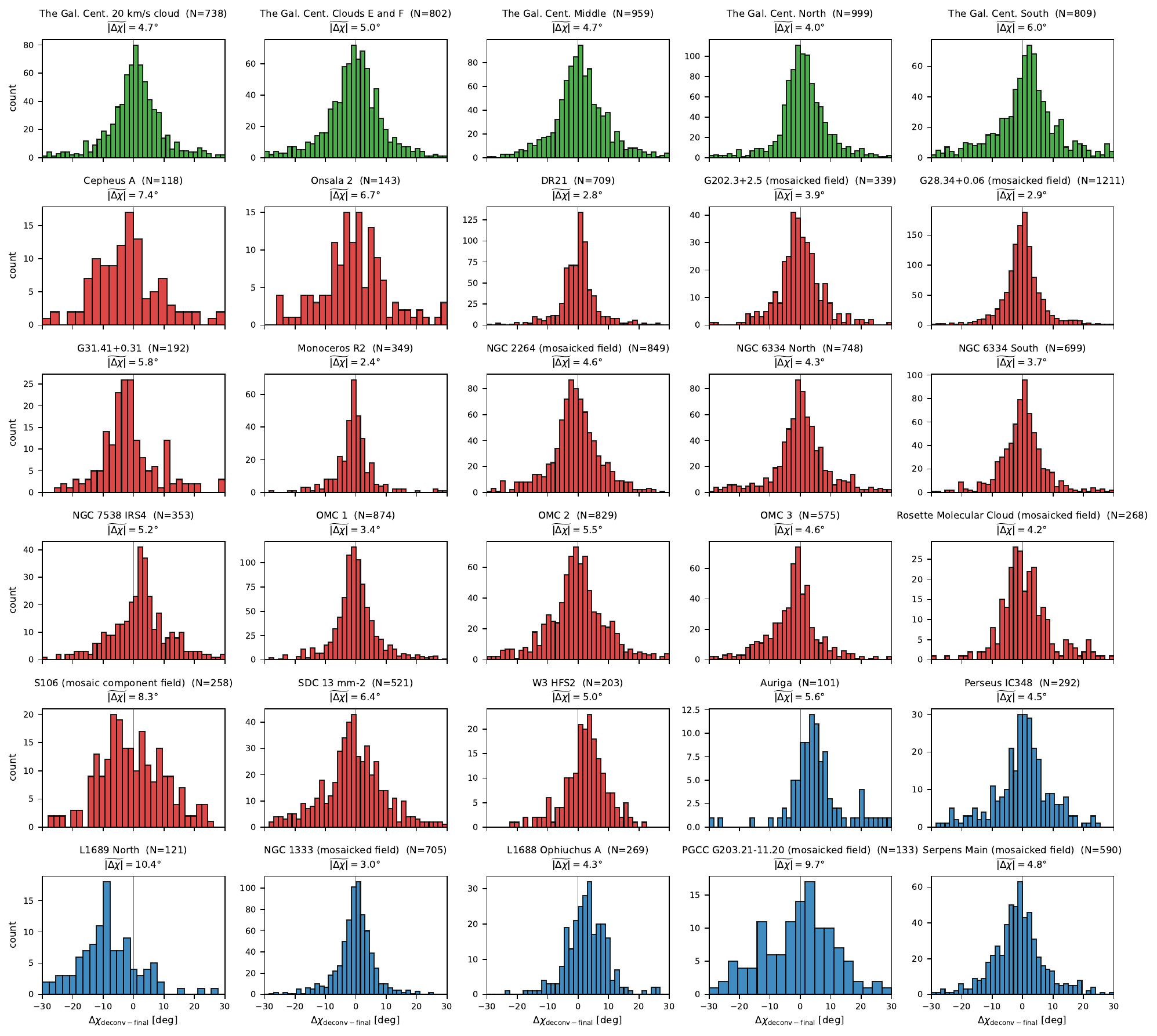}
\caption{
Histograms of the $\Delta\chi$ residuals between the deconvolve--reconvolve and final products at $I/\sigma_I > 30$ for each of the $30$ sources used in the validation across the sample in figure~\ref{fig:beam_validation_survey}.
Panels are grouped by environment (CMZ in green, HMSF in red, LMSF in blue), with $\widetilde{|\Delta\chi|}$ for each source printed at the top of its panel.
Histogram bin widths use the Freedman--Diaconis rule.
}
\label{fig:beam_validation_persource}
\end{figure*}

We also applied the same comparison to an extended archival sample.
All three reductions were available for the $147$ POL-2 fields, of which $92$ satisfied the matched-sample requirement.
The $92$ fields comprise the $30$ sources analyzed above and $62$ additional archival fields (mosaic-component fields and other archival sources).
The archival sample of $92$ fields agrees quantitatively with the subsample of $30$ sources used for beam validation: the median of the residuals for individual sources is $6\fdg2$ before and $4\fdg9$ after matching, the improvement factor per source is $1.31$, and $95\%$ reach $\widetilde{|\Delta\chi|}_{\rm deconv} < 10\degr$.
The performance of harmonization is therefore not specific to the subset used for the scientific analyses.
The histograms for the $30$ individual sources in the subsample used for beam validation are shown in figure~\ref{fig:beam_validation_persource}.

These summaries characterize the range of sensitivity to processing among the eligible clouds.
They do not assign the same error bound to every pixel or to sources that lack the required comparison products.
 
\subsubsection{Implications for inter-band polarization-angle science}
\label{ss:beam_validation_implications}
The three 450\,$\mu$m reductions examined here constrain how the processing changes the recovered $\vec{B}$-field structure while keeping the underlying observations fixed.
A difference between the 450 and 850\,$\mu$m polarization angles can arise either from the weighting of that structure at different wavelengths or from the way the two maps recover it.
 
The L1688 example, the comparison on both sampling grids, and the tests across the survey demonstrate a statistically significant reduction in the angle differences between the processed products.
The two comparisons separate the two processing changes.
Comparison of (i) with (iii) gives a larger difference, so most of the change in the recovered angle field follows the beam matching rather than the restriction of angular scales.
Comparison of (ii) with (iii) isolates the restriction of the angular scales, which leaves a smaller difference.
The two differences are measured against the same final product and are not independent error terms to be added or subtracted.

The median residuals for individual sources after harmonization define a typical floor of $\widetilde{|\Delta\chi|} \sim 4\degr\hbox{--}6\degr$ in high-quality regions (figure~\ref{fig:beam_validation_survey}c).
The extended archival sample of $92$ fields yields quantitatively consistent floors, so the result is not driven by source selection.
 
The angle differences that remain when the final 450 and 850\,\micron\ maps are compared are similar in size to the processing sensitivity measured here.
They therefore do not by themselves require intrinsically different $\vec{B}$-field geometries between the two bands.
We adopt the floor $\widetilde{|\Delta\chi|} \sim 4\degr\hbox{--}6\degr$ as the technical reference for assessing the agreement between polarization angles in the two bands in Paper~S1.
Residuals well above this floor remain candidates for an astrophysical explanation specific to each source.

The 4--6 degree scale describes the sensitivity to the 450\,$\mu$m reduction choices tested, rather than a direct measurement of the angle difference between 450 and 850\,$\mu$m.
We recommend interpreting discrepancies between the bands together with this processing sensitivity, the angle uncertainties of individual pixels, and the evidence for structure that depends on wavelength in each source.

\subsubsection{Summary of Technical Validation 5}
\label{ss:beamval_summary}
Across eligible sources, we compared the three products to measure how the recovered 450\,$\mu$m angle field responds to beam transformation and spatial filtering.
\textit{This quantification across the survey completes Pillar~3 of the technical framework.}
We interpret the local agreement in angles, distributions of angle differences, $\vec{B}$-field morphology, and statistics of angle dispersion in Papers~S1 and~S2 against the quantified instrumental floor $\widetilde{|\Delta\chi|} \sim 4\degr\hbox{--}6\degr$.
After harmonization, we find $\widetilde{|\Delta\chi|} < 10\degr$ in $29$ of the $30$ dual-band sources at $I/\sigma_I > 30$.
In section~\ref{sec:discussion}, we combine this validation with Pillar~1 (subsection~\ref{ss:flux}: photometric consistency) and Pillar~2 (subsections~\ref{ss:noiseI}--\ref{ss:noisemodel}: noise characterization) to quantify the uncertainties in the principal dual-band observables of Papers~S1 and~S2.
These uncertainties include the terms in $P_{450}/P_{850}$ and $\gamma-\alpha$ from the calibration and from residual zero-level offsets in $Q/U$, and the sensitivity of the $\vec{B}$-field morphology to the tested processing choices.
We assess an observed angle difference between the bands by considering this processing sensitivity together with the Stokes~$Q,U$ measurement errors and residuals specific to each source.
A stable angle pattern under the tested reductions does not, on its own, establish an equivalent bound on errors in polarization fraction or spectral index.

\section{Discussion}
\label{sec:discussion}
 
\subsection{Integrated Systematic Uncertainty Budget for the Dual-Band Derived Quantities}
\label{sec:uncertainty}
\label{app:systematics}%
We assess whether the quantified calibration errors and residual Stokes-$Q/U$ offsets can account for the difference between the polarization fractions measured in the two bands.
We now combine the results of the technical validations from the three pillars (figure~\ref{fig:T_pillar_overview} and table~\ref{tbl:T_pillar_index}) to estimate the systematic uncertainties in two quantities derived from the two bands, $P_{450}/P_{850}$ and $\gamma-\alpha$ (equation~\ref{eq:spectral_indices}).
These estimates define the technical precision floor used in Papers~S1 and~S2.
Subsections~\ref{ss:flux}--\ref{ss:beamval} address five requirements for interpreting polarization measurements in the two bands.
The first three are consistent calibration of surface brightness (subsection~\ref{ss:flux}), empirically justified thresholds in Stokes~$I$ (subsection~\ref{ss:noiseI}), and propagation of Stokes~$Q,U$ noise into $P$ and polarization angle (subsection~\ref{ss:noisemodel}).
The remaining two are correction of additive zero-level offsets in the 450\,\micron\ Stokes~$Q/U$ intensities (subsection~\ref{ss:offset}) and validation of residual angle differences between the two bands after beam harmonization (subsection~\ref{ss:beamval}).
Paper~S2 subsection~\ref{S2-ss:S2_P_ratio} defines the three statistical estimators of the polarization-fraction ratio.
The two quantities are related through equation~(\ref{eq:Pratio_SpIdx}).
Here we consider systematic contributions.
Papers~S1 and~S2 report statistical uncertainties from pixel-to-pixel scatter separately.
 
\subsubsection{Absolute Flux Calibration}
\label{sss:discussion_fcf}
The flux conversion factors for extended sources have uncertainties of $\pm 7$\% at 850\,$\mu$m and $\pm 17$\% at 450\,$\mu$m \citep{Mairs2021}, as applied in the data preparation (Paper~S1 section~\ref{S1-sec:obs}).
For an individual spectral index of total intensity, $\alpha = \log(I_{450}/I_{850})/\log(\nu_{450}/\nu_{850})$ (equation~\ref{eq:spectral_indices}), the combined uncertainty of the FCF contributes
\begin{equation}
\sigma_\alpha^{\rm cal}
\simeq
\frac{\sqrt{0.17^2+0.07^2}}{\ln(\nu_{450}/\nu_{850})}
\simeq 0.29 .
\end{equation}
The polarized-intensity spectral index $\gamma$ carries the same contribution, because wavelength-dependent FCF errors multiply Stokes $I$ and polarized intensity $PI$ by the same factors.
Thus, the calibration contribution cancels in the difference:
\begin{equation}
\delta(\gamma - \alpha)^{\rm FCF}
=
\delta\gamma^{\rm FCF}
-
\delta\alpha^{\rm FCF}
=
0 .
\end{equation}
The polarization fraction $P = PI/I$ is likewise not affected by FCF errors, because both the numerator and the denominator scale with the same flux conversion factor at a given wavelength.
The absolute FCF uncertainty therefore cancels in $P_{450}/P_{850}$.
 
\subsubsection{Residual 450\,$\mu$m Stokes-$Q/U$ Zero-Level Offsets}
\label{sss:discussion_offset}
The three-step regression procedure (sub-subsection~\ref{ass:3step}) leaves residual offsets whose combined amplitude has a median of $0.07$~mJy~arcsec$^{-2}$, with an upper quartile of $0.13$~mJy~arcsec$^{-2}$ (sub-subsection~\ref{ass:ResidualOffsets}; equations~\ref{eq:residual_mean}--\ref{eq:residual_uncertainty}).
Unlike multiplicative FCF errors, these residuals are additive and perturb $Q_{450}$, $U_{450}$, and hence $PI_{450}$ directly, while leaving the total-intensity scale largely unchanged.
Among the systematic terms considered here, the residual offset is the only contribution that enters the derived quantities non-trivially.
 
\begin{figure*}[tp]
\centering
\includegraphics[width=\textwidth]{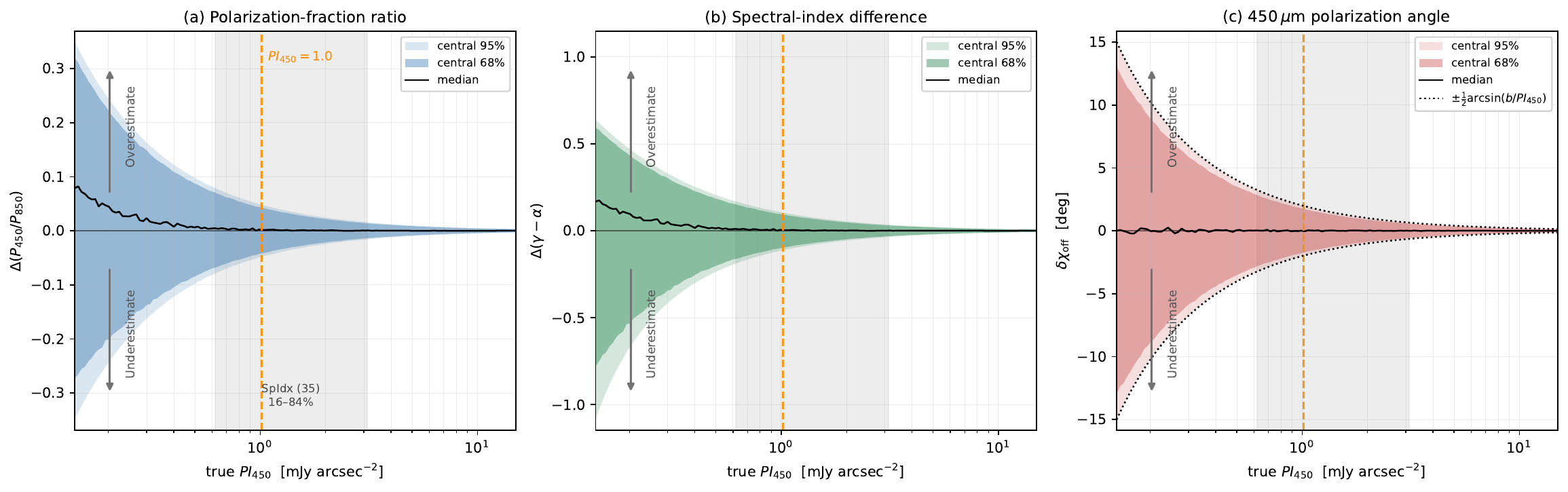}
\caption{Monte Carlo propagation of the residual 450\,$\mu$m Stokes-$Q/U$ zero-level offset.
We add to $(Q_{450}, U_{450})$ a residual Stokes vector of representative amplitude $b = 0.07$\,mJy\,arcsec$^{-2}$ (sub-subsection~\ref{ass:ResidualOffsets}) and random orientation, and apply it uniformly across the field, with $10^{4}$ realizations at each true $PI_{450}$.
The panels show the resulting change in (a) the polarization-fraction ratio $P_{450}/P_{850}$, (b) the spectral-index difference $\gamma - \alpha$ (equation~\ref{eq:spectral_indices}), and (c) the 450\,$\mu$m polarization position angle.
Shaded bands are the central 68\% (dark) and 95\% (light) intervals of the realizations, and the black line is their median.
The bands and the line describe the simulated distribution and are not confidence intervals on a measurement.
The vertical axes show the \textit{change} from the true value, so that negative values indicate underestimation of the plotted quantity.
Negative values do not imply negative ratios of polarization fractions, which cannot occur by definition.
The gray band marks the 16th--84th percentile range of $PI_{450}$, using the median for each source in the 35-source {\sc SpIdx} sample of table~\ref{tbl:T_sample_inventory}.
The orange dashed line marks the sample median $PI_{450} = 0.68$\,mJy\,arcsec$^{-2}$, the intensity at which the uncertainties for individual sources in sub-subsection~\ref{sss:discussion_offset} are quoted.
For $PI_{450}>b$, the dotted curves in panel~(c) show the analytic envelope $\pm\frac{1}{2}\arcsin(b/PI_{450})$, the largest angular shift attainable for the adopted residual amplitude.
We do not add measurement noise or apply debiasing in the simulation (sub-subsection~\ref{sss:discussion_offset}).
\label{fig:offset_mc}}
\end{figure*}

We simulated the effect of the residual in the Stokes $Q$--$U$ plane to quantify its contribution to the derived quantities (figure~\ref{fig:offset_mc}).
The residual is additive in this plane, so we applied it there instead of directly perturbing the polarized intensity.
For the true polarized intensity $PI_{450}$ and position angle $\chi_0$, we set $(Q_0, U_0) = PI_{450}\,(\cos 2\chi_0,\ \sin 2\chi_0)$.
We added a residual vector $(\delta Q_{450}, \delta U_{450}) = b\,(\cos\theta,\ \sin\theta)$ with fixed amplitude $b$ and uniformly random orientation $\theta$ over $[0, 2\pi)$, using $10^{4}$ realizations at each $PI_{450}$.
Within each realization, we applied the same $(\delta Q, \delta U)$ to every position because the three-step correction gives one pair of intercepts for each field.
We recalculated polarized intensity and position angle from the perturbed Stokes values.
We characterized the changes in $P_{450}/P_{850}$, $\gamma - \alpha$ (equation~\ref{eq:spectral_indices}), and the 450\,$\mu$m polarization angle, $\delta\chi_{\rm off}$, which enter the analyses of Papers~S1 and~S2.
 
We adopt $b = 0.07$~mJy~arcsec$^{-2}$ as a representative residual amplitude.
This is the median combined residual amplitude over the sample of individual fields, and the upper quartile extends to $0.13$~mJy~arcsec$^{-2}$ (sub-subsection~\ref{ass:ResidualOffsets}); it is not an upper bound.
The simulated uncertainty decreases with increasing $PI_{450}$, because the additive residual becomes a smaller fractional perturbation.
We therefore evaluated the uncertainty over the polarized intensities represented in the sample.
Taking the median polarized intensity of each source in the 35-source {\sc SpIdx} sample (table~\ref{tbl:T_sample_inventory}), we find a sample median of $PI_{450} = 0.68$~mJy~arcsec$^{-2}$, with a Maritz--Jarrett standard error of $0.35$~mJy~arcsec$^{-2}$ and a 16th--84th percentile range of $0.15$--$2.4$~mJy~arcsec$^{-2}$.
The same quantity measured over the wider set of fields with matched data in both bands agrees with the {\sc SpIdx} value within its uncertainty, so we quote the {\sc SpIdx} value alone.
At $PI_{450}=0.68$~mJy~arcsec$^{-2}$, representative of the
{\sc SpIdx} sample, the simulations give, for an individual source,
\begin{equation}
\sigma(P_{450}/P_{850}) \lesssim 0.05,
\qquad
\sigma(\gamma - \alpha) \lesssim 0.11,
\qquad
\sigma(\chi_{450}) \lesssim 2\fdg1 ,
\end{equation}
where each $\sigma$ is the standard deviation over the realizations.
The central 68\% intervals drawn in figure~\ref{fig:offset_mc} are about 20\% wider than these values, because the component of a randomly oriented residual projected onto a fixed direction follows an arcsine rather than a Gaussian distribution.
 
Papers~S1 and~S2 report medians over 30--35 fields, so we next evaluate how the uncertainties of individual sources propagate to the sample medians.
We fitted the three-step correction separately in each field and modeled the residual offsets as independent between fields.
We repeated the simulation with an independent residual vector for each field at its measured $PI_{450}$ and took the median of the perturbed values for individual sources.
The repeated simulation gives $\sigma \simeq 0.008$ for the sample median of $P_{450}/P_{850}$ and $\sigma \simeq 0.018$ for the sample median of $\gamma - \alpha$.
Both values are smaller by a factor of $6.5$ than the values for individual sources given above, close to the expected $\sqrt{35}$ reduction.
For uncertainties in the sample medians, we adopt in table~\ref{tbl:systematics} the conservative bounds of $\lesssim 0.02$ for $P_{450}/P_{850}$ and $\lesssim 0.04$ for $\gamma-\alpha$.
These bounds are about 2.6 and 2.3 times the corresponding
standard deviations obtained from the simulations.
 
\begin{deluxetable}{lccc}
\tablecaption{Systematic uncertainty contributions to the dual-band ratio diagnostics\label{tbl:systematics}}
\tablehead{
  \colhead{Systematic source} &
  \colhead{Adopted magnitude} &
  \colhead{Effect on $P_{450}/P_{850}$} &
  \colhead{Effect on $\gamma - \alpha$}
}
\startdata
Absolute flux calibration --- 850\,$\mu$m\tablenotemark{a}
& $\pm 7$\%
& $\approx 0$\tablenotemark{d}
& $\approx 0$\tablenotemark{d} \\
Absolute flux calibration --- 450\,$\mu$m\tablenotemark{a}
& $\pm 17$\%
& $\approx 0$\tablenotemark{d}
& $\approx 0$\tablenotemark{d} \\
450\,$\mu$m $Q/U$ zero-level offset residual\tablenotemark{b}
& $\approx 0.07$~mJy~arcsec$^{-2}$ (median)
& $\lesssim 0.02$
& $\lesssim 0.04$ \\
Stokes-$I$ intensity threshold\tablenotemark{f}
& $I/\sigma_I > 50$ in both bands (Method~3)
& filtered upstream\tablenotemark{g}
& filtered upstream\tablenotemark{g} \\
Beam morphology mismatch\tablenotemark{c}
& \parbox[t]{4.5cm}{\raggedright Wiener deconvolution to the modeled 850\,$\mu$m beam\strut}
& ---\tablenotemark{e}
& ---\tablenotemark{e} \\
Beam spatial-frequency truncation\tablenotemark{c}
& $300\arcsec$ common-scale cutoff
& ---\tablenotemark{e}
& ---\tablenotemark{e} \\
\hline
Practical systematic floor (quantified terms)
& ---
& $\lesssim 0.02$
& $\lesssim 0.04$ \\
\enddata
 
\tablenotetext{a}{We adopt the extended-source FCF uncertainties of \citet{Mairs2021}.
Subsection~\ref{ss:flux} describes the calibration of surface brightness, and sub-subsection~\ref{sss:discussion_fcf} gives the contribution of the uncertainty in this calibration.}
\tablenotetext{b}{The three-step procedure for correcting the 450\,$\mu$m offsets is described in subsection~\ref{ss:offset} and sub-subsection~\ref{ass:3step}.
The residual level is characterized in sub-subsection~\ref{ass:ResidualOffsets} and propagated in sub-subsection~\ref{sss:discussion_offset}.
The quoted residual amplitude is the sample median, not an upper bound.
The two propagated entries refer to the sample medians of Papers~S1 and~S2, assuming independent residuals between fields.
For an individual source at $PI_{450}=0.68$\,mJy\,arcsec$^{-2}$, the same simulation gives $\sigma(P_{450}/P_{850})\lesssim0.05$, $\sigma(\gamma-\alpha)\lesssim0.11$, and $\sigma(\chi_{450})\lesssim2\fdg1$ (sub-subsection~\ref{sss:discussion_offset}).}
\tablenotetext{c}{The procedure for beam harmonization is described in subsection~\ref{ss:beamval}; its contribution to the systematic budget is discussed in sub-subsection~\ref{sss:discussion_beam}.
The two rows concerning the beam describe successive steps in harmonization.
The first replaces the morphology of the native 450\,$\mu$m beam with the modeled 850\,$\mu$m compound-beam response via Wiener deconvolution.
The second truncates the largest recoverable scale at 450\,$\mu$m to the common $12\farcs6$--$300\arcsec$ range matched to the standard 850\,$\mu$m reduction.}
\tablenotetext{d}{FCF errors are multiplicative and cancel in $P = PI/I$.
They also enter $\alpha$ and $\gamma$ in common mode and therefore cancel in $\gamma - \alpha$.}
\tablenotetext{e}{Common multiplicative response factors cancel in the ratio diagnostics.
Residual errors in beam shape and spatial filtering can change the relative spatial weighting of total and polarized emission.
Their contributions to these diagnostics are not quantified by the polarization-angle comparisons alone.}
\tablenotetext{f}{The Method~3 Stokes-$I$ mask requires $I/\sigma_I > 50$ and selection from the brightest 5\% of the pixels of each source at both wavelengths, with no threshold on the signal-to-noise ratio of polarized intensity (Paper~S1 subsection~\ref{S1-ss:P_ratio}).
This mask defines the parent set of pixels for the systematic budget.}
\tablenotetext{g}{Above the intensity threshold, we debias polarized intensity with the modified asymptotic estimator (MAS) of \citet{Plaszczynski2014} and \citet{Montier2015}.
We report the remaining statistical scatter separately as $\delta X_{\rm stat}$ in Papers~S1 and~S2, rather than as a systematic floor in this table.}
 
\tablecomments{
Systematic contributions to the polarization-fraction ratio $P_{450}/P_{850}$ and the spectral-index difference $\gamma-\alpha$ (subsection~\ref{sec:uncertainty}).
Paper~S2 subsection~\ref{S2-ss:S2_P_ratio} defines the three estimators of the ratio.
Equations~(\ref{eq:spectral_indices}) and~(\ref{eq:Pratio_SpIdx}) define the spectral indices and relate the two diagnostics.
The practical floor for $\gamma-\alpha$, $\sigma\lesssim0.04$, is about an order of magnitude below the observed median displacement $|\gamma-\alpha|\simeq0.51$ (Paper~S1 F2).
This comparison concerns the quantified terms.
Note~(e) discusses the contributions of residual errors in beam shape and spatial filtering.
The table does not include a difference in the modulation efficiency of POL-2 between the two bands (sub-subsection~\ref{ss:fcf_calibration}).
}
\end{deluxetable}
 
The angular response follows from the geometry of the Stokes $Q$--$U$ plane.
For small residuals, the component parallel to the intrinsic polarization vector changes mainly the polarized intensity, whereas the perpendicular component changes mainly the polarization angle.
We define the residual amplitude as $b=\sqrt{(\delta Q_{450})^2+(\delta U_{450})^2}$ and adopt $b=0.07$~mJy~arcsec$^{-2}$ as a representative value.
When $PI_{450}>b$, the maximum angular shift is
\begin{equation}
|\delta\chi_{\rm off}|_{\rm max}
=
\frac{1}{2}\arcsin\!\left(\frac{b}{PI_{450}}\right).
\end{equation}
The dotted curves in panel~(c) of figure~\ref{fig:offset_mc} show this upper bound as a function of $PI_{450}$ for the adopted $b=0.07$~mJy~arcsec$^{-2}$.
At the sample median $PI_{450}=0.68$~mJy~arcsec$^{-2}$, the maximum angular shift is $3\fdg0$.
The maximum shift is smaller than the median angle residual between 450\,$\mu$m reductions after beam harmonization, $4\fdg7$ over the subsample of 30 sources used for beam validation (subsection~\ref{ss:beamval}).
Sources with larger residuals, or with fainter polarized emission, can lie above the value quoted here because $b$ is a representative amplitude rather than a bound.

We corrected the observed polarized intensities with the modified asymptotic estimator of \citet{Plaszczynski2014}.
We did not apply this estimator in the simulations.
At fixed noise parameters, the debiasing transformation $PI_{\rm db}=f(PI)$ gives $\delta PI_{\rm db}\simeq f'(PI)\,\delta PI$.
Omitting the correction therefore rescales the simulated intensity response by the local slope $f'$ and leaves the polarization angles unchanged.
 
The simulations isolate the contribution from residual Stokes $Q_{450}$ and $U_{450}$ offsets and do not include random measurement noise.
We therefore recommend including the resulting uncertainties alongside the statistical uncertainties reported in Papers~S1 and~S2 when assessing the total uncertainty.
Subsection~\ref{ss:noisemodel} describes how measurement noise propagates into polarized intensity and angle.
 
The Monte Carlo calculation estimates the contribution of the residual offset for individual pixels and is therefore most directly relevant to Method~3, which weights pixels equally, in Paper~S2 subsection~\ref{S2-ss:S2_P_ratio}.
Paper~S2 measures $P_{450}/P_{850}$ with three complementary estimators: an intensity-weighted mean (Method~1), a weighted correlation fit (Method~2), and a distribution of pixel values with equal weights (Method~3).
The last is the standard estimator used to report results for individual sources.
The residual offset is shared by all positions within a field, so its contribution depends on the polarization structure and the weighting used by each estimator.
For all three estimators of Paper~S2 (subsection~\ref{S2-ss:S2_P_ratio}), we adopt the result of this calculation as a common representative uncertainty in $P_{450}/P_{850}$ due to residual offsets.
This is a reference estimate because we have not propagated the common residual vector through the specific weighting procedure of each estimator.

The residual amplitude $b$ is measured from the three-step regressions and therefore describes the maps from which the three-step intercepts are subtracted (sub-subsection~\ref{ass:ResidualOffsets}).
For the 14 Category~1 fields of the {\sc SpIdx} sample that use maps reduced with optimized thresholds without subtraction, the offset measured at the selected threshold is the minimum $a_0^{\rm tot}$ over the threshold grid (sub-subsection~\ref{ass:offset_significance}; examples in appendix figures~\ref{afig:PCAcutoff_cat1_p1} and~\ref{afig:PCAcutoff_cat1_p2}).
The Monte Carlo calculation applies the same $b$ to all 35 sources, and table~\ref{tbl:systematics} gives the resulting bounds for the sample medians.
The tables of $P_{450}/P_{850}$ and spectral indices in Paper~S2 do not include a separate term for this systematic uncertainty in each entry.
 
\subsubsection{Beam Mismatch}
\label{sss:discussion_beam}
Beam convolution acts as a linear operator on the intrinsic sky brightness.
Comparing extended emission observed with different effective beam profiles therefore requires explicit beam transformation rather than simple smoothing (e.g., \citealt{Furuya2006}, appendix~A).
We transform the 450\,$\mu$m maps to the 850\,$\mu$m angular response with a Wiener deconvolution kernel (subsection~\ref{ss:beamval}):
\begin{equation}
K =
\mathcal{F}^{-1}
\left[
\frac{\tilde{B}_{850}}{\tilde{B}_{450}}
\right],
\end{equation}
where $\tilde{B}_{450}$ and $\tilde{B}_{850}$ are the Fourier transforms of the two-component beam models of \citet{Mairs2021}.
This procedure transforms the effective 450\,$\mu$m beam into the 850\,$\mu$m beam response rather than simply broadening the map.
 
For the ratio quantities considered here, we distinguish uncertainties in a common multiplicative scale from uncertainties in the spatial response.
A multiplicative factor common to Stokes $I$, $Q$, and $U$ at a given wavelength cancels exactly in $P=\sqrt{Q^2+U^2}/I$.
The corresponding contributions to $\alpha$ and $\gamma$ are identical and therefore cancel in $\gamma-\alpha$.
This cancellation applies to uncertainties in the beam-area conversion insofar as they act as a common multiplicative factor.
 
Errors in beam shape and spatial filtering have a different effect.
For a beam response $B$, the measured polarization fraction is
\begin{equation}
P_B =
\frac{\sqrt{(B*Q)^2+(B*U)^2}}{B*I},
\end{equation}
where $*$ denotes convolution.
The response mixes emission with different polarization fractions and angles, even though it is applied to all three Stokes maps.
The resulting changes in polarized and total intensity are not necessarily proportional.
Residual errors in spatial response therefore do not necessarily cancel in $P_{450}/P_{850}$, nor, through equation~(\ref{eq:Pratio_SpIdx}), in $\gamma-\alpha$.
 
Harmonization of the beam establishes a common angular response and a range of recovered angular scales for comparison.
The quantities analyzed here refer to this common response.
Uncertainties caused by departures from it in either band depend on the spatial structure of the source.
The comparisons in subsection~\ref{ss:beamval} quantify the sensitivity of polarization angles to the tested choices for beam matching and filtering.
The corresponding sensitivity of ratios of polarization fractions and differences in spectral indices is not evaluated here.
This sensitivity can be assessed by recomputing these quantities under variations of the adopted spatial response.
The simulations of residual offsets in sub-subsection~\ref{sss:discussion_offset} quantify the additive contribution for the adopted maps, separately from this dependence on spatial response.
 
\subsection{Survey-Wide Technical Precision Floor and Implications for Papers~S1 and~S2}
\label{sec:precisionfloor}
The preceding analysis identifies which systematic contributions remain in the ratios used to describe the survey and quantifies the calibration and residual $Q/U$ zero-level offset terms.
Table~\ref{tbl:systematics} summarizes the resulting budget of systematic uncertainties.
We must assess the errors that remain in the ratio before interpreting a lower polarization fraction at 450\,$\mu$m as a change in polarized dust emission.

The dominant multiplicative uncertainties in flux calibration cancel in the quantities used for the scientific interpretation.
FCF errors cancel in $P = PI/I$ and hence in $P_{450}/P_{850}$.
Errors in the FCF and in the normalization by beam area contribute identically to $\gamma$ and $\alpha$ and therefore cancel in $\gamma - \alpha$.
Among the terms quantified here, the only one that contributes appreciably is the residual additive 450\,$\mu$m Stokes-$Q/U$ zero-level offset.
Even this term remains small over the range of polarized intensities spanned by the statistical sample.
 
The resulting practical floors for the three dual-band observables are
\begin{equation}
\sigma(\widetilde{P_{450}/P_{850}}) \lesssim 0.02,
\qquad
\sigma(\widetilde{\gamma - \alpha}) \lesssim 0.04,
\qquad
\widetilde{|\Delta\chi|} \sim 4\degr\hbox{--}6\degr ,
\label{eq:precision_floors}
\end{equation}
where a tilde denotes a median.
We conclude that errors in flux calibration or residual zero-level offsets in Stokes $Q/U$ at 450\,$\mu$m do not account for the detection of $\gamma < \alpha$.
The first two floors are substantially smaller than the signals reported for the sample as a whole in Papers~S1 and~S2.
The floor for $\gamma - \alpha$ is an order of magnitude below the observed median difference between the spectral indices, $|\gamma - \alpha| \approx 0.51$ (Paper~S1 F2).
Similarly, the systematic floor on $P_{450}/P_{850}$ is far below the observed departure of the polarization-fraction ratio from unity (Paper~S1 F3; Paper~S2 subsection~\ref{S2-ss:S2_P_ratio}).

The first two floors are conservative bounds on the uncertainties of the medians calculated over sources and reported in Papers~S1 and~S2.
These bounds cover the calibration and residual $Q/U$ offsets quantified here.
We have not evaluated how beam matching and spatial filtering affect these two medians, so these effects are not included in the bounds.
The bounds also do not include a difference in the modulation efficiency of POL-2 between the two bands (sub-subsection~\ref{ss:fcf_calibration}).
These bounds are smaller than the values for individual sources in sub-subsection~\ref{sss:discussion_offset}, $\lesssim 0.05$ and $\lesssim 0.11$ at the sample median $PI_{450} = 0.68$~mJy~arcsec$^{-2}$.
This difference arises because the residual $Q/U$ offsets, modeled as independent between fields, partially average out in a median over sources.

The third floor is the typical value of the median angle residual for individual sources, measured between 450\,$\mu$m reductions after beam harmonization (subsection~\ref{ss:beamval}).
This floor sets the reference scale for differences in polarization angle between the two bands.
The residual $Q/U$ zero-level offset adds $\lesssim 2\fdg1$ to this scale for each source (sub-subsection~\ref{sss:discussion_offset}).
 
The cancellation of common multiplicative factors is central to the design of subsequent scientific analyses.
The paired diagnostics $P_{450}/P_{850}$ and $\gamma - \alpha$ are more robust because the leading multiplicative systematics cancel.
Individual measurements of $P_{450}$, $PI_{450}$, $\alpha$, and $\gamma$ can have larger uncertainties from calibration or the beam.
We therefore use ratio diagnostics with controlled technical floors for the principal analyses in Papers~S1 and~S2.
 
Equation~(\ref{eq:precision_floors}), which includes the floor of the uncertainty in polarization angle from subsection~\ref{ss:beamval}, completes the technical framework of three pillars.
The present paper quantifies technical limits for the three principal dual-band observables used in Papers~S1 and~S2: the calibration and residual $Q/U$ zero-level offset terms in $P_{450}/P_{850}$ and $\gamma - \alpha$, and the sensitivity of the morphology of polarization angles in the two bands to processing.
We have explicitly propagated the quantified systematics from calibration, noise, and offsets into the observables used for the survey and shown that they are smaller than the astrophysical signals traced by those observables.
Our scientific interpretation in Papers~S1 and S2 rests on these results.
Table~\ref{tbl:T_dependency} summarizes the technical results of this paper that are used in Papers~S1 and~S2.
The cancellation of common multiplicative calibration factors does not remove additive baselines or changes in spatial weighting.
The physical interpretation therefore requires an uncertainty budget appropriate to the source, its polarized intensity, and the map products being compared.
 
\begin{longrotatetable}
\begin{deluxetable*}{lll}
\tablecaption{Paper~T technical results used in Papers~S1 and~S2\label{tbl:T_dependency}}
\tablewidth{0pt}
\tablehead{
  \colhead{Technical result} &
  \colhead{Established here} &
  \colhead{Used in Papers~S1 and~S2}
}
\startdata
\parbox[t]{5.4cm}{\raggedright Pillar~1: extended-source FCF + $\beta = \alpha - C(T_{\rm d})$ (equation~\ref{eq:planck_correction})\strut} &
Subsection~\ref{ss:flux} &
\parbox[t]{6.6cm}{\raggedright Paper~S1 subsection~\ref{S1-ss:SpIdx} ($\alpha, \gamma$ for F2); Paper~S2 subsection~\ref{S2-ss:S2_SpIdx} (F4 spectral-index inventory)\strut} \\
\parbox[t]{5.4cm}{\raggedright Pillar~2: Stokes-$I$ + $Q,U$ analytic noise model\strut} &
Subsections~\ref{ss:noiseI}--\ref{ss:noisemodel} &
\parbox[t]{6.6cm}{\raggedright Paper~S1 subsections~\ref{S1-ss:bfield_consistency}--\ref{S1-ss:P_ratio} (F1, F3 detection masks); Paper~S2 subsection~\ref{S2-ss:S2_HGBS} (F6 per-pixel masks)\strut} \\
\parbox[t]{5.4cm}{\raggedright Pillar~3: 450\,$\mu$m $Q/U$ zero-level offset correction + beam harmonization\strut} &
Subsections~\ref{ss:offset}--\ref{ss:beamval} &
\parbox[t]{6.6cm}{\raggedright matched-resolution dual-band products throughout Papers~S1 and~S2\strut} \\
\parbox[t]{5.4cm}{\raggedright Systematic-uncertainty floor (quantified terms) $\sigma(P_{450}/P_{850}) \lesssim 0.02$, $\sigma(\gamma{-}\alpha) \lesssim 0.04$\strut} &
Subsection~\ref{sec:uncertainty} (table~\ref{tbl:systematics}) &
\parbox[t]{6.6cm}{\raggedright $\delta X_{\rm sys}$ entries throughout Paper~S1 section~\ref{S1-sec:results} and Paper~S2 section~\ref{S2-sec:S2_results}\strut} \\
\parbox[t]{5.4cm}{\raggedright Sample inventory (174/36/30/9 sources; 417 pixel pairs for Method~3)\strut} &
Subsection~\ref{ss:obs_sample} (table~\ref{tbl:T_sample_inventory}) &
\parbox[t]{6.6cm}{\raggedright Paper~S1 35-source {\sc SpIdx} + 21-source dual-band-angle + 9-region HGBS subsets;\\ Paper~S2 69-source {\sc PolSrc} + 7-source CMZ + 417-pixel HGBS\strut} \\
\enddata
\tablecomments{
The five technical results listed here support the scientific analyses of Papers~S1 and~S2.
The derivation of these results does not depend on the scientific conclusions of those papers.
Corresponding lists of inputs from Paper~T appear in Paper~S1 table~\ref{S1-tbl:S1_dependency} and Paper~S2 table~\ref{S2-tbl:S2_dependency}.
}
\end{deluxetable*}
\end{longrotatetable}

\section{Summary}
\label{sec:summary}
 
We have developed and tested a methodology to calibrate, select, and compare JCMT SCUBA-2 / POL-2 polarization
measurements at 450 and 850\,$\mu$m.
We summarize the results first as five technical validations and then as the three methodological pillars they establish, following figure~\ref{fig:T_pillar_overview} and table~\ref{tbl:T_pillar_index}.
 
\subsection{Principal results of the five technical validations}
\label{ss:summary_technical_validations}
 
\begin{enumerate}
 
\item \textit{TV1: Surface-brightness calibration and spectral indices.}
Calibrating extended emission with FCF$_{\rm arcsec}$ in mJy\,arcsec$^{-2}$ avoids the deterministic shift of $\Delta\beta=-1.05$ introduced by applying FCF$_{\rm peak}$ to these data.
The relation $\beta=\alpha-C(T_{\rm d})$ (equation~\ref{eq:planck_correction}) provides the corresponding basis for interpreting dust spectral indices.
Our comparison of the calibrations demonstrates the importance of a consistent surface-brightness scale for thermal dust emission in molecular clouds (subsection~\ref{ss:flux}).
 
\item \textit{TV2: Stokes-$I$ noise and data selection.}
Across 174 POL-2 fields, we separated additive noise from fluctuations that depend on signal strength, using a two-component model and a hybrid estimator.
The median ratio of surface-brightness noise is $\sigma_{450}/\sigma_{850}\simeq25$, and the adopted fit gives $\sigma_{450}\propto\sigma_{850}^{1.94\pm0.16}$.
At both wavelengths, the decrease in noise with integration time is consistent with the ideal $t^{-1/2}$ scaling (subsection~\ref{ss:noiseI}).
The measured noise properties provide an empirical basis for selection based on intensity and for estimates of sensitivity.
 
\item \textit{TV3: Stokes-$Q,U$ noise and polarization uncertainties.}
The two-component model extends to $Q$ and $U$, separating additive noise from gain fluctuations that depend on signal strength.
The additive noise scales $\sigma_Q$ and $\sigma_U$ are closely matched, with typical absolute logarithmic ratios $|\log_{10}(\sigma_Q/\sigma_U)|\simeq0.03$.
Blank-sky measurements and differences between repeated observations independently support weak correlations between the errors in $Q$ and $U$.
These tests support the approximately isotropic additive-noise model used to propagate uncertainties in the polarized intensity and polarization angle (subsection~\ref{ss:noisemodel}).
We examined running medians of $\alpha$ and $\gamma$ against their formal $S/N$ and tested the effect of imposing a threshold on $S/N(\gamma)$.
These tests show that the sign of the population offset $\gamma-\alpha$ does not depend on selection by $S/N(\gamma)$, whereas its amplitude does (sub-subsections~\ref{ss:spidx_snrbias}--\ref{ss:spidx_snrbias_sens}).
 
\item \textit{TV4: Additive Stokes-$Q/U$ zero-level offsets and preparation of matched maps.}
PCA settings chosen for individual fields and a three-step regression procedure reduce additive 450\,$\mu$m Stokes-$Q/U$ zero-level offsets of order $1$\,mJy\,arcsec$^{-2}$.
The combined residual amplitude has a median of approximately $0.07$\,mJy\,arcsec$^{-2}$.
The maps correspond to the 850\,$\mu$m angular response and the common $12\farcs6$--$300\arcsec$ range of angular scales.
Independent SOFIA/HAWC+ comparisons support the correction in strongly affected fields, including Ophiuchus~A (subsection~\ref{ss:offset}).
 
\item \textit{TV5: Sensitivity of polarization angles to beam
matching and spatial filtering.}
Comparisons of three reductions of the same 450\,$\mu$m observations quantify the response of the inferred $\vec{B}$-field orientations to these processing choices.
Taking the median angle difference from the final product for each of the 30 sources, we find that the sample median decreases from $6\fdg3$ to $4\fdg7$ after beam matching; 29 sources have median differences below $10\degr$ at $I/\sigma_I>30$.
The comparisons of Ophiuchus~A on both $12\arcsec$ and $4\arcsec$ grids support the same trend.
The resulting scale of $4\degr$--$6\degr$ characterizes sensitivity to the tested 450\,$\mu$m reduction choices and provides a reference for interpreting angle differences between the bands (subsection~\ref{ss:beamval}).
 
\end{enumerate}
 
\subsection{Three methodological pillars for dual-band analysis}
\label{ss:summary_pillars}
 
The five validations establish three complementary requirements for interpreting polarization measurements in both bands.
 
\begin{enumerate}
 
\item \textit{Pillar~1: Photometric consistency.}
To relate the observed dust spectral slopes to grain properties, we establish a consistent calibration of surface brightness and account for dust temperature in interpreting the spectral indices (TV1).
 
\item \textit{Pillar~2: Noise characterization.}
To investigate spatial variations in grain alignment and $\vec{B}$-field structure within molecular clouds, we relate the data selection and polarization uncertainties to the measured noise properties of Stokes $I$, $Q$, and $U$ (TV2 and TV3).
These analyses provide the statistical basis for assessing the significance of variations in polarization fraction and field orientation.
 
\item \textit{Pillar~3: Map-domain comparability.}
To investigate how polarization fractions and projected $\vec{B}$-field structures depend on wavelength, we prepare maps with compatible zero levels and spatial responses (TV4).
We then quantify the sensitivity of the recovered polarization angles to the processing choices (TV5), providing a technical reference for interpreting the observed differences between the bands.
 
\end{enumerate}
 
The combined uncertainty assessment in subsection~\ref{sec:uncertainty} translates these results
into the derived quantities used in Papers~S1 and~S2.
Multiplicative FCF errors common to $I$ and $PI$ within each band cancel in $P_{450}/P_{850}$ and $\gamma-\alpha$.
Using the model of residual offsets (sub-subsection~\ref{sss:discussion_offset}), we adopt $\sigma(P_{450}/P_{850})\lesssim0.02$ and $\sigma(\gamma-\alpha)\lesssim0.04$ for the sample medians.
These two floors cover only the calibration and residual offset terms quantified in subsection~\ref{sec:precisionfloor}.
This assessment treats the residual vectors as independent between fields.
We use these estimates together with the statistical uncertainties and the sensitivity of polarization angles to processing established by TV5.
The quantified errors in flux calibration and residual Stokes-$Q/U$ offsets do not account for the lower polarization fraction at 450\,$\mu$m relative to 850\,$\mu$m reported for the sample as a whole in Papers~S1 and~S2.
Effects of beam matching and spatial filtering on this ratio, and any difference in modulation efficiency between the two bands, remain outside this budget.
 
The methodology provides calibration choices, noise diagnostics, and correction procedures for future observations and archival analyses of POL-2 data.
We apply it in Paper~S1 to tests of grain alignment and $\vec{B}$-field structure, and in Paper~S2 to the census of polarization in both bands for 69 sources and comparisons among molecular cloud environments.
The products of the technical validations, including the 450\,\micron\ Stokes-$Q/U$ maps with matched beams and the parameters for offset correction in appendix table~\ref{tab:PCAparams}, will be made available electronically as supplementary material.

\begin{acknowledgments}
RSF is deeply grateful to the medical staff who have continuously supported his recovery from cancer.
Their dedicated care made this work possible.
The authors thank Paul Ho, the former Director of the East Asian Observatory (EAO), for his full and unwavering support for this project, as well as the EAO staff for their continued assistance over a decade.
We also express our sincere gratitude to Dr.~Jessica Dempsey, former Deputy Director of the EAO, for her continuous and decisive leadership of the international instrument team throughout the commissioning and validation phase of POL-2.
Dr.~Dempsey sustained institutional support for the extensive test observations and performance assessments needed to establish POL-2 as a reliable scientific instrument, particularly to characterize instrumental polarization (IP).
She also enabled the participating teams to contribute their full capabilities.
In particular, she approved and supported dedicated IP validation experiments conducted with the JCMT dome cover removed, thereby enabling critical verification of the IP correction models developed independently by DB and PB.
We acknowledge Takeshi Inagaki, a UNIX systems engineer at the Subaru Telescope, who visited RSF for approximately two weeks each year from 2015 to 2020 and made substantial contributions to improving the efficiency and reliability of the data-analysis workflow.
We also thank the successive Directors of the Subaru Telescope, who not only approved Mr.~Inagaki's long-term annual visits, but also provided extensive in-kind support during JCMT observing runs from 2015 to 2019.
We gratefully acknowledge the Astronomy Data Center (ADC) of the National Astronomical Observatory of Japan (NAOJ) for long-term support and access to shared computing resources, including 32 UNIX machines and more than 100~TB of storage.
In 2022, Mr.~Inagaki also made \texttt{pol2map} run efficiently on all 32 of these machines, including the handling of jobs under the security settings of the ADC.
The authors are grateful to Drs.~Dan Bintley and Shaoliang Li (EAO), who generously shared their in-depth knowledge of SCUBA-2 instrumentation and the JCMT optics through close collaboration as members of the SCUBA-3 development team.
Their critical insights into instrumental characteristics and systematic effects informed the uncertainty analysis presented in this work.
We thank Prof.\ Doug Johnstone for helpful advice on this manuscript.
Discussions of observational uncertainties associated with submillimeter telescopes and bolometer cameras greatly benefited from interactions with Prof.~Ryohei~Kawabe and Dr.~Tai~Oshima (NAOJ).
We thank Prof.~Yoshito~Shimajiri (Kyushu Kyoritsu University, Japan) for discussions of bolometer imaging algorithms, Dr.~Hiroyuki~Hirashita (ASIAA, Taiwan) for discussions of dust properties, and Dr.~Riccardo~Cesaroni (INAF Arcetri, Italy) for advice on \emph{Herschel} data handling.
Discussions with Profs.~Hideki Takami and Yutaka Hayano (Subaru Telescope), drawing on their experience in adaptive optics development and operations, helped us understand atmospheric fluctuations at Maunakea relevant to submillimeter observations.
RSF expresses sincere gratitude to Prof.~Akira~Obuchi (Tokushima University, Japan) for extensive guidance and invaluable discussions on mathematical aspects relevant to this work.
Finally, we thank Mike Chen, Qilao Gu, Jennifer Hatchell, Tsuyoshi Inoue,
Masato I.~N.\ Kobayashi, Jungmi Kwon, Chi-Yan Law,
Yusuke Tsukamoto, Jintai Wu, Tetsuya Zenko, and Yapeng Zhang
for helpful discussions.
 
These observations were obtained by the James Clerk Maxwell Telescope, operated by the East Asian Observatory on behalf of The National Astronomical Observatory of Japan; Academia Sinica Institute of Astronomy and Astrophysics; the Korea Astronomy and Space Science Institute; the National Astronomical Research Institute of Thailand; Center for Astronomical Mega-Science (as well as the National Key R\&D Program of China with No.~2017YFA0402700).
Additional funding support is provided by the Science and Technology Facilities Council of the United Kingdom and participating universities and organizations in the United Kingdom and Canada.
The POL-2 data analyzed here were obtained under JCMT programs M16AL004, M17BL011, M20AL006 and M20AL018 (BISTRO Large Programs), together with E21AC002, E21BJ001, E21BJ002, M16AD003, M17AP019, M17AP067, M17AP073, M17AP074, M17BP037, M17BP041, M17BP046, M17BP050, M17BP070, M18AH27C, M18BD002, M18BP019, M18BP030, M18BP041, M18BP055, M19AP003, M19AP009, M19AP014, M19AP016, M19AP032, M19AP034, M19AP038, M19BP015, M19BP058, M19BP061, M20AP022, M20AP023, M20AP040, M20BP007, M20BP022, M20BP043, M20BP046, M21AH20A, M21AP016, M21AP021, M21BF004, M21BF005, M21BP004, M21BP045, M21BP083, M22AP018, M23BP011, R20XP002.
Programs M21BF004 and M21BF005, which observed fields in Orion~A, used observing time purchased through the EAO Individually Funded Program with JSPS KAKENHI Grant Numbers JP19H01938 and JP21H00045.
This work is based in part on archival observations made with the NASA/DLR Stratospheric Observatory for Infrared Astronomy (SOFIA), retrieved from the NASA/IPAC Infrared Science Archive (IRSA).
SOFIA was jointly operated by the Universities Space Research Association, Inc.\ (USRA), under NASA contract NNA17BF53C, and the Deutsches SOFIA Institut (DSI) under DLR contract 50~OK~2002 to the University of Stuttgart.
The SOFIA/HAWC+ observations used here were obtained under plan IDs 09\_0054, 09\_0203, 09\_0215, 70\_0509, 70\_0511 and 70\_0609.
 
\textit{Use of AI tools.}
This study began in early 2016, and AI tools participated in it for roughly the last two years, from the autumn of 2024.
Their role changed as both the research and the tools developed.
We describe it here in some detail because AI tools contributed substantially to the analysis code and to the text of all three papers of this series.
Before that period, until about 2023, RSF used the AI-based language tools Grammarly and DeepL under paid subscriptions.
He corrected the English of his drafts with Grammarly and checked their meaning by translating them back into Japanese with DeepL.
Some of the wording corrected with Grammarly probably remains in the present text, but it can no longer be identified.
 
Before AI tools were introduced, the observations and the basic data reduction had been completed by people.
All raw data were transferred from the JCMT Science Archive between 2017 March and 2024 February and reduced to Stokes $I$, $Q$, and $U$ maps (steps~1--3 of \texttt{pol2map}) by RSF on the computers of the NAOJ Astronomy Data Center.
No command or parameter of these steps was produced by an AI tool.
This reduction rests on the research of DB, who implemented \texttt{pol2map} in the Starlink SMURF package and advised on its use.
RSF produced the polarization-vector catalogs using his procedure; their scripts were later tidied with ChatGPT and refined with Claude Code.
The core analysis code was developed by RSF from 2015, together with Takeshi Inagaki until 2022, and RSF's notebooks of 2018--2024 underlie much of the present analysis.
From the autumn of 2024, AI tools took over part of the role that Mr.~Inagaki had played, helping RSF to debug this code and to run it in new computing environments.
Development of the analysis code proceeded iteratively between RSF and the AI tools, including substantial rewriting or re-implementation of parts of RSF's earlier code.
 
The principal tools were ChatGPT (OpenAI; GPT-4o from 2024 October, later GPT-5 and, from 2026 September, GPT-6 Astra) and Claude (Anthropic).
Claude was used through the AI assistant of the Warp terminal from the autumn of 2024 (Claude 3.5-series and other models; the model that answered cannot always be identified), through Claude Cowork, and through Claude Code.
The session logs of Claude Code, which survive from 2026 June onward, record the models Claude Opus 4.8, Claude Fable 5, Claude Opus 5, Claude Fable 5.1, and Claude Opus 5.5, with Claude Haiku 4.5 for auxiliary tasks.
The models used with Claude Code before 2026 June were not recorded.
The periods of use of these tools overlapped, and we do not assert the order in which they were adopted.
OpenAI Codex was used in 2026 September to review the work records and to draft working instructions.
In 2026 October, Codex also checked independently whether the cited papers support the statements for which we cite them (the model was not recorded in either period).
Gemini (Google) was used occasionally, including the summaries shown automatically with search results and Gemini Code Assist in a code editor.
Perplexity was used occasionally until 2026 May; it suggested points to check in the manuscripts, and text or arguments from it may have entered the papers.
 
The AI tools carried out work packages that could not have been completed by people in a reasonable time, and some work packages were led by them.
They shared the management of the large survey data set and the evaluation of calibration observations, for example, of quasars, used as references when inspecting the science data.
They helped RSF to read and interpret DB's implementation in the Starlink software, the Starlink source code, and the accompanying documents, including e-mail correspondence.
They converted RSF's notebooks into stand-alone scripts and packages, wrote new analysis and plotting code to RSF's specifications, and modified and accelerated RSF's existing code.
They transcribed RSF's handwritten notes, articulated, organized, and consolidated RSF's ideas, and drafted and rewrote much of the English text.
In choosing the English language, RSF adopted many of their recommendations to match the strength of each claim to the statistical result.
The three papers were first planned as one, and their division was considered from late 2024 onward.
During this period the AI tools consolidated the many comments received in several rounds from more researchers than there are co-authors.
They were instructed to weigh each comment on its scientific merit alone, irrespective of who had made it, and this also eased communication in English among non-native speakers.
They kept the source counts and the numbers consistent among the text, tables, and figures of the three papers, a task that could not have been done by hand within realistic limits.
In the final stage they read the three manuscripts adversarially as simulated referees set up by RSF.
They also checked the cited statements against the cited papers, which RSF had obtained.
In 2026 October, they revised the English toward RSF's own style, as learned from his hand edits and his published papers.
The AI tools mainly handled short-term writing plans, while RSF set the medium- and long-term plans.
 
RSF set the astronomical motivation and the questions, specified the analyses, and verified the results from the astronomical point of view.
The verification from the data-science point of view was done mainly by the AI tools.
The statistical results were interpreted by RSF together with the AI tools.
RSF decided the outline of the astronomical conclusions, the AI tools then tested this outline adversarially, and RSF made the final decision on what to adopt.
RSF wrote the original versions of much of the analysis code, examined substantial parts of the code written or modified by the AI tools, and checked their outputs.
All visual checks, including separate displays of the original data, were made by RSF.
These checks found errors both in AI-written code and in RSF's earlier code, and the errors were corrected.
For example, RSF noticed that the statistics of one comparison of polarization angles were more tightly aligned than the maps suggested, which revealed a cut on the polarization-fraction ratio inherited from code written in 2021.
RSF also identified the cause of an error in which polarization-angle maps had been read in place of angle-dispersion maps.
 
In sub-subsection~\ref{ss:noiseI_motivation}, the atmospheric model of equations~(\ref{eq:Iobs})--(\ref{eq:Itrans}) was transcribed from RSF's handwritten notes and typeset with GPT-4o, without change of content.
The comparison of Ricean formulations in appendix~\ref{app:ricean} developed over several years in discussions among several of the authors, in the course of which RSF also used ChatGPT.
The contributions of the individuals and of the tool to these parts cannot now be separated, and we have not tried to reconstruct them from memory.
We assess the reliability of these parts from the analyses presented here and the authors' judgment, rather than their origin.
Whether, and how far, the co-authors used AI tools in preparing their own contributions and comments is not known to the first author.
The handling of unpublished data, manuscripts, and correspondence with these tools was judged comprehensively in light of the regulations of Tokushima University and the laws of Japan.
Under this judgment, RSF shared this material with the providers of these tools, mainly Anthropic, OpenAI, and Google.
The data were astronomical maps and catalogs that RSF had derived from raw data that are publicly available in the JCMT Science Archive.
RSF regards such sharing as a premise of using AI tools in research, and this study may have contributed modestly to the development of their models.
 
The figures in this paper were produced with Matplotlib or TikZ between 2025 November and 2026 October, that is, after AI tools had entered this work.
We divide these figures into three groups according to the origin of their plotting code.
Figures~\ref{afig:Q8U8vsQ4U4_initialEstimate}--\ref{fig:450_singlefield_before_after} and \ref{afig:PCAcutoff_cat1_p1}--\ref{afig:PCAcutoff_cat3} were made with analysis code that RSF wrote between 2020 and 2024, before AI tools were used, and extended with AI tools from late 2025 onward.
Figure~\ref{afig:OphA_multiwavelength_Bfield} was made with the same code after an overlay of SOFIA data had been added with AI tools in 2025 November.
Figures~\ref{afig:sensitivity_time}, \ref{afig:f_smooth}, \ref{afig:QUnoise_DR21}--\ref{fig:SpIdx_BiasCheck_examples}, \ref{afig:SOFIA_CDF}, and \ref{fig:QU_W51_case2}--\ref{afig:AzEl_offset_maps} were made with notebooks and scripts that RSF developed from 2025 October onward while using AI tools interactively.
The contributions of RSF and of the tools to this code cannot be separated.
Figures~\ref{fig:T_pillar_overview}--\ref{afig:sig450vs850}, \ref{afig:nefd_transmission}, \ref{fig:T_S6_roadmap}, and \ref{fig:beam_validation_main}--\ref{fig:offset_mc} were made with scripts written by Claude-based coding tools to RSF's specifications, in several cases re-implementing RSF's notebook code.
Figures~\ref{fig:T_pillar_overview}, \ref{fig:T_S4_roadmap}, and \ref{fig:T_S6_roadmap} are TikZ diagrams drafted in the same way.
For submission, Claude Code typeset the TikZ diagrams and the figures with several panels as single figure files, without changing their content.
RSF inspected every figure against the underlying data.
 
No AI tool is an author of this paper.
The authors take full responsibility for its content, including every part prepared with the help of AI tools.
No output of these tools entered the paper without RSF's judgment, and RSF has read the whole text.
Without the commercially available models of OpenAI and Anthropic that took part in this work, these survey papers could not have been brought to submission before RSF's retirement.
\end{acknowledgments}
 
\section*{Funding}
RSF was supported by the Visiting Scholars Program provided by the NAOJ Research Coordination Committee, NINS (\mbox{NAOJ-RCC-23DS-050}).
RSF was supported by Grants-in-Aid for Scientific Research of Japan (JSPS KAKENHI Grant Numbers JP19H01938 and JP21H00045).
Team BISTRO-J gratefully acknowledges partial financial support from 260 individual contributors through a crowdfunding campaign.
 
\section*{Data Availability}
The raw SCUBA-2/POL-2 observations analyzed in this paper are available from the JCMT Science Archive hosted at the Canadian Astronomy Data Centre (CADC).
The SOFIA/HAWC+ Level~4 products used for the independent validation at 154 and 214\,\micron\ (table~\ref{tbl:SOFIA_validation}) are public and were retrieved from the NASA/IPAC Infrared Science Archive (IRSA).
 
\facilities{JCMT (SCUBA-2, POL-2), SOFIA (HAWC+)}
 
\software{Starlink \citep{Currie2014}, SMURF \citep{Chapin2013}, \texttt{pol2map} \citep{Parsons2021}, \textsc{polpack}, SOFIA Redux/\texttt{hawc\_drp} \citep{Clarke2023}, Astropy \citep{Astropy2013,Astropy2018,Astropy2022}, NumPy \citep{Harris2020}, SciPy \citep{Virtanen2020}, Matplotlib \citep{Hunter2007}, \texttt{emcee} \citep{Foreman-Mackey2019}, \texttt{reproject}}
 
\appendix

\section{Comparison of Ricean Expectation Formulations}
\label{app:ricean}
\label{as:bessel_vs_laguerre}%

\subsection{Exact Expectation}
In the Ricean analyses used elsewhere in this series, we evaluate the exact expectation with modified Bessel functions $I_0$ and $I_1$.
Under the condition of isotropic polarization noise validated in subsection~\ref{ss:noisemodel} ($\sigma_Q \simeq \sigma_U \equiv \sigma_{QU}$, ${\rm Cov}(Q,U)\approx 0$), this expectation is
\begin{equation}
\langle PI_{\rm obs} \rangle = \sigma_{QU} \sqrt{\frac{\pi}{2}}\,
\exp\!\left(-\frac{\xi^2}{4}\right)
\left[
\left(1 + \frac{\xi^2}{2}\right) I_0\!\left(\frac{\xi^2}{4}\right)
+ \frac{\xi^2}{2} I_1\!\left(\frac{\xi^2}{4}\right)
\right],
\label{eq:ricean_main}
\end{equation}
where $\xi \equiv PI_{\rm true}/\sigma_{QU}$ is the signal-to-noise ratio of the true polarized intensity.
Equation~(\ref{eq:ricean_main}) is exact and valid for all $\xi$, from the regime dominated by noise ($\xi \ll 1$) through intermediate signal-to-noise ratios ($\xi \sim 1$) to the limit of high signal-to-noise ratio ($\xi \gg 1$).
This expression is identical to the form written with the Laguerre polynomial $L_{1/2}$, $\langle PI_{\rm obs} \rangle = \sigma_{QU}\sqrt{\pi/2}\,L_{1/2}(-\xi^2/2)$.
\citet{Pattle2019} adopted this form as the Ricean-mean model of the polarization fraction in the Ophiuchus molecular cloud.

\subsection{High Signal-to-Noise Expansion}
Even at a high signal-to-noise ratio, noise in $Q$ and $U$ slightly increases the expected value of the measured polarization amplitude.
For $\xi \gg 1$, equation~(\ref{eq:ricean_main}) reduces to
\begin{equation}
\langle PI_{\rm obs} \rangle \simeq PI_{\rm true}
\left(1 + \frac{1}{2\xi^2}\right).
\label{eq:ricean_laguerre}
\end{equation}
This expansion differs from equation~(\ref{eq:ricean_main}) by at most 3.3\% for $\xi \geq 0.7$, but it diverges as $\xi \to 0$.
The exact expectation instead approaches the value for pure Gaussian noise in $Q$ and $U$,
\begin{equation}
\lim_{\xi \to 0} \langle PI_{\rm obs} \rangle =
\sigma_{QU} \sqrt{\frac{\pi}{2}} \approx 1.25\,\sigma_{QU}.
\label{eq:bessel_lowSN}
\end{equation}
The expansion therefore cannot replace equation~(\ref{eq:ricean_main}) for the faint lines of sight that occupy much of a 450\,$\mu$m map.
The non-zero amplitude expected from an unpolarized line of sight is a measurement effect, not a detection of aligned dust.

Debiasing estimators work in the opposite direction: they estimate the true amplitude of an individual measurement from the measured one.
For example, the estimator $PI_{\rm db} = (PI_{\rm obs}^2 - \sigma_{QU}^2)^{1/2}$ of \citet{Wardle1974} has the first-order form $PI_{\rm obs}\,(1 - \sigma_{QU}^2/2PI_{\rm obs}^2)$, with the sign opposite to that of equation~(\ref{eq:ricean_laguerre}) \citep[see also][]{Montier2015,Vidal2016}.
We use such estimators to debias individual measurements (appendix~\ref{app:mas_debias}), not as models of the expected measured amplitude.

\subsection{Recommendation}
We adopt the exact expectation (equation~\ref{eq:ricean_main}) throughout this work; modern numerical libraries evaluate it at negligible cost.
Many useful 450\,$\mu$m lines of sight lie outside the limit of strong polarization, so a model for their measured amplitudes must remain valid as the intrinsic signal approaches the noise.
Using the exact expectation removes an avoidable approximation in modeling weak polarization, provided that the assumed noise model for $Q$ and $U$ is appropriate.
The exact expectation does not recover an intrinsic polarization law of the cloud where the data lack constraining power.
The uncertainty in such a law must remain part of the physical interpretation.

\section{Stokes $Q,U$ Reliability Diagnostics for Bright 850\,\micron\ Fields (W51 and DR21)}
\label{app:QU_diagnostics}
 
In this appendix, we examine the uncertainties in Stokes~$Q$ and~$U$ in two exceptionally bright 850\,\micron\ fields, W51 and DR21, to supplement the characterization of the Stokes-$I$ noise in subsection~\ref{ss:noiseI}.
The $\Delta Q$ and $\Delta U$ diagnostics identify algorithmic systematics that the Stokes-$I$ analysis (subsection~\ref{as:IinvDI}) cannot fully reveal on its own.
They also refine the criteria for selecting pixels in the polarization noise model of subsection~\ref{ss:noisemodel}.
 
The measurement principle of POL-2 explains why the polarization uncertainties can diagnose the reliability of the total intensity.
In POL-2 observations, the signal recorded by each bolometer can be idealized as
\begin{equation}
d(t) = I(t) + Q(t)\cos\{4\gamma(t)\} + U(t)\sin\{4\gamma(t)\} + n(t),
\label{eq:pol2_signal_app}
\end{equation}
where $\gamma(t)$ is the rotation angle of the half-wave plate and $n(t)$ represents noise and systematics.
We extract all three Stokes parameters, $I$, $Q$, and $U$, by fitting a sinusoid to the same time-series data.
In this fit, $Q$ and $U$ correspond to the amplitude and phase of the modulation $4\gamma$, while $I$ corresponds to the mean signal level.
Thus, total intensity is not measured separately before polarization is derived.
The quality of the extraction of $Q$ and $U$ directly reflects the fidelity of the time-series data from which $I$ is also derived.
 
Detector saturation or strong non-linearity can severely compromise sinusoidal demodulation, which cleanly separates the polarized signal from the total intensity when the bolometer operates in its linear regime.
Signal clipping truncates the sinusoidal waveform, causing errors in the amplitude and phase of the fitted $Q$ and $U$.
Gain variations that depend on signal level introduce waveform distortions that generate spurious harmonics.
These distortions can leak intensity fluctuations into the $4\gamma$ component, creating artificial polarization signals through the leakage of Stokes~$I$ into Stokes~$Q$ and~$U$.
Even minor distortions in the strong $I$ signal can produce proportionally large errors in the derived polarization parameters because $Q$ and $U$ represent small modulation amplitudes relative to $I$.
Anomalies in $\Delta Q$ and $\Delta U$ therefore provide a sensitive diagnostic of errors in the time-stream data.
Such errors necessarily imply that the corresponding Stokes~$I$ values are also unreliable, even if the problem is not apparent from $I$ alone.
 
Our analysis uses the final maps rather than the raw time-stream data.
We therefore infer the quality of the sinusoidal fitting from the statistical properties of the reconstructed Stokes parameters, without testing that process directly.
If saturation or map-making artifacts compromise the time-series fitting, we should see correlated anomalies in $\Delta Q$ and $\Delta U$.
 
\begin{figure*}[tp]
\centering
\includegraphics[width=0.95\textwidth]{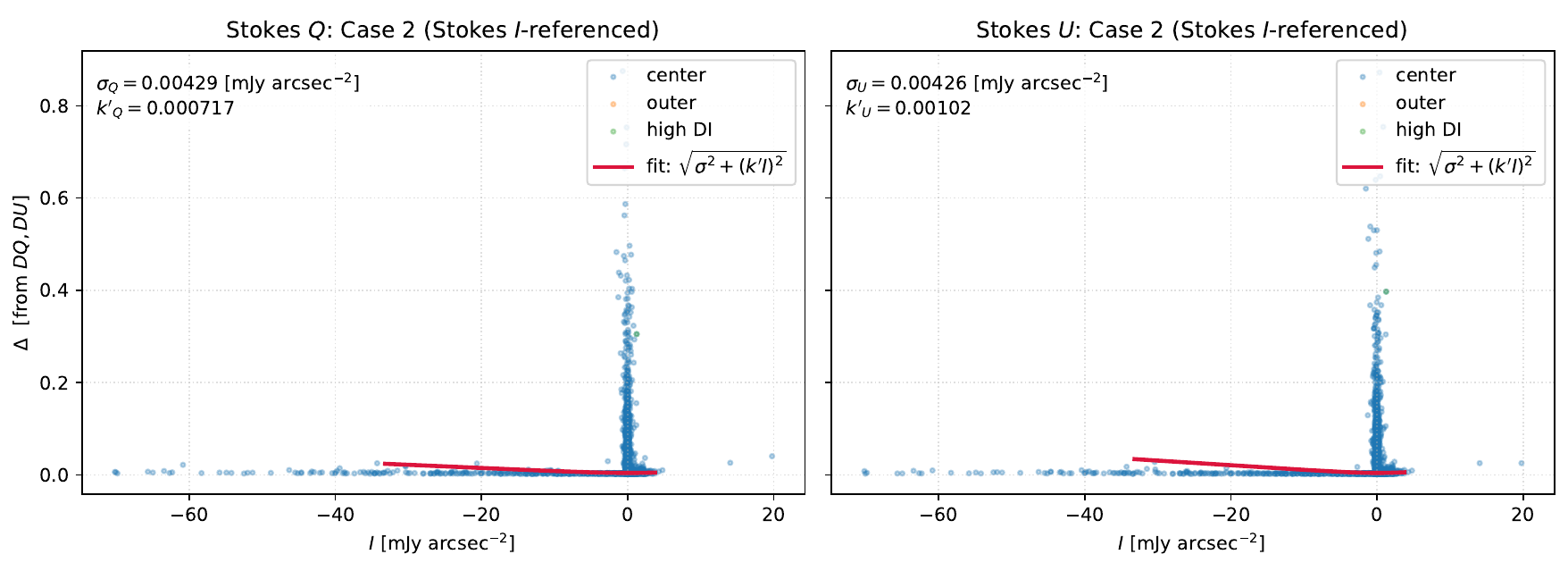}
\caption{
Diagnostics of Stokes $Q$ and $U$ uncertainties for the 850\,$\mu$m observations of W51 (Case~2; referenced to Stokes $I$).
The red curves show the best-fit empirical noise model, $\Delta_{Q,U}(I)=\sqrt{\sigma_{Q,U}^2 + (k'_{Q,U} I)^2}$.
The presence of large negative Stokes $I$ values toward the bright center (extending to $I \simeq -60$~mJy~arcsec$^{-2}$) provides direct evidence for ``negative bowling'' artifacts.
Such values indicate algorithmic overcorrection in the iterative map making of \texttt{makemap}, because thermal dust emission is intrinsically non-negative.
Negative $I$ values in the map outskirts, where noise dominates, are not by themselves a diagnostic of this failure.
The coincidence of negative $I$ values with elevated $\Delta Q$ and $\Delta U$ demonstrates that map-making artifacts affect the full Stokes vector simultaneously.
All Stokes parameters are derived from the same time-series data via sinusoidal fitting (equation~\ref{eq:pol2_signal_app}).
We reduced the data using Starlink version 2021 July 23.
}
\label{fig:QU_W51_case2}
\end{figure*}

Figure~\ref{fig:QU_W51_case2} presents Case~2 noise diagnostics referenced to Stokes $I$ for the 850\,$\mu$m observations of W51, following the analysis of the two regimes established in subsection~\ref{as:QU_noise_model}.
The horizontal axis extends to \textit{negative} Stokes~$I$ values, reaching $I \simeq -60$~mJy\,arcsec$^{-2}$ in the central region.
We interpret Stokes~$I$ values this far below zero as evidence of negative bowls introduced by the iterative map maker \texttt{makemap} \citep{Chapin2013}, since thermal dust emission is intrinsically non-negative ($I \geq 0$).
These artifacts arise when spatial filtering and common-mode subtraction overcorrect the time-stream data in the vicinity of bright, compact sources.
This overcorrection produces localized negative residuals in the reconstructed map.
 
We interpret the concurrent elevation of $\Delta Q$ and $\Delta U$ (reaching $\sim$0.2--$0.8$~mJy\,arcsec$^{-2}$) in these pixels with negative $I$ as evidence that the systematics in the map making propagate into all three Stokes parameters simultaneously.
Such propagation is precisely the expected behavior given equation~(\ref{eq:pol2_signal_app}).
Any systematic affecting the quality of the time-stream data---whether detector non-linearity, clipping, or algorithmic artifacts---will degrade $I$, $Q$, and $U$ concurrently, since all three are extracted from the same corrupted data.

The 850\,$\mu$m observations of DR21, by contrast, do not reproduce this signature.
Within $60\arcsec$ of the DR21 emission peak, not one pixel carries a negative reconstructed intensity, against 96\% of the pixels in the same aperture for W51.
Within $120\arcsec$, 21 of 305 pixels (6.9\%) in DR21 have negative intensities, with a median of $-0.03$~mJy\,arcsec$^{-2}$.
This fraction falls inside the range spanned by fields that show no anomaly of any kind (0--29\% across the seven fields other than W51, with Ophiuchus~A at 11.5\% and NGC~1333 South at 28.9\%).
The median Stokes-$Q$ uncertainty in the central $30\arcsec$ of DR21 is 0.5 times the noise measured far from its center, rather than elevated.
 
The fitted additive noise floors for both fields remain comparable to those of other survey fields when evaluated using only the well-behaved outer regions: $\sigma_Q \simeq \sigma_U \simeq 0.003$--$0.004$~mJy\,arcsec$^{-2}$.
The comparable noise floors show that the instrumental performance is nominal outside the affected central pixels.
We take these findings as independent confirmation that the central pixels of the 850\,$\mu$m map of W51 are affected by algorithmic systematics.
This confirmation validates the exclusion of these pixels from quantitative polarization analysis.
A threshold on the Stokes-$I$ signal-to-noise ratio $I/\sigma_I$, where $\sigma_I$ is the per-pixel intensity uncertainty, automatically rejects the affected pixels because of how the variance is defined.
In the $12\arcsec$ products, the Stokes-$I$ variance is defined only where the reconstructed intensity is positive.
Across the eight fields examined here (W51, DR21, OMC~1, Sagittarius~B2, L1688 Ophiuchus~A, NGC~6334 North, Monoceros~R2, and NGC~1333 South), at both wavelengths, every pixel with $I < 0$ lacks a usable $\sigma_I$ and every pixel with $I > 0$ has one, with no exceptions.
A pixel with non-positive intensity therefore has no defined $I/\sigma_I$ and is rejected by every selection used in this paper series.
This condition applies identically in all fields and preserves valid data in the surrounding regions.
 
OMC-1 and Sgr~B2, which are among the brightest sources in our sample and whose 850\,$\mu$m peak intensities are comparable to that of W51, do not exhibit the anomalous pixels with negative $I$ seen toward the W51 center.
Within $60\arcsec$ of their emission peaks, both fields maintain $I/\sigma_I \gg 10$ throughout (median 567 and 940 respectively in the central $30\arcsec$).
In these regions, Stokes~$I$ remains strictly positive, and the noise characteristics are consistent with the two-component model established in subsection~\ref{as:QU_noise_model}.
The exclusion therefore does not apply to all sources with high intensities.
The contrast indicates that the artifact is specific to the 850\,$\mu$m map of W51 rather than a general limitation of POL-2 observations at high intensity.
This specificity likely reflects differences in observing conditions, scan patterns, or source morphology.
 
The 450\,$\mu$m observations do not show a counterpart of this failure in either field.
Within $120\arcsec$ of the emission peak, W51 has 17 of 305 pixels (5.6\%) with negative intensity at 450\,$\mu$m and DR21 has 8 (2.6\%), against 11 (3.6\%) for OMC~1 and 21 (6.9\%) for Ophiuchus~A.
The failure is therefore specific to W51 at 850\,$\mu$m, and is not a wavelength-independent consequence of source brightness.
 
The contrast is clear when we examine the regions around the emission peaks.
Across all eight fields examined here and both wavelengths---sixteen combinations of field and band---the number of pixels with negative intensity within $60\arcsec$ of the emission peak is zero in fourteen cases.
The two exceptions are W51 at 850\,$\mu$m, with 66 of 69 pixels (96\%), and NGC~1333 South at 850\,$\mu$m, with 22 of 69 (32\%).
The latter is the faintest field in the set: its 850\,$\mu$m peak intensity of 8~mJy\,arcsec$^{-2}$ places its central pixels in the regime dominated by noise.
Here the reconstructed intensity is expected to scatter about zero, rather than show the negative bowls that we seek to identify.
Within $30\arcsec$ of the W51 peak, 18 of 21 pixels (86\%) have negative intensity, with a median of $-12$~mJy\,arcsec$^{-2}$.
The intensity reaches $I \simeq -60$~mJy\,arcsec$^{-2}$, as shown in figure~\ref{fig:QU_W51_case2}.
This negative intensity is an order of magnitude larger in absolute value than the peak intensity of any pixel that survives in the same aperture.
Within $120\arcsec$ of the W51 peak, 282 of 305 pixels (93\%) have negative intensities; these pixels occupy $\sim$5\% of the mapped area.
The negative region is the entire footprint of the defect.
 
The fraction of pixels with negative intensity over an entire map does not diagnose map quality.
Across all finite pixels, this fraction is 38--46\% in every field examined here, at both wavelengths, with one exception.
This range includes OMC~1 (41\%, 43\%) and Sgr~B2 (39\%, 41\%), which the diagnostics above identify as clean, as well as DR21 (40\%, 42\%).
Only W51 at 850\,$\mu$m has a higher fraction, and the difference is modest (51\%).
The fraction is determined by the map periphery, where noise dominates, the reconstructed intensity scatters about zero, and the negative fraction approaches one half in every field.
This fraction does not measure the behavior of the bright central emission that we seek to diagnose, and we do not use it.
 
\subsection{The adopted pixel quality mask}
\label{ass:qcmask}
 
The W51 map at 850\,$\mu$m contains localized regions toward the bright central emission in which the reconstructed Stokes parameters are affected by systematics in the map making.
The diagnostics do not identify a comparable failure in DR21.
We excluded these affected pixels from the quantitative analysis of both wavelengths using the quality criteria defined above.
Outside of the affected regions, the fitted Stokes-$Q$ and Stokes-$U$ noise floors are consistent with those of the broader survey, indicating nominal instrumental performance.
W51 remains in the 69-source {\sc PolSrc} census and in the technical diagnostics presented here.
We exclude it from the scientific data set analyzed in Papers~S1 and~S2 because its default 850\,$\mu$m reduction is affected by this failure.
DR21 is retained subject to the standard quality criteria.
 
We use the same definition of the mask throughout the paper series.
We include a pixel in an analysis of both wavelengths only if the following conditions hold in both bands.
The reconstructed Stokes~$I$ must be positive and have a defined variance, and $I/\sigma_I$ must exceed the threshold appropriate to the measurement: $>10$ for the fits of the noise model in subsection~\ref{ss:noisemodel}, $>30$ for comparisons of polarization angles (subsection~\ref{ss:beamval}), and $>50$, together with brightness cuts for each source, for the quantities derived from both bands in the systematic budget of subsection~\ref{sec:uncertainty}.

This mask has the following two important properties.
First, the parent mask is defined by Stokes~$I$ and its uncertainty alone.
We do not use polarization fraction, polarized intensity, polarization angle, spectral index, or other scientific quantities to define this parent mask.
Additional requirements for particular diagnostics define subsets of these pixels, such as the $S/N(PI)>10$ subset used for the direct fit of the noise model in sub-subsection~\ref{sss:QU_validation}.
Second, we apply this parent mask identically to every field, W51 and DR21 included.
We do not use a mask specific to an individual field or drawn by hand anywhere in this series.
 
The two-component noise model of subsection~\ref{ss:noisemodel} provides a further constraint on the construction of such masks.
The diagnostic signature of a map-making failure is the \textit{joint} occurrence of non-positive reconstructed intensity and inflated uncertainty at the same pixels, as observed toward the center of W51 at 850\,$\mu$m; neither condition on its own is sufficient.
An elevation of the polarization uncertainties toward bright emission is not by itself evidence of a defect: the multiplicative term $k_\lambda I_\lambda$ guarantees that $\sigma_Q$ and $\sigma_U$ rise with intensity in every field.
In the central $30\arcsec$ of OMC~1 at 850\,$\mu$m the median $\sigma_Q$ is 4.3 times its value far from the center.
This elevation is larger than DR21 shows anywhere (0.5 times, i.e.\ none), and OMC~1 is nonetheless clean according to the above criteria.
 
\section{Comparison with the JCMT POL-2 Sensitivity Calculator}
\label{app:itc_noise_comparison}
 
The empirical scaling in figure~\ref{afig:sensitivity_time} shows that the noise in the final BISTRO maps decreases with integration time consistently with the ideal $t^{-0.5}$ expectation for uncorrelated white noise at both wavelengths.
To check this result against an external estimate, we compare the same Stokes-$I$ noise floors with the official JCMT SCUBA-2/POL-2 sensitivity calculator (Calculator~v3.2.0; \citealt{EAOSCUBA2ITC}).
We express all quantities on the same surface-brightness scale, mJy\,arcsec$^{-2}$, following the convention for flux density used throughout this paper series.

The calculator estimates sensitivity for observing proposals, whereas the BISTRO products are reduced science maps of bright, spatially structured star-forming regions.
Their noise includes residuals from map making, coupling to source structure, masking choices, and systematics from later processing steps that the calculator does not represent.
This comparison does not recalibrate the integration-time calculator.
 
We define the empirical-to-ITC ratio as
\begin{equation}
\mathcal{R}_{\rm ITC} = \frac{\Delta I_\lambda^{\rm emp}}{\Delta I_\lambda^{\rm ITC}} ,
\label{eq:R_ITC}
\end{equation}
where $\Delta I_\lambda^{\rm emp}$ is our empirical power-law fit (sub-subsection~\ref{ass:NoiseResults}) evaluated at a representative integration time, and $\Delta I_\lambda^{\rm ITC}$ is the RMS predicted by the calculator and expressed on the matching surface-brightness scale.
 
For a representative total exposure time of $t = 13.5$\,hr, our empirical fit gives $\Delta I_{850}^{\rm emp}=0.0073$\,mJy\,arcsec$^{-2}$ and $\Delta I_{450}^{\rm emp}=0.205$\,mJy\,arcsec$^{-2}$.
The ITC prediction for Band~1 conditions (POL-2 daisy map of $\sim$3\,arcmin, $12''$ pixel sampling, median in four representative declinations $\delta = \{+45^\circ, +30^\circ, -5^\circ, -25^\circ\}$) gives
$\Delta I_{850}^{\rm ITC}=0.00124$\,mJy\,arcsec$^{-2}$ at 850\,$\mu$m.
The 850\,$\mu$m map defines the reference resolution and is not re-smoothed during harmonization.
We therefore use the area of the native compound beam, $\Omega^{\rm FCF}_{850}=239$\,arcsec$^{2}$, for this unit conversion.
The quoted surface-brightness value corresponds to a calculator RMS of $0.296$\,mJy\,beam$^{-1}$ at 850\,$\mu$m.
At 450\,$\mu$m, we obtained $\Delta I_{450}^{\rm ITC}=0.0156$\,mJy\,arcsec$^{-2}$ by dividing the calculator output of $1.90$\,mJy\,beam$^{-1}$ by the area of the native compound beam ($\Omega^{\rm FCF}_{450}=122$\,arcsec$^{2}$), consistent with the calculator's beam normalization.
Dividing the same output by the Gaussian reference area $\Omega^{\rm mat}_{450}=180$\,arcsec$^{2}$ instead gives $0.0106$\,mJy\,arcsec$^{-2}$.
Table~\ref{tab:itc_noise_comparison} lists this value to illustrate the reference normalization; it is not a second estimate of the same surface brightness.
The resulting ratios are $\mathcal{R}_{\rm ITC}=5.9$ at 850\,$\mu$m and $\mathcal{R}_{\rm ITC}=13.1$ at 450\,$\mu$m ($19.3$ on the illustrative Gaussian reference normalization), summarized in table~\ref{tab:itc_noise_comparison}.
 
\begin{deluxetable*}{lccc}
\tablecaption{
Comparison of the empirical noise model with the JCMT SCUBA-2/POL-2 sensitivity calculator, in mJy\,arcsec$^{-2}$.
\label{tab:itc_noise_comparison}
}
\tablehead{
\colhead{Band / convention} &
\colhead{Empirical $\Delta I_\lambda^{\rm emp}$} &
\colhead{ITC $\Delta I_\lambda^{\rm ITC}$} &
\colhead{$\mathcal{R}_{\rm ITC}$}
}
\startdata
850\,$\mu$m, native beam ($\Omega^{\rm FCF}_{850}=239$\,arcsec$^2$) & 0.0073 & 0.00124 & 5.9 \\
450\,$\mu$m, native beam ($\Omega^{\rm FCF}_{450}=122$\,arcsec$^2$) & 0.205 & 0.0156 & 13.1 \\
450\,$\mu$m, Gaussian reference normalization ($\Omega^{\rm mat}_{450}=180$\,arcsec$^2$) & 0.205 & 0.0106 & 19.3 \\
\enddata
\tablecomments{
We evaluated Calculator~v3.2.0 for Band~1 weather, a POL-2 daisy map of $\sim$3\,arcmin, $12\arcsec$ pixel sampling, and $t=13.5$\,hr.
The entries are medians over declinations $\delta=+45\degr,+30\degr,-5\degr,-25\degr$.
Both 450\,$\mu$m entries use the same calculator RMS of $1.90$\,mJy\,beam$^{-1}$.
The calibrated native-beam conversion divides by $\Omega^{\rm FCF}_{450}=122$\,arcsec$^2$, consistent with the calculator's beam normalization.
The alternative divides by the Gaussian reference area $\Omega^{\rm mat}_{450}=1.133\,\theta_{\rm eff}^2=180$\,arcsec$^2$ for $\theta_{\rm eff}=12\farcs6$, the two-component effective FWHM of equation~(12) in \citet{Mairs2021} adopted for the common resolution (subsection~\ref{ss:beamval}).
This second value illustrates a reference normalization; it does not predict the RMS after beam harmonization.
At 850\,$\mu$m, we divided the calculator RMS of $0.296$\,mJy\,beam$^{-1}$ by $\Omega^{\rm FCF}_{850}=239$\,arcsec$^2$.
The 850\,$\mu$m map is not re-smoothed during harmonization.
}
\end{deluxetable*}
 
The relations between NEFD, atmospheric transmission, and exposure time used in the ITC were calibrated on Stokes $Q$ and $U$ maps of bright compact calibrators \citep{EAOPOL2obs}.
The magnitude of $\mathcal{R}_{\rm ITC}$ is on the same broad scale as the independently inferred efficiency factors for smoothing and map making, with the median $f_{\rm smooth}\simeq 4.4$ at 850\,$\mu$m and $\simeq 3.9$ at 450\,$\mu$m (figure~\ref{afig:f_smooth}).
We therefore regard the ITC comparison as an external consistency check on the empirical noise model and retain the model of subsection~\ref{ss:noisemodel} for all subsequent polarization analyses.
We do not interpret these ratios as a failure of the ITC prediction or use the calculator as a replacement for the empirical model.

\section{Null Hardware Tests for the 450\,\micron\ Stokes-$Q/U$ Zero-Level Offset}
\label{app:hardware_nulltests}
 
This appendix presents the three null hardware tests summarized in sub-subsection~\ref{as:HardwareInvestigation}: telescope pointing and attitude, dependence on source brightness, and the IP correction model.
These tests required observations of exceptionally bright fields to achieve the $S/N$ ratio necessary to characterize the offset in individual Minimum Schedulable Blocks (MSBs), typically 40--42 minutes per observation.
 
\subsection{Telescope Pointing and Attitude Parameters}
\label{ass:TelescopeAttitude}
\begin{figure*}[tp]
\centering
\includegraphics[width=\textwidth]{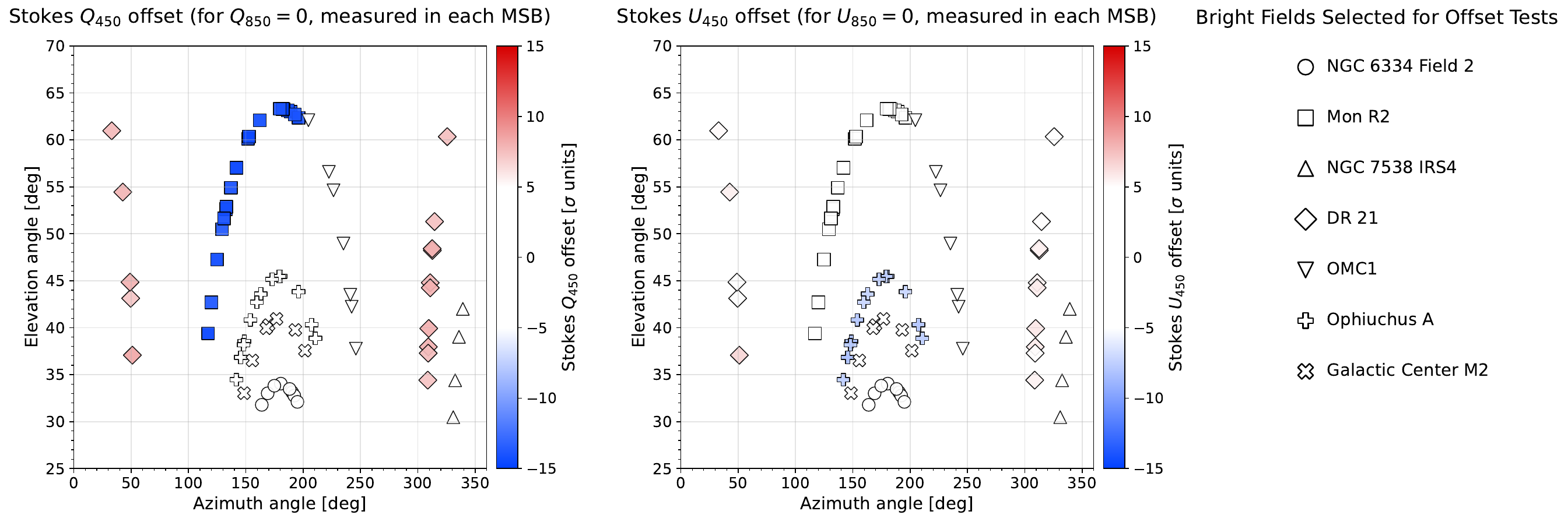}
\caption{
Azimuth--elevation distribution of the residual 450\,$\mu$m zero-level offsets in Stokes~$Q$ (left panel) and $U$ (right panel), measured when the corresponding 850\,$\mu$m component equals zero.
Each point represents a single MSB (typically 40--42 minutes of observation) from one of seven bright fields selected for their exceptional $S/N$ ratios.
The offsets for individual MSBs were measured on an earlier (2020) reduction of these data at the default threshold, not on the final products used elsewhere in this paper.
The color scale represents the offset amplitude in units of the uncertainty for each field ($\sigma$), with a white plateau for values within $\pm5\sigma$ and a diverging blue--red color scale (blue for negative, red for positive) beyond this range.
Neither Stokes component shows a systematic dependence on azimuth or elevation, and the distributions of their offsets are comparable.
These observations support the null result of the tests of telescope geometry for the seven fields examined.
}
\label{afig:AzEl_offset_maps}
\end{figure*}

We first tested whether the 450\,\micron\ Stokes-$Q/U$ offsets correlate with the geometry of the telescope pointing or the attitude parameters.
Figure~\ref{afig:AzEl_offset_maps} shows the azimuth--elevation distribution of the residual Stokes~$Q$ and $U$ offsets, evaluated for individual MSBs in seven bright star-forming regions.
We selected these regions from the larger survey sample because their exceptional brightness allowed us to measure offsets with sufficient precision in individual 40--42 minute observations.
The offsets are presented in units of the uncertainties for each field ($\sigma$), with values within $\pm 5\sigma$ rendered in white and larger deviations transitioning to blue (negative) or red (positive).
The azimuth--elevation distribution shows no systematic structure, gradient, or clustering that would indicate telescope-dependent effects.
Both Stokes components show similar amplitude distributions and spatial symmetry.
Additional tests confirmed the absence of correlations with declination, parallactic angle, or hour angle in the analyzed fields.
We therefore do not find evidence that telescope pointing or tracking geometry contributes to the offset.
 
\subsection{Source Brightness Dependence}
\label{ass:IntensityDependence}
We investigated whether the 450\,\micron\ $Q/U$ offsets might scale with source brightness, which could indicate detector non-linearity or a bolometer response that depends on signal strength.
The signal-to-noise requirements restricted the analysis of individual MSBs to bright fields.
The dynamic range accessible in our sample spans peak intensities from approximately 100\,mJy\,beam$^{-1}$ to several Jy\,beam$^{-1}$.
Across this range, the offsets remained constant within measurement uncertainties when presented in units of surface brightness (mJy\,arcsec$^{-2}$).
We interpret this insensitivity to source flux as evidence that saturation effects, non-linear detector response, and calibration errors that depend on flux do not contribute to the observed phenomenon.
The most luminous targets in our sample (including OMC-1 and NGC\,6334) had offsets comparable in magnitude to those of fainter sources when normalized by their respective noise levels.
 
\subsection{Instrumental Polarization Correction}
\label{ass:InstrumentalPol}
The JCMT POL-2 data-reduction pipeline applies an elevation-dependent IP correction following the model of \citet{Friberg2018}.
We adopt the AUG2019 IP model, which improves the consistency of 450\,\micron\ maps taken at different elevations \citep{EAO2019IPmodels}.
This model has been the default in \texttt{pol2map} since \texttt{SMURF}~1.7.0 \citep{EAO2020IPdefault}.
To test the sensitivity of the data to this correction, we reprocessed them with a constant offset added to the modeled IP, stepping from $-0.6\%$ to $+0.6\%$ in increments of $0.15\%$.
This range extends to twice the $0.3\%$ that the observatory quotes as a reasonable estimate of the uncertainty in the IP level \citep{EAO2019IPuncertainty}.
The \texttt{IPOFFSET} parameter raises or lowers the IP level by the given percentage of the total intensity rather than rescaling the model.
Across this range, the resulting changes in Stokes~$Q$, $U$, and the polarized intensity $PI = \sqrt{Q^2 + U^2}$ do not show a strong dependence on the IP correction above the map noise.
However, some fields respond by an amount that is not negligible, and the response is largest for the brightest sources, so we recommend treating results for very bright targets with caution.
These tests measure how sensitive the data are to the IP correction.
Because the true uncertainty of the IP model is not known independently, they neither establish nor exclude the IP model as a contributor to the observed $Q/U$ offsets.
 
\subsection{Summary of Null Hardware Tests}
\label{ass:HardwareSummary}
These three null tests identify no clear origin in the hardware for the 450\,\micron\ Stokes-$Q/U$ zero-level offset in the fields and parameter ranges examined.
They motivate the investigation of a software origin in sub-subsections~\ref{ass:PhysicalEvidence}--\ref{ass:PhysicalConstraints}.
However, the IP test constrains the sensitivity of the data to the IP correction, rather than the accuracy of the IP model itself.
The subsequent reduction experiments (sub-subsection~\ref{ass:offset_significance}) show that the recovered offsets respond to the treatment controlled by the \texttt{pca.zero\_snr} parameter in the \texttt{SMURF}/\texttt{pol2map} pipeline.
This response implicates the map making by \texttt{makemap} without establishing it as the sole contributor to the offsets.
 
\section{Origin of the 450\,\micron\ Stokes-$Q/U$ Zero-Level Offset}
\label{app:pca_origin}
We linked the 450\,\micron\ offsets to the construction of low-frequency background models in the \texttt{pol2map} pipeline by systematically testing the \texttt{makemap} algorithm \citep{Chapin2013}.
The FLT, PCA, and COM stages of \texttt{makemap} replace masked time-stream samples with linearly interpolated values before filtering in the Fourier domain.
The 450\,\micron\ time streams are noisier than those at 850\,\micron.
Under the driest conditions, the sky background power at 450\,\micron\ is about ten times that at 850\,\micron.
In good observing conditions, the typical NEFDs are 400 and 90\,mJy\,s$^{1/2}$ at 450 and 850\,\micron, respectively \citep{Holland2013}.
The interpolated segments therefore differ systematically between bands and introduce zero-level biases into the fast Fourier transform (FFT) that depend on wavelength.
High-pass filtering in \texttt{makemap} introduces ringing \citep{Mairs2021}.
Sharp edges in the interpolated data can produce ringing that propagates as negative bowls or spatial ripples in the final map.
 
The biases do not depend monotonically on the area of the PCA mask: fields with comparable mask sizes can exhibit offsets differing by more than an order of magnitude.
The lack of a monotonic dependence indicates that the offset reflects the {\em interaction} between mask geometry, interpolation noise, and the edge response of the FFT during the early filtering iterations, rather than mask size alone.
Reducing the spatial extent of the FLT, PCA, and COM masks through the (\texttt{pca.zero\_snr}, \texttt{pca.zero\_snrlo}) pair limits artificial interpolation and extrapolation in poorly constrained regions.
Bright targets retain non-zero masks to suppress ringing, while faint or smoothly varying fields reach minimal offsets with minimal masking.
Table~\ref{tab:PCAparams} in appendix~\ref{app:pca_grid_atlas} summarizes the parameter grid.
The default configuration $(\texttt{pca.zero\_snr}, \texttt{pca.zero\_snrlo}) = (3, 2)$ is one point on this grid.
\texttt{pol2map} uses this configuration in Steps~2 and~3 when the configuration file does not set these parameters (appendix~\ref{app:reduction_workflow}).
Our results show that the optimal threshold depends on the field.

\section{Physical Assumptions Behind the Zero-Intercept Condition}
\label{app:zero_intercept_assumptions}
 
This appendix describes the astrophysical assumptions and considerations of data quality underlying the zero-intercept criterion used in sub-subsection~\ref{ass:PhysicalConstraints}.
A proportional relation between the two submillimeter bands is expected when both trace a sufficiently coherent polarization pattern with compatible spatial weighting and an approximately constant ratio of polarized intensities over the fitted region.
Optical thinness alone does not establish these conditions.
In particular, components with different temperatures or emissivity indices may contribute unequally at the two wavelengths.
Averaging along the line of sight or within the beam may then make the polarization depend on wavelength, if the components also differ in polarization efficiency or $\vec{B}$-field orientation.
We summarize the relevant assumptions and potential sources of contamination below.
 
\begin{enumerate}
 
\item \textit{Reliability of the 850\,\micron\ reference polarization.}
We adopt the 850\,\micron\ maps as the reference for the comparison between the bands, supported by the characterization of instrumental polarization in \citet{Friberg2018}.
Attributing a measured intercept specifically to the 450\,\micron\ maps requires residual additive errors in the 850\,\micron\ reference to be sufficiently small.
This assumption about the reference is separate from the assumptions about the intrinsic polarization geometry discussed below.
\label{assumption:850ref}
 
\item \textit{Potential contamination at 450\,\micron.}
The calibrated 450\,\micron\ maps may retain instrumental, atmospheric, or processing contributions to the measured Stokes parameters.
Such contributions are possible explanations for an additive offset and must be assessed using observational and data-reduction tests.
They are not astrophysical premises from which a zero-intercept relation is derived.
\label{assumption:450contamination}
 
\item \textit{Optically thin dust emission.}
We assume that the dust emission is optically thin at both wavelengths, so that absorption and self-absorption have negligible effects on the measured polarization.
The validity of this approximation depends on the column density and the dust opacity adopted in the regions analyzed \citep{Hildebrand1983, Ossenkopf1994}.
Using the brightest native-beam pixel of each field, we estimate 450\,$\mu$m dust optical depths of up to $\tau_{450}=0.56$, assuming a single temperature of 40\,K and a beam filling factor of unity.
The optical depth exceeds 0.2 only toward the brightest peaks of NGC~6334, Sgr~B2, OMC~1, W51, and G34.257+0.155, and it is below 0.15 in the other fields.
We do not expect self-absorption to produce the zero-level offsets, because they also appear in fields where the optical depth is below 0.15, such as Ophiuchus~A and OMC~3 (sub-subsection~\ref{ass:offset_significance}).
Optical thinness permits the emission from different locations along the line of sight to be added, but does not prevent cancellation between their Stokes contributions.
\label{assumption:opticallythin}
 
\item \textit{Line-of-sight integration and beam averaging.}
In the optically thin approximation, the observed Stokes parameters are integrals of the corresponding emissivities along the line of sight, subsequently averaged over the observing beam \citep{Draine2003}.
Different $\vec{B}$-field orientations can produce partial cancellation within these integrals or within the beam \citep{Jones2015}.
A common polarization pattern at the two wavelengths therefore requires sufficiently coherent polarization geometry and a compatible relative weighting of the emitting components.
Matching the angular resolution and spatial-frequency coverage is necessary for the comparison, but does not by itself ensure identical weighting of physically distinct components.
\label{assumption:radiativetransfer}
 
\item \textit{Stokes parameters as projections of polarized intensity.}
The intrinsic linear Stokes parameters are related to the polarized intensity and polarization position angle by
\begin{equation}
Q_{\lambda}=PI_{\lambda}\cos(2\chi_{\lambda}),
\qquad
U_{\lambda}=PI_{\lambda}\sin(2\chi_{\lambda}),
\end{equation}
following the standard description of emission polarimetry \citep{Hildebrand1988}.
Where the two bands share the same polarization angle, their Stokes vectors differ only in amplitude.
However, this local relation does not require the amplitude ratio to be constant across a map.
\label{assumption:stokesprojection}
 
\item \textit{Wavelength-independent alignment geometry and inter-band scaling.}
We assume, as a working model, that the directions of grain alignment do not depend on wavelength for the two POL-2 bands \citep[see discussions of grain alignment and polarized dust emission in][]{Andersson2015, Planck2020-XII}.
A common local alignment direction does not guarantee a common observed polarization angle by itself because the relative contributions of different emitting components may vary with wavelength.
The conditions on coherence and weighting described above are therefore also required.
For a single proportional relation over the fitted region, we further assume that the ratio of polarized intensities is approximately constant.
This ratio can depend on the dust emission spectra and polarization efficiencies of the contributing components.
\label{assumption:alignmentdirection}
 
\end{enumerate}
 
Under these conditions, the intrinsic Stokes~$Q,U$ maps can be approximated by
\begin{equation}
Q_{450}^{\rm sky}(\boldsymbol{x})
\simeq R\,Q_{850}^{\rm sky}(\boldsymbol{x}),
\qquad
U_{450}^{\rm sky}(\boldsymbol{x})
\simeq R\,U_{850}^{\rm sky}(\boldsymbol{x}),
\end{equation}
where $R>0$ is the approximately constant ratio of polarized intensities in the fitted region.
In the exact proportional limit, a position satisfying
$Q_{850}^{\rm sky}=U_{850}^{\rm sky}=0$
also satisfies
$Q_{450}^{\rm sky}=U_{450}^{\rm sky}=0$.
The corresponding intrinsic linear relations pass through the origin.
Thus, zero intercepts are expected within this model, subject to measurement and fitting uncertainties.
 
The approximately constant ratio is an essential additional condition when interpreting a regression across multiple pixels.
If the ratio varies with position, each local Stokes relation can retain a multiplicative form, while a single fitted line acquires a non-zero intercept.
Similarly, changes with wavelength in the weighting of components with different polarization directions can invalidate the approximation of a common polarization pattern.
A non-zero fitted intercept therefore indicates a departure from the proportional model, but does not uniquely identify its cause.
 
We regard residual processing effects as the preferred explanation for the offsets, on the basis of the combined evidence from multiple fields and the observational and processing tests.
The recurrence of the observed intercepts (equation~\ref{aeq:OphA_intercepts}) across many fields favors a common non-astrophysical origin.
The hardware investigation in sub-subsection~\ref{as:HardwareInvestigation} and appendix~\ref{app:hardware_nulltests} constrains effects associated with telescope orientation and tracking geometry.
The analysis in sub-subsection~\ref{ass:3step} examines the dependence on the data-reduction procedure.
The intercepts must be interpreted together with this evidence.
 
The three-step correction described in sub-subsection~\ref{ass:3step} is intended to remove the additive contamination supported by these tests.
Consistency with a zero-intercept relation provides a diagnostic of the corrected maps within the stated assumptions.
Spatial variations in the spectrum of the polarized emission and possible cancellation along the line of sight or within the beam must also be considered when assessing whether a correction is warranted.

\section{PCA-Threshold Grid Search for the 450\,\micron\ Stokes-$Q/U$ Offset Optimization}
\label{app:pca_grid_atlas}
 
In this appendix, we show how the 450\,\micron\ Stokes-$Q/U$ zero-level offset in each field responds to changes in the PCA threshold, which underlie the three-step correction of sub-subsection~\ref{ass:3step}.
We first describe the grid search that converts the offsets for individual fields into a quantitative significance test (appendix~\ref{app:pca_optimization}), based on their empirical origin in the time-stream interpolation and Fourier filtering of \texttt{makemap} (appendix~\ref{app:pca_origin}).
We then examine the limitations of 450\,\micron\ mosaicking that restrict the optimization to reductions of individual fields (appendix~\ref{app:pca_mosaicking}) and the ranges of the optimal thresholds (appendix~\ref{app:pca_optimal_ranges}).
Table~\ref{tab:PCAparams} lists the grid of $(\texttt{pca.zero\_snr}, \texttt{pca.zero\_snrlo})$ combinations.
Figures~\ref{afig:PCAcutoff_cat1_p1}--\ref{afig:PCAcutoff_cat3} show the signed intercepts $a_0^Q$ and $a_0^U$ versus \texttt{pca.zero\_snr} for 30 of the 119 fields, chosen as representative of Categories~1--3.
The interval around zero that is consistent with the noise corresponds to $|a_0^Q| < \sigma_{\rm eff}$ and $|a_0^U| < \sigma_{\rm eff}$ (equation~\ref{eq:sigmaeff_def}).
This interval is left white, and gray shading marks the range outside it.
 
\subsection{Optimization Procedure}
\label{app:pca_optimization}
To assess the choice of reduction settings against the polarization uncertainty, we examine how the additive polarization baseline changes with the PCA masking thresholds.
This baseline has its largest fractional effect where the cloud emission is faint, so selecting a reduction solely for producing a small absolute offset is insufficient.
We reprocessed the data for each field in the survey (both individual pointings and the component tiles of mosaicked targets) across the grid of \texttt{pca.zero\_snr}/\texttt{pca.zero\_snrlo} values listed in table~\ref{tab:PCAparams}.
We then measured the baseline terms $(a_0^Q, a_0^U)$ by weighted regression.
We define the combined offset and the effective noise as
\begin{equation}
a_0^{\rm tot} = \sqrt{(a_0^Q)^2 + (a_0^U)^2} ,
\label{eq:a0tot_def}
\end{equation}
\begin{equation}
\sigma_{\rm eff} = \sqrt{\sigma_Q \, \sigma_U} ,
\label{eq:sigmaeff_def}
\end{equation}
where $\sigma_Q \simeq \sigma_U$ and ${\rm Cov}(Q, U) \approx 0$ are empirically validated in equation~(\ref{aeq:CovQUzero}) of sub-subsection~\ref{ass:QU_isotropy_validation}.
This formulation expresses a residual baseline directly in units of the polarization-noise amplitude.
We use it to test the significance of the offsets and assign the field categories summarized in sub-subsection~\ref{ass:offset_significance}.
The threshold sweep identifies where the measured baseline is sensitive to the reduction settings and where it remains large relative to the adopted noise scale.
The comparison with the noise scale guides the treatment of faint polarization in the tested map product; it does not by itself demonstrate that the preferred setting recovers the true sky baseline.
 
\subsection{Limitations of Mosaicking for 450\,\micron\ Offset Analysis}
\label{app:pca_mosaicking}
The optimization procedure is reliable only for reductions of individual fields.
For the 450\,\micron\ mosaicked and mosaic-component products, $(a_0^Q, a_0^U)$ becomes nearly independent of \texttt{pca.zero\_snr}: the curves flatten and the minima disappear.
We nevertheless classify these products with the same diagnostics and select their thresholds by the same rule.
The three-step intercepts are subtracted from the two mosaicked fields whose maps at the default threshold have three-step fits, Taurus~B211/213 and G28.34+0.06 (sub-subsection~\ref{ass:ResidualOffsets}).
Diagnostic case studies of failures in mosaicking for OMC~1, NGC~1333, and NGC~2264 are presented in appendix~\ref{app:mosaicking_diagnostics}.
 
\subsection{Optimal Threshold Ranges}
\label{app:pca_optimal_ranges}
The thresholds of smallest $a_0^{\rm tot}$ span the whole grid of table~\ref{tab:PCAparams}.
For the 61 single fields whose minimum is not at the default, the 16th, 50th, and 84th percentiles of \texttt{pca.zero\_snr} are 17, 100, and 340.
The minimum lies, for example, at 200 for OMC~1, 50 for Sgr~B2, 50 for OMC~2, and 150 for OMC~3.
The minima are generally broad, with $\Delta\texttt{pca.zero\_snr} \sim 50$--100 giving near-optimal performance.
For the 22 single Category~1 fields reduced with an optimized threshold, the smallest $a_0^{\rm tot}$ has a median of $0.16$\,mJy\,arcsec$^{-2}$ (16th--84th percentiles $0.03$--$0.74$\,mJy\,arcsec$^{-2}$) and is below $0.1$\,mJy\,arcsec$^{-2}$ in 8 of them.
These results have two practical implications.
The significance of $(Q, U)$ baselines must be evaluated against $\sigma_{\rm eff}$ rather than against a criterion on the absolute amplitude.
We optimize the PCA threshold for each field with a significant 450\,\micron\ offset, but this is currently feasible only for reductions of individual fields.
 
\begin{table*}[t]
\centering
\caption{
Tested combinations of the \texttt{pca.zero\_snr} and \texttt{pca.zero\_snrlo} parameters in the grid search over the masking thresholds.
When the configuration file does not set them, \texttt{pol2map} (version 2021 July 23) uses \texttt{pca.zero\_snr} = 3 and \texttt{pca.zero\_snrlo} = 2 for the PCA mask in Steps~2 and~3 (appendix~\ref{app:reduction_workflow}).
}
\label{tab:PCAparams}
\begin{tabular}{cc}
\hline
\hline
\texttt{pca.zero\_snr} & \texttt{pca.zero\_snrlo} \\
\hline
    2   &   1.2  \\
    3   &   2    \\
    5   &   3    \\
   10   &   6    \\
   17   &  10    \\
   25   &  15    \\
   50   &  30    \\
  100   &  60    \\
  150   &  90    \\
  200   & 120    \\
  300   & 180    \\
  400   & 240    \\
  800   & 480    \\
 1600   & 960    \\
\hline
\end{tabular}
\end{table*}
 
\begin{figure*}[tp]
\centering
\includegraphics{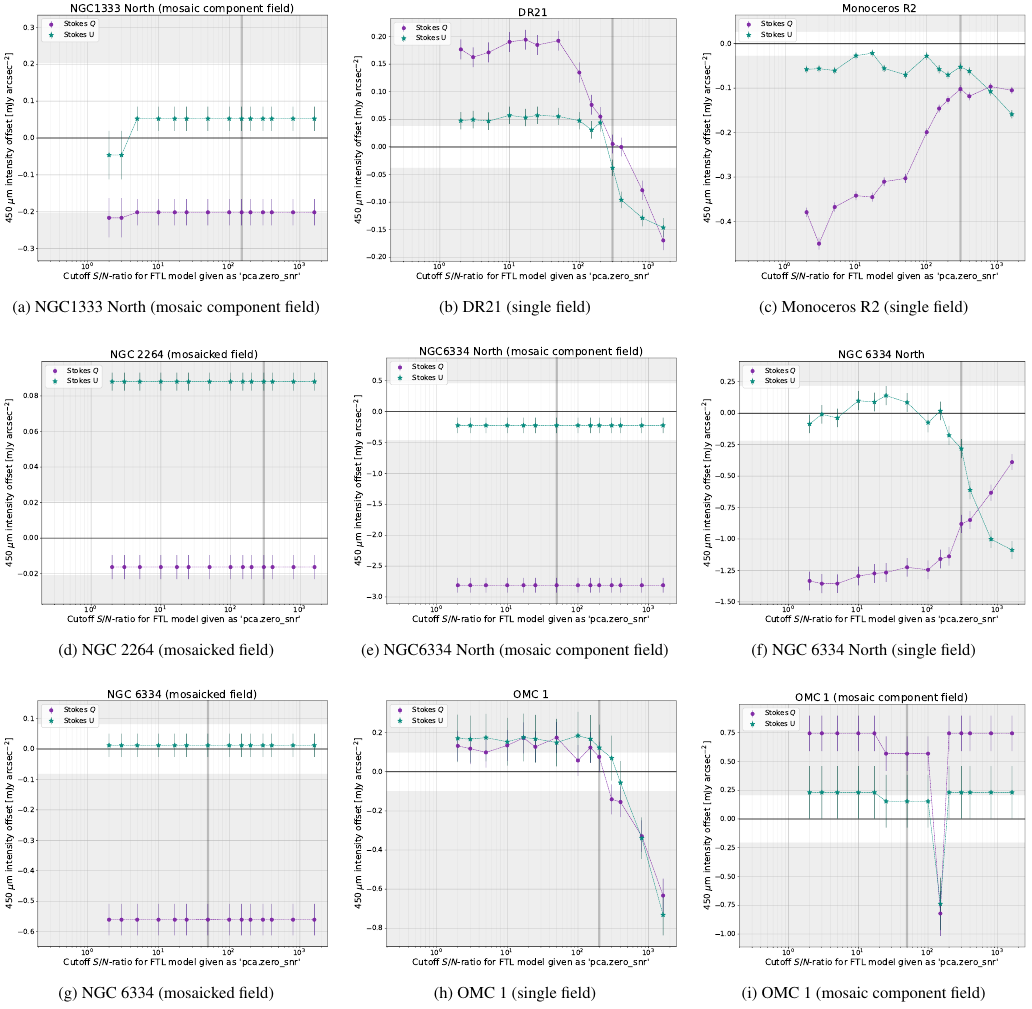}
\caption{
Category~1.
Representative fields where the 450\,\micron\ Stokes-$Q$ or $U$ zero-level offset exceeds the effective noise $\sigma_{\rm eff}$ at the default \texttt{pca.zero\_snr}=3, or where the minimum $a_0^{\rm tot}$ over the tested thresholds exceeds $\sigma_{\rm eff}$.
We reduced these fields with the threshold of smallest $a_0^{\rm tot}$ (sub-subsection~\ref{ass:offset_significance}).
Each panel plots the signed intercepts $a_0^Q$ and $a_0^U$, with their uncertainties, against \texttt{pca.zero\_snr}.
The white band around zero marks $|a_0^Q| < \sigma_{\rm eff}$ and $|a_0^U| < \sigma_{\rm eff}$, and gray shading lies outside it.
The vertical gray line marks the threshold of smallest $a_0^{\rm tot}$.
Labels indicate the field type.
A ``single field'' denotes an individual POL-2 pointing, a ``mosaic component field'' refers to a single tile within a multi-tile mosaic, and a ``mosaicked field'' designates the combined product of multiple component fields.}
\label{afig:PCAcutoff_cat1_p1}
\end{figure*}
 
\begin{figure*}[tp]
\centering
\includegraphics{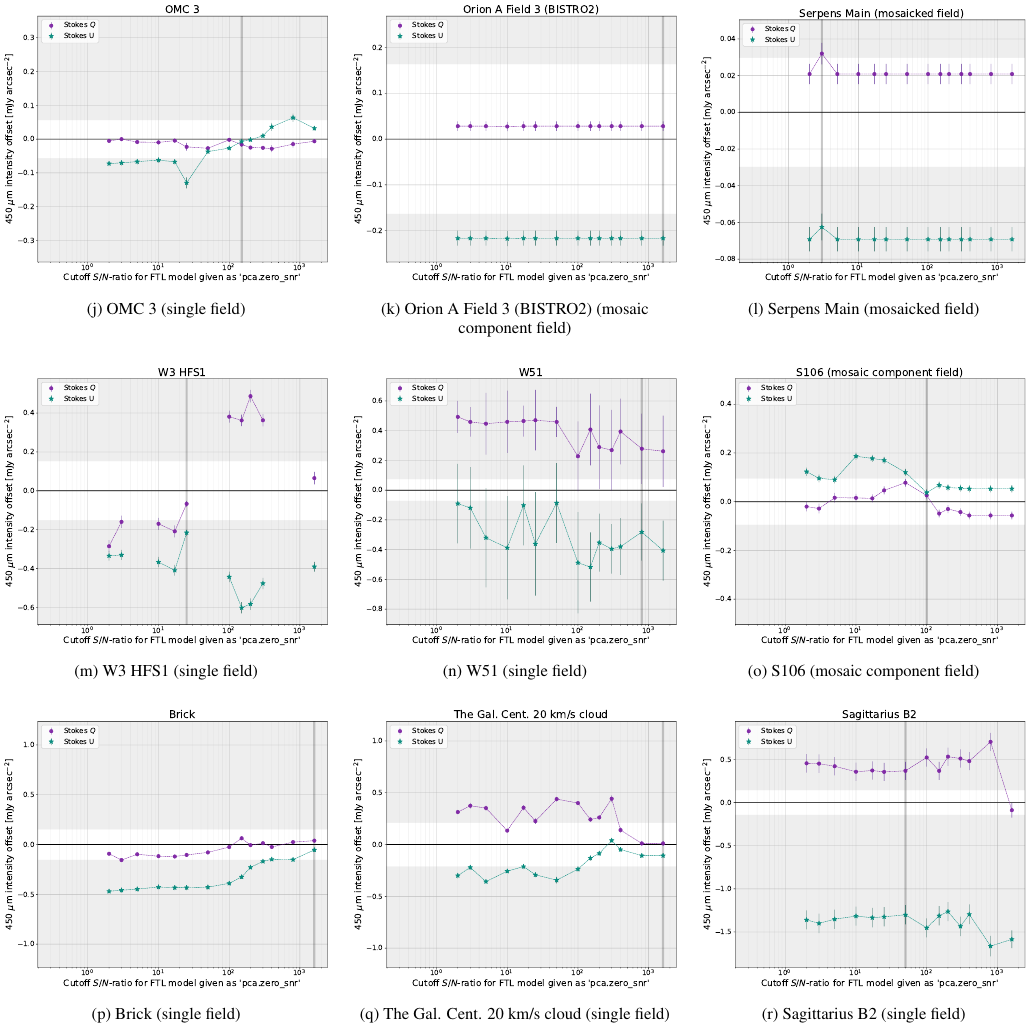}
\caption{
Category~1 (continued).
Additional representative Category~1 fields, defined as in figure~\ref{afig:PCAcutoff_cat1_p1}.
These targets show that bright, structured regions such as W51, the Central Molecular Zone (CMZ), and Orion~A tend to have 450\,\micron\ offsets that exceed the effective noise at the default threshold.
}
\label{afig:PCAcutoff_cat1_p2}
\end{figure*}
 
\begin{figure*}[tp]
\centering
\includegraphics{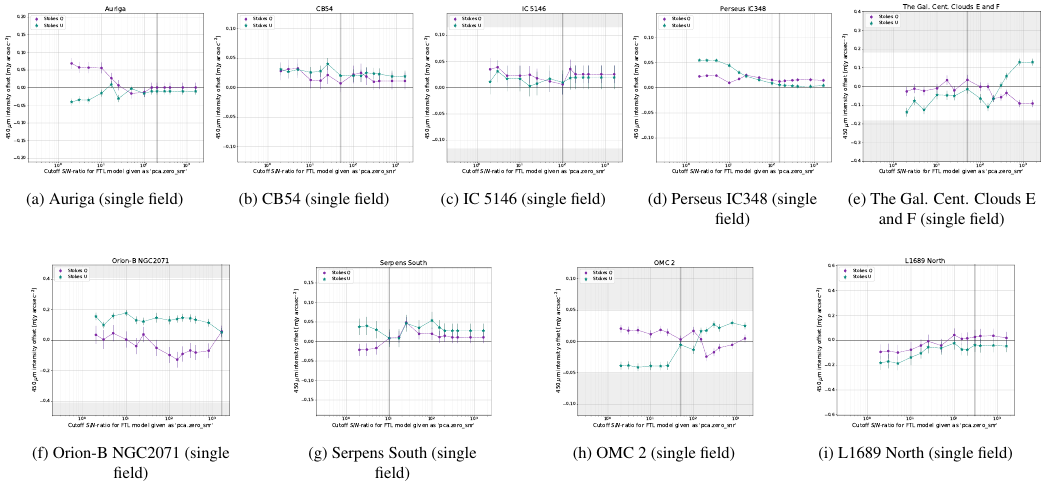}
\caption{
Category~2.
Representative fields whose combined amplitude $a_0^{\rm tot}$ of the 450\,\micron\ Stokes-$Q/U$ zero-level offsets is smallest at a \texttt{pca.zero\_snr} other than the default value of three.
However, neither the offsets at the default threshold nor the minimum $a_0^{\rm tot}$ exceed the effective noise level $\sigma_{\rm eff}$ in these fields (see sub-subsection~\ref{ass:offset_significance}).
For these targets, the default \texttt{pca.zero\_snr = 3} remains appropriate for the subsequent analysis.
In several panels no gray shading appears, because the offsets are small enough that the whole plotted range lies within $\pm\sigma_{\rm eff}$.}
\label{fig:PCAcutoff_cat2}
\end{figure*}
 
\begin{figure*}[tp]
\centering
\includegraphics{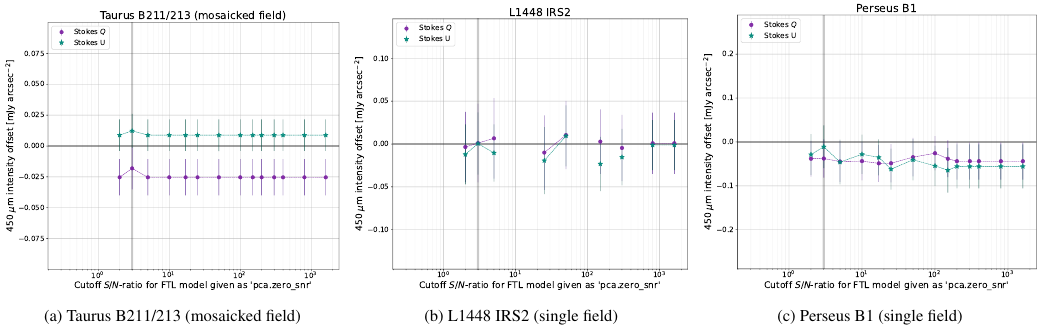}
\caption{
Category~3.
Representative fields whose combined amplitude $a_0^{\rm tot}$ of the 450\,\micron\ Stokes-$Q/U$ zero-level offsets (equation~\ref{eq:a0tot_def}) is smallest at the default \texttt{pca.zero\_snr}=3 and does not exceed the effective noise scale $\sigma_{\rm eff}$ there (sub-subsection~\ref{ass:offset_significance}).
These fields keep the default \texttt{pca.zero\_snr}=3.
Where a three-step fit exists, its intercepts are subtracted (sub-subsection~\ref{ass:ResidualOffsets}).
No gray shading appears in these panels, because the offsets are small enough that the whole plotted range lies within $\pm\sigma_{\rm eff}$.
These small offsets are consistent with the dominance of noise in Category~3.}
\label{afig:PCAcutoff_cat3}
\end{figure*}
 
\section{Mosaicking Failure Modes for the 450\,\micron\ Stokes-$Q/U$ Zero-Level Offset Diagnostics}
\label{app:mosaicking_diagnostics}
 
We compare (i)~reductions of individual fields, (ii)~mosaic-component reductions, and (iii)~mosaicked products for several POL-2 targets observed with overlapping tiles.
These comparisons support the conclusion in sub-subsection~\ref{ass:offset_significance} that the 450\,\micron\ offset diagnostics fail under mosaicking, and they document the signatures of these failures.
 
For reductions of individual 450\,\micron\ fields, the offset amplitude $a_0^{\rm tot}$ depends on \texttt{pca.zero\_snr}.
Bright fields (e.g., OMC~1, Mon~R2, OMC~2, NGC~6334 North) exhibit clear minima at finite thresholds, so that a threshold can be selected and the three-step correction applied (sub-subsection~\ref{ass:3step}).

Processing the same data as mosaic-component fields changes the behavior of the 450\,\micron\ offsets qualitatively.
The baselines $(a_0^Q, a_0^U)$ become nearly independent of \texttt{pca.zero\_snr}: the curves flatten, the minima disappear, and the amplitudes can increase relative to reductions of the individual fields.
For example, the OMC~1 component maintains offsets of $(a_0^Q, a_0^U) \approx (+0.7, +0.2)$\,mJy\,arcsec$^{-2}$ for $\texttt{pca.zero\_snr} = 2$--$25$ and then transitions abruptly to $\approx (-0.8, -0.7)$\,mJy\,arcsec$^{-2}$ at larger thresholds with little variation thereafter.
NGC~1333 and NGC~2264 show analogous behavior: their individual tiles exhibit modest threshold dependence when reduced separately, whereas the mosaic-component reductions hold nearly constant offsets across the full parameter grid.
 
A second failure mode is the propagation and amplification of residual offsets in the final mosaicked products.
For NGC~1333 the northern component shows individual offsets of only $\sim 0.05$--$0.10$\,mJy\,arcsec$^{-2}$, but yields $(a_0^Q, a_0^U) \approx (-0.20, +0.05)$\,mJy\,arcsec$^{-2}$ under mosaic-component processing.
The final mosaic inherits the larger values, which places it in Category~1.
NGC~2264 shows an even stronger effect: the northern component maintains $a_0^U \approx +0.34$\,mJy\,arcsec$^{-2}$ across nearly all tested thresholds, a factor of two larger than any offset seen in its single-field reduction.
 
Finally, for two of the ten mosaicked and mosaic-component fields whose selected threshold is not the default, the polarization products (\texttt{polvecat}) generated with $\texttt{pca.zero\_snr} = 3$ are identical to those generated with the selected threshold.
These two fields are NGC~1333 (mosaicked field) and NGC2264 North (mosaic component field).
Papers~S1 and~S2 nevertheless use seven mosaicked fields and one mosaic-component field.
Sub-subsection~\ref{ass:ResidualOffsets} states how we corrected each field.
 
\section{Reproducible \texttt{pol2map} Reduction Configuration for the 450\,\micron\ Single-Field Products}
\label{app:reduction_workflow}
 
In this appendix, we provide the \texttt{pol2map} commands and modifications for individual fields used in the 450\,\micron\ reductions throughout this work (summarized in sub-subsection~\ref{ass:offset_significance}).
We follow the standard three-step \texttt{pol2map} procedure with two modifications: (i)~\texttt{flt.filt\_edge\_largescale=300}, which enforces the common range of angular scales of 12\farcs6--300\arcsec, and (ii)~$(\texttt{pca.zero\_snr}, \texttt{pca.zero\_snrlo})$ thresholds chosen for each field from the grid in table~\ref{tab:PCAparams}.
 
\subsection{pol2map Step~1: auto-masked total intensity map}
We first added \texttt{flt.filt\_edge\_largescale=300} to the configuration file, limiting sensitivity to structures larger than $\sim$300\arcsec.
The large-scale filter, combined with \texttt{smooth450=yes}, ensures that both wavelengths sample identical angular scales (12\farcs6--300\arcsec).
This smoothing option deconvolves the 450\,\micron\ double-Gaussian beam and reconvolves to the modeled 850\,\micron\ compound beam (12\farcs6 effective FWHM).
Here \texttt{ref=iext850.sdf} guarantees that the 450 and 850\,\micron\ maps are matched pixel by pixel, allowing direct regression of the pixel values.
We produced the auto-masked reference map with:
\begin{verbatim}
pol2map config=^conf in=^infiles iout=iauto qout=! uout=! \
        ref=iext850.sdf mapdir=maps qudir=qudata pixsize=4 smooth450=yes
\end{verbatim}
 
\subsection{pol2map Step~2: externally masked intensity map}
The \texttt{normalise=yes} setting stabilizes the large-scale background across multiple MSBs.
We refined the masks and applied the normalization between observations with:
\begin{verbatim}
pol2map config=^conf in=qudata/* iout=iext qout=! uout=! mapdir=maps mask=iauto \
        maskout1=astmask maskout2=pcamask BINSIZE=12 mapvar=yes skyloop=yes \
        obsweight=no normalise=yes smooth450=yes retain=yes
\end{verbatim}
 
\subsection{pol2map Step~3: final $Q$ and $U$ maps with optimized PCA thresholds}
For Category~1 fields (sub-subsection~\ref{ass:offset_significance}), we modified the configuration file to include the pair $(\texttt{pca.zero\_snr}, \texttt{pca.zero\_snrlo})$ from table~\ref{tab:PCAparams} that gives the smallest $a_0^{\rm tot}$.
Category~2 and~3 fields keep the default values.
In figures~\ref{afig:PCAcutoff_cat1_p1}--\ref{afig:PCAcutoff_cat3}, the vertical gray line marks the threshold of the smallest $a_0^{\rm tot}$ for each field.
We generated the final Stokes $Q$ and $U$ maps with:
\begin{verbatim}
pol2map config=^conf2 in=qudata/* iout=! qout=qext uout=uext mapdir=maps mask=iauto \
        maskout1=astmask maskout2=pcamask ipref=iext mapvar=yes skyloop=yes \
        obsweight=no normalise=yes smooth450=yes cat=polvecat.FIT BINSIZE=12 PIXSIZE=4 \
        debias=yes JY=no retain=yes
\end{verbatim}
 
\subsection{Note on debiasing}
The \texttt{debias=yes} option applies the bias estimator selected by the \texttt{pol2map} parameter \texttt{debiastype}, which by default is the asymptotic estimator \citep{Montier2015}.
For the final products, we applied the Modified Asymptotic Estimator \citep{Plaszczynski2014} after the offset correction; appendix~\ref{app:mas_debias} gives the explicit form of this estimator.
Users who require alternative statistical treatments of the polarization amplitude should run Step~3 with \texttt{debias=no} and apply their preferred correction during post-processing.
 
\subsection{Note on flux calibration}
We use \texttt{JY=no} to retain native units (pW) appropriate for extended emission.
The alternative \texttt{JY=yes} applies the flux conversion for point sources, which is unsuitable for molecular clouds.
We applied the extended-emission FCFs of \citet{Mairs2021} and the POL-2 attenuation factors ($1.96 \pm 0.023$ at 450\,$\mu$m and $1.35 \pm 0.008$ at 850\,$\mu$m) during post-processing.
 
\section{Modified Asymptotic Debiasing of the 450\,\micron\ Polarimetric Products}
\label{app:mas_debias}
 
In this appendix, we give the explicit form of the MAS of \citet{Plaszczynski2014} and \citet{Montier2015}.
We adopt this estimator for the final 450\,\micron\ polarimetric products $(PI_{\rm MAS}, p_{\rm MAS}, \chi)$ summarized in sub-subsection~\ref{ass:ResidualOffsets}.
We use these products in our scientific analysis and in Papers~S1 and~S2, with the polarization angle $\chi$ defined in equation~(\ref{eq:chi_def}).

We apply the MAS estimator after the $Q/U$ zero-level offset correction of equations~(\ref{aeq:Qfit_initial})--(\ref{aeq:Qfinal}) in sub-subsection~\ref{ass:3step}, so that the statistics of the corrected Stokes parameters satisfy its assumptions.
We express the corrected Stokes parameters as in equation~(\ref{eq:Q450corr_definition}).
We calculate the polarized intensity as $PI^{\rm corr} = \sqrt{(Q_{450}^{\rm corr})^2 + (U_{450}^{\rm corr})^2}$, but Gaussian noise introduces a positive bias in this raw amplitude.
The MAS estimator removes this bias in a way that is stable and robust to noise, which we consider well suited to POL-2 data.
 
Let $\sigma_Q$ and $\sigma_U$ denote the uncertainties in the corrected Stokes $Q$ and $U$ parameters.
Defining $S^2 = (Q_{450}^{\rm corr})^2 + (U_{450}^{\rm corr})^2$, the noise term entering the Rice distribution is
\begin{equation}
\sigma_{PI}^2 =
\frac{(Q_{450}^{\rm corr})^2 \sigma_Q^2 + (U_{450}^{\rm corr})^2 \sigma_U^2}
     {(Q_{450}^{\rm corr})^2 + (U_{450}^{\rm corr})^2} .
\label{eq:MAs_sigmaP}
\end{equation}
This noise term has the form of the variance (equation~31 of \citealt{Plaszczynski2014}; equation~20 of \citealt{Montier2015}) rather than that of the noise bias (their equations~30 and 15, respectively).
The two coincide when $\sigma_Q = \sigma_U$ and the $Q$ and $U$ noise are uncorrelated.
The polarized intensity debiased with the MAS becomes
\begin{equation}
PI_{\rm MAS} = PI^{\rm corr} - \frac{\sigma_{PI}^2}{2 PI^{\rm corr}}
\left[ 1 - \exp\!\left( -\frac{S^2}{\sigma_{PI}^2} \right) \right] ,
\label{eq:MAS_app}
\end{equation}
where $PI^{\rm corr} = \sqrt{S^2}$ is the raw amplitude.
The corresponding debiased polarization fraction is
\begin{equation}
p_{\rm MAS} = \frac{PI_{\rm MAS}}{I} ,
\label{eq:pMAS_def}
\end{equation}
where $I$ is the corresponding Stokes $I$ intensity.

\bibliography{450um1G}{}
\bibliographystyle{aasjournal}
 
\end{document}